\documentclass[fleqn,usenatbib]{mnras}

\usepackage{newtxtext,newtxmath}

\usepackage[T1]{fontenc}

\DeclareRobustCommand{\VAN}[3]{#2}
\let\VANthebibliography\thebibliography
\def\thebibliography{\DeclareRobustCommand{\VAN}[3]{##3}\VANthebibliography}

\usepackage{graphicx}	
\usepackage{amsmath}	
\usepackage{afterpage}
\usepackage{color}
\usepackage[flushleft]{threeparttable}
\usepackage{booktabs} 

\title[VLA observations toward OMC-2/3]{Multifrequency deep radio continuum VLA observations toward OMC-2/3}

   \author[A. K. Diaz-Rodriguez et al.]
   {Ana K. Diaz-Rodriguez,$^{1}$
	Mayra Osorio,$^{1}$\thanks{Corresponding author: osorio@iaa.csic.es}
	Guillem Anglada,$^{1}$
	S. Thomas Megeath,$^{2}$
	John J. Tobin,$^{3}$
	\newauthor
	Elise Furlan,$^{4}$
	M. Dolores Hurtado-Espejo,$^{1}$
	A. Florin Placinta-Mitrea,$^{1}$
	Jos\'e F. G\'omez,$^{1}$
	\newauthor
	Guillermo Bl\'azquez-Calero,$^{1}$
	William J. Fischer,$^{5}$$^\dag$
	P. Manoj,$^{6}$
	Dan M. Watson,$^{7}$
	Amelia M. Stutz,$^{8}$
	\newauthor
	Luis F. Rodr{\'i}guez$^{9}$
	\\
	\\ 
	$^{1}$Instituto de Astrof{\'i}sica de Andaluc{\'i}a (CSIC) Glorieta de la Astronom{\'i}a s/n E-18008 Granada, Spain\\
	$^{2}$Department of Physics and Astronomy, University of Toledo 2801 West Bancroft Street, Toledo, OH 43606, USA\\
	$^{3}$National Radio Astronomy Observatory, 520 Edgemont Rd., Charlottesville, VA 22903, USA\\
	$^{4}$IPAC, Mail Code 314-6, Caltech 200 E. California Blvd., Pasadena, CA 91125, USA\\
	$^{5}$Space Telescope Science Institute 3700 San Martin Drive, Baltimore, MD 21218, USA\\
	$^{6}$Department of Astronomy and Astrophysics, Tata Institute of Fundamental Research Homi Bhabha Road, Colaba, Mumbai 400005, India\\
	$^{7}$University of Rochester, Rochester, NY, USA\\
	$^{8}$Departamento de Astronom\'{i}a, Universidad de Concepci\' on, Casilla 160-C, Concepci\'{o}n,
	Chile\\
	$^{9}$Instituto de Radioastronom{\'i}a y Astrof{\'i}sica, UNAM Apartado Postal 3-72 (Xangari), 58089 Morelia, Michoac{\'a}n, Mexico}

\date{Accepted XXX. Received YYY; in original form ZZZ}

\pubyear{2026}

\begin{document}
\label{firstpage}
\pagerange{\pageref{firstpage}--\pageref{lastpage}}
\maketitle

\begin{abstract}

We present multiwavelength (0.7-5 cm) Very Large Array continuum observations of high sensitivity (3 $\mu$Jy beam$^{-1}$), reaching an angular resolution of $0.4''$ (154 au) toward the Integral Shaped Filament in Orion. We image a field of $12.5'$ in diameter, containing the whole OMC-2 region and the southernmost $2.5'$ of OMC-3. We report the positions, flux densities and spectral indices of 47 radio sources, and discuss their nature.  We conclude that, out of the 57 radio sources in the field of view, 36 (63\%) are very likely tracing YSOs, 4 are tracing jet/outflow knots, and 6 are extragalactic sources; the nature of the remaining sources is unclear. We identify the driving sources of the 9 previously reported molecular outflows in our field of view.  Four of them show a clear elongated radio jet morphology. We detect free-free emission from two potential photoevaporating discs associated with Class III YSOs. We find variability in 11 radio sources, 7 of which are also variable at shorter wavelengths. We find 7 pairs of radio sources with projected separations of 150-3000 au, which are potential binary systems; we identify a total of 17 potential multiple systems within this separation range when taking into account sources detected at other wavelengths. This work highlights the potential of deep VLA observations to uncover a large population of weak radio sources and study their diversity, serving as a guide for future surveys with the SKA and ngVLA.
 
\end{abstract}

\begin{keywords}
ISM: jets and outflows --- radio continuum: stars --- stars: binaries --- stars: formation --- stars: protostars
\end{keywords}

\section{Introduction} \label{sec:intro}

The Orion Molecular Cloud complex, with reported distances ranging from 380 to 450 pc \citep{Menten2007,Kim2008,Stutz2016,Kounkel2017,Kounkel2018,Grossschedl2018}, is one of the nearest active star-forming regions, where young stellar objects (YSOs) of both low and high luminosity coexist. The complex is composed of two elongated giant molecular clouds, Orion A and B, with a total mass exceeding $2 \times 10^5 M_{\odot}$ \citep{Wilson2005}, roughly aligned in the north-south direction (see fig. 1 of \citealt{Megeath2012}). Toward the northern end of the Orion A cloud, \citet{Mezger1988,Mezger1990} and \citet{Chini1997} identified several bright millimeter sources, aligned along an Integral Shaped Filament \citep[ISF,][]{Bally1987,Peterson2008,Stutz2016,Kainulainen2017}. Most of these sources were later associated with YSOs (e.g., \citealt{Adams2012}, \citealt[and references therein]{Furlan2014,Furlan2016, Osorio2017}). Outstanding among them are the bright FIR 3 and FIR 4 sources, which are associated with the Class I HOPS 370 and the Class 0 HOPS 108 YSOs, respectively. In this paper, we analyze the protostellar content toward the Orion Molecular Clouds 2 and 3 (OMC-2/3), two of the richest molecular clouds in protostars and prestellar cores, located in the northern part of the ISF.

The Orion complex has been intensively observed in the infrared spectral range with surveys carried out with the \textit{Spitzer} Space Telescope \citep{Megeath2012,Megeath2016} and the \textit{Herschel} Space Observatory. In particular, 330 protostars have been studied in great detail as part of the open-time key program Herschel Orion Protostar Survey (HOPS, e.g., \citealt{Stanke2010,Fischer2010,Fischer2017,Fischer2020,Manoj2013,Stutz2013,Furlan2014,Furlan2016,Gonzalez2016}). For these objects, identified with HOPS numbers, \citet{Furlan2016} modeled their spectral energy distributions (SEDs) from near-IR to submillimeter wavelengths allowing them the determination of fundamental properties such as their luminosity and the envelope mass, as well as their classification in terms of evolutionary stage. However, in some cases, the SED was contaminated by emission from nearby sources, due to the relatively low angular resolution of those observations. Therefore, observations at other wavelength ranges and with higher angular resolution can help to better understand the nature of these sources. In this sense, the VLA/ALMA Nascent disc and Multiplicity (VANDAM) survey of Orion \citep{Tobin2020I}, targeting 328 protostars with ALMA at 870~$\mu$m and including observations with the VLA at 9 mm toward 148 protostars, has helped to identify at high angular resolution ($\sim 0\farcs$1, much higher than the data used to construct the SED), possible discs associated with the sources, and clarify their nature and multiplicity (see HOPS catalog of discs, jets and envelopes in Orion\footnote{Available at \url{http://planetstarformation.iaa.es}.}).

In particular, radio observations at cm wavelengths can detect the most embedded, youngest YSOs, even those of very low luminosity. These radio sources trace the base of collimated outflows of partially ionized gas (radio jets), providing not only an extinction-free census of outflow (and thereby accretion) activity but also a tool to pinpoint the position of YSOs and identify embedded multiple systems \citep{Anglada2018}.

The OMC-2/3 region has been observed by different programs at radio wavelengths with the Very Large Array (VLA). \citet{Reipurth1999} observed it at 3.6 cm in the VLA D-configuration (angular resolution of $\sim8\arcsec$ and rms of 30 - 40 $\mu$Jy beam$^{-1}$) to investigate its protostellar content, detecting 14 radio sources toward the ISF, 11 of which are likely related to the star-forming region according to these authors. As part of a study of several star-forming regions with outflow activity, \citet{Reipurth2004} observed a small fraction of OMC-3 (a region of $\sim3\arcmin\times3\arcmin$ around the bipolar reflection nebula Haro 5a/6a), using the VLA A-configuration at 3.6 cm (angular resolution of $\sim 0\farcs3$ and rms of 7 $\mu$Jy beam$^{-1}$), detecting two additional YSOs\footnote{\citet{Reipurth2004} reported originally four sources from their observations, but sources VLA 15 and 17 in that paper correspond to sources VLA 5 and 6, respectively, from \citet{Reipurth1999}. On the other hand, source VLA 15 from \citet{Reipurth2004} is a different radio source than source VLA 15 in \citet{Osorio2017}.}. \citet{Kounkel2014} surveyed the whole Orion complex at 5 cm in A-configuration (sub-arcsecond angular resolution and rms of 60 $\mu$Jy beam$^{-1}$) looking for non-thermal emitters in order to use them later in VLBI observations to measure the distance to the region. They increased by 25 the known radio sources in the OMC-2/3 region, 17 of which are likely YSOs according to those authors. In summary, these three papers report a total of 41 radio sources toward the OMC-2/3 region, and propose 30 of them as likely YSOs.

\citet{Osorio2017} presented VLA multifrequency observations of OMC-2 in the C-configuration from 0.7 to 5 cm (with angular resolution from $\sim 0\farcs4$ to $\sim$ 4\arcsec), as well as very sensitive (rms = 3 $\mu$Jy beam$^{-1}$) observations at 5 cm in the A-configuration, more than 10 times deeper than previous surveys. A $\sim40\arcsec\times30\arcsec$ region around FIR 4, the brightest submillimeter source in OMC-2 \citep{Furlan2014}, was analysed. The paper was focused in discening the nature of the FIR 4 object and nearby sources. One of the main results was the detection of the thermal radio jet driven by the Class I YSO HOPS 370 (associated with FIR 3) and a non-thermal lobe associated with it. Since the Class 0 YSO HOPS 108 (associated with FIR 4) is located in projection toward the non-thermal jet lobe, the authors suggested that the interaction of this jet with the ambient could have triggered the formation of HOPS 108.

The vicinity of FIR 3/FIR 4 has also been studied recently by \citet{Tobin2019} through VLA continuum observations at 9 mm and ALMA continuum and line observations at 870 $\micron$. They characterized the continuum sources in the region on $\sim$ 40 au scales, and found emission from high-excitation transitions of complex organic molecules (COMs) associated with both HOPS 108 (FIR 4) and HOPS 370 (FIR 3). In this study, 10 radio sources were reported, 7 of them tracing YSOs, and 3 showing large proper motions and tracing non-thermal knots of the HOPS 370 radio jet.

In this paper we present a comprehensive analysis of the sources in the entire field of view of the \citet{Osorio2017} observations, a region of $12\farcm$5 in diameter, comprising the southernmost $2\farcm5$ of OMC-3 and the whole OMC-2 regions.  Our aim is to combine our radio observations with the available multiwavelength information to investigate the nature of the detected sources, distinguishing YSOs associated with the OMC-2/3 star-forming region from background extragalactic objects. We also seek to investigate the physical (thermal or non-thermal) mechanisms responsible for the radio emission from YSOs, particularly the relative contributions of thermal and non-thermal processes, through the analysis of their radio spectral indices and flux density variability. Together with the study of source multiplicity and the comparison with observations at other wavelengths, this approach will allow us to characterise individual radio sources in OMC-2/3 and compare the properties of the YSO population with those reported in other star-forming regions observed through similar radio interferometric surveys.

More broadly, this study aims to provide a reference for interpreting the increasingly detailed observations of young stellar populations obtained with current and forthcoming facilities. In particular, our results may help motivate and guide future investigations of radio emission from YSOs with next-generation centimetre-wavelength interferometers, such as the next-generation Very Large Array (ngVLA) and the Square Kilometre Array (SKA).

We adopt an average distance to OMC-2/3 of 390 pc \citep{Kounkel2018, Grossschedl2018, Tobin2020I,Tobin2020II}, slightly different from the value of 414 pc \citep{Menten2007} used in \citet{Osorio2017}. The use of this updated value for the distance would decrease the luminosities derived by \citet{Furlan2016}, who used 420 pc, by 16\%. 

This paper is structured as follows. In Section \ref{sec:observations} we describe the observations. In Section \ref{sec:results} we present our results and discuss the global and individual properties of the detected radio sources. Finally, in Section \ref{sec:conclusions} we summarize our conclusions. Most of the work presented in the paper was included in the PhD thesis of Ana K. Diaz-Rodriguez \citep{Diaz-Rodriguez2021}

\section{Observations and Data Reduction} \label{sec:observations}

The observations were carried out with the Karl G. Jansky VLA of the National Radio Astronomy Observatory (NRAO) in the C configuration in October 2014 (Project 14B-296) at C (5 cm), X (3 cm), K (1.3 cm), and Q (0.7 cm) bands, and in the A configuration in August-September 2015 (Project 15A-369) at C band (5 cm), with angular resolutions ranging from $\sim0\farcs4$ to $\sim4\arcsec$ (see Table \ref{tab:maps}). The observational bandwidth was 4 GHz at C and X bands, and 8 GHz at K and Q bands. The largest angular scale that can be imaged reasonably well in our VLA observations\footnote{\url{https://science.nrao.edu/facilities/vla/docs/manuals/oss/performance/resolution}} is $9''$ (A-configuration in C band) and ranges from $32''$ to $240''$ in the C-configuration observations, which is enough to image typical YSOs without a significant loss of extended emission. The phase centre was similar in both projects, with coordinates $\alpha$(J2000) = 05$^{\rm h}$35$^{\rm m}$27$\fs$070, $\delta$(J2000) = $-$05\degr 10\arcmin00\farcs60 for the 2014 runs and $\alpha$(J2000) = 05$^{\rm h}$35$^{\rm m}$27$\fs$080, $\delta$(J2000) = $-$05\degr 10\arcmin00\farcs30 for the 2015 ones. The data were edited and calibrated using the Common Astronomy Software Applications package (CASA; version 4.2.2). For further details of the observational setup and calibration see table 1 in \citet{Osorio2017}, where these observations were first reported.

\begin{table*}
	\setlength{\tabcolsep}{11pt}
	\caption[]{Characteristics of the VLA maps.}
	\label{tab:maps}
	\centering
	\begin{tabular*}{\linewidth}{ccccccccc}
		\hline\hline
		& &Central &Frequency&Field & & \multicolumn{2}{c}{Synthesized beam}\\
		\cline{7-8}
		& & wavelength$^a$& range & of view $^b$& & Size & PA & rms$^d$ \\
		Configuration &Band &  (cm)& (GHz) & ($'$) & Weighting$^c$ & ($''\times''$)  &($^\circ$)  & ($\mu$Jy\, beam$^{-1}$)\\
		\hline
		A & C & 5 & 4.162-8.160   & 12.5 & Natural  & 0.42$\times$0.40 & 14.9 & 3  \\
		C & C & 5 & 3.912-7.960   & 12.5 & Robust 0 & 4.76$\times$2.48 & 46.3 & 10  \\
		C & X & 3 & 7.912-11.960  & 7.6  & Robust 0 & 2.68$\times$1.73 & 36.1 & 9  \\
		C & K & 1.3 & 17.976-25.960 & 3.4  & Robust 0 & 0.84$\times$0.71 & 5.2  & 10\\
		C & Q & 0.7 & 40.039-48.023 & 1.7  & Robust 0 & 0.46$\times$0.36 & $-$20.5 & 15\\
		\hline
	\end{tabular*}
	\begin{tablenotes}
		\footnotesize
		\item{$^a$Central wavelength of the bandwidth. $^b$Diameter of the field of view where radio sources have been searched for, corresponding to the region where the primary beam response is above 10\% of that at the field centre. At C band, the 13 arcmin$^2$ region where extended emission from M43 dominates (shown in Fig. \ref{fig:filament}) is excluded from the analysis. $^c$The robust parameter \citep{Briggs1995} is defined as implemented in CASA, taking values between $-$2.0 (close to uniform weighting) to 2.0 (close to natural). $^d$At the centre of the map; the rms at the border of the map is 10 times higher.}
	\end{tablenotes} 
\end{table*}

All the images were made with the CASA (version 4.2.2 for the 2014 data, and versions 5.0.0 and 5.1.1 for the 2015 data) task CLEAN using multi-frequency synthesis \citep{Conway1990}. For the A-configuration map, the wide-field algorithm w-projection (with the number of w-projection planes for convolution automatically determined by CLEAN) was used to correct for the curvature of the sky plane, as noted in the CASA User Reference \& Cookbook\footnote{Release 4.7.2, available at \url{https://casa.nrao.edu/casa_cookbook.pdf}.}. Not using this correction results in a shift of the positions of the sources away from the phase centre, which reaches $\sim$ 0$\farcs$01 at 6$\arcmin$. The extended emission from the nearby bright HII region M43 (also known as NGC 1982), $\sim5\arcmin$ to the south of the phase centre, makes the imaging at C band difficult. As in \citet{Osorio2017}, to properly clean the radio sources in our field of view, we downweighted the effects of the extended emission from M43, by using only baselines $>$ 5 k$\lambda$ (suppressing extended emission at scales larger than $\sim$ 40$\arcsec$) for the C-configuration maps and $>$ 25 k$\lambda$ (suppressing extended emission at scales larger than $\sim$ 8$\arcsec$) for the A-configuration map. To obtain the C-band image in the C configuration, where the sensitivity to extended structures is higher, we additionally used the multiscale deconvolution algorithm. All the maps presented in this work have been corrected for the primary beam response.

\begin{figure*}
\centering
\includegraphics[width=0.75\hsize]{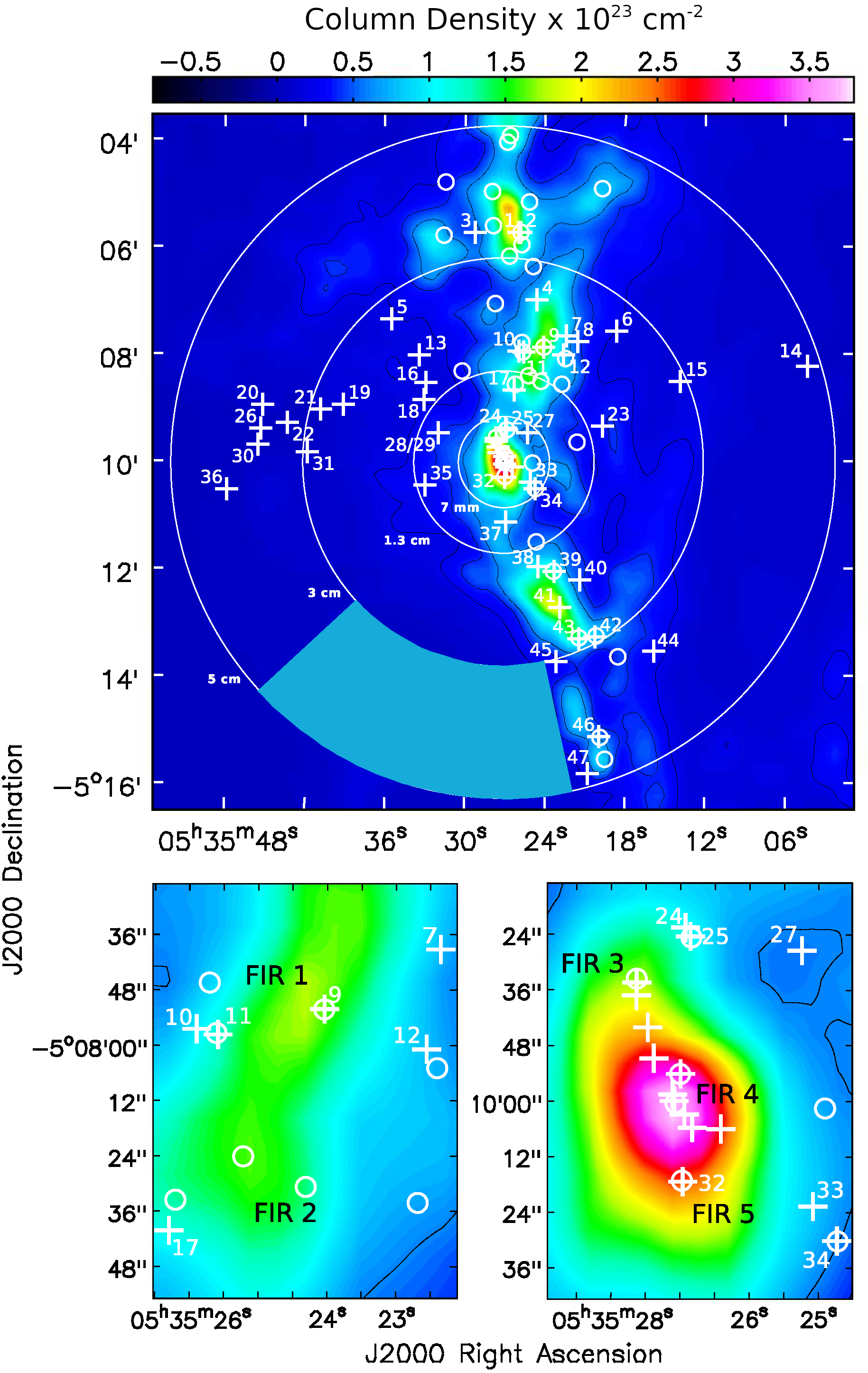}
\caption{Location of the detected radio sources marked with plus signs. The background is the column density map of the ISF obtained by \citet{Stutz2018} from {\it Herschel} dust emission data (160 - 500 $\micron$); contour levels are 2.3, 4 and 6 times $10^{22}$ cm$^{-2}$ \citep[see their Appendix A]{Hacar2018}.  Numbered sources are reported in this work, while those unnumbered were reported by \citet{Osorio2017}. Open circles mark the position of HOPS sources \citep{Furlan2016,Stutz2013,Tobin2015}. Concentric circles mark the region where the primary beam response is $>10\%$ of that at the field centre for each frequency. The light blue annular sector to the south in the main panel marks the region where extended emission from the bright HII region M43 (NGC 1982) is detected, which is not analyzed in this work. The panels at the bottom show a close-up of the most crowded regions.}
\label{fig:filament}
\end{figure*}

Given the high sensitivity reached, it has been possible to image the region where the primary beam response is above 10\% of that at the field centre, where we still have enough sensitivity to detect sources of a few tenths of mJy. The diameter of the imaged region changes with frequency, ranging from $\sim2\arcmin$ at 0.7 cm to $\sim13\arcmin$ at 5 cm. The largest of these fields covers the ISF in the southernmost 2$\farcm$5 of OMC-3 and in the whole OMC-2 region. The four concentric circles in Fig. \ref{fig:filament} mark the size of the imaged region in each band. At C band (5 cm), the extended emission from M43 could be masking the emission from individual radio sources in the S-SE edge of the field, making it difficult the analysis of the maps. Therefore, we excluded from our analysis the 13 arcmin${^2}$ annular sector shown in Fig. \ref{fig:filament}, where the extended emission from M43 dominates. Table \ref{tab:maps} summarizes the characteristics of the maps obtained.

\section{Results and discussion} \label{sec:results}

In \citet{Osorio2017} we discussed 10 radio sources detected in a $\sim40\arcsec\times30\arcsec$ region around FIR 3/4 (see table 2 of that paper). There are 47 additional radio sources detected in a field of $\sim12\farcm5$ in diameter (Table \ref{tab:maps}), whose positions, deconvolved sizes, and flux densities at the observed bands are listed in Tables \ref{tab:positions} and \ref{tab:fluxes}. We numbered the sources according to their declination, from north to south. We consider as detected those sources with peak intensity above a 3$\sigma$ level in more than one band or epoch, or above a 6$\sigma$ level if detected only in one band and epoch.

In Fig. \ref{fig:filament} we show the positions of the 57 radio sources detected in our observations of the region, overlaid on a column density map derived from {\it Herschel} dust emission data at 160 $\micron$, 250 $\micron$, 350 $\micron$, and 500 $\micron$ (adapted from \citealt{Stutz2018}). For the 47 radio sources discussed in this paper, we show in Fig. \ref{fig:sources} our high angular resolution A-configuration (5 cm) maps. In Fig. \ref{fig:SEDs} we present the spectra in the cm wavelength range, constructed using all the available data \citep[this work]{Reipurth1999,Reipurth2004,Kounkel2014} for 41 of these sources (for 6 of them we only have data at a single frequency). Spectral indices are listed in Table \ref{tab:fluxes} (see details in Section \ref{subsec:alpha}) and additional properties of the sources, compiled from the literature, are presented in Table \ref{tab:sources}. We searched unsuccessfully for sources with large circular polarization.

\begin{table*}
			\caption{Association with emission at other wavelengths and nature of the detected radio sources$^a$.}
			\label{tab:sources}		
					\setlength{\tabcolsep}{1.5pt}
					\scriptsize
					\centering
						\begin{tabular*}{\linewidth}{@{\extracolsep{\fill}}lccccccc}
								\hline\hline
								\multicolumn{8}{c}{}\\
									{Source} &  {Alternative}& {Compact/unresolved} & {Refs.} & {Outflow/SF} & {Ref.} & Nature/ & {Ref.}\\
									{Number} & {Names} & Emission$^b$ & & {Signatures} & & Evol. Stage$^c$ & \\
								\hline
								1 & HOPS 78  & SMM,\,FIR,\,MIR$^d$& 1,\,2,\,3 & mm core,\,H$_2$ knots,\,CO outf. & 4,\,5,\,6,\,7 & Class 0 & 2\\
								2 &...& MIR?$^d$ & 8 & mm core? & 4 & YSO or Source 1 knot & 8,\,9\\ 
								3 &...&...&...&...&...& EG & 10\\ 
								4 &...& NIR & 11 & MIR nebulosity  & 9 & HII & 10\\ 
								5 &...& MIR,\,NIR,\,XR & 12,\,9,\,13 &...&...& disc galaxy?& 9,\,13\\ 
								6 &...&...&...&...&...& ? & \\ 
								7 &...& MIR,\,NIR,\,XR$^\ast$&  3,\,11,\,13 &...&...& Class I/II & 3,\,8\\ 
								8 &...&...&...&...&...& ? & \\ 
								9 & PBR 019003,\,HOPS 394,\,FIR 1a & SMM,\,FIR?,\,XR? & 1,\,14,\,8 & mm core?$^e$ ,\,CO outf. & 4,\,21 & Class 0?,\,PBR? & 2,\,14\\
								10 & ...& SMM,\,MIR,\,NIR,\,OP,\,XR$^\ast$& 1,\,8,\,11,\,15,\,13    & ...& ...& Class II & 8,\,9\\ 
								11 & HOPS 71 & SMM,\,FIR,\,MIR,\,NIR & 1,\,2,\,3,\,11  & mm core,\,CO outf.  & 4,\,6 & Class I & 2\\
								12 & V0492 Ori & MIR,\,NIR,\,OP$^\ast$,\,XR & 3,\,11,\,15,\,13,\,16 & mm core? & 4 & Class I/II & 3,\,8\\ 
								13 & ...& XR & 13 &...&...& EG & 10,\,9\\ 
								14 & V1321 Ori & MIR,\,NIR,\,OP$^\ast$,\,XR & 8,\,11,\,15,\,13,\,16 & ...& ...& Class III & 8,\,13\\
								15 & ...&...&...& ...& ...& ?& \\
								16 &...& NIR &   9 & ...&...& ?& \\ 
								17 & NP Ori  & MIR,\,NIR,\,OP$^\ast$,\,XR$^\ast$& 8,\,11,\,15,\,13,\,16 &...&...& Class III & 8,\,13\\ 
								18 &...& NIR &   9 & ...& ...& ? & \\
								19 & V1576 Ori & MIR,\,NIR,\,OP$^\ast$,\,XR &  8,\,11,\,15,\,13,\,16     &...&...& Class III & 8\\
								20 & ...& ...& ...& ...& ...&  Source 26 knot?    & 9\\
					 			21 & V0363 Ori & MIR,\,NIR,\,OP$^\ast$,\,XR & 8,\,11,\,15,\,13,\,16 & ...& ...& Class II/III  & 8,\,13\\ 
								22 & V0569 Ori & MIR,\,NIR,\,OP$^\ast$,\,XR$^\ast$ & 8,\,11,\,15,\,13,\,16  &...&...& Class III & 8,\,13\\
								23 & ...& NIR &    9 & ...& ...& EG or Class II/III & 9\\
								24 & HOPS 66,\,SOF 1,\,FraSCO-4 & SMM,\,FIR$^f$  & 1,\,2 & mm core?
								& 4 & YSO& 9\\ 
								25 & HOPS 66,\,SOF 1,\,V2455 Ori,\,FraSCO-6 & SMM,\,FIR$^f$ ,\,MIR,\,NIR$^\ast$ & 1,\,2,\,3,\,11,\,16 & mm core,\,IR cavity & 4,\,17 & Flat spectrum$^g$  & 2\\
								26 &...& MIR,\,NIR &  12,\,9 &...& ...& Radio galaxy? & 9\\ 
								27 & V1344 Ori & MIR,\,NIR,\,OP$^\ast$,\,XR & 8,\,11,\,15,\,13,\,16&...&...& Class III & 8\\ 
								28 & V0360 Ori,\,FraSCO-9 & MIR,\,NIR,\,OP$^\ast$,\,XR$^\ast$ $^h$ & 3,\,11,\,15,\,13,\,16 &...&...& Class I/II$^i$ & 3,\,8\\ 
								29 & V0360 Ori & MIR,\,NIR,\,OP$^\ast$,\,XR$^\ast$$^h$ & 3,\,11,\,15,\,13,\,16 &...& ...& Class I/II$^i$  & 3,\,8\\ 
								30 &...& ...& ...&...&...& Source 26 knot?    & 9\\ 
								31 &...& NIR & 9 & ...&...& ? & \\ 
								32 & HOPS 369,\,SOF 4, FraSCO-45 & SMM,\,FIR,\,MIR,\,NIR  & 1,\,2,\,3,\,11 & mm core? & 4,\,17 & Flat spectrum & 2\\
								33 & ...&...&...& ...& ...& HII & 9\\ 
								34 & HOPS 368,\,SOF 5,\,FraSCO-49 & SMM,\,FIR,\,MIR,\,NIR  & 1,\,2,\,3,\,11 & mm core,\,CO outf.,\,IR cavity & 4,\,6,\,17 & Class I & 2\\
								35 & ...& MIR,\,NIR,\,XR?   & 12,\,9,\,8 & ...&...& EG or Class II/III & 9\\ 
								36 & NSV 16575 & MIR,\,NIR,\,OP$^\ast$,\,XR & 8,\,11,\,15,\,13,\,16  & ...& ...& Class III & 8,\,13\\
								37 & SOF 6,\,AI Ori,\,FraSCO-51 & MIR,\,NIR,\,OP$^\ast$,\,XR & 3,\,11,\,15,\,13,\,16 & mm core? & 4 & Class I/II  & 3,\,8\\ 
								38 & ...& MIR,\,NIR,\,OP,\,XR?   & 9,\,18,\,15,\,8 & ...& ...& YSO & 9\\ 
								39 & HOPS 60,\,SOF 7,\,FIR 6b & SMM,\,FIR,\,MIR & 1,\,2,\,3 & mm core,\,CO outf.,\,IR cavity & 4,\,6,\,19,\,17 & Class 0 & 2\\ 
								40 & ...& MIR,\,NIR,\,OP,\,XR   & 8,\,11,\,15,\,13  & ...& ...& Class III & 8,\,13\\ 
								41 & ...& ...& ...& mm core? & 4 & ? &\\ 
								42 & HOPS 59, SOF 8, FIR 6d, V2364 Ori & SMM,\,FIR,\,MIR,\,NIR$^\ast$,\,OP,\,XR & 1,\,2,\,3,\,11,\,15,\,13,\,16 & mm core,\,CO outf.,\,IR cavity & 4,\,19,\,17 & Flat spectrum & 2\\
								43 & PBR 135003,\,HOPS 409,\,FIR 6c & SMM,\,FIR & 1,\,20 & mm core,\,CO outf. & 4,\,6,\,19,\,21 & Class 0,\,PBR  & 2,\,20\\
								44 & ...& ...& ...& ...&...& ? & \\ 
								45 & ...& MIR,\,NIR,\,OP,\,XR   & 8,\,11,\,15 & ...& ...&  Class II/III & 8,\,13\\
								46 & HOPS 57 & SMM,\,FIR,\,MIR,\,NIR,\,XR  & 1,\,2,\,3,\,11,\,13 & mm core & 4 & Flat spectrum & 2,\,16\\
								47 & ...& MIR,\,NIR,\,OP,\,XR$^\ast$ & 8,\,11,\,15,\,13 & ...& ...& Class III & 8\\	
								\hline										 																
							\end{tabular*}
			\vspace{3mm}		
			\begin{tablenotes}
					\scriptsize
					\item{$^a$For sources VLA 11 (HOPS 370), HOPS 108, HOPS 64, MIPS 2301, MIPS 2297, VLA 15 and VLA 16, see \citealt{Osorio2017}.
							$^b$Point source detected at sub-millimeter (SMM), far-IR (FIR), mid-IR (MIR), near-IR (NIR), optical (OP) or X-ray (XR) wavelengths; the * indicates variability in that wavelength range.
							$^c$EG: Extragalactic, HII: Ionized region, PBR: PACS Bright Red source, YSO: Young stellar object.
							$^d$Emission detected as a single source with an intermediate position between our sources 1 and 2. We favor the association with Source 1 (see Section \ref{1and2}).
							$^e$Note that the source is associated with the mm core \#11 of \citet{Kainulainen2017}, not with \#12 as mistakenly reported in their table 2.
							$^f$Sources 24 and 25 appear as a single far-IR source due to insufficient angular resolution.
							$^g$This classification is based on a SED that probably contains emission from both sources 24 and 25.
							$^h$Sources 28 and 29 have each one a different optical counterpart; due to insufficient angular resolution, a single source is detected al at mid-IR, near-IR and X-ray (likely with contribution from both sources 28 and 29).
							$^i$This classification is based on a SED that contains emission from both sources 28 and 29.}
						{\bf References:} (1) \citealt{Tobin2020I}, (2) \citealt{Furlan2016}, (3) \citealt{Megeath2012},  (4) \citealt{Kainulainen2017}, (5) \citealt{Stanke2002},  (6) \citealt{Takahashi2008}, (7) \citealt{Williams2003}, (8) \citealt{Getman2017}, (9) This paper, (10) \citealt{Reipurth1999}, (11) \citealt{Skrutskie2006}, (12) T. Megeath, private comunication, (13) \citealt{Tsujimoto2002}, (14) \citealt{Stutz2013}, (15) \citealt{Gaia2023}, (16) \citealt{Samus2017}, (17) \citealt{Habel2021}, (18) \citealt{DaRio2009}, (19) \citealt{Shimajiri2009}, (20) \citealt{Tobin2015},  (21) \citealt{Tobin2016outf}.
				\end{tablenotes}
			\end{table*}

We point out that there are five previously reported radio sources that fall within our field of view, but are not discussed here. Three of these sources, GBS-VLA J053529.00-050909.0, J053517.06-051423.7 and J053525.65-050757.2 \citep{Kounkel2014} are not detected in our observations of higher sensitivity, which suggests they are variable; the first two sources were reported as probably extragalactic, and the third one as a Class I YSO. The source OMC-2/3 VLA 4, a proposed radio jet \citep{Reipurth1999,Reipurth2004}, falls at the edge of our field of view, and only a hint of C-band emission is observed in our data; the detection is not clear enough for the source to be included in our discussion. Finally, we note that \citet{Reipurth1999} reported a source (VLA 14, that they associate with the star-forming region) at a position located $\sim5\farcs5$ from our Source 40, and we cannot discard the possibility that both are the same object given the poor angular resolution (8$\arcsec$) of those observations.

\subsection{Distribution of the detected radio sources and their association with the star-forming region} \label{subsec:distribution}

As a first step, we analyze the association of the detected radio sources with the star-forming region. In our maps, we expect a number of extragalactic sources that increases with the wavelength of the observation and the field size \citep[see their Appendix A]{Anglada1998}. Therefore, in our set of maps, the largest contamination of extragalactic sources is expected to occur in the A-configuration C-band map since it is the most sensitive, at the longest wavelength and with the largest field of view. Using equation A11 of \citet{Anglada1998} for a field of view of 12$\farcm$5 in diameter but excluding the solid angle of the annular sector affected by the extended emission of M43, we estimate that 13$\pm4$ of the 54 sources detected in this C-band map above a 6$\sigma$ threshold could be extragalactic. Thus, we expect that $41\pm4$ of the sources detected in this map are galactic and likely associated with the star-forming region. On the other hand, in the X-band map we report a total of 31 sources, and we expect that only 2$\pm$1 ($\sim6\%$) are extragalactic, strongly suggesting that the 3 sources detected at X-band but not detected in the A-configuration C-band map are galactic. In summary, we conclude that 44$\pm$4 ($\sim77\%$) of the 57 radio sources detected in the field are galactic and probably associated with the star-forming region.

As illustrated in Fig. \ref{fig:filament}, a total of 42 sources out of 57 ($74\%$) are observed toward the ISF (with a column density $>2.3\times10^{22}$ cm$^{-2}$, marked as medium and high column density in Table \ref{tab:positions}), and 30 of them ($53\%$ of the total) lie in the densest part of the filament (with a column density $>6\times10^{22}$ cm$^{-2}$, marked as high column density in Table \ref{tab:positions}). Since the ISF contains the phase centre, from the ratio between the areas of the ISF and the field of view, we can estimate the number of expected extragalactic sources toward the ISF. Assuming a width of 2$\arcmin$ for the filament, we obtain that only $3\pm2$ of the 42 sources observed toward the ISF ($1\pm1$ of those observed toward the densest part) are expected to be extragalactic.

In order to further investigate the membership to the star-forming region of the 42 radio sources detected toward the ISF, we analyzed their association with continuum clumps in the 3 mm ALMA map obtained by \citet{Kainulainen2017} with an angular resolution of $\sim$3$\arcsec$, the 870 $\micron$ ALMA maps obtained by \citet{Tobin2020I} with an angular resolution of $\sim 0\farcs11$, and by \citet{Federman2023} with an angular resolution of $\sim4\arcsec$, and the HOPS sources \citep{Furlan2016} detected in the far-IR with {\it Herschel} with an angular resolution of $\sim$5$\arcsec$-7$\arcsec$ (see Section \ref{subsec:associations}). We find that 18 ($\sim 43\%$) of the 42 radio sources detected toward the ISF are associated with compact emission at 870 $\micron$, likely tracing a disc and we consider them as ``bona fide'' YSOs (Table \ref{tab:sources}). Of these 18 sources, 11 fall very close (within $\le 1''$) to the peak of a 3 mm compact core (nine of them are also associated with HOPS sources) suggesting they are deeply embedded objects (Class 0/I); the remaining seven sources are probably somewhat more evolved YSOs\footnote{Four of them (all associated with HOPS sources) are detected toward the border of a 3 mm core (within $\lesssim 3''$ of the peak), one (associated with a HOPS source) is close (within $\le 1''$) to a faint 3 mm peak, and two are not associated with any 3 mm compact core or HOPS source.}. Among the remaining 24 radio sources toward the ISF without a reported association at 870 $\micron$, two fall in the proximity of 3 mm dust emission and we consider them as likely YSOs. In summary, we find that for 20 out of the 42 radio sources detected toward the ISF there is additional evidence of their specific association with dust clumps at arcsec/sub-arcsec scales ($\sim$40 - 2700 au), as traced by mm/sub-mm emission. Since we estimated that only $\sim$3 extragalactic sources should be detected toward the ISF, we expect that a large fraction of the remaining 22 radio sources are also YSOs, likely in a relatively evolved stage. In fact, 15 of them are detected as near-IR or optical point-like sources (see Table \ref{tab:sources}). In Section \ref{subsec:individual} and Appendix~\ref{subsec:individual2} we comment on the nature of the individual sources using additional information.

\begin{figure}
	\includegraphics[width=\hsize]{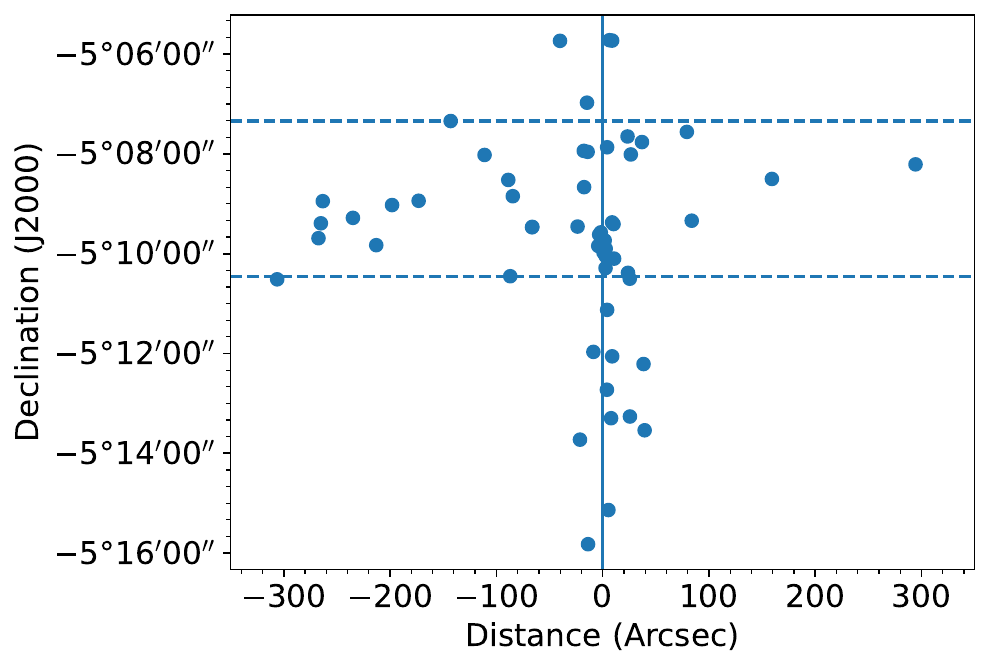}
	\caption{Declination of the radio sources versus their distance to the ISF ridgeline. The farthest sources are concentrated in a region with a declination between $-05\degr10\arcmin31\arcsec$ and $-05\degr07\arcmin20\arcsec$ (the region bounded by dashed lines). The ridgeline of the ISF is formed by finding the right ascension of the pixel with peak emission at each declination in the column density map \citep{Stutz2018} shown in Fig. \ref{fig:filament}.}
	\label{fig:distVSdec}
\end{figure}

\begin{figure}
	\includegraphics[width=0.48\textwidth]{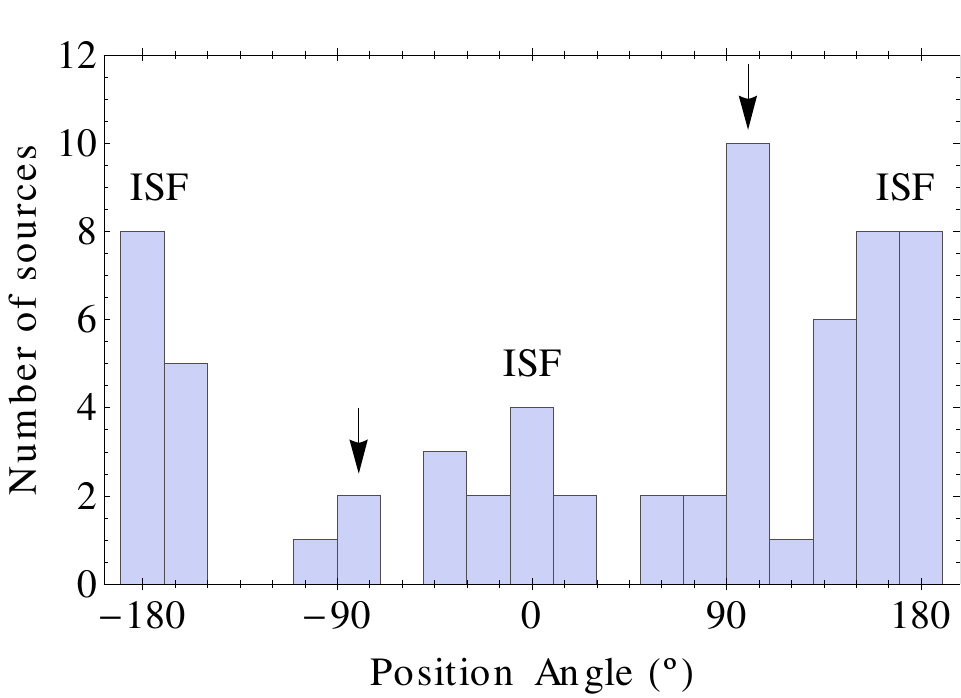}
	\caption{Distribution of the position angles of the sources, measured from N to E, taking the origin as the intersection point between the ISF and the E-W stripe. We excluded the only source at a distance $\le30\arcsec$ from the origin since its position angle is not well determined. Most of the sources have a position angle with an absolute value close to 0$\degr$ or 180$\degr$ (i.e., along the ISF, which is roughly aligned N-S), but a significant number of sources appear clustered close to a position angle with an absolute value of $\sim90\degr$ (i.e., perpendicular to the ISF), indicated by arrows in the histogram.}
	\label{fig:hist-all}
\end{figure}

Interestingly, we note that a number of sources appear clustered along an east-west stripe almost perpendicular to the ISF (see Fig. \ref{fig:filament}). This E-W stripe has a width of $\sim2\arcmin$ and extends in right ascension up to the limits of our field of view (a total extent of $12\farcm5$). The distribution of the sources along the E-W stripe is not uniform, with the eastern side appearing more populated than the western side. This is further illustrated in Fig. \ref{fig:distVSdec}, where we plot, for the detected sources, their declination and distance to the ISF ridgeline (formed by finding the right ascension of the pixel with peak emission at each declination in the column density map shown in Fig. \ref{fig:filament}). The sources farther from the ISF are concentrated in a region with a declination between $-05\degr10\arcmin31\arcsec$ and $-05\degr07\arcmin20\arcsec$ (Fig. \ref{fig:distVSdec}), where the dispersion ($\sigma$) of the distances to the ISF ridgeline reaches $114\arcsec$. Outside this declination range, the positions of the sources fall very close to the ISF ridgeline, with a dispersion of the distances of $\sim20\arcsec$. The two preferred orientations (along the ISF and along the E-W stripe) can be clearly seen in Fig. \ref{fig:hist-all}, where we present the distribution of the position angles of the sources, measured from N to E and taking the origin as the intersection point between the ISF and the E-W stripe (approximately  1\farcm3 north from the phase centre). Most of the sources have a position angle with an absolute value close to 0$\degr$ or 180$\degr$ (i.e., along the ISF, which is roughly aligned N-S), but a significant number of sources appear clustered close to a position angle with absolute value of $\sim90\degr$ (i.e., the sources in the E-W stripe roughly perpendicular to the ISF).

The alignment of sources along the E-W stripe suggests that the majority of them are not extragalactic since one would expect unrelated extragalactic sources to be randomly distributed (unless we are detecting a cluster of radio emitting galaxies). In any case, as we will see in Section \ref{stripe}, most of these objects are classified as YSOs. Furthermore, following a similar approach as in the ISF, we estimate that only 2$\pm1$ of the $\sim$ 16 sources that fall toward the stripe, but outside the ISF filament, are extragalactic. We consider, then, that most of the sources in the E-W stripe are also probably associated with the star-forming region. We note, however, that there is an apparent scarcity (or, at least, an irregular distribution) of extragalactic sources, since there are no detected radio sources outside the declination range $-05\degr10\arcmin31\arcsec<$ Dec. $< -05\degr07\arcmin20\arcsec$ at distances $>50''$ of the ISF ridgeline, where we would expect to detect about $\sim6$ extragalactic sources. We do not have an obvious explanation for this apparently anomalous distribution. However, this result deserves a more robust confirmation since we are dealing with relatively small numbers with poor statistical significance.

The column densities toward the E-W stripe, typically $\sim 1\times10^{22}$ cm$^{-2}$, are significantly lower than toward the filament ($\sim$ (2.5-40)$\times10^{22}$ cm$^{-2}$), suggesting that most of the sources in this stripe could correspond to an older stellar population, in agreement with the fact that most of them have been previously classified as Class III YSOs (see further details in \ref{stripe}). To be consistent with their observed alignment, these objects should have formed in a structure almost perpendicular to the ISF. Recently, \citet{Hacar2018} found that the ISF is composed of several velocity-coherent fibers, fertile in star formation. While most of the fibers have their axis aligned with the filament, some deviate from this direction, and a few of them are almost perpendicular to it, like the branches of a tree (see their fig. 4). Therefore, it is conceivable that the radio sources located in the E-W stripe may have previously formed in perpendicular fibers whose material has already been depleted.

In this Section we made, essentially, statistical considerations on the expected number of extragalactic sources, and the association of the detected radio sources to the star-forming region. However, there is additional information on the individual sources  (spectral index,  emission at other wavelengths, etc), that allows  us to ascertain which particular sources belong to the star-forming region. Using these complementary data, in Section \ref{subsec:individual} and Appendix~\ref{subsec:individual2} we discuss in detail on the nature of the individual sources.

\subsection{Spectral indices of the detected radio sources}\label{subsec:alpha}

A useful parameter to investigate the nature of the radio sources is the spectral index $\alpha$, where $S_{\nu} \propto \nu^\alpha$. However, the interpretation of spectral indices, especially when they are restricted to a limited range of wavelengths, is not unambiguous. There are uncertainties in the spectral index measurement (due to source variability, differences in angular resolution and/or intrinsic uncertainties in the flux measurements). Additionally, different emission mechanisms (or a combination of them) can produce similar values of the spectral index, while the same emission mechanism can produce different values of the spectral index. Despite these uncertainties, in general, a flat or positive spectral index ($\alpha\ge-0.1$) usually indicates thermal emission, while a negative one ($\alpha<-0.1$) is indicative of non-thermal emission. Below we summarize the possible astrophysical objects that we can detect with our cm observations, and characterize their emission in terms of $\alpha$. 

Massive young stars can photoionize the surrounding gas, creating the so called photoionized HII regions, with thermal free-free emission. The free-free spectrum of a homogeneous HII region is characterized by optically thick emission at low frequency (with $\alpha\simeq2$) and optically thin emission at high frequency ($\alpha\simeq-0.1$). The turnover frequency at which the emission becomes optically thin depends on the emission measure ($n_e^2 L$, where $n_e$ is the electron density and $L$ is the size) of the region. Very young HII regions can be very compact and dense ($n_e>10^6$ cm$^{-3}$), in contrast to the more evolved, classical HII regions that can be much more extended, reaching sizes of the order of several pc, and less dense ($n_e\simeq10^3$ cm$^{-3}$).  Because of the very high $n_e$ of the youngest HII regions, their turnover frequency occurs at high frequencies and their cm emission is (moderately) optically thick, with typical spectral indices $\gtrsim$1.

Thermal radio jets from YSOs \citep{Anglada1995,Anglada1996,Anglada2018} trace the base (a region very close to the exciting source) of the larger ionized jets wavelengths observed at infrared and optical. These radio jets are observed through their thermal free-free emission arising from shock-ionized gas. The emission is expected to be optically thick ($\alpha\simeq2$) in the region nearest to the star (where densities are higher) and optically thin ($\alpha\simeq-0.1$) at larger distances. The combination of these two components results in a total emission with a constant spectral index with an intermediate value over most of the radio spectrum. For a conical radio jet with constant temperature, velocity, and ionization fraction, $\alpha=0.6$ is expected \citep{Reynolds1986}. In general, a typical value of the  spectral indices of the observed radio jets is $\alpha\sim0.4$ \citep{Anglada2018}. Frequently, radio jets appear just as weak compact sources, which, in sensitive sub-arcsec angular resolution observations, can show an elongated morphology in the direction of the large-scale outflow  (see \citealt{Anglada2018} for a complete review on protostellar radio jets and their properties). They are useful to pinpoint the location of YSOs, particularly those most deeply embedded.

Photoionized HII regions can be stronger radio sources than thermal radio jets, but only for objects of high bolometric luminosity. In the low-luminosity regime ($L_{\rm bol} \la 100~L_\odot$), shock-ionized radio jets are dominant since stellar photoionization is negligible in these objects. Radio emission from both thermal radio jets and photoionized HII regions increases with bolometric luminosity but with a much steeper slope for HII regions so that they become comparable for a bolometric luminosity $L_{\rm bol}\simeq 1500~L_\odot$, where $S_\nu d^2 \simeq 1$ mJy kpc$^2$, corresponding to $S_\nu \simeq 6$ mJy at 3.6 cm at the distance of Orion (see fig. 8 in \citealt{Anglada2018}). Therefore, among the radio sources with $-0.1\le\alpha<2$, we will tend to identify as radio jets the weak ($<$ 1 mJy), elongated or unresolved radio sources, while we will tend to consider as photoionized HII regions the strongest ones (several mJy), either compact or extended with a cometary or irregular morphology, provided the observed cm flux density is consistent with the photoionization rate expected from the source luminosity.

Photoionized discs are also free-free emitters, and thus, they are expected to present radio spectral indices rather similar to those of radio jets \citep{Lugo2004,Anglada2018}. If angularly resolved, these sources can appear elongated in the direction of the disc plane \citep{Hoare2006,Macias2016}, making it difficult to distinguish them from the radio jets when no additional information on the orientation of the disc or outflow is available. The mechanism proposed to ionize and heat the disc surface in low-mass YSOs is high-energy radiation (far-UV, extreme-UV, and X-ray) from the stellar chromosphere of the Class II/III young stars \citep{Clarke2001,Gorti2009,Owen2010}.

Thermal emission from dust, either from a disc or from a dusty envelope, also has a positive spectral index, but steeper than that of free-free (i.e., $\alpha\ge2$). However, in observations with insufficient angular resolution the dust emission from the disc and the free-free emission from the jet cannot be spatially separated, resulting in emission with an intermediate spectral index between free-free and dust. This confusion will usually occur around 1 cm, since for most sources the thermal dust emission does not contribute significantly to the radio spectra at wavelengths $\ga$ 2 cm.

Young stars with active magnetospheres (Class II/III YSOs) can be sources of non-thermal gyrosynchrotron emission. These radio sources are very compact, can be highly variable on timescales of days, are expected to present negative spectral indices, and can be circularly polarized \citep{Andre1996,Garay1996,Rodriguez1999,Kounkel2014,Ortiz2015,Diaz-Rodriguez2022}. Detection of circular polarization strongly suggests gyrosynchrotron emission, but the lack of detectable polarization does not exclude this kind of emission.

In several cases, negative spectral indices, suggestive of non-thermal synchrotron emission, have been measured toward the lobes of YSO radio jets, at relatively large distances from the star, while the core of the jet, closer to the star, presents a positive spectral index, indicative of thermal emission (e.g., HH 80-81: \citealt{Marti1993,Rodriguez2005}, Serpens: \citealt{RodKam2016}, HOPS 370: \citealt{Osorio2017}, NGC 1333 IRAS 2B: \citealt{Tychoniec2018nonthermal}; in this last case, the non-thermal emission is observed quite close to the star).

The detection of linear polarization in the radio emission with negative spectral index observed in the HH 80-81 jet lobes \citep{Carrasco2010} confirmed its non-thermal synchrotron nature, suggesting that the emission originates from relativistic particles accelerated by strong shocks in the jet. Since the degree of polarization in YSO jets is low, linear polarization is hard to detect (see discussion in \citealt{Anglada2018}) and, in general, a negative spectral index is considered an indication of a synchrotron jet lobe.

Extragalactic sources also present non-thermal synchrotron emission, typically with a spectral index $\alpha\sim-0.7$ for optically thin emission \citep{Condon1984}. Optically thick synchrotron emission has a flat or positive spectral index and could be confused with thermal emission, but the turnover frequency in extragalactic sources is low (30 MHz - 1 GHz, \citealt{Odea1998}) and falls outside the frequency range of our observations. Extragalactic sources can be highly variable. They are usually very compact but can also show a lobe-core-lobe morphology that can be confused with a YSO jet. In some cases, proper motions can be measured and, thus, an extragalactic origin can be discarded.  

Thus, radio sources with negative spectral indices are candidates to be either a relatively evolved YSO, an extragalactic source or a non-thermal knot of a radio jet. Additional information (e.g., morphology, proper motions and polarization studies) can help to discern their nature. We will analyze their association with features at other wavelengths (see Section \ref{subsec:associations}) as an additional piece of information in an attempt to properly classify them.

For all the 47 sources reported here, we measured the flux densities at each observed band (5, 3, 1.3 and 0.7 cm) and constructed their spectra at radio wavelengths using our data, as well as data compiled from the literature (\citealt{Reipurth1999,Reipurth2004,Kounkel2014}; see Table \ref{tab:fluxes} and Fig. \ref{fig:SEDs}). The spectral index is obtained from a least-squares fit (in log-log scale) to the data points at the different frequencies where we detected the sources, taking into account upper limits when they are useful to constrain the fit. For the spectral index calculations we used only our data, which were obtained within a relatively short time interval, $<1$ yr (2014-2015), in order to minimize variability problems that would result in spurious values of $\alpha$. For sources 10, 12, 19, 21 and 35, only the 2014 epoch data is used since their 2015 flux density is significantly different, indicating they could be variable sources (see Section \ref{subsec:variab}). There are 13 sources with data only at 5 cm wavelength (see Table \ref{tab:fluxes}), and a spectral index cannot be obtained\footnote{We do not report an in-band spectral index (derived from maps made with the lower and higher frequency half-band) because errors in image processing, which are particularly significant in the case of relatively faint sources (SNR $<$ 50, such as those in our sample), have a considerable impact on the calculation of their spectral indices \citep{Rau2016} and may result in unreliable values. Actually, for each source, we derived 5 cm in-band spectral indices from the A and the C configuration data. We compared these values with each other, and, when possible, with the spectral index derived using several bands, finding clear inconsistencies in a number of cases.}. In Table \ref{tab:fluxes} we list the values of the spectral index obtained from our observations. For completeness, for a few sources, when our data were insufficient to obtain the spectral index, we list in the table values reported by \citet{Kounkel2014} which were derived from data obtained at very close epochs (3 epochs within 1 month); we note that only for one of these sources (Source 36) those values provide a useful constrain.

As mentioned before, the sign of the spectral index gives information about the nature of the radio emission (flat or positive indicates thermal emission, and negative non-thermal emission). Of the whole sample of 57 radio sources, consisting of the 10 radio sources reported by \citet{Osorio2017} and the 47 additional sources reported here, we are able to assess the sign of the spectral index for 27 of them (47\%): 19 sources with $-0.1\le\alpha<2$ that we identify as likely thermal free-free emission, and 8 sources with $\alpha<-0.1$, suggesting non-thermal emission (values given in Table \ref{tab:fluxes} of this work and in table 3 of \citealt{Osorio2017}). Most of the sources toward the ISF (17/24, or 71\%) have a positive spectral index, and 76\% (13/17) of them are associated with compact mm/submm emission, either at core \citep{Kainulainen2017} or disc \citep{Tobin2020I} scales (see Section \ref{subsec:distribution}), indicating they are probably YSOs. Notably, all the sources detected toward the peak of a 3 mm core (likely tracing deeply embedded protostars) have a positive spectral index. In Table  \ref{tab:alpha}  we  show  a  summary  of  the  spectral  indices and association with dust of the 57 detected radio  sources.

\begin{table}
	\setlength{\tabcolsep}{15.5pt}	
	\caption{Summary of the spectral indices and association with dust tracers.\label{tab:alpha}}
	\begin{tabular*}{\linewidth}{lcccc}
		\hline\hline
		Spectral & \multicolumn{3}{c}{Toward ISF $^b$}& Outside\\
		\cline{2-4}
		Index$^a$& \multicolumn{2}{c}{Compact Dust?$^c$} & Total & ISF$^b$\\
		\cline{2-3}
		& Yes & No &\\
		\hline
		$\ge-0.1$  & 13 & 4 & 17 & 2\\
		$<-0.1$ & 1 & 6 & 7 & 1\\
		?   & 8 & 10 & 18 & 12\\
		\hline
	\end{tabular*}
	\begin{tablenotes}
		\footnotesize
		\item{$^a$From Table \ref{tab:fluxes} of this work and table 3 of \citet{Osorio2017}. $^b$We consider that a source is located toward the Integral Shaped Filament (ISF) when the column density at its position, as estimated from Herschel 100 - 500 $\mu$m data \citep{Stutz2018}, is >$2.3\times10^{22}$ cm$^{-2}$; otherwise it is considered to be outside of it. $^c$Association with compact dust mm/sub-mm emission, either at core \citep{Kainulainen2017} or disc \citep{Tobin2020I} scales. See Section \ref{subsec:distribution}.}
	\end{tablenotes} 
\end{table}

\subsection{Association of the radio sources with other star-formation activity tracers}\label{subsec:associations}

Emission in other wavelength ranges can give us valuable additional information about our radio sources. For example, protostars (Class 0/I) are usually associated with sub-mm and far-IR thermal dust emission coming from the envelope of infalling gas and dust. Signs of mass ejection such as molecular outflows, collimated jets and cavities in the envelope are also often observed toward protostars. On the other hand, more evolved YSOs (Class II/III) are generally associated with near-IR and/or optical emission, coming from the accretion disc and the star itself. Class III YSOs are often associated with X-ray emission from magnetic reconnection flaring near the stellar surface, although a fraction of Class I and II YSOs also present detectable X-ray emission \citep{Getman2017}.

We searched for counterparts of our detected radio sources in point source catalogs in the far-IR \citep{Stutz2013,Furlan2016}, mid-IR \citep{Megeath2012,Megeath2016,Getman2017}, near-IR \citep{Skrutskie2006,Hillenbrand1997,Hillenbrand2013,Kounkel2016}, optical \citep{Gaia2023}, and X-ray \citep{Tsujimoto2002,Tsujimoto-Err,Getman2017}. We also searched the literature for signposts and features potentially associated with them, such as mm dust cores \citep{Chini1997,Kainulainen2017}, dust discs \citep{Tobin2020I}, jets and outflows \citep{Williams2003,Takahashi2008,Shimajiri2009, Sato2023} and outflow cavities \citep{Habel2021}. Additionally, we looked for extended emission associated with our radio sources in images obtained with the following facilities: HST ($0.8 - 2.5 \micron$, \citealt{Habel2021}), NEWFIRM (1.662 and 2.159 $\micron$, T. Megeath, private communication), {\it Spitzer} (3.6, 4.5, 5.8, 8 and 24 $\micron$, \citealt{Megeath2012}), {\it Herschel} (70 and 160 $\micron$, \citealt{Stutz2013}), APEX (350 and 870 $\micron$, \citealt{Stutz2013}). Our results are summarized in Table \ref{tab:sources}, where we report the counterparts and other interesting features observed at other wavelengths, and potentially associated with our radio sources. Among the IR counterparts, we report six NEWFIRM and one {\it Spitzer} sources that are identified here for the first time (see further details in Sections \ref{weakKs} and \ref{38}).

\begin{figure}
	\includegraphics[width=0.99\hsize]{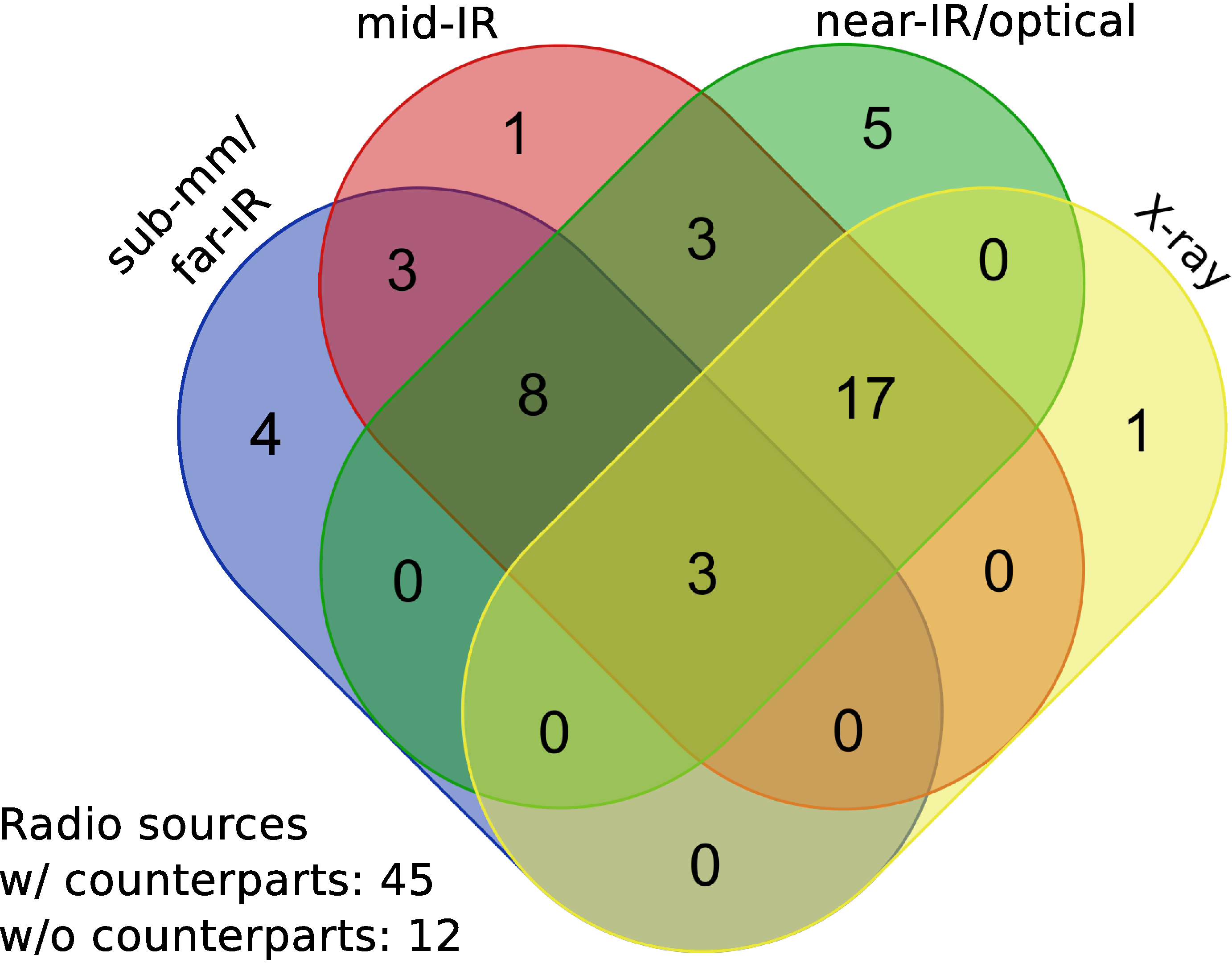}
	\caption{Venn diagram of the sub-mm/far-IR (18), mid-IR (35), near-IR/optical (36), and X-ray (21) counterparts of the 57 detected radio sources. The total radio sources with counterparts are 45. The few (5) doubtful identifications marked with ``?'' in Table \ref{tab:sources} are not included in this statistics. Diagram generated at \href{http://bioinformatics.psb.ugent.be/webtools/Venn/}{Bioinformatics \& Evolutionary Genomics}.} 
	\label{fig:counterparts}
\end{figure}

Of the 57 radio sources reported here and in \citet{Osorio2017}, we find that 45 (79\%) have possible counterparts. For a better visualization we show in Fig. \ref{fig:counterparts} a Venn diagram of the different types of counterparts. As can be seen in the figure, 21 radio sources have X-ray, 36 have near-IR or optical, 35 have mid-IR and 18 have far-IR or sub-mm proposed counterparts\footnote{We do not include in these statistics the few doubtful identifications marked with ``?'' in Table \ref{tab:sources}.}. Out of the 21 radio sources with an X-ray counterpart only 3 (sources 10, 42, and 46, see Table \ref{tab:sources}) also have a far-IR/sub-mm counterpart, in agreement with the general belief that X-ray emission is probably not associated with the youngest protostars \citep{Getman2017}. Furthermore, Source 42, which is the only one with counterparts at all the wavelengths inspected, is an embedded protostar with an irregular cavity (see further details in Section \ref{42}) through which we are probably seeing the short-wavelength emission from the inner envelope and disc. Four radio sources only have far-IR/sub-mm counterparts (sources 9, 24, 43 from this work, and VLA 15 from \citealt{Osorio2017}), suggesting they are very young. As expected, the radio sources with far-IR/sub-mm counterparts are located in the densest part of the ISF, while radio sources with counterparts at shorter wavelengths are systematically located farther from the ISF (see Table \ref{tab:sources} and Fig. \ref{fig:filament}). We do not find a trend for the spatial distribution of the radio sources with X-ray counterparts.

Of the same 57 radio sources considered above, we find that 9 (16\%) have known outflows and/or cavities associated with them (Table \ref{tab:sources}). These are the sources 1, 9, 11, 25, 34, 39, 42, 43 (discussed in this paper) and VLA 11 (discussed in \citealt{Osorio2017}). These sources are also associated with far-IR/sub-mm emission and are observed toward compact dust cores (see Section \ref{subsec:distribution}). They are probably among the youngest YSOs in our sample.

In Table \ref{tab:sources} we include information on the proposed nature of the radio sources, and the evolutionary stage of the YSOs, when known (see Section \ref{subsec:individual} and Appendix~\ref{subsec:individual2} for further details). We make a suggestion for the nature of 46 out of 57 radio sources, with 36 of them probably being YSOs.

In the future,  proper motions may be measured from multi-epoch observations, thereby providing a way to distinguish between jet knots (with proper motions of $\sim$100 km/s), emission pinpointing the position of a star (with proper motions of a few km/s) or extragalactic sources, which  exhibit no measurable proper motion.

\subsection{Variability of the radio sources}\label{subsec:variab}

In order to investigate the radio variability of the sources reported in this work and \citet{Osorio2017}, we compared our flux density measurements with those of previous works (\citealt{Reipurth1999,Reipurth2004,Kounkel2014}; see Table \ref{tab:fluxes} and Fig. \ref{fig:SEDs}) for the 16 sources previously reported at cm wavelengths. Because the flux values given by \citet{Reipurth2004} for sources 1 and 3 are systematically higher than values in other works, we recalibrated their data following the more recent recommendations for the VLA calibration process and obtained the more consistent values shown in our spectra (Fig. \ref{fig:SEDs}). We consider a source as probable radio variable (Table \ref{tab:fluxes}) if the flux density at the same band changes by a factor greater than 1.5 between different epochs, taking into account the uncertainties. We select this threshold to exclude sources with small amplitude flux density fluctuations ($\sim$ 10–20\%) caused by short-timescale ejections of ionised condensations. We are instead interested in finding sources with stronger variability driven by either non-thermal emission or singular events (see discussion in Section 3.3 of \citealt{Anglada2018}).
We have paid particular attention when inferring variability from data with significant differences in uv coverage. In such cases, the existence of variability has only been asserted when the flux density observed with the longest baseline observations is greater than that observed in the more compact array, in order to ensure that the variability is not the result of an extended emission that has been filtered out due to the lack of short baselines.

\begin{figure}
	\includegraphics[width=1\hsize]{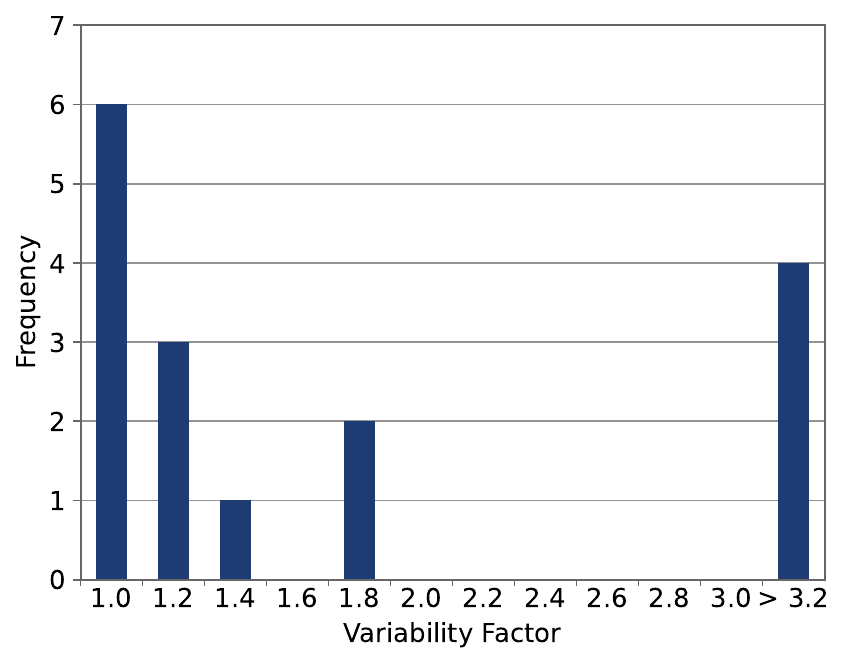}
	\caption{Distribution of the variability factor (defined as $F =  \left ( {S_{\mathrm{max}}-\sigma_{S_\mathrm{max}}} \right )/ \left ( {S_\mathrm{min}+\sigma_{S_{\mathrm{min}}}} \right )$, where $S_\mathrm{max(min)}$ is the maximum (minimum) flux density and $\sigma_{S_\mathrm{max(min)}}$ its uncertainty) between our flux density measurements and previous reports in the same band. Values $F\le1$ are all consistent with $F=1$ within the uncertainties and therefore represented together; values $F > 1$ should be interpreted as lower limits.} 
	\label{fig:var_old}
\end{figure}

\begin{figure}
	\includegraphics[width=1\hsize]{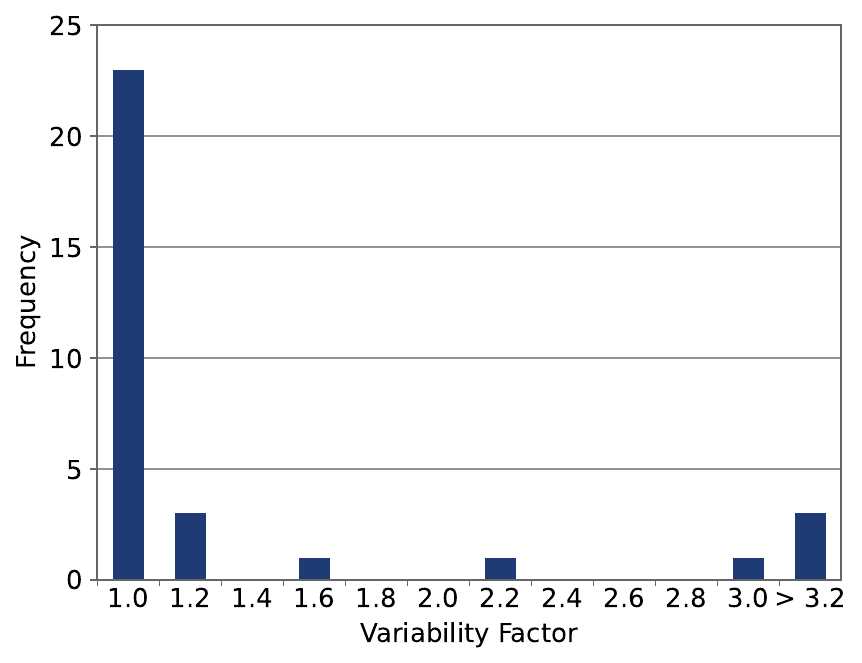}
	\caption{Same as Fig. \ref{fig:var_old} but between our 5 cm (2014 and 2015) flux density measurements.} 
	\label{fig:var_5cm}
\end{figure}

We find that six out of these 16 sources meet our variability criterion (sources 11, 12, 14, 21, 36, and 45; see Fig. \ref{fig:var_old}). Additionally, comparing our two 5 cm (2014 and 2015) observations taken 10-11 months apart, we find that six out of the 32 radio sources for which we can derive a restrictive variability factor show signs of variability in this period (sources 10, 12, 19, 21, 27, and 35; see Fig. \ref{fig:var_5cm}).  

In total, we find that 10 sources seem to be variable at radio wavelengths, with changes by a factor up to 30 (the most extreme case is Source 12) on time-scales of a few years. All of these 10 sources seem to be YSOs, mostly evolved ones (Class II/III), except the Class I source 11 (see further details in Section \ref{subsec:individual} and Appendix~\ref{subsec:individual2}); Source 35 could be either an extragalactic source or an evolved YSO. In none of these cases, the reported radio variability is obviously due to differences in the array configuration of the observations (but in some cases there could be residual extended emission that slightly modifies our results).

Since 37 radio sources have counterparts at near-IR, optical and/or X-rays (see Table \ref{tab:sources} and \citealt{Osorio2017}), we searched the literature for variability at those wavelengths \citep{Tsujimoto2002,Tsujimoto-Err,Samus2017}, finding that 19 (51\%) are reported as variable\footnote{This includes sources MIPS 2301 and HOPS 64, which are variable in the near-IR, and MIPS 2297, which is variable in the optical, whose radio counterparts were reported in \citet{Osorio2017}.} (see Table \ref{tab:sources}). All these 19 sources seem to be evolved YSOs (Class II/III, see further details in Section \ref{subsec:individual} and Appendix~\ref{subsec:individual2}). Interestingly, seven of them are also among the suspected radio variables (sources 10, 12, 14, 19, 21, 27, and 36).

\subsection{Multiplicity of the radio sources}\label{subsec:multip}

The high angular resolution reached in our observations at 5 cm allows us to separate emission from radio sources as close as 0\farcs4 (154 au at the adopted distance to Orion). In this way, we have resolved into multiple different sources emission that was previously associated with a single source in lower angular resolution observations (\citealt{Reipurth1999,Furlan2016}; see Tables \ref{tab:positions} and \ref{tab:sources}). In particular, we resolve the radio sources VLA 5 into sources 1 and 2, VLA 9 into sources 10 and 11, and VLA 13 into sources 33 and 34, and the far-IR source HOPS 66 into sources 24 and 25.

Previous works \citep[e.g., ][]{Reipurth1993,Duchene2004,Tobin2016mult} have searched for multiple systems with a projected separation  $<$2000-3000 au, which is an order of magnitude smaller that the typical star-to-star separation (between $\sim $10000 - 20000 au, depending on the density of the star-forming region, according to \citealt{Reipurth1993}). These authors argued that companions found within these separations are likely to be gravitationally bound. However, using this approach to identify multiple systems has some drawbacks. First, the number of multiple systems identified depends considerably on the adopted range of separations, which is somewhat arbitrary. Secondly, projected separations are smaller than the real distances, and in some cases, unrelated objects may appear as close companions. Finally, unresolved systems, with components closer than the observational resolution limit, will be missed.

Nevertheless, if we consider projected separations between 154 au (our angular resolution limit) and 3000 au ($\sim$ 8$\arcsec$), we find 7 possible resolved binary systems among the 53 sources reported here and in \citet{Osorio2017}\footnote{We exclude of this analysis sources VLA 12N, VLA 12C, VLA 12S (reported by \citealt{Osorio2017}) and Source 33 (see Section \ref{33&34}) since they are not tracing YSOs.}. These are the ones formed by sources 28 and 29 (separation = 232 au = $0.6''$), sources 24 and 25 (separation = 849 au = $2.2''$), HOPS 370 and MGM 2301 (separation = 1042 au = $2.7''$), sources 1 and 2 (separation = 1081 au = $2.8''$), sources 10 and 11 (separation = 1853 au = $4.8''$), HOPS 64 and HOPS 108 (separation = 2316 au = $5.9''$), and VLA 15 and VLA 16 (separation = 2393 au = $6.1''$). All of these objects are probable YSOs (see Section \ref{subsec:individual}, Appendix~\ref{subsec:individual2}, and \citealt{Osorio2017}), except for Source 2 that could be a YSO or a knot of Source 1 (see Section \ref{1and2}). No higher order multiple systems are found at this separation range.

In order to establish the significance of the number of proposed binaries, we can make a simple estimate of the probability of a chance alignment of the positions in the plane of the sky of unrelated objects. As can be seen in Fig.~\ref{fig:filament} and discussed in Section \ref{subsec:distribution}, the sources we have detected are not uniformly distributed across the entire observed field of view, but appear to be concentrated towards the ISF (42 sources) and along an east–west stripe (16 sources), with one source in the overlapping central part. We shall therefore conservatively estimate the probability of a chance alignment amongst the potential binaries in each of these two regions, where this probability is highest. For a given source, the probability that another source will fall by chance within a solid angle of radius $a$ around the first one, thus simulating a false binary with angular separation $\le$$a$, can be estimated from the probability of the opposite event (i.e. that all other detected sources lie outside that solid angle), which yields:
\begin{equation} \label{eq:Pfake}
	P({\rm false~binary})=1- {\left(\frac{\Omega - \pi a^2}{\Omega}\right)}^{N-1},
\end{equation}
where $\Omega$ is the solid angle of the region considered, and $N$ the total number of sources detected toward this region. The solid angle is $\Omega\simeq 12.5'\times2' = 90,000$ arcsec$^2$ for both the ISF region (where $N$ = 42 sources) and the E-W stripe (where $N$ = 16 sources). Thus, for the observed binary separations listed above, the values of the probability of chance alignment are $P({\rm false~binary})$ = 0.007-0.052 for the six proposed binaries toward the ISF, and $P({\rm false~binary})$ = 0.0002 for the proposed binary toward the E-W stripe (sources 28 and 29). Therefore, the individual probabilities of being true binaries are $P({\rm true~binary})$ = 1- $P({\rm false~binary})$ = 0.993-0.948 for the proposed binaries towards the ISF and $P({\rm true~binary})$ = 0.9992 for the proposed binary towards the stripe, and the expected number of binaries is $\sum_{i=1}^{7} P({\rm true~binary})_i$ = 6.838, which is very close to 7. 

The probability that all the proposed 7 binaries are true binaries is given by:
\begin{equation} \label{eq:Pfinal}
	P({\rm 7~true~binaries})=\prod_{i=1}^{7} P({\rm true})_i = 0.848, 
\end{equation}
and the probability of finding at least one false positive among the seven proposed binaries is $1-0.848$ = 0.152, which is small. Thus, we conclude that the expected number of binaries is 7 with a high probability. This result is consistent with a more detailed analysis performed by \citet{Tobin2022} over a larger region, encompassing the Orion A and B molecular clouds, who concluded that pairs of sources with projected separations $\la 8''$are mostly true binaries.

Two parameters broadly used to characterize the multiplicity of a sample are the multiplicity frequency or multiplicity fraction ({\em MF}), that gives the fraction of non-single systems, and the companion fraction ($CF$), which quantifies the average number of stellar companions per system. They are defined as $MF = (B+T+Q+...)/(S+B+T+Q+...)$, and $CF = (B+2T+3Q+...)/(S+B+T+Q+...)$, where $S$, $B$, $T$, $Q$, ..., are the number of single, binary, triple, and quadruple ..., systems \citep{Reipurth1993}. The value of $MF$ is always between 0 and 1; $CF = MF$ in samples with only single stars or binary systems, while $CF > MF$ in samples with higher-order multiple systems (see \citealt{Duchene2013}). For our sample of 53 radio sources, we find $MF = CF = 0.15^{+0.06}_{-0.04}$\footnote{Like in \citet{Tobin2022}, $1\sigma$ uncertainties of $MF$ and $CF$ are calculated assuming binomial statistics, specifically using the Wilson score interval \citep{Wilson1927}, whose edges are given by \[
	\sigma_{P}
	=
	\left(1 + N_{\mathrm{sys}}^{-1}\right)^{-1}
	\left[
	P + \tfrac{1}{2} N_{\mathrm{sys}}^{-1}
	\right]
	\pm
	\left(1 + N_{\mathrm{sys}}^{-1}\right)^{-1}
	\left[
		P(1 - P) N_{\mathrm{sys}}^{-1}
		+ \tfrac{1}{4} N_{\mathrm{sys}}^{-2}
	\right]^{ \tfrac{1}{2}},
	\] where $P$ is $MF$ or $CF$, and $N_\mathrm{sys}$ is the number of systems.}. This is a ``local'' result, since our sample represents a small fraction of the more than 300 known protostars in Orion. 

\citet{Tobin2022} explored the multiplicity of their target sample of 289 protostars in Orion for different separation ranges between 20 and 10000 au. They found a bi-modal distribution of companion separation for their whole sample, with peaks at $\sim100$ au and $\sim3000$ au, suggesting that multiple star formation and fragmentation is a multi-scale process. The shorter peak (below our resolution limit) would characterize the proto-binary separation, while the larger peak, the core-to-core separation. This bi-modal distribution has been also observed in the Perseus star-forming region \citep{Tobin2016mult,Tobin2022}. For the northern ISF (comprising OMC-1, OMC-2 and OMC-3 regions), \citet{Tobin2022} found $MF = 0.18^{+0.05}_{-0.04}$ and $CF = 0.22^{+0.06}_{-0.05}$ in a separation range from 20 to 1000 au. In our sample we have only two potential binaries with separation below 1000 au (see above), giving $MF = CF = 0.04^{+0.04}_{-0.02}$ for a separation range between 154 and 1000 au. However, we cannot make a direct comparison since, first, we are analyzing a smaller portion of the ISF (OMC-2 and part of OMC-3) and, second, we are probably loosing an important fraction of close multiples due to our poorer angular resolution limit (154 au).

It is possible that our radio sources have companions with no detectable radio emission but that have been detected at other wavelengths. In our inspection of the NEWFIRM deep H and K${_s}$  images, we find 10 near-IR sources at a projected separation $<3000$ au from the radio sources 12, 14, 17, 36, 38, 42 and 45; \citet{Kounkel2016} reported 4 near-IR sources between 100 and 1000 au from the sources 7, 11 and 46; and \citet{Gaia2023} reported 1 optical source at $\sim100$ au from Source 27. Based on the accurate distances reported by \citet{Gaia2023} and/or on the analysis of the spacial and velocity structure of the region done by \citet{Kounkel2018}, we conclude that sources 14, 27, 36 and 38 probably do not form binary systems with their respective nearby (in projection) near-IR/optical sources (see further details in Sections \ref{14}, \ref{stripe}, \ref{27} and \ref{38}), leaving 11 possible near-IR companions. Also, there are 6 additional sources detected with the VLA at 9 mm and/or ALMA at 870 $\micron$ \citep{Tobin2020I} within 3000 au of sources 1 and 2, Source 12, and VLA 15 and VLA 16. On the other hand, sources 19, 22 and 37 are spectroscopic binaries \citep{Tobin2009}, and Source 40 is an astrometric binary with components detected by \citet{Kounkel2017} through Very Long Long Baseline Array (VLBA) observations. If we include these 21 possible companions detected at other wavelengths, we find that three of the previously potential binaries become higher multiple systems (one triple, one quadruple, and one quintuple), and that there are, additionally, seven binary systems, one triple, and two quadruple. Therefore, we identify a total of 29 singles, 11 binaries, 2 triples, 3 quadruples and one quintuple candidates. We note that in one of the possible quadruple systems probably not all the sources are gravitationally bound (see further information in Section \ref{10and11}). After the inclusion of the 21 additional possible companions, we obtain $MF=0.37^{+0.06}_{-0.05}$ and $CF = 0.61^{+0.05}_{-0.06}$.

In summary, we are detecting components of 17 potential multiple YSO systems (see Section \ref{subsec:individual} and Appendix~\ref{subsec:individual2}). Further analysis, e.g., proper motion studies, will be necessary to determine if these sources are indeed gravitationally bound. For example, if the components of a multiple system candidate present a common proper motion distinguishable of that of the nearby sources (the proper motions in the region have a dispersion of up to $\sim3$ km s$^{-1}$ \citep{Dzib2017,Kounkel2018} this would suggest that they are indeed gravitationally bound. On the other hand, if the relative proper motion of the components of the binary candidate exceeds the expected escape velocity, this would be suggestive of a gravitationally unbound system.

\subsection{On the correlation of the observed radio emission with the bolometric luminosity and the outflow momentum rate}\label{subsec:nature}

The existence of empirical correlations between the observed centimeter luminosity ($S_\nu d^2$) of protostars and both the bolometric luminosity ($L_{\rm bol}$) and the outflow momentum rate ($\dot{P}$) is fairly well established. These correlations (sometimes referred as the `Anglada correlations') have been inferred from observations of a large sample of sources covering a wide range of bolometric luminosities \citep{Anglada1992,Anglada1995,Anglada1996,Anglada2015,Anglada2018,Rosero2019} and have been interpreted as a result of the connection between accretion and outflow processes in star formation. As discussed in the above references (see also, Section \ref{subsec:alpha}), protostars are frequently associated with relatively weak cm continuum emission which is attributed to free-free radiation from partially ionized jets. The mechanism that produces the necessary ionization of these jets is not photoionization by stellar UV photons (as in classical HII regions) but shocks associated with the outflowing gas itself, and a correlation of the radio luminosity and the outflow momentum rate is expected if this mechanism is at work \citep{Curiel1987,Curiel1989,Anglada1996}. On the other hand, since the bolometric luminosity of protostars is largely dominated by the accretion luminosity, and because outflows are driven by accretion, protostars with higher bolometric luminosities are expected to be associated with more powerful outflows, and consequently with the stronger (shock-ionized) cm continuum emission; thus, a positive correlation between bolometric and cm radio luminosities is also expected.

These correlations, with recently updated parameters derived from a selected sample of 81 YSOs reliably associated with outflows (with an estimate of $\dot{P}$ for 73 of them) and covering a range of eight orders of magnitude in bolometric luminosity ($10^{-2} L_\odot \lesssim L_{\rm bol} \lesssim 10^6 L_{\odot}$), are given by equations (28) and (31) of \citet{Anglada2018}:
\begin{equation} \label{eq:Lbol}
	\left(\frac{S_\nu d^2}{{\rm mJy \: kpc}^2}\right) = 10^{-1.90\pm0.07} \left(\frac{L_{{\rm bol}}}{L_\odot}\right)^{0.59\pm0.03}, 
\end{equation}
\begin{equation} \label{eq:Ppto}
	\left(\frac{S_\nu d^2}{{\rm mJy \: kpc}^2}\right) = 10^{2.97\pm0.27} \left(\frac{\dot{P}}{M_\odot \: {\rm year^{-1} \: km \: s^{-1}}, }\right)^{1.02\pm0.08}
\end{equation}
where $S_\nu$ is the flux density at 3.6 cm wavelength.

In these equations, the outflow momentum rate, $\dot{P}$, is the parameter that is most poorly determined by observations. In general, it is estimated from the mass, velocity, and dynamical time of the CO molecular outflow using simple assumptions for the geometry and outflow kinematics, with an important dependence on the assumed CO molecular abundance, opacity effects, and inclination angle. Uncertainties in this parameter are often of an order of magnitude or greater. Our knowledge of the distance, $d$, to star-forming regions has been much improved recently, so this parameter is generally not a substantial cause of uncertainty. In addition, equation (\ref{eq:Lbol}) is not affected by distance, as long as the same value of this parameter has been used for radio and bolometric luminosities. The uncertainties in the bolometric luminosity, $L_{\rm bol}$, are mainly due to the poor angular resolution of the far-IR observations, where the SED of the youngest objects reaches its maximum, causing it to include the emission of several objects in crowded regions, so that in some cases these values can only be used as upper limits. Often the structure of the sources is complex (with important asymmetries due to the presence of cavities, etc.), so that the emission is anisotropic and the geometry and orientation of the object strongly influence the observed emission, introducing additional uncertainties in the derived bolometric luminosity, even when this is obtained through detailed modeling. Thus, in general, the parameter that is better determined by the observations is the cm flux density, $S_\nu$; in this case, the main sources of uncertainty are the possible variability in the radio emission and the uncertain behavior of the spectral index when data at different wavelengths are compared. Because of these characteristics of the observational uncertainties, the least-squares fitting has been performed taking $S_\nu d^2$ as the independent variable (i.e., contrary to the way equations \ref{eq:Lbol} and \ref{eq:Ppto}  are displayed), and minimizing the errors in $L_{\rm bol}$ and $\dot{P}$. Uncertainties shown in equations (\ref{eq:Lbol}) and (\ref{eq:Ppto}) correspond to the formal errors of the fits (residual standard deviation) and partially reflect the actual dispersion in these correlations.

Although the general trend traced by these correlations is clear, suggesting the existence of a connection between accretion, outflow and radio emission, the precise relationship between these three phenomena depends on complex interactions in the outflow ejection through an accretion disc, and in the shock-ionization of the outflow that originates the free-free emission observed at radio wavelengths, which are unlikely described by the simple power-law correlations shown in equations  \ref{eq:Lbol} and \ref{eq:Ppto}. For example, as shown by \citet{Anglada1996}, a power-law relationship like equation (\ref{eq:Ppto}) is derived from the shock-induced ionization model of \citep{Curiel1987,Curiel1989} under the simplifying assumptions of plane-parallel shocks, optically thin emission and equaling the momentum rates in the molecular outflow (tracing larger scales and dynamical times) and in the radio jet (which traces smaller scales and dynamical times). Thus, the specific relationship between $S_\nu d^2$, $L_{\rm bol}$, and $\dot{P}$ for a given source will certainly depend on a complex way on many factors, such as the individual properties of the source, environment and orientation with respect to the observer. Consequently, it is expected that in these correlations, in addition to the dispersion of data points due to the observational errors of the measurements, there is an additional intrinsic scatter of the data points due to the particular conditions of each source \citep{Anglada2018}. Systematic studies, focused on individual regions (where distance uncertainties are minimized), performed under uniform conditions, and with accurate determinations of both the bolometric luminosity and outflow parameters, can help to better characterize this intrinsic scatter of the data points. For example, \citet{Tychoniec2018corr} analyzed the Anglada correlations for a large and homogeneous sample of YSOs, consisting of all known protostars in Perseus molecular cloud, but covering a narrow range of luminosities $10^{-1.5} L_{\odot}\lesssim L_{\rm bol}\lesssim10^{1.5} L_{\odot}$. These authors conclude that, while the general trends are robust when including additional sources with $L_{\rm bol}$ up to $10^4 L_{\odot}$, the correlations are weakened, or even disappear when data points were restricted to the Perseus sample, revealing the intrinsic scatter of data points.

The field of view covered by our observations includes the positions of 31 far-IR HOPS sources (see Tables \ref{tab:sources} and \ref{tab:HOPS}, and Fig. \ref{fig:filament}), with 14 of them being associated with detected radio sources. The HOPS sources have been observed in a homogeneous way, obtaining their SEDs from near-IR to submillimeter wavelengths. Therefore, their $L_{\rm bol}$ have been very well determined in most cases by \citet{Furlan2016} using {\it Herschel} bands at 70 and 160 $\micron$, which constrain the SED peak, very sensitive to the luminosity. We use this valuable data to investigate the correlation with radio emission.

\begin{table}
	\caption{Bolometric and radio luminosities of the HOPS YSOs in the field of view\label{tab:HOPS}.}
	\centering
	\resizebox{\columnwidth}{!}{
	\begin{tabular}{l c c c c c}
	\hline
	\multicolumn{1}{c}{HOPS} &
	\multicolumn{1}{c}{Source} &
	\multicolumn{1}{c}{$L_{\rm bol}$$^a$} &
	\multicolumn{1}{c}{$\dot{P}$$^b$} &
	\multicolumn{1}{c}{$S_{\nu}d^2$$^c$} &
	\multicolumn{1}{c}{Evolutionary} \\
	ID & name/number & (L$_\odot$) & (M$_\odot$ km s$^{-1}$ yr$^{-1}$) & (mJy kpc$^2$) & Stage$^a$ \\
		\hline	
		
		56  & - & 20 & ... & $<$0.05$^d$ & Class 0\\
		57  & 46 & 3 & ... & 0.03$^d$ & Flat Spectrum\\
		58  & - & 4 & ... & $<$0.019$^d$ & Flat Spectrum\\
		59  & 42 & 42 & ... & 0.11$^d$ & Flat Spectrum\\
		60  & 39 & 19 & 9.8$\times10^{-5}$ & 0.019$^d$ & Class 0\\
		62  & - & 0.6 & ... & $<$0.009 & Flat Spectrum\\
		63  & - & 0.4 & ... & $<$0.007 & Flat Spectrum\\
		64  & HOPS 64 & 13 &... & 0.0016$^d$ & Class I\\
		65  & - & 0.30 & ... & $<$0.008 & Class I\\
		66  & 24 & $<$18$^e$ & ... & 0.006$^d$ & Flat Spectrum\\
		66  & 25 & $<$18$^e$ & ... & 0.013$^d$ & Flat Spectrum\\
		68  & - & 5 & ... &  $<$0.009$^d$ & Class I\\
		69  & - & 2 & ... & $<$0.009 & Flat Spectrum\\
		70  & - & 6 & ... & $<$0.009$^d$ & Flat Spectrum\\
		71  & 11 & 5 & 4.6$\times10^{-5}$ & 0.08$^d$ & Class I\\
		73  & - & 1.4 & ... & $<$0.012$^d$ & Class 0\\
		74  & - & 1.0 & ... & $<$0.015$^d$& Flat Spectrum\\
		75  & - & 3 & ... & $<$0.016$^d$ & Class 0\\
		76  & - & 1.6 & ... & $<$0.018$^d$ & Class I\\
		77  & - & 11 & ... & $<$0.021 & Flat Spectrum\\
		78  & 1 & 8 & 4.5$\times10^{-4}$ & 0.06$^d$ & Class 0\\
		79  & - & 0.07 & ... & $<$0.021 & Flat Spectrum\\
		80  & - & 0.07 & ... & $<$0.03 & Flat Spectrum\\
		81  & - & 1.1 & ... & $<$0.03$^d$ & Class 0\\
		82  & - & 2 & ... & $<$0.04 & Flat Spectrum \\
		84  & - & 42 & ... & $<$0.06$^d$ & Class I\\
		108  & HOPS 108 & 33 & ... & 0.003$^d$ & Class 0\\
		368  & 34 & 59 & 8.4$\times10^{-5}$ & 0.07$^d$ & Class I\\
		369  & 32 & 30 & ... & 0.005 & Flat Spectrum\\
		370  & VLA 11 & 305 & ... & 0.3$^d$ & Class I\\
		394  & 9 & 5 & 4.8$\times10^{-6}$ & 0.04$^d$ & Class 0\\
		409  & 43 & 7 & 1.9$\times10^{-5}$ & 0.009$^d$ & Class 0\\
		\hline
	\end{tabular} 
}
	\begin{tablenotes}
		\footnotesize
		\item{$^a$From \citet{Furlan2016}, with the luminosities scaled to our adopted distance of 390 pc. $^b$From \citet{Takahashi2008} and \citet{Tobin2016outf}.$^c$Radio luminosity obtained from C-configuration 5 cm data (from Table \ref{tab:fluxes} in this work and table 3 in \citealt{Osorio2017}). A-configuration data have been used for HOPS 66, 78 and 108 (that are blended with nearby sources in the C-configuration data) and HOPS 57 and 64 (undetected in the C-configuration map). Upper limits are 5$\sigma$.$^d$Source detected at Ka band (9 mm) by \citet{Tobin2020I}.$^e$The far-IR source HOPS 66 includes both radio sources 24 and 25.}
	\end{tablenotes} 
\end{table}

\begin{figure*}
	{\includegraphics[width=0.49\hsize]{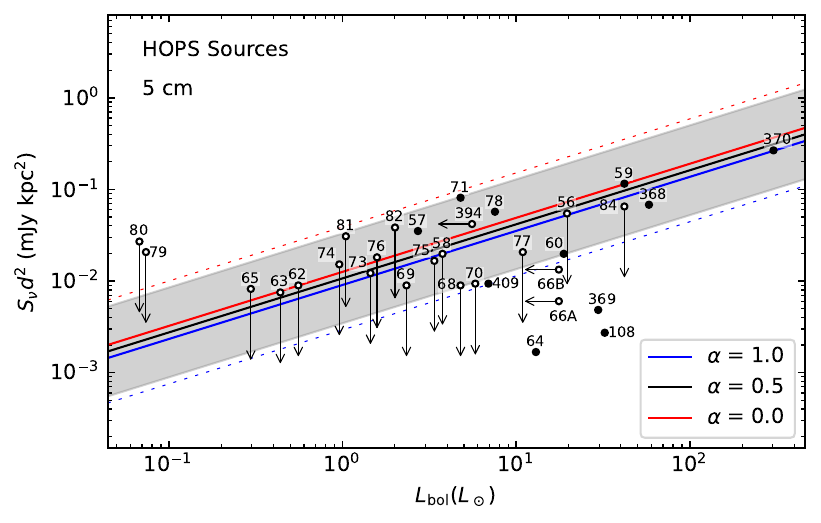}}
	\hfill
	{\includegraphics[width=0.49\hsize]{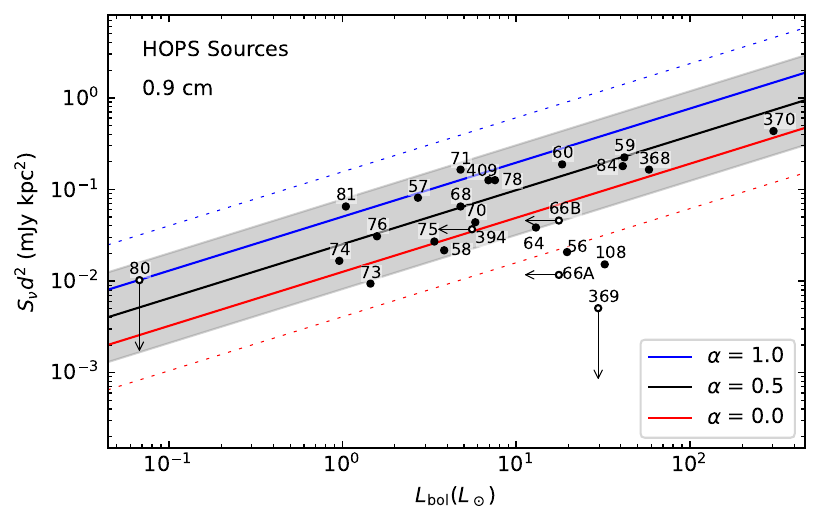}}
	\caption{Radio luminosity, $S_{\nu}d^2$, at 5 cm (left panel, Table \ref{tab:HOPS}) and 0.9 cm (right panel, \citealt{Tobin2020I}) versus the bolometric luminosity, $L_{\rm bol}$ (from \citealt{Furlan2016}, after correction to our adopted distance of 390 pc) for the HOPS YSOs observed in the analyzed field of view. Data points are labelled with their HOPS ID number. The black solid line represents the extrapolation to 5 cm (left panel) and 0.9 cm (right panel), assuming a typical average spectral index $\alpha = 0.5$, of the empirical correlation obtained at 3.6 cm (equation~\ref{eq:Lbol}) by \citet{Anglada2018} from a fit to a large sample of sources. The shadowed gray area indicates the residual standard deviation of this fit. To illustrate how the adopted spectral index affects the extrapolation to other wavelengths, we also plot the results assuming a spectral index $\alpha = 0.0$ (red solid line) and $\alpha = 1.0$ (blue solid line). The dotted lines indicate the variation of the limits of the shadowed area when the red and blue line correlations are considered.
Vertical arrows indicate 5$\sigma$ upper limits. Horizontal arrows indicate $L_{\rm bol}$ upper limits for radio sources 24 and 25 (both associated with HOPS 66 but unresolved in the far-IR data, see Section \ref{24&25}).
		\label{fig:Lbol}}
\end{figure*}

In Fig. \ref{fig:Lbol} (left) we plot the 5 cm radio luminosity ($S_{\nu} d^2$) versus $L_{\rm bol}$ (Table \ref{tab:HOPS}) for the 31 HOPS YSOs with known $L_{\rm bol}$ in our field of view. For comparison, we plot equation (\ref{eq:Lbol}), extrapolated from 3.6 cm to 5 cm assuming a spectral index $\alpha =0.5$, which is a typical value for radio jets (see \citealt{Anglada2018}). The uncertainty interval (residual standard deviation of the fit from which equation \ref{eq:Lbol} was derived) is plotted as a gray shaded region. To illustrate how the adopted spectral index affects the extrapolation to other wavelengths, we have also plotted the results assuming spectral indices of $\alpha = 0.0$ (no correction with respect to 3.6 cm) and $\alpha = 1.0$. As can be seen in the figure, the correction from 3.6 to 5 cm is negligible, given the small difference in wavelength and the relatively flat spectral indices of thermal radio jets. We find that most of the HOPS sources in our field of view have bolometric luminosities consistent with the Anglada correlation. There is no source with a radio luminosity significantly higher than expected from equation (\ref{eq:Lbol}), suggesting that shock-induced ionization is the mechanism at work, since photoionization by stellar photons for sources in this bolometric luminosity range, 0.1-300 $L_{\odot}$, is expected to produce radio luminosities in the range $<$$10^{-16}$ to $\sim$$10^{-2}$ mJy kpc$^{-1}$, respectively (fig. 5 in \citealt{Anglada1995}), well below the observed values. However, although the large majority of 5 cm data points are consistent with the correlation, most of them (18) correspond to non-detections, and only upper limits for the radio luminosity are available. At least three out of the 14 radio sources detected at 5 cm have radio luminosities clearly lower than expected, namely the ones associated with HOPS 64, HOPS 108 and HOPS 369; HOPS 66 coincides with two radio sources, which are not angularly resolved in the far-IR data (see discussion below).

Out of the 18 HOPS sources not detected at 5 cm, 11 were observed at 9 mm by \citet{Tobin2020I},  detecting 10 of them. In Fig. \ref{fig:Lbol} (right) we plot the observed radio luminosities at 9 mm, together with the Anglada correlation extrapolated to that wavelength. As can be seen in the figure, most of the 9 mm sources appear to follow the correlation. It should be noted that, in some cases, the emission at 9 mm can have a significant contribution of dust thermal emission from the associated disc or envelope; therefore, the actual free-free contribution in some sources can be somewhat lower than the values plotted. Also, the extrapolation of the Anglada correlation from 3.6 cm to 9 mm is more uncertain than at 5 cm (see Fig.~\ref{fig:Lbol}), which increases the area of uncertainty and therefore the number of sources that are potentially consistent with the correlation. These results suggest that the 9 mm emission is dominated by free-free emission. In any case, it would be interesting to perform still more sensitive observations at 5 cm, a wavelength where the dust contribution should be negligible, to unveil the free-free radio emission from this undetected population of weak radio sources. We estimate that future 5 cm observations reaching a rms sensitivity of $\sim$ 3 $\mu$Jy beam$^{-1}$ across the whole field of view (similar to the sensitivity reached toward the centre of the field in our current observations) should be able to detect the large majority of these sources above a 5-$\sigma$ level, if they have a flux density similar to that expected. This will unambiguously clarify the degree of agreement with the predictions. Nevertheless, we identify three sources, HOPS 56, HOPS 108 and HOPS 369, that have radio luminosities at 9 mm lower than expected.

In summary, we find that most of the HOPS sources in our field of view follow the Anglada correlation of radio emission with $L_{\rm bol}$, and that our non-detections at 5 cm are mostly due to sensitivity issues. However, a few protostars (HOPS 56, HOPS 64, HOPS 108 and HOPS 369; and perhaps HOPS 66) clearly do not follow the correlation and appear as underluminous at radio wavelengths.

The correlation between radio and bolometric luminosities is usually interpreted as a consequence of the connection between outflow (traced by the cm radio emission) and accretion (traced by the bolometric luminosity which, in protostars, is assumed to be dominated by the accretion luminosity). Therefore, for the sources that appear as underluminous in radio (or, equivalently, overluminous in bolometric luminosity), one would expect that either the mechanism of shock-induced ionization is less effective than in the other sources, or that the bolometric luminosity has a significant contribution from mechanisms other than accretion. In more evolved YSOs, where the accretion has decreased significantly, it has been suggested that the stellar luminosity (which is not expected to be directly related with accretion and outflow) can have an important contribution to the bolometric luminosity, making theses objects to appear radio-underluminous in the Anglada correlation \citep{Rodriguez2014,Rodriguez2016}. However, our sources are younger and we suggest, alternatively, that the contribution to the bolometric luminosity additional to accretion could be due to external heating. Actually, these sources are located in a crowded region (see Fig. \ref{fig:filament}), where contamination from nearby sources is not unlikely. Additionally, we note that HOPS 64, HOPS 108, and HOPS 369 fall very close to the non-thermal (likely synchrotron) emitting lobes of the powerful jet driven by HOPS 370 (\citealt{Osorio2017}) where a significant heating due to the these strong shocks is expected \citep{Shimajiri2008,Manoj2013,Gonzalez2016,Osorio2017}.

In the case of HOPS 66, the low-resolution far-IR emission is associated with two radio sources (sources 24 and 25; see Fig. \ref{fig:24&25}) but we do not really know which part of the estimated $L_{\rm bol}$ corresponds to each one. However, we note that the bolometric luminosity required to account for the observed 5 cm radio luminosity of Source 24 is of the order of only $\sim$ 0.3 $L_{\odot}$, which is a very small fraction of the overall bolometric luminosity of HOPS 66 (18 $L_{\odot}$). Thus, a distribution of luminosities where Source 24 has a bolometric luminosity of the order of $\sim$ 0.3 $L_{\odot}$ and Source 25 a bolometric luminosity of the order of $\sim$ 18 $L_{\odot}$ seems a likely possibility, in which both radio sources would follow the Anglada correlation with $L_{\rm bol}$ (see Section \ref{24&25}).

\begin{figure}
	{\includegraphics[width=1.0\hsize]{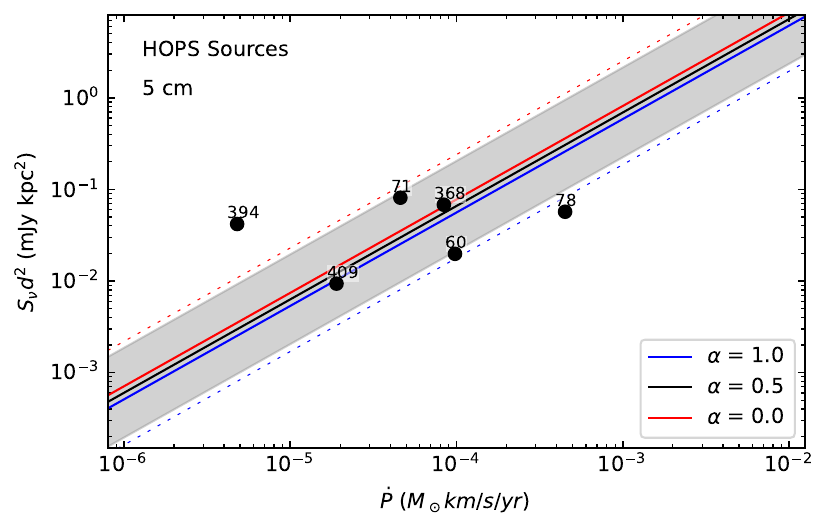}}
	\caption{Radio luminosity, $S_{\nu}d^2$, at 5 cm (Table \ref{tab:HOPS}) versus the outflow momentum rate, $\dot{P}$ (from \citealt{Takahashi2008,Tobin2016outf}; see Table \ref{tab:HOPS}) for the six HOPS YSOs in our field of view with known molecular outflow parameters. Data points are labelled with their HOPS ID number. 
The black solid line represents the extrapolation to 5 cm, assuming a typical average spectral index $\alpha = 0.5$, of the empirical correlation obtained at 3.6 cm (equation~\ref{eq:Ppto}) by \citet{Anglada2018} from a fit to a large sample of sources. The shadowed gray area indicates the residual standard deviation of this fit. To illustrate the dependence of the extrapolation to other wavelengths on the assumed spectral index, we also plot the results assuming a spectral index $\alpha = 0.0$ (red solid line) and $\alpha = 1.0$ (blue solid line). The dotted lines indicate the variation of the limits of the shadowed area when the red and blue line correlations are considered.
		\label{fig:Ppto}}
\end{figure}

For six of the 14 HOPS sources (HOPS 60, 71, 78, 368, 394 and 409), we also have an estimate of the molecular outflow force (or outflow momentum rate), $\dot P$, obtained from the literature (\citealt{Takahashi2008,Tobin2016outf}; see Table \ref{tab:HOPS}). All these sources have values of the radio luminosity and outflow momentum rate consistent, within the uncertainties, with the expected correlation described by equation (\ref{eq:Ppto}) (see Fig. \ref{fig:Ppto}), in agreement with the suggestion that the ionization necessary to account for the observed radio emission is produced by shocks associated with the outflow. Note that the physical parameters of molecular outflows are more scarce than the bolometric luminosities of the YSOs, since they are more difficult to obtain, and here the number of data points is insufficient to trace the trend of the relationship between radio emission and outflow force, but just the coincidence in the expected values.

\subsection{Comparison with radio studies of other star-forming regions}

In order to place our results in a broader context, we have identified in the literature similar radio studies in other star-forming regions. In the following, we therefore compare our results on OMC-2/3 (\citealt{Osorio2017}; this paper) with those of the VLA studies performed toward the star-forming regions of Perseus \citep{Pech2016}, Serpens \citep{Ortiz2015}, Ophiuchus \citep{Dzib2013}, Taurus-Auriga \citep{Dzib2015}, and IRDC G14.2-N and G14.2-S \citep{Diaz-Marquez2024}. 

\subsubsection{Evolutionary-class composition of the radio samples}

From Table~\ref{tab:fluxes}, we obtained the spectral index for those sources where the evolutionary stage was available (Table~\ref{tab:sources}). We grouped in a single class (hereafter referred as Class 0/I) the sources in the protostellar stages (Class 0, I, Flat spectrum, and PBR), as defined, e.g., in \citet{Furlan2016}. We note, however, that a number of sources are listed in Table \ref{tab:sources} with an uncertain classification, falling between two classes. For these sources, we have searched for complementary information in the literature to refine their classification; as a result, we favoured a classification of sources 7, 12, 21, 28, and 37 as Class II, and of Source 45 as Class III. Given the small number of Class II sources in our sample, this uncertainty regarding the classification of most of them could potentially disguise some evolutionary trends in our sample. For the radio sources obtained from the literature we equate the standard classification of YSOs in the class categorisation based on the IR spectral index (e.g., \citealt{Lada1987,Andre1993,Furlan2016}), and the stage categorisation proposed by \citep{Robitaille2006} from SED modelling. 

\begin{figure}
\centering
\includegraphics[width=0.48\textwidth]
{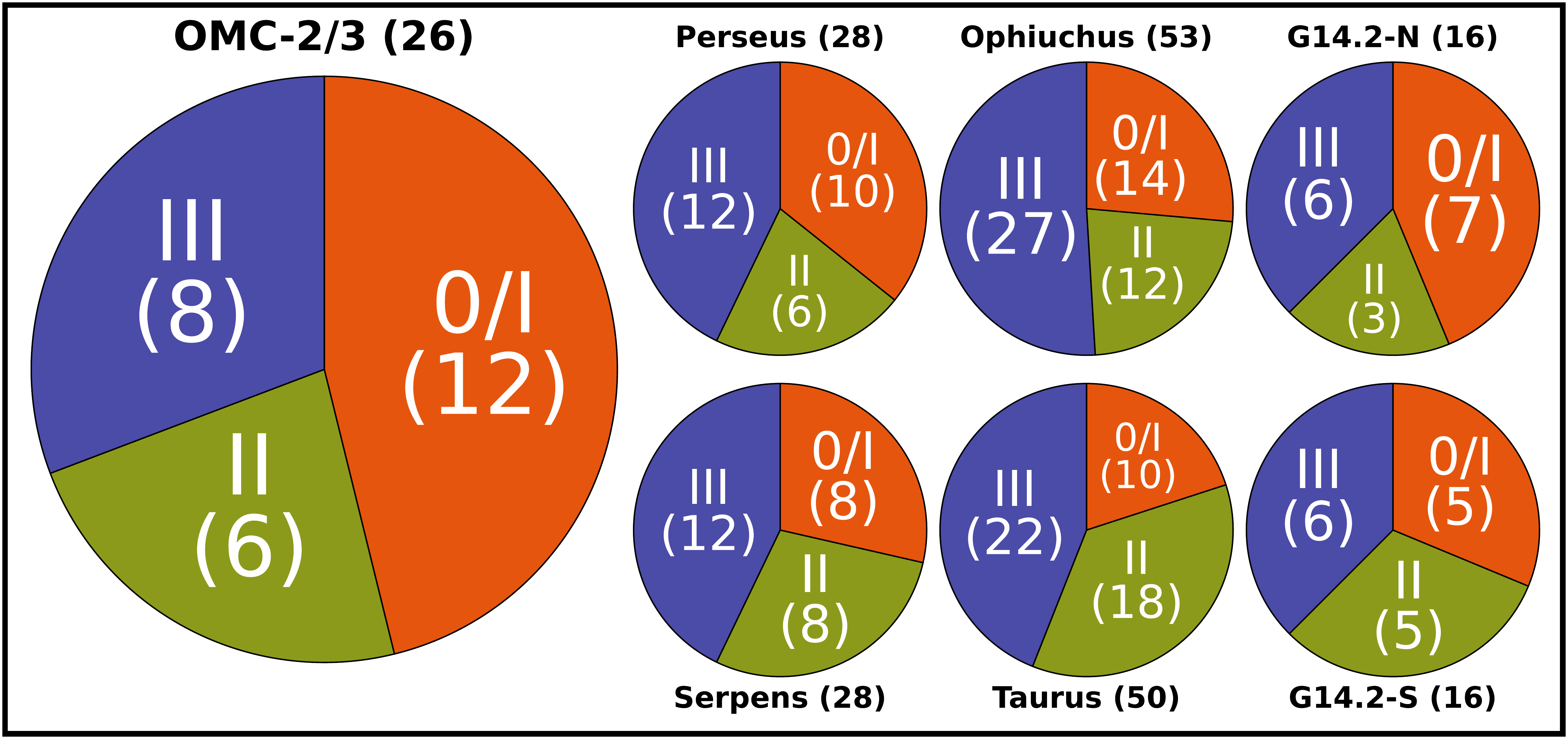}
\caption{Relative distributions of Class 0/I, II, and III sources in OMC-2/3 (\citealt{Osorio2017} and this paper; see Table~\ref{tab:sources}) and in six star-forming regions with VLA surveys reported in the literature (Perseus: \citealt{Pech2016}, Serpens: \citealt{Ortiz2015}, Ophiuchus: \citealt{Dzib2013}, Taurus-Auriga: \citealt{Dzib2015}, and G14.2-N and G14.2-S: \citealt{Diaz-Marquez2024}). The total number of sources in each region is indicated in parentheses, while the numbers within the sectors correspond to the individual evolutionary classes. A chi-squared test of homogeneity does not reveal statistically significant differences in the class distributions among the regions.
}
\label{fig:pies}
\end{figure}

Figure~\ref{fig:pies} shows a pie chart diagram of the distribution of the radio sources detected toward OMC-2/3 with known spectral classification (Tables~\ref{tab:fluxes} and  \ref{tab:sources}) and comparison with the results of radio surveys from the literature. Some differences are apparent, with OMC-2/3 and G14.2-N showing relatively large proportions of Class 0/I sources and Ophiuchus exhibiting a larger proportion of Class III sources. To analyse whether these differences are significant, we applied a Pearson's chi-squared test of homogeneity, implemented using the \texttt{scipy.stats.chi2\_contingency} function in Python, to compare the evolutionary-class distributions among the seven regions. The test yielded $\chi^2=10.24$, with (7$-$1)$\times$(3$-$1)=12 degrees of freedom and a $p$-value of 0.595. Since some expected source counts are below five, we also assessed the statistical significance through $10^5$ Monte Carlo simulations. In each simulation, sources were randomly redistributed among the evolutionary classes and regions, while preserving both the total number of sources in each region and the overall number of sources in each evolutionary class. The Monte Carlo $p$-value, $p_{\rm MC}$, was calculated as the fraction of simulated distributions yielding a chi-squared statistic equal to or greater than the observed value of 10.24. We obtained $p_{\rm MC}\simeq0.607$, consistent with the $p$-value of 0.595 derived from the theoretical chi-squared distribution. Both methods indicate that the observed differences in evolutionary-class composition among the seven regions are not statistically significant.

Thus, the evolutionary-class compositions of the seven radio samples are statistically compatible with a common distribution. However, this result should not be interpreted as evidence that the underlying YSO populations have identical evolutionary compositions, since the samples are subject to rather different observational sensitivities, spatial coverage, and selection criteria. Nevertheless, the absence of a statistically significant difference in the overall class composition provides useful context for interpreting the spectral-index comparisons presented below.

\subsubsection{Radio spectral indices and comparison of OMC-2/3 with other star-forming regions}
\label{sec:spectral_indices_comparison}

The radio spectral index provides useful information on the emission mechanisms operating in YSOs. Positive or approximately flat spectral indices are commonly associated with thermal free-free emission from ionised gas, whereas negative indices may indicate a contribution from non-thermal processes. Although the spectral index alone does not uniquely determine the emission mechanism (see Section \ref{subsec:alpha}), as a simplified analysis tool, we will henceforth adopt a spectral index $\alpha < -0.1$ as an indicator of non-thermal emission, whilst spectral indices $\alpha \ge -0.1$ will be considered indicative of thermal emission.

\begin{figure*}
\centering
\includegraphics[width=0.98\textwidth]
{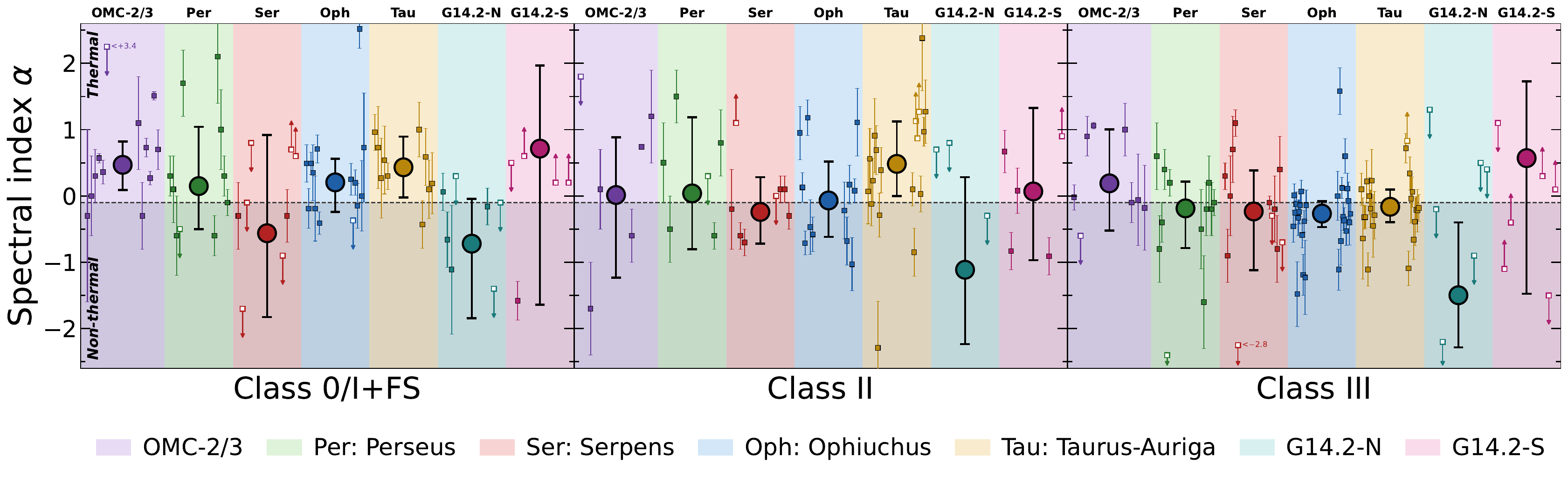}
\caption{Spectral indices (from Table~\ref{tab:fluxes}) and their median as a function of the YSO evolutionary stage (from Table~\ref{tab:sources}) for the radio sources in OMC-2, and comparison with the results obtained from radio surveys toward Perseus \citep{Pech2016}, Serpens \citep{Ortiz2015}, Ophiuchus \citep{Dzib2013}, Taurus-Auriga \citep{Dzib2015}, and G14.2-N and G14.2-S \citep{Diaz-Marquez2024}  
star-forming regions. Error bars in the median correspond to 95\% percentile intervals (see Table~\ref{tab:spectral_medians_fractions_boot} and Appendix~\ref{app:spectral_statistics}). The black dashed line at $\alpha = -0.1$ draws the boundary between mostly thermal emission and non-thermal emission (see Section~\ref{subsec:alpha}).
}
\label{fig:indices}
\end{figure*}

To investigate whether the radio spectral-index distributions of YSOs in OMC-2/3 resemble those observed in other star-forming regions, we compared the radio spectral-index distributions of the YSOs detected in OMC-2/3 with those reported in VLA surveys of Perseus, Serpens, Ophiuchus, Taurus, and G14.2 \citep{Pech2016, Ortiz2015, Dzib2013, Dzib2015, Diaz-Marquez2024}. The comparisons were performed separately for Class 0/I, Class II, and Class III sources. 
In Fig.~\ref{fig:indices}, we show the measured spectral indices, by class and region, for the entire dataset considered in our study. Table~\ref{tab:spectral_medians_fractions_boot} presents an overview of the statistics, together with the median values of the spectral indices and the fractions of sources considered as likely non-thermal ($\alpha < -0.1$).

\begin{table}
\centering
\caption{Median radio spectral indices and fractions of sources with $\alpha$$<$$-$$0.1$ by region and evolutionary class. 
}
\label{tab:spectral_medians_fractions_boot}
\setlength{\tabcolsep}{4.9pt}
\begin{tabular}{llrrrrcc}
\toprule
Region & Class & $N$ & Det. & U.l. & L.l. & $\alpha_{\rm med}$ & $f(\alpha$$<$$-0.1)$ \\
\midrule
OMC-2/3 & 0/I & 12 & 11 & 1 & 0 & $+0.47^{+0.35}_{-0.38}$ & $+0.20^{+0.22}_{-0.20}$ \\
OMC-2/3 & II & 6 & 5 & 1 & 0 & $+0.01^{+0.87}_{-1.25}$ & $+0.47^{+0.36}_{-0.31}$ \\
OMC-2/3 & III & 8 & 7 & 1 & 0 & $+0.19^{+0.81}_{-0.71}$ & $+0.36^{+0.39}_{-0.36}$ \\
\addlinespace
Perseus & 0/I & 10 & 9 & 1 & 0 & $+0.15^{+0.89}_{-0.65}$ & $+0.38^{+0.32}_{-0.28}$ \\
Perseus & II & 6 & 5 & 1 & 0 & $+0.04^{+1.15}_{-0.84}$ & $+0.47^{+0.36}_{-0.31}$ \\
Perseus & III & 12 & 11 & 1 & 0 & $-0.19^{+0.40}_{-0.60}$ & $+0.55^{+0.28}_{-0.30}$ \\
\addlinespace
Serpens & 0/I & 8 & 2 & 4 & 2 & $-0.56^{+1.48}_{-1.26}$ & $+0.63^{+0.37}_{-0.38}$ \\
Serpens & II & 8 & 6 & 1 & 1 & $-0.24^{+0.52}_{-0.48}$ & $+0.59^{+0.29}_{-0.34}$ \\
Serpens & III & 12 & 9 & 3 & 0 & $-0.23^{+0.62}_{-0.89}$ & $+0.55^{+0.28}_{-0.30}$ \\
\addlinespace
Ophiuchus & 0/I & 14 & 13 & 1 & 0 & $+0.21^{+0.35}_{-0.44}$ & $+0.33^{+0.24}_{-0.26}$ \\
Ophiuchus & II & 12 & 12 & 0 & 0 & $-0.07^{+0.59}_{-0.55}$ & $+0.48^{+0.27}_{-0.23}$ \\
Ophiuchus & III & 27 & 27 & 0 & 0 & $-0.27^{+0.19}_{-0.20}$ & $+0.66^{+0.15}_{-0.18}$ \\
\addlinespace
Taurus & 0/I & 10 & 10 & 0 & 0 & $+0.43^{+0.46}_{-0.45}$ & $+0.19^{+0.31}_{-0.19}$ \\
Taurus & II & 18 & 15 & 0 & 3 & $+0.48^{+0.64}_{-0.49}$ & $+0.25^{+0.19}_{-0.19}$ \\
Taurus & III & 22 & 21 & 0 & 1 & $-0.16^{+0.26}_{-0.23}$ & $+0.55^{+0.22}_{-0.19}$ \\
\addlinespace
G14.2-N & 0/I & 7 & 4 & 3 & 0 & $-0.72^{+0.68}_{-1.13}$ & $+0.78^{+0.22}_{-0.35}$ \\
G14.2-N & II & 3 & 0 & 3 & 0 & $-1.11^{+1.40}_{-1.12}$ & $+0.82^{+0.18}_{-0.49}$ \\
G14.2-N & III & 6 & 0 & 6 & 0 & $-1.50^{+1.10}_{-0.79}$ & $+0.88^{+0.12}_{-0.38}$ \\
\addlinespace
G14.2-S & 0/I & 5 & 1 & 1 & 3 & $+0.71^{+1.25}_{-2.35}$ & $+0.36^{+0.44}_{-0.36}$ \\
G14.2-S & II & 5 & 4 & 0 & 1 & $+0.07^{+1.26}_{-1.04}$ & $+0.46^{+0.34}_{-0.46}$ \\
G14.2-S & III & 6 & 0 & 2 & 4 & $+0.57^{+1.16}_{-2.05}$ & $+0.34^{+0.33}_{-0.34}$ \\
\bottomrule
\end{tabular}
\par\smallskip
\begin{minipage}{0.49\textwidth}
\footnotesize
\textit{Notes.} 
Det., U.L., and L.L. denote the numbers of spectral-index detections, upper limits, and lower limits, respectively. Data from \citet{Osorio2017,Pech2016, Ortiz2015, Dzib2013, Dzib2015, Diaz-Marquez2024}; and Tables~\ref{tab:fluxes} \ref{tab:sources} of this paper. The quoted uncertainties represent model-dependent 95\% Monte Carlo–bootstrap percentile intervals (see Appendix~\ref{app:spectral_statistics} for details). The criterion $\alpha$$<$$-0.1$ identifies sources with spectral indices consistent with non-thermal emission.
Estimates for G14.2-N Classes II and III are particularly model-dependent, as all their spectral indices are upper limits. 
\end{minipage}
\end{table}

The data includes both measured spectral indices and upper or lower limits. To account for measurement uncertainties and upper and lower limits,  we generated 10\,000 Monte Carlo realisations, adopting a physically motivated interval $-2.5\leq\alpha\leq2.5$ for the underlying spectral indices (see Appendix~\ref{app:spectral_statistics}). We then calculated the two-sample Kolmogorov--Smirnov (KS) statistic, $D$, defined for each realisation. 

The median value of $D$ was used as a descriptive measure of the separation between the distributions. Table~\ref{tab:ks_omc_regions} shows the KS distances between OMC-2/3 and other regions for each evolutionary class. The largest KS distances are found for G14.2-N, with median values of $D_{\rm KS}=0.750$, 0.667, and 0.708 for Classes 0/I, II, and III, respectively. These values suggest substantial differences between the spectral-index distributions of the two regions. However, the small sample sizes and the high fraction of censored measurements in G14.2-N, particularly for Classes II and III, which contain only upper limits, make these results strongly dependent on the adopted treatment of censored data. Therefore, these large KS distances should be interpreted with caution and cannot, by themselves, establish statistically significant differences between the underlying populations. 

Taurus shows the smallest median KS separation from OMC-2/3 for Class 0/I sources, with $D_{\rm KS}=0.267$ (see Table~\ref{tab:ks_omc_regions}). Their median spectral indices are also similar, with $\alpha_{\rm med}=0.47$ for OMC-2/3 and $\alpha_{\rm med}=0.43$ for Taurus (Table~\ref{tab:spectral_medians_fractions_boot}). This similarity is consistent with broadly comparable radio spectral-index distributions among the youngest objects in these two samples. In contrast, the Class 0/I distributions of OMC-2/3 and Serpens show a larger separation, with a median $D_{\rm KS}=0.542$ (Table~\ref{tab:ks_omc_regions}). Their estimated median spectral indices are $0.47$ and $-0.56$, respectively (Table~\ref{tab:spectral_medians_fractions_boot}). We note, however, that the Serpens sample contains a substantial fraction of censored measurements. 

For Class III objects, the comparison between OMC-2/3 and Ophiuchus gives a median KS statistic of $D_{\rm KS}=0.463$, larger than the corresponding values for Perseus, Serpens, and Taurus, which are approximately $D_{\rm KS}=0.375$ (see Table~\ref{tab:ks_omc_regions}). The estimated median spectral indices are $0.19$ for OMC-2/3 and $-0.27$ for Ophiuchus (Table~\ref{tab:spectral_medians_fractions_boot}). This difference suggests a possible variation in the spectral-index distributions of the radio-detected Class III populations.

\begin{figure}
\centering
\includegraphics[width=0.48\textwidth]{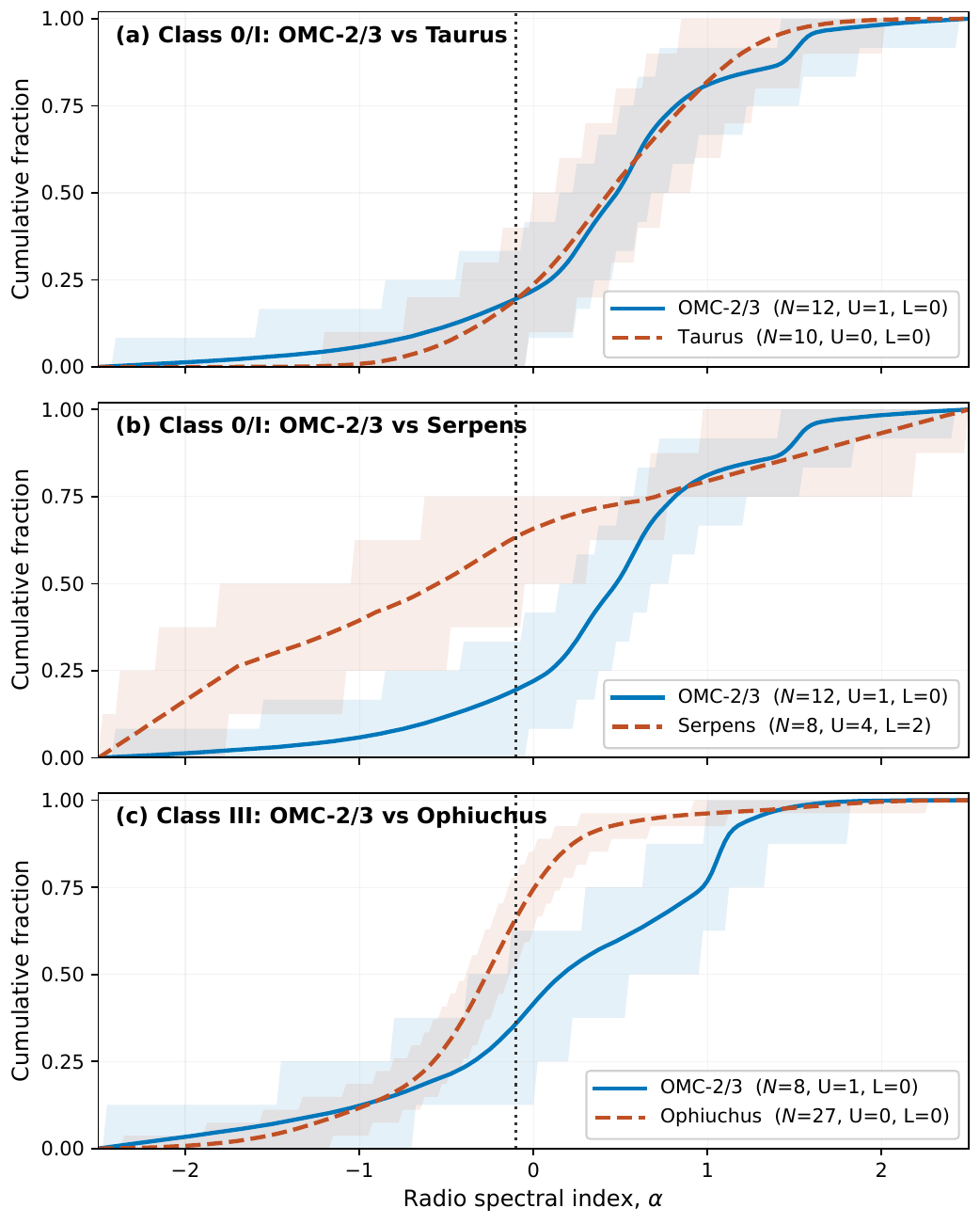}
\caption{Cumulative distribution functions (CDFs) of the radio spectral indices for selected comparisons between OMC-2/3 and other star-forming regions: (a) Class 0/I sources in OMC-2/3 and Taurus; (b) Class 0/I sources in OMC-2/3 and Serpens; and (c) Class III sources in OMC-2/3 and Ophiuchus. The curves represent the mean CDFs obtained from 10,000 Monte Carlo realizations, accounting for measurement uncertainties and censored data. The shaded regions indicate the 2.5th–97.5th percentile ranges of the simulated CDFs, calculated with the source catalogues held fixed (see Appendix~\ref{app:spectral_statistics}~for details). The vertical dotted line marks $\alpha=-0.1$, below which spectral indices are consistent with non-thermal emission. The number of detections (N), upper limits (U) and lower limits (L) are indicated in the insets. 
}
\label{fig:omc_cdf_comparison}
\end{figure}

Figure~\ref{fig:omc_cdf_comparison} presents the cumulative distribution functions (CDFs) for three representative comparisons of OMC-2/3 Class 0/I sources versus Taurus (Figure~\ref{fig:omc_cdf_comparison}a) and Serpens (Figure~\ref{fig:omc_cdf_comparison}b), and for OMC-2/3 Class III sources versus Ophiuchus (Figure~\ref{fig:omc_cdf_comparison}c), illustrating the same trends.

The main trends are relatively stable under alternative assumptions for the censored measurements, although bootstrap resampling reveals substantial uncertainty associated with the limited sample sizes (Appendix~\ref{app:spectral_statistics}). However, because the surveys differ in sensitivity, frequency coverage, and source-selection criteria, these comparisons cannot be interpreted as direct tests of differences between the complete YSO populations. Moreover, the KS statistics are used here descriptively, without assigning formal significance probabilities to the censored samples.

\subsubsection{Evolutionary-class dependence of radio spectral indices}

The dependence of radio spectral-index distributions on evolutionary class may provide insights into changes in the dominant radio emission mechanisms during YSO evolution. 

In Taurus, the median spectral indices, $\alpha_{\rm med}$, of Class 0/I and Class II sources are similar ($0.43$ and $0.48$, respectively), whereas Class III sources have a more negative median ($-0.16$) (see Table~\ref{tab:spectral_medians_fractions_boot}). The corresponding KS separations are $D_{\rm KS}=0.518$ for Class 0/I versus III, and $D_{\rm KS}=0.465$ for Class II versus III  (Table~\ref{tab:ks_omc_regions}). In Ophiuchus, the median spectral index decreases from $\alpha_{\rm med}$=$0.21$ for Class 0/I to $\alpha_{\rm med}$=$-0.07$ for Class II and $\alpha_{\rm med}$=$-0.27$ for Class III, with $D_{\rm KS}=0.460$ between Class 0/I and III (see Tables~\ref{tab:spectral_medians_fractions_boot} and \ref{tab:ks_omc_regions}). These trends are consistent with an increasing contribution from emission components producing negative radio spectral indices in the more evolved, radio-detected samples of Taurus and Ophiuchus. This is in agreement with the findings of \citep{Dzib2013,Dzib2015}, who suggested that the emission of the youngest protostars (Class 0/I and flat spectrum sources) is dominated by thermal free-free emission from dense ionised stellar winds and thermal jets, whilst in more evolved (Class II/III) YSOs the observed emission is generally non-thermal gyrosynchrotron related to magnetic activity near the stellar surface.
 
In OMC-2/3, however, the dependence on evolutionary class is less clearly defined. In OMC-2/3, the median spectral index is highest for Class 0/I sources ($\alpha_{\rm med}=0.47$), followed by Class III ($0.19$) and Class II ($0.01$) (see Table~\ref{tab:spectral_medians_fractions_boot}). Thus, the radio sources in OMC-2/3 do not exhibit a monotonic decrease in median spectral index with evolutionary class. 

This behaviour is further illustrated by the KDE and CDF distributions for OMC-2/3, Taurus, and Ophiuchus shown in Figure~\ref{fig:regions_kde_cdf}.

Some observational and sample-related limitations should also be considered for OMC-2/3. The bright H II region M43 (e.g., \citealt{Rashid2026}), located towards the southeastern corner of the OMC-2 field, produced artifacts in the radio images, requiring the exclusion of an approximately $2'\times4'$ area from the source search in the C-band maps. Although external photoionisation and photoevaporation may affect the circumstellar environments of young stars near an H II region, the exclusion of this area prevents us from assessing whether such effects influence their radio properties. In addition, the evolutionary classification of a few sources required complementary information to resolve initial ambiguities (see above). Given the small number of sources in some evolutionary classes, particularly Class II, the statistical results should be interpreted with appropriate caution.

To assess the statistical significance of the evolutionary trends in radio spectral indices, in Appendix~\ref{app:negative_fractions} we compared the change in the fraction of sources with $\alpha<-0.1$ from Class 0/I to Class III within OMC-2/3, Taurus, and Ophiuchus, as well as the differences in these trends among the three regions. We find tentative evidence for an increase in this fraction in Taurus and Ophiuchus, where the Monte Carlo–bootstrap 95\% percentile intervals marginally exclude zero. However, the corresponding increase in OMC-2/3 is not statistically significant, and neither are the differences in evolutionary trends among the regions.

In summary, the comparison of radio spectral-index distributions across seven star-forming regions suggests an increasing prevalence of negative spectral indices towards more evolved YSOs, particularly in Taurus and Ophiuchus. Although this trend is less evident in OMC-2/3, we find no statistically significant evidence that the evolutionary trends differ among regions. These results are consistent with evolutionary changes in the relative contributions of thermal and non-thermal radio emission mechanisms, but the limited sample sizes, censored measurements, and observational selection effects prevent firmer conclusions.

\subsection{Notes on the nature of the individual sources}\label{subsec:individual}

In this Section and Appendix~\ref{subsec:individual2} we discuss the nature of the 47 reported radio sources by analyzing their radio spectral index, morphology, emission at other wavelength ranges, position relative to the ISF, association with dust clumps, etc. For simplicity, in several cases, we discuss together sources that are close to each other, or those that have similar observational characteristics or evolutionary stages. We are able to determine quite confidently the nature of 36 of these sources (see Table \ref{tab:sources}).

We conclude that, out of the total of 57 total radio sources in the field of view, 41 (72\%) are very likely associated with the star-forming region, 6 are extragalactic background sources, and we are unsure about the association with the star-forming region of the remaining 10 sources. Of the 41 sources associated with the star-forming region, 36 are tracing the position of YSOs, 3 are tracing non-thermal emission from jet knots (discussed in \citealt{Osorio2017}), one is tracing the shock-ionized wall of an outflow cavity, and one could be either a YSO or a jet knot.

\subsubsection{Source 1 (VLA 5, HOPS 78) and Source 2 }\label{1and2}

Source 1 is located in the densest part of the ISF (Fig. \ref{fig:filament}), toward the 1.3 mm dust condensation MMS 9 \citep{Chini1997}. The source, also known as VLA 5, was previously detected at 3.6 cm by \citet{Reipurth1999,Reipurth2004}\footnote{The source is called VLA 15 in \cite{Reipurth2004}.} and at 5 cm by \citet{Kounkel2014}. This source has been proposed as the driving source  of a powerful bipolar CO outflow \citep{Williams2003,Takahashi2008} that extends almost 1 pc in the E-W direction. The blue lobe of the outflow, to the W of Source 1, is associated with a prominent chain of collimated H$_2$ knots \citep{Yu1997,Stanke2002}.

In our inspection of the NEWFIRM (T. Megeath, private communication), {\it Spitzer} \citep{Megeath2012} and {\it Herschel} \citep{Stutz2013} images from 2.2 to 70 $\micron$, we noticed that the peak of the IR emission shifts from a position compatible with Source 1, to positions up to 3\farcs8 farther to the W, as we go from long to short wavelengths. This could be interpreted as implying that the IR emission is not tracing Source 1, but rather a cavity opened by the blue lobe of the outflow driven by Source 1, seeing closer to the protostar at longer wavelengths. 

\begin{figure}
	\includegraphics[width=0.99\hsize]{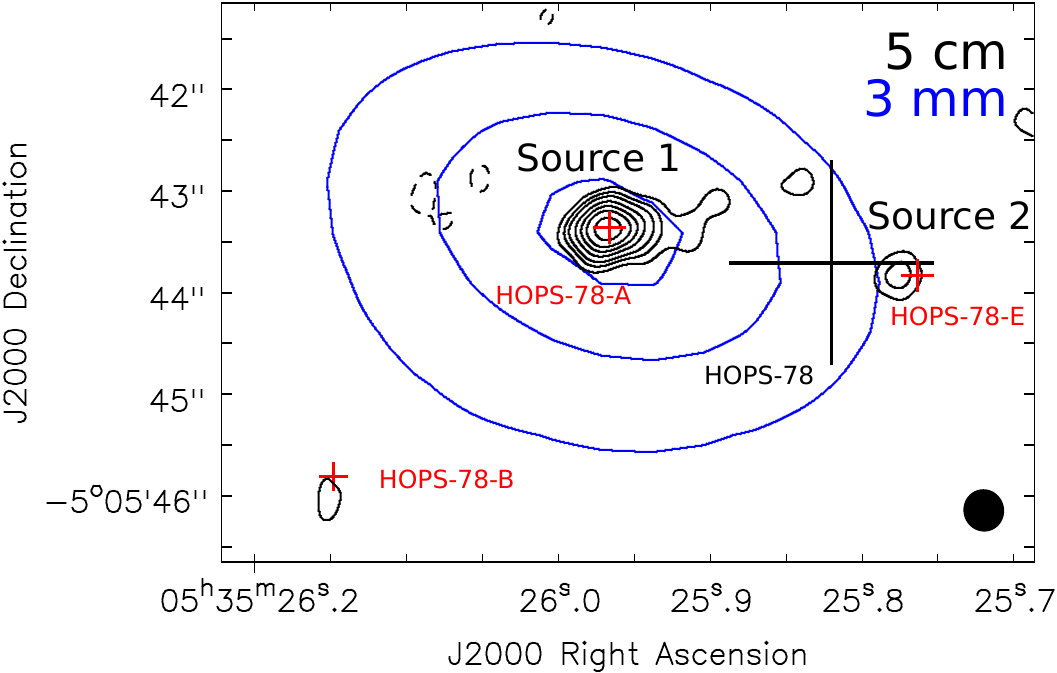}
	\caption{Sources 1 and 2. VLA 5 cm A-configuration map with natural weighting (black contours) overlaid on the ALMA 3 mm map (blue contours, adapted from \citealt{Kainulainen2017}). Plus signs mark the position of the HOPS source (black, \citealt{Furlan2016}, derived at 4.5 $\micron$) and the 9 mm/870 $\micron$ sources (red, \citealt{Tobin2020I}) in the region. Black contour levels are $-$3, 3, 5, 7, 10, 13, 16, and 20 times 10 $\mu$Jy beam$^{-1}$, the rms of the 5 cm map. Blue contour levels are 0.3, 0.6, and 0.9 times 19.6 mJy beam$^{-1}$, the peak of the 3 mm core. The 5 cm VLA synthesized beam is shown in the bottom right corner of the image. The 3 mm ALMA synthesized beam (HPBW) is $3\farcs8\times2\farcs3$, P.A. = 71$\degr$.\label{fig:1&2}}
\end{figure}

Source 1 seems to be very young since it is located toward the peak of a 3 mm compact dust core (\citealt{Kainulainen2017}, Fig. \ref{fig:1&2}), indicating that it is deeply embedded. We associate Source 1 with the far-IR source HOPS 78 (see Table \ref{tab:sources} for the counterparts), which is tracing a Class 0 YSO \citep{Furlan2016}. The coordinates of HOPS 78 were derived from the {\it Spitzer} 4.5 $\micron$ image, and therefore are shifted to the W with respect to our Source 1, which better traces the YSO.

In our 5 cm high-resolution map, Source 1 appears elongated at P.A. = $99\pm5\degr$ (Fig. \ref{fig:1&2}), consistent with the direction of the E-W outflow. This source follows the Anglada correlations with $L_{\rm bol}$ and \.{P} (see Section \ref{subsec:nature}), suggesting it is a radio jet. Additionally, \citet{Tobin2020I} detected with ALMA at 870 $\micron$ with $\sim0\farcs11$ resolution, a structure nearly perpendicular to the radio source that appears to be a circumstellar dust disc (see Fig. \ref{fig:sources}). They also detected the source with the VLA at 9 mm with a flux density of $0.83\pm0.02$ mJy. Although we cannot derive a spectral index only from our observations, using this value together with the flux densities shown in Fig. \ref{fig:SEDs} we find a positive spectral index $\alpha = 0.4\pm0.3$ for this source, consistent with thermal free-free emission. Therefore, we confirm this source as the driving source of the molecular outflow.

Source 2 is located $2\farcs8$ ($\sim1000$ au) in projection to the W of Source 1 (Fig. \ref{fig:1&2}). It has also been detected with the VLA at 9 mm with a flux density of $0.046\pm0.017$ mJy \citep{Tobin2020I}, which, together with our 5 cm flux density (Table \ref{tab:fluxes}), gives a flat spectral index $\alpha = -0.2\pm0.3$, whose uncertainty does not allow us to determine the nature of the emission. The source is located toward the border of the 3 mm compact dust core (\citealt{Kainulainen2017}, Fig. \ref{fig:1&2}), suggesting it is more evolved than Source 1, under the assumption that it is an independent object.

Another possibility is that Source 2 were a knot ejected from the Source 1 outflow. This could be hinted at from the VLA maps at 9 mm since the position of the source slightly moves to the W in those observations, taken $\sim$ 1 year after ours. Additionally, in the previous 3.6 cm high-resolution observations of \citet{Reipurth2004} taken in 2002 with an rms of 10 $\mu$Jy beam$^{-1}$ in the vicinity of these sources, Source 1 is resolved with a prominence to the W side while Source 2 is not detected (unpublished map). A reliable measure of the proper motions of Source 2 should be done to clarify its nature.

Interestingly, Source 1 could be member of a quadruple system (or quintuple, if Source 2 were indeed a YSO), since \citet{Tobin2020I} detected three additional objects at less than $5\arcsec$ ($\sim1900$ au) from Source 1, through VLA observations at 9 mm and ALMA observations at 870 $\micron$.

\subsubsection{Source 4 (VLA 7)}\label{4}

\begin{figure}
	\includegraphics[width=0.99\hsize]{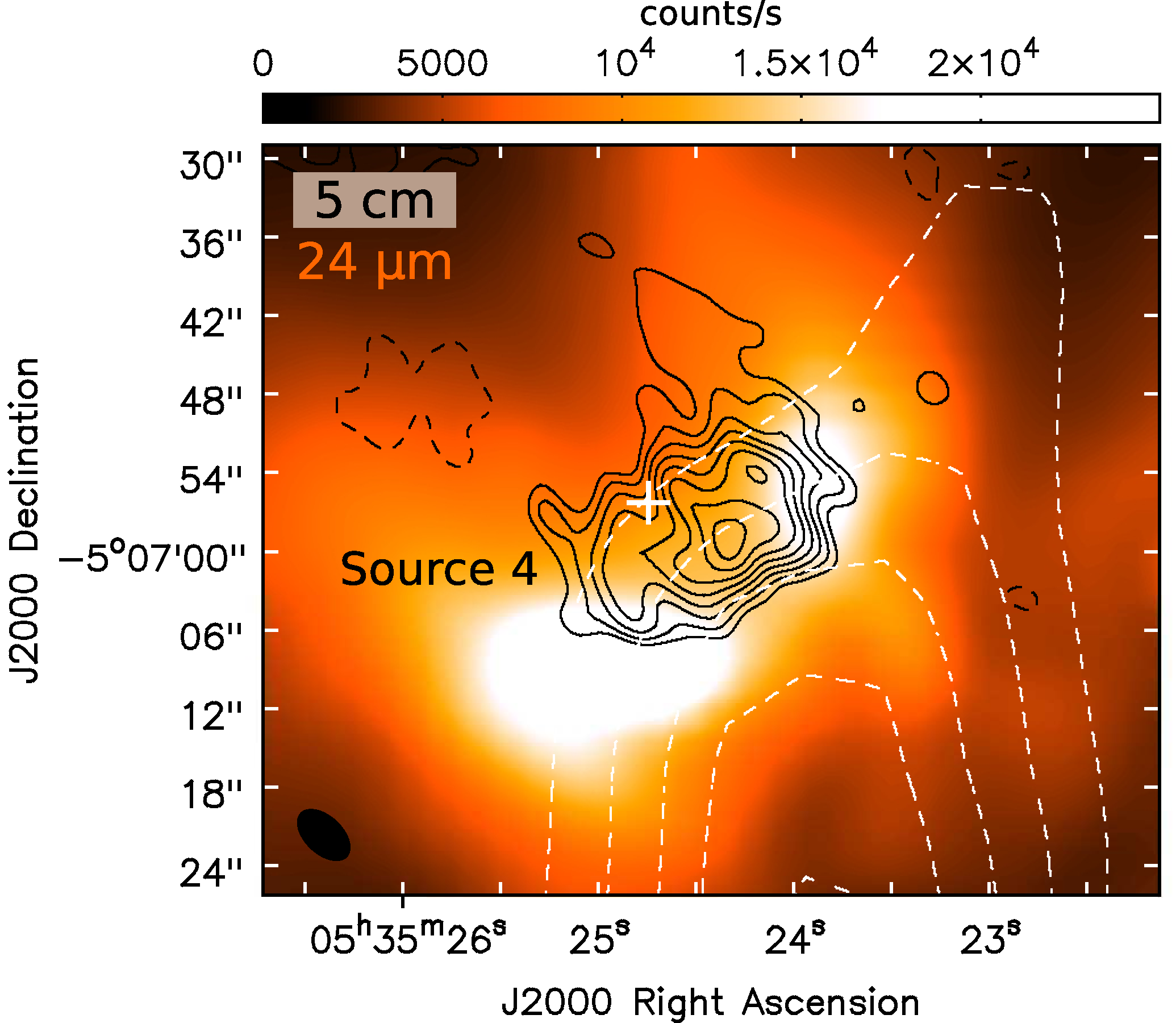}
	\caption{Source 4. VLA 5 cm C-configuration map with robust 0 weighting and uvtaper = 50 k$\lambda$ (black contours) and column density of the ISF (dashed white contours, adapted from \citealt{Stutz2018}) overlaid on the Spitzer 24 $\micron$ image (color scale, adapted from \citealt{Megeath2012}). The plus sign marks the position of 2MASS J05352474-0506563. Black contour levels are $-$3,$-$2, 2, 3, 4, 5, 6, 7, 8, and 9 times 21 $\mu$Jy beam$^{-1}$, the rms of the 5 cm map. Dashed white contour levels are 0.5, 0.6, 0.7, 0.8, and 0.9 times 1.77$\times10^{23}$ cm$^{-2}$, the local maximum of the column density. The VLA synthesized beam is shown in the bottom left corner of the image.
	}
	\label{fig:4}
\end{figure}

This extended source, previously detected at 3.6 cm and named VLA 7 \citep{Reipurth1999}, is located toward the densest part of the ISF, at $\sim15\arcsec$ from its ridgeline (see Fig. \ref{fig:filament}), but is not associated with any 3 mm compact dust core \citep{Kainulainen2017}. The source shows a cometary morphology at 5 cm (Fig. \ref{fig:4}), with the SW part (located toward a higher column density, traced with white contours in Fig. \ref{fig:4}) being more intense than the NE one. This asymmetry is also observed at 3 cm, although the map has a lower quality because the source is far from the phase centre and the primary beam is smaller at this band. There is extended emission at 24 $\micron$ tracing heated dust, which shows a morphology similar to that of the radio emission but reaching larger scales (Fig. \ref{fig:4}).

We obtain a spectral index $\alpha = -0.3\pm0.3$ which, although uncertain, is consistent with an optically thin HII region. In the optically thin regime, the flux density at 5 cm could be produced by a B3 zero-age main-sequence star with a luminosity of $L_{\rm bol} \sim 1000$ $L_{\sun}$ \citep{Thompson1984}, which is in agreement with \citet{Reipurth1999}. These authors proposed IRAS 05329-0508 (V* V415 Ori) as the ionizing source, but this star is located $\sim 40 \arcsec$ from Source 4, well outside the HII region. We propose instead 2MASS J05352474-0506563, a near-IR point source better centred in the HII region, $\sim$ 6$\arcsec$ from the 5 cm emission peak (see Fig. \ref{fig:4}). According to the Spitzer and Newfirm photometry, this source does not present an IR excess and shows an extinction of A$_K = 10$ mag, with a de-reddened absolute magnitude of $-$4.1, consistent with an O star with $L_{\rm bol} \sim 10000$ $L_{\sun}$, higher than what was inferred from the radio continuum. A possible reason for this discrepancy could be that not all the photons are being used for the photoionization of the gas, e.g., due to the non-homogeneity of the medium, or that we are underestimating the radio continuum flux density since the source is very extended, and our observations probably are not recovering the whole emission of the source.

\subsubsection{Source 9 (VLA 8, HOPS 394)}\label{9}

\begin{figure}
	\center\includegraphics[width=0.98\hsize]{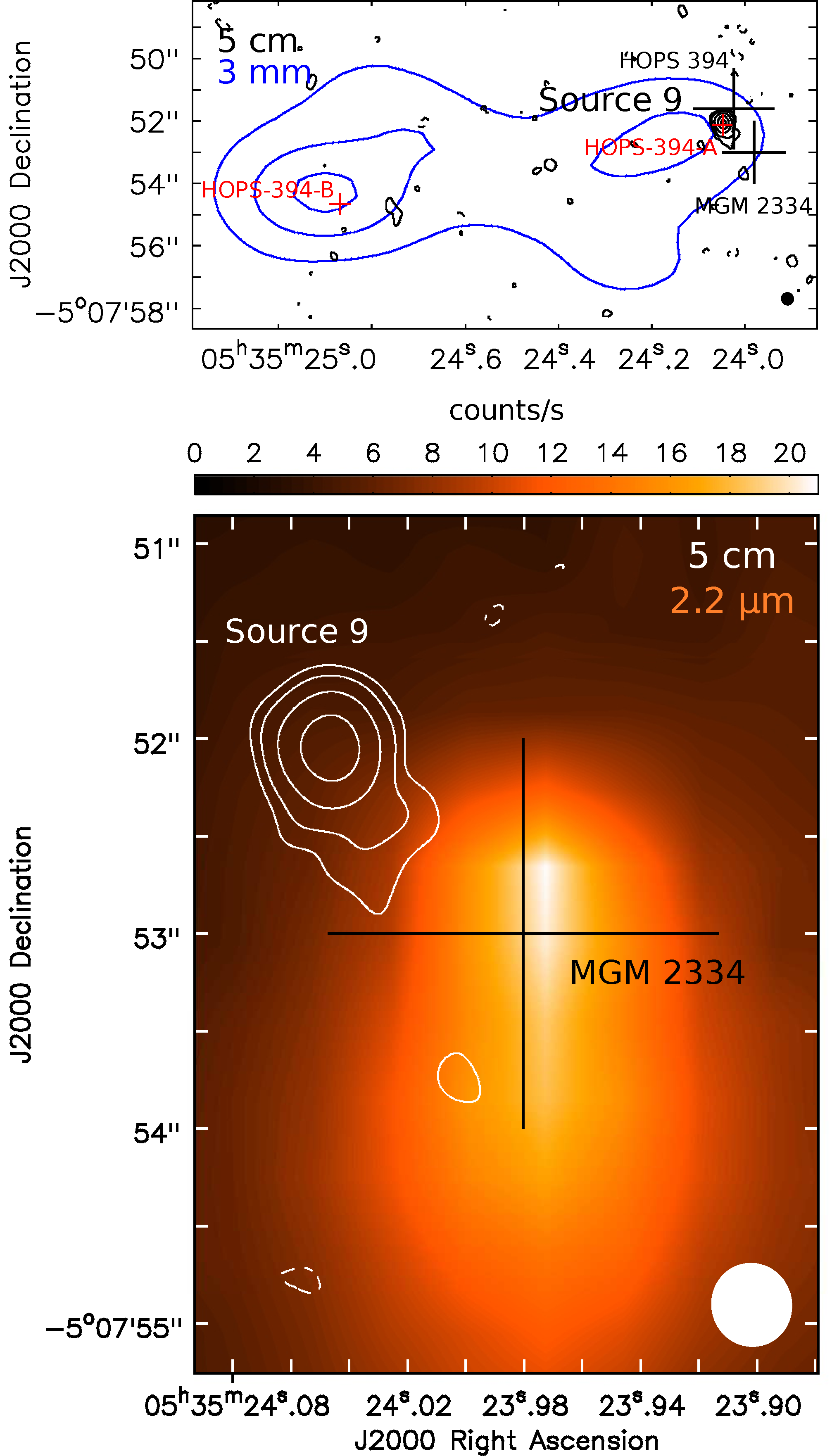}
	\caption{Source 9. {\it Top}: VLA 5 cm A-configuration map with natural weighting (black contours) overlaid on the ALMA 3 mm map (blue contours, adapted from \citealt{Kainulainen2017}). Black plus signs mark the positions of the far-IR source HOPS 394 (derived at 70 $\micron$ from \citealt{Stutz2013}, after applying an offset correction, as indicated in the text; the position uncertainty is twice the size of the sign) and the mid-IR source MGM 2334 (derived at 4.5 $\micron$; \citealt{Megeath2012}). The red plus signs mark 9 mm/870 $\micron$ sources HOPS-394-A and B (\citealt{Tobin2020I}). {\it Bottom}: VLA 5 cm A-configuration map with natural weighting (white contours) overlaid on the NEWFIRM 2.2 $\micron$ image (color scale, T. Megeath, private communication). Black plus sign marks the position of the MGM 2334 source \citep{Megeath2012}. The black (top) and white (bottom) contour levels  are $-$3, 3, 5, 10, 20, and 30 times 4.7 $\mu$Jy beam$^{-1}$, the rms of the 5 cm map. Blue contour (top) levels are 0.3, 0.6, and 0.9 times 12.4 mJy beam$^{-1}$, the peak of the 3 mm core. The 5 cm VLA synthesized beam is shown in the bottom right corner of the images. The 3 mm ALMA synthesized beam (HPBW) is $3\farcs8\times2\farcs3$, P.A. = 71$\degr$.}
	\label{fig:9}
\end{figure}

This source, previously detected at 3.6 cm and named VLA 8 \citep{Reipurth1999}, is located in the densest part of the ISF (Fig. \ref{fig:filament}), toward the 1.3 mm dust condensation named FIR 1a \citep{Chini1997}. In this region, the 3 mm ALMA map \citep{Kainulainen2017} revealed two compact dust cores with peaks separated about $\sim10\arcsec$ in projection, joined by a filamentary structure (see Fig. \ref{fig:9}, top). The eastern core is associated with a weak source detected at 9 mm and 870 $\micron$ \citep{Tobin2020I}, and the western core is associated with Source 9. Additionally, the mid-IR source MGM 2334, a YSO with a disc \citep{Megeath2012}, and the far-IR source HOPS 394, an extremely young Class 0 object (source PBR 019003\footnote{The map used to derive the coordinates of PBR 019003 reported in Table 4 of \citet{Stutz2013} has an offset of $\Delta\alpha = 1\farcs4$, $\Delta\delta = -1\farcs9$ with respect to the 24 $\micron$ sources (A. Stutz, private communication), which better trace the position of the YSOs. We use in this work the corrected position of HOPS 394.} in \citealt{Stutz2013, Furlan2014}), have been reported within 2$\arcsec$ of Source 9 (see Fig. \ref{fig:9}, top). Although previous works have considered MGM 2334 and HOPS 394 to be the same source, there are hints that suggest they are different objects (see below). On the other hand, there is a proposed bipolar CO outflow NE-SW (P.A. $\sim 50\degr$, or equivalently $\sim 230\degr$; \citealt{Tobin2016outf}) in the region that has been associated with HOPS 394.

In our 5 cm high-resolution map, Source 9 is angularly resolved (Fig. \ref{fig:9}, bottom), showing an extension to the SW at P.A. $\sim 35\degr$ (equivalent to $\sim 215\degr$), although the inner part of the source appears marginally elongated roughly in the N-S direction (PA = $16\degr\pm17\degr$, equivalent to $196\degr\pm17\degr$; Table \ref{tab:positions}); at 9 mm, the source is elongated at P.A. $= 27\degr\pm10\degr$ (equivalent to $= 207\degr\pm10\degr$) \citep{Tobin2020I}. These elongations are consistent (within the errors) with the direction of the CO outflow. We derive a flat spectral index $\alpha = 0.0\pm0.6$ (Table \ref{tab:fluxes} and Fig. \ref{fig:SEDs}) at cm wavelengths, consistent with thermal free-free emission despite the large uncertainty. Additionally, the source follows the Anglada correlation with $L_{\rm bol}$ and \.{P} (see Section \ref{subsec:nature}) suggesting it is a radio jet. Thus, we propose that this source is the driving source of the outflow reported in its vicinity. On the other hand, ALMA 870 $\micron$ observations detected extended emission at scales of $\sim$ 300 au associated with this source \citep{Tobin2020I}, which could be tracing a dust disc or its inner envelope (see Fig. \ref{fig:sources}).

The mid-IR source MGM 2334 is located 1$\farcs 4$ ($\sim540$ au) to the SW of Source 9 (Fig. \ref{fig:9}, bottom), and appears to be a different object given that the separation is somewhat larger than the IR positional uncertainty ($1\arcsec$ at 4.5 $\micron$). On the other hand, both NEWFIRM (1.7 and 2.2 $\micron$) and HST (0.8 - 2.5 $\micron$) images show a scattered light cone with the vertex at MGM 2334 and extending toward the S, indicating the presence of a cavity associated with this source (Fig. \ref{fig:9}, bottom). This is inconsistent with the direction of the outflow detected in the vicinity of the sources (P.A. $\sim50\degr$, \citealt{Tobin2016outf}), that we associate with Source 9.

The far-IR source HOPS 394 was detected in observations at 70 and 160 $\micron$ (\citealt{Stutz2013}) with an angular resolution of 5$\arcsec$ and $7\farcs8$, respectively. It is located $\sim0\farcs6$ in projection from Source 9 and could be associated with it, but also with MGM 2334 due to its positional uncertainty (Fig. \ref{fig:9}, top). \citet{Stutz2013} classified it as an extremely young object because of its strong far-IR emission and faint mid-IR emission, although they warned that it is located in a complex field where the photometry may be strongly contaminated. We favor the association of HOPS 394 with Source 9 since the radio source appears to be younger than MGM 2334. However, we note that extremely young objects (like HOPS 394) are deeply embedded and our accurate position of Source 9 reveals that the radio source does not appear toward the peak but toward the border of the 3 mm compact dust core \citep{Kainulainen2017}, $\sim2\arcsec$ ($\sim800$ au) from the peak (see Fig. \ref{fig:9}, top). Additionally, although there is ammonia emission (tracing dense gas) in the surroundings of Source 9, this source is not associated with any ammonia peak (D. Li, private communication).

In summary, we propose that Source 9 is a radio jet and suggest it as the driving source of the molecular outflow reported in its vicinity. We show that Source 9 is a different object than MGM 2334 and favor the radio source as the counterpart of HOPS 394. New observations of this complex region should be carried out to fully reveal its protostellar content.

\subsubsection{Sources 10 and 11 (VLA 9, HOPS 71)}\label{10and11}

\begin{figure}
	\includegraphics[height=1.16\hsize]{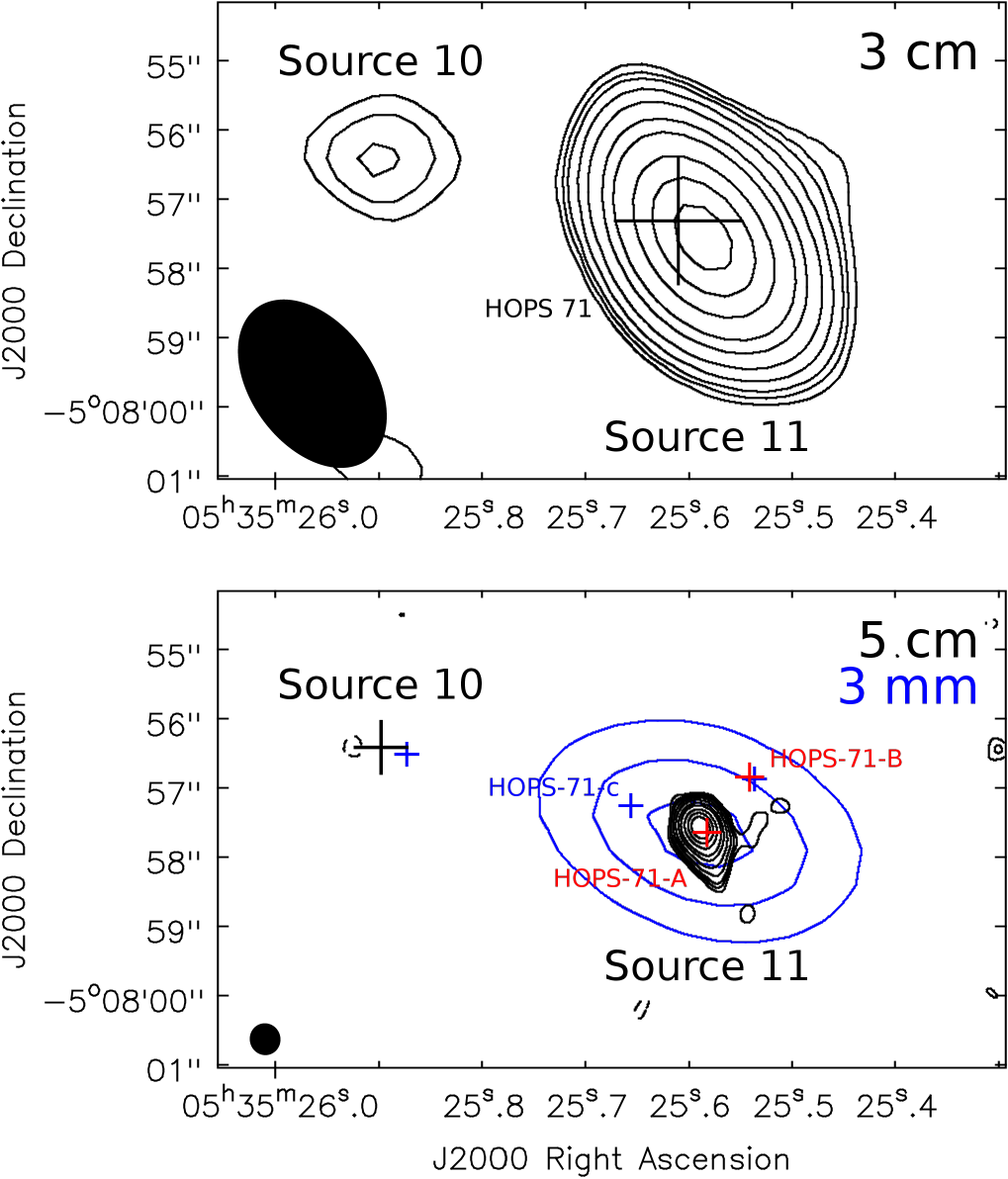}
	\caption{Sources 10 and 11. {\it Top}: VLA 3 cm C-configuration map with robust 0 weighting. Plus sign marks the position of the HOPS source (\citealt[derived at 4.5 $\micron$]{Furlan2016}). {\it Bottom}: VLA 5 cm A-configuration map with natural weighting (black contours) overlaid on the ALMA 3 mm map (blue contours, adapted from \citealt{Kainulainen2017}). Plus signs mark the position of the 5 cm Source 10 (black, this work), the 9 mm/870 $\micron$ (red, named HOPS-71-A and B; \citealt{Tobin2020I}), and the near-IR/optical objects in the region (blue, \citealt{Gaia2023}). Blue contour levels are 0.3, 0.6, and 0.9 times 8.6 mJy beam$^{-1}$, the peak of the 3 mm core. In both panels, black contour levels are $-$3, 3, 4, 5, 7, 10, 15, 20, 30, 40, and 50 times the rms of each map, which is 10 $\mu$Jy beam$^{-1}$ (top) and 4.5 $\mu$Jy beam$^{-1}$ (bottom). The VLA synthesized beams are shown in the bottom left corner of the maps.  The 3 mm ALMA synthesized beam (HPBW) is $3\farcs8\times2\farcs3$, P.A. = 71$\degr$.}
	\label{fig:10&11} 
\end{figure}

Sources 10 and 11, separated $4\farcs8$ ($\sim 1900$ au) in projection (Fig. \ref{fig:10&11}, top), are located toward the densest part of the ISF, $\sim 15\arcsec$ ($\sim 5800$ au) from its ridgeline (see Fig. \ref{fig:filament}). They were previously detected as a single source, VLA 9, in the 3.6 cm lower angular resolution observations of \citet{Reipurth1999}. VLA 9 was later associated by \citet{Takahashi2008} with a probable monopolar outflow at P.A. $\sim40\degr$.

Source 11 is located toward the peak of a 3 mm compact dust core \citep{Kainulainen2017}, suggesting it is deeply embedded (see Fig. \ref{fig:10&11}, bottom). It is also associated with the far-IR source HOPS 71 (Fig. \ref{fig:10&11}, top; see Table \ref{tab:sources} for the counterparts), classified as a Class I YSO \citep{Furlan2016}. In our 5 cm high-resolution map, the source is elongated at P.A. $= 24\pm5\degr$ (Fig. \ref{fig:10&11}, bottom), roughly consistent with the direction of the outflow (P.A. $\sim40\degr$) reported in its vicinity \citep{Takahashi2008}. The source has a positive flat spectral index, $\alpha=0.3\pm0.4$ (Table \ref{tab:fluxes} and Fig. \ref{fig:SEDs}), consistent with thermal emission, and follows the Anglada correlations with $L_{\rm bol}$ and \.{P} (see Section \ref{subsec:nature}), suggesting it is a radio jet. All these facts lead us to propose that Source 11 is a thermal radio jet, driving the large scale outflow. We note that the flux densities measured by us for Source 11 are $\sim$ 2 times higher than the flux density reported by \citet{Reipurth1999} at lower angular resolution (corresponding to the combination of sources 10 and 11), which could indicate some variability for Source 11 (Fig. \ref{fig:SEDs}).

\begin{figure}
	\includegraphics[width=0.48\textwidth]{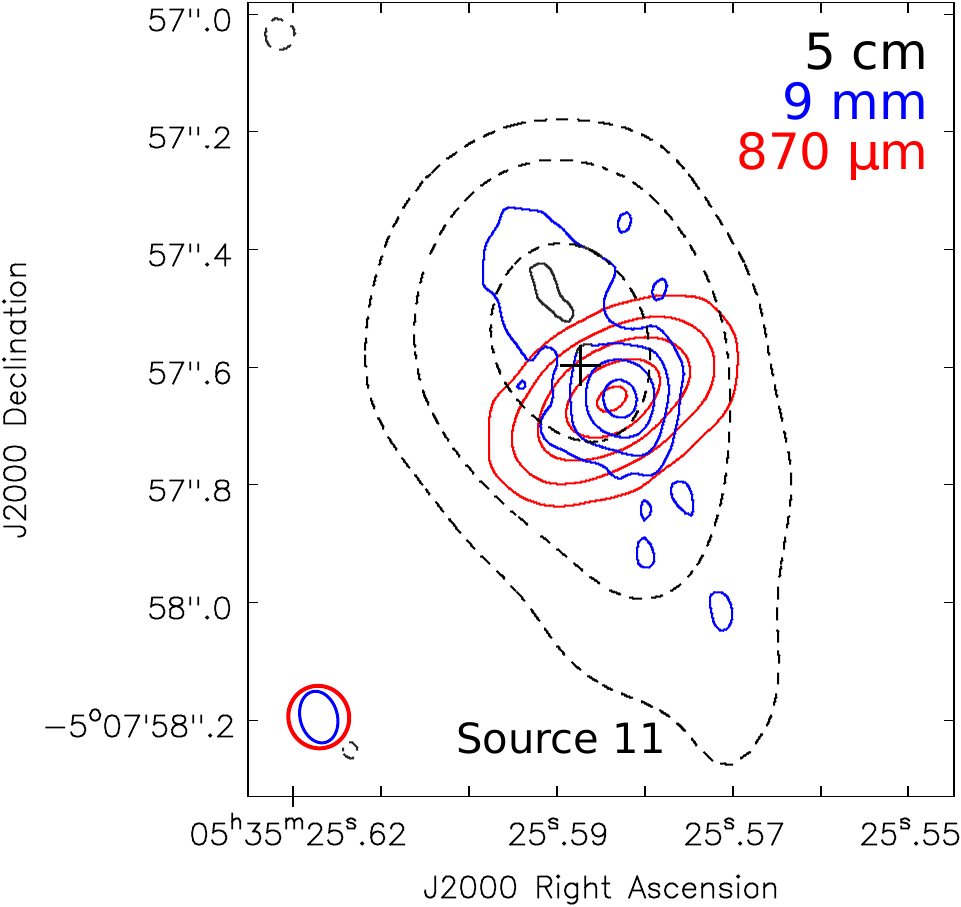}
	\caption{Source 11. Emission at 9 mm (blue contours) and 870 $\micron$ (red contours) adapted from \citealt{Tobin2020I} associated with Source 11. The plus sign marks the position derived from a Gaussian fit to the 5 cm emission (dashed contours, this work). The core at 9 mm coincides well with the 870 $\micron$ emission tracing the disc, while the 9 mm lobe extends in the direction of the 5 cm elongated source. Red contour levels are 5, 20, 50, 100, and 160 times 310 $\mu$Jy beam$^{-1}$, the rms of the 870 $\micron$ map. Blue contour levels are $-$3, 3, 8, 20, and 45 times 9.55 $\mu$Jy beam$^{-1}$, the rms of the 9 mm map. Black dashed contour levels are 4, 10 and 30 times 5.5 $\mu$Jy beam$^{-1}$, the rms of the 5 cm map with robust 0.5. The 9 mm (blue) and 870 $\micron$ (red) synthesized beams are shown in the bottom left corner of the map. The 5 cm synthesized beam (HPBW) is $0\farcs34\times0\farcs32$, P.A. = $-6\degr$.
		\label{fig:11-disc}} 
\end{figure}

\citet{Tobin2020I} detected Source 11 (HOPS-71-A in their nomenclature) at 9 mm as a compact core ($S_\nu$ = 1.106 $\pm$ 0.024 mJy) with a lobe ($S_\nu$ $\sim$ 0.3 mJy) extending to the NW, coinciding with the elongation we detect at 5 cm.  At 870 $\micron$ the source is detected by \citet{Tobin2020I} as an elongated structure ($S_\nu$ = 130.1 $\pm$ 1.1 mJy), consistent with a circumstellar disc of dust nearly perpendicular to the 9 mm lobe (see their fig. 32 and Fig. \ref{fig:sources}). The peak of the 9 mm core coincides very well with the 870 $\micron$ emission peak, while our 5 cm emission peak is displaced $\sim0\farcs1$ to the NW, at an intermediate position (plus sign in Fig. \ref{fig:11-disc}) between the 9 mm compact core and its lobe. From the 9 mm and 870 $\micron$ observations, we derive a spectral index $\alpha = 2.04\pm0.01$ for the compact core, which appears too flat for a dust-only source (for which a spectral index between 2.5 and 3 would be expected), suggesting a free-free contribution of $\sim$0.7-1 mJy at 9 mm in the compact core.  Assuming that the observed $\sim$0.3 mJy flux density of the NE lobe at 9 mm is dominated by free-free, then we would expect a total free-free flux density at 9 mm of  $\sim$1-1.3 mJy, which is consistent with the value of $\sim$0.4-1.5 mJy derived by extrapolating our data in the cm regime with a spectral index $\alpha=0.3\pm0.4$ (Table \ref{tab:fluxes} and Fig.~\ref{fig:SEDs}).

Our interpretation is that the protostar underwent a one-sided ejection of ionized plasma, which caused an asymmetric distortion in the shape of the source and, consequently, a shift in the centroid position in our 5-cm map. This ejected material is best identified at 9 mm, where the NE lobe, dominated by free-free emission, is well differentiated from the compact core, which has contributions from both dust and free-free emission. We note that in this particular case, the position of the protostar is better traced by the circumstellar disc \citep{Tobin2020I}.

Source 10 is located close ($\sim5\arcsec$) to the 3 mm compact dust core associated with Source 11 (see Fig. \ref{fig:10&11}, bottom). It was classified as a YSO with a disc (Class I/II) by \citet{Getman2017} using the SED in the 1.6 - 8 $\micron$ range (see Table \ref{tab:sources} for the counterparts), but this source has not been detected with ALMA at 870 $\micron$ \citep{Tobin2020I}. The source has an M spectral type according to \citet{Hillenbrand2013} and is located at a distance of $391\pm4$ pc \citep{Gaia2023}. According to \citet{Kounkel2018}, the source belongs to the ONC-6 cluster, with a mean age of $0.76^{+1.01}_{-0.76}$ Myr.  We derive a negative spectral index $\alpha=-1.7\pm0.7$ (Table \ref{tab:fluxes} and Fig. \ref{fig:SEDs}), indicating non-thermal emission, suggesting it is an evolved YSO (Class II/III).

There are two additional objects close in projection to sources 10 and 11 that we do not detect with our observations (Fig. \ref{fig:10&11}, bottom). \citet{Kounkel2016} detected them at near-IR wavelengths and called them HOPS 71-b (coincident with HOPS-71-B from \citealt{Tobin2020I}) and HOPS 71-c. These possible companions appear more evolved than Source 11 according to these authors. HOPS 71-c has also been detected as a variable source in radio (at  5 cm, \citealt{Kounkel2014}). \citet{Gaia2023} reported distances of $360\pm16$ pc for HOPS 71-b, and $433\pm13$ pc for HOPS 71-c, confirming their association with the star-forming region but separated by a considerable distance (73 $\pm$ 21 pc). Indeed, \citet{Kounkel2018} report, based on their 3D velocities, that Source 10 and HOPS 71-c belong to different ONC clusters (ONC-6 and ONC-12, respectively). Therefore, the possibility of a quadruple system composed of sources 10, 11, HOPS 71-b, and HOPS 71-c does not seem plausible. A triple system composed of Source 11, HOPS 71-b, and Source 10 could be possible.

In summary, we are resolving VLA 9 into two radio sources called sources 10 and 11 in this work, both YSOs. Source 10 is more evolved than Source 11, and we suggest it is a Class II/III based on its observed properties. Source 11, which could be a member of a triple system, is a radio jet and we propose it is the driving source of the molecular outflow reported in its vicinity.

\subsubsection{Source 12}\label{12}

\begin{figure}
	\includegraphics[width=0.99\hsize]{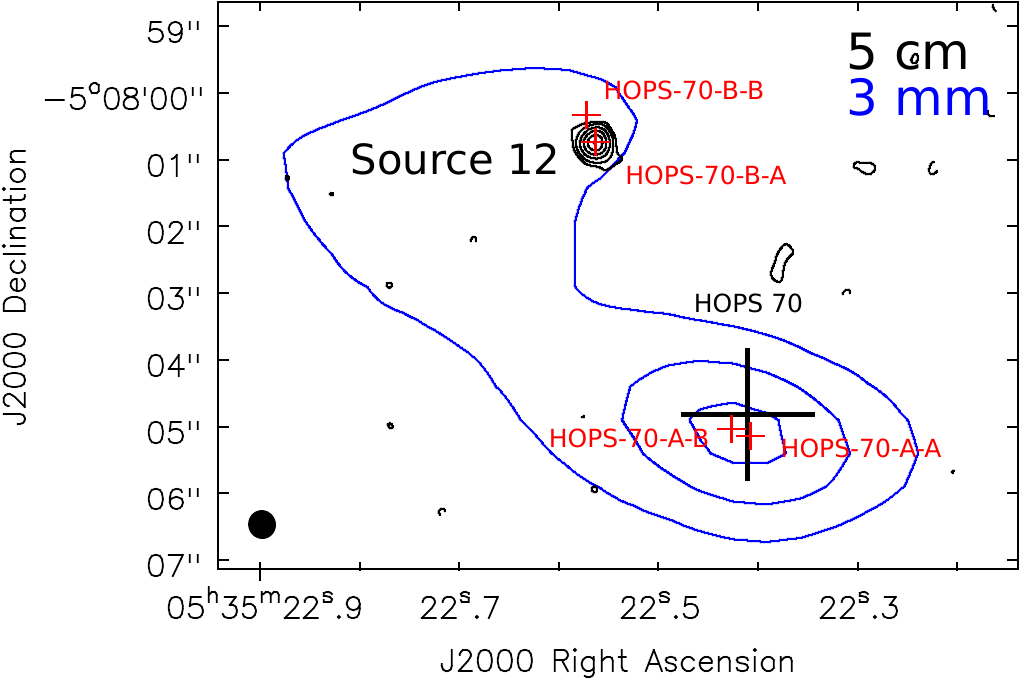}
	\caption{Source 12. VLA 5 cm A-configuration map with robust 0.5 weighting (black contours) overlaid on the ALMA 3 mm map (blue contours, adapted from \citealt{Kainulainen2017}). Plus signs mark the position of the HOPS source (black, \citealt{Furlan2016}, derived at 4.5 $\micron$), and the 9 mm/870 $\micron$ sources (red, \citealt{Tobin2020I}) in the region. Black contour levels are $-$3, 3, 6, 12, 20, 30, 40, and 50 times 5 $\mu$Jy beam$^{-1}$, the rms of the 5 cm map. Blue contour levels are 0.3, 0.6, and 0.9 times 3.1 mJy beam$^{-1}$, the peak of the 3 mm core.  The VLA synthesized beam is shown in the bottom left corner of the map.  The 3 mm ALMA synthesized beam (HPBW) is $3\farcs8\times2\farcs3$, P.A. = 71$\degr$.
		\label{fig:12}}
\end{figure}

Source 12 is located toward the densest part of the ISF (see Fig. \ref{fig:filament}), close ($\sim0\farcs6$) to a faint 3 mm peak (see Fig. \ref{fig:12}) and has mid-IR, near-IR, X-ray and variable optical counterparts (Table \ref{tab:sources}). This source was classified as a YSO with a disc based on IR emission \citep{Megeath2012,Getman2017}, and its spectral type is K according to \citet{Hillenbrand2013}.  Source 12 is located at a distance of $310\pm15$ pc \citep{Gaia2023} and, according to \citet{Kounkel2018}, belongs to the ONC-12 cluster, with a mean age of $0.7^{+1.6}_{-0.7}$ Myr.

This source, previously detected at 5 cm with the VLA by \citet{Kounkel2014}, presents signs of high radio variability (see Fig. \ref{fig:SEDs} and Section \ref{subsec:variab}). The source was also detected with the VLBA (milli-arcsec angular resolution) at 6 cm with a flux density of $0.80\pm0.06$ mJy \citep{Kounkel2017}, indicating that it has a very compact and therefore non-thermal core. In our 5 cm high resolution map the source is marginally resolved (Table \ref{tab:positions} and Fig. \ref{fig:sources}), indicating that some extended emission is also present.

Source 12 (HOPS-70-B-A in \citealt{Tobin2020I}) could have a close companion, since there is another source (HOPS 70-B-B, detected at 9 mm and 870 $\micron$, \citealt{Tobin2020I}) located $0\farcs4$ ($\sim$ 150 au) in projection to the NE, at a  P.A. = 19$\degr$ (see Fig. \ref{fig:12}). Our 5 cm high-resolution natural weighting map shows a marginal elongation along a P.A. = $45\degr \pm 17\degr$ (Table \ref{tab:positions} and Fig. \ref{fig:sources}), which is somewhat different from the P.A. of the potential binary, suggesting that the observed elongation is not due to the presence of a marginally resolved binary. Indeed, in a higher-resolution 5 cm map (robust 0.5) of Source 12, the elongation persists and the emission appears to be associated only with source HOPS 70-B-A (see Fig. \ref{fig:12}). 
Due to the proximity of these two sources it is unclear if the near- and mid-IR emission (Table \ref{tab:sources}) is associated with both components, or with only one of them, and therefore the previous classification of Source 12 as a YSO with a disc should be taken with caution. The fact that it is detected with the VLBA suggests it is a relatively evolved YSO (Class II/III). 

Additionally, the far-IR source HOPS 70 (large black plus sign in Fig. \ref{fig:12}), which \citet{Tobin2020I} resolved into two sources at 9 mm and 870 $\micron$, is located at $\sim5\arcsec$ ($\sim1900$ au) of projected separation to the SW of Source 12. Therefore, Source 12 could be a member of a quadruple system. 

\subsubsection{Source 14}\label{14}

Source 14, previously reported at 5 cm by \citet{Kounkel2014}, is located $\sim5\arcmin$ from the ISF ridgeline (see Fig. \ref{fig:filament}). This source has been proposed as a runaway star \citep{Neuhaeuser1998} that appears to have been ejected $\sim12\times10^4$ years ago from the vicinity of the star $\theta^1$ Ori C, in the Orion Trapezium Cluster \citep{Kounkel2017}. It has been previously classified as a Class III YSO \citep{Tsujimoto2002,Getman2017} with a K spectral type \citep{Strassmeier2009}. The radio source is separated $4\farcs3$ ($\sim1700$ au) in projection from the K0 star JW 242 \citep{Hillenbrand2013}, which we do not detect in radio. Since \citet{Gaia2023} reported a distance of $557\pm22$ pc for JW 242 and $393\pm2$ pc for the optical counterpart of Source 14 (see Table \ref{tab:sources} for the counterparts), we discard that the two objects constitute a binary system.

This is one of the strongest sources in our sample ($\sim 1$ mJy) and presents signs of high radio variability (see Fig. \ref{fig:SEDs} and Section \ref{subsec:variab}). We cannot derive a spectral index from our observations since the source was observed only at C band. \citet{Kounkel2014} reported a flat spectral index $\alpha=-0.02\pm0.19$, unable to constrain the nature of the emission. However, the source was also detected with the VLBA (milli-arcsec angular resolution) at 6 cm with a flux density of 1.6 mJy \citep{Kounkel2017}, indicating that it has a very compact and therefore non-thermal core. This is in agreement with the source being a Class III YSO. 

Source 14 is marginally resolved in our 5 cm high-resolution map (Fig. \ref{fig:sources}). This could suggest that it is composed of a non-thermal compact core (detected by the VLBA) surrounded by a more extended emission (detected by us with the VLA). 

\subsubsection{Sources 17, 19, 21, 22, 28, 29, and 36}\label{stripe}

We discuss all these sources together since all of them are located toward the E-W stripe (Fig. \ref{fig:filament}), have mid-IR, near-IR, X-rays, and variable optical counterparts (Table \ref{tab:sources}), and appear to be evolved YSOs (Class II or Class III, see below) with late spectral types K-M \citep{Hillenbrand1997,Strassmeier2009,Hillenbrand2013,Daemgen2012}. We suggest that these sources are part of a population older than those of the filament, which may have formed in structures perpendicular to the ridgeline of the ISF, now depleted of material (see Section \ref{subsec:distribution}).

Source 17 is reported here at radio wavelengths for the first time. It is located $\sim17\arcsec$ from the ISF ridgeline (Fig. \ref{fig:filament}) and has been previously classified as a Class III YSO based on IR observations \citep{Tsujimoto2002,Getman2017}. We derive a negative spectral index $\alpha\le-0.6$ (Table \ref{tab:fluxes} and Fig. \ref{fig:SEDs}) for the radio source, consistent with non-thermal emission.

In our 5 cm high-resolution map, the source is elongated in the NE-SW direction (see Fig. \ref{fig:sources}), which could indicate a marginally resolved binary system, with a separation of a few tenths of an arcsecond, or a (non-thermal) radio jet. The jet interpretation seems less plausible, since evolved Class III YSOs exhibit weak or undetectable radio jets and are not expected to have non-thermal lobes. Sensitive higher angular resolution observations would clarify the nature of this source. The source has several possible companions, since it is separated in projection $\sim4\arcsec$ ($\sim1500$ au) from the spectroscopic binary and variable star JW 785 \citep{Tobin2009,Samus2017}, and $\sim7\arcsec$ ($\sim2700$ au) from HOPS 349, none of which we detect in radio. Since \citet{Gaia2023} reported distances of $397\pm5$ pc for Source 17 and $380\pm12$ pc for JW 785, it is plausible that they belong to a multiple system.

Source 19, located $\sim3\arcmin$ from the ISF ridgeline (see Fig. \ref{fig:filament}), is reported at radio wavelengths for the first time here and presents signs of high radio variability (see Fig. \ref{fig:SEDs} and Section \ref{subsec:variab}). It was previously classified as a Class III YSO \citep{Getman2017}, and is a spectroscopic binary \citep{Tobin2009}.  Source 19 is located at a distance of $389\pm2$ pc \citep{Gaia2023} and, according to \citet{Kounkel2018}, belongs to the ONC-21 cluster, with a mean age of $0.8^{+0.8}_{-0.8}$ Myr.
We derive a positive spectral index $\alpha=0.9\pm0.3$ for this source (Table \ref{tab:fluxes} and Fig. \ref{fig:SEDs}), although it should be taken with caution due to the high variability of the source.

Source 21, previously reported at 5 cm by \citet{Kounkel2014}, is located $\sim3\farcm3$ from the ISF ridgeline (see Fig. \ref{fig:filament}). \citet{Tsujimoto2002} classified the source as a Class II YSO, while \citet{Getman2017} classified it as Class III. Source 21 is located at a distance of $363\pm13$ \citep{Gaia2023} and, according to \citet{Kounkel2018}, belongs to the ONC-12 cluster, with a mean age of $0.7^{+1.6}_{-0.7}$ Myr.
We derive a negative spectral index $\alpha = -0.6\pm0.4$ (Table \ref{tab:fluxes} and Fig. \ref{fig:SEDs}), consistent with non-thermal emission. This is further supported by the detection of the source at 6 cm with the VLBA (milli-arcsec angular resolution) with a maximum flux density of $0.47\pm0.12$ mJy \citep{Kounkel2017}, indicating that it has a very compact and therefore non-thermal core. Those authors noted that the source appears to be marginally extended in RA, which could point to a marginally resolved binary system. In our 5 cm high-resolution map, the source is marginally resolved (Table \ref{tab:positions} and Fig. \ref{fig:sources}), indicating that some extended emission is also present.

Source 22 is reported here at radio wavelengths for the first time. It is located $\sim4\arcmin$ from the ISF ridgeline (see Fig. \ref{fig:filament}). The source has been previously classified as a Class III YSO \citep{Tsujimoto2002,Getman2017}, and is a spectroscopic binary according to \citet{Tobin2009}. According to \citet{Kounkel2018}, this source belongs to the ONC-16 cluster\footnote{The source was erroneously associated with ONC-9 by \citet{Diaz-Rodriguez2021}}, which has a mean age of $1.5^{+1.7}_{-1.5}$ Myr and is located at a mean distance of $387\pm35$ pc. We cannot derive a spectral index from our observations since the source was only observed at C band. However, in our inspection of the 3 GHz (10 cm) VLASS map \citep{Lacy2020}, we find that the source is detected with a flux density of $1.04\pm0.23$ mJy. This value, along with our 5 cm flux densities (Table \ref{tab:fluxes}), suggests a negative spectral index. The source is marginally resolved in our A-configuration, natural weighting 5 cm map. Additionally, in a higher-resolution map made with robust $-$0.5, the source also appears elongated but the P.A. of the elongation is very uncertain. Similarly to Source 17, Source 22 could be a marginally resolved subarcsecond binary. Sensitive, higher angular resolution observations are needed to clarify the nature of this source.

Sources 28 and 29, separated only $0\farcs6$ ($232$ au) in projection, are reported at radio wavelengths for the first time here. They are located $\sim66\arcsec$ ($\sim25500$ au) from the ISF ridgeline (see Fig. \ref{fig:filament}). In our 5 cm high-resolution map, we separate the emission from the two sources (see Fig. \ref{fig:sources}). The two emission peaks in that map --- that correspond to our radio sources 28 and 29 --- coincide within $0\farcs1$ with the position of two optical sources \citep{Gaia2023}, which are components of a binary system \citep{Kohler2006,Daemgen2012}. Using the C-configuration maps, in which the emission from sources 28 and 29 is blended, we derived a positive spectral index $\alpha=0.74\pm0.03$ for the combination of the two components (Table \ref{tab:fluxes} and Fig. \ref{fig:SEDs}). Emission in the IR is detected at low resolution as a single source that has been previously classified as a YSO with a disc \citep{Megeath2012,Getman2017}. According to \citet{Kounkel2018}, Source 28 belongs to the ONC-12 cluster, which has a mean age of $0.7^{+1.6}_{-0.7}$ Myr and is located at a mean distance of $402\pm15$ pc.

Source 36 was previously reported at 5 cm by \citet{Kounkel2014}, who derived a positive spectral index $\alpha=1.0\pm0.4$ (Table \ref{tab:fluxes} and Fig. \ref{fig:SEDs}), consistent with thermal free-free emission. It has been previously classified as a Class III YSO \citep{Tsujimoto2002,Getman2017}. Source 36 is located at a distance of $384\pm4$ pc \citep{Gaia2023} and, according to \citet{Kounkel2018}, belongs to the ONC-12 cluster, with a mean age of $0.7^{+1.6}_{-0.7}$ Myr.
The source is separated $6\farcs2$ ($\sim2400$ au) in projection from the optical variable V574 Ori \citep{Samus2017}, located at $401\pm4$ pc \citep{Gaia2023}, which we do not detect in radio. Since according to \citet{Kounkel2018} V574 Ori belongs to a different cluster than Source 36, we discard that they are components of a binary system.

\subsubsection{Sources 20, 26, and 30}\label{20&26&30}

\begin{figure}
	\includegraphics[width=0.84\hsize]{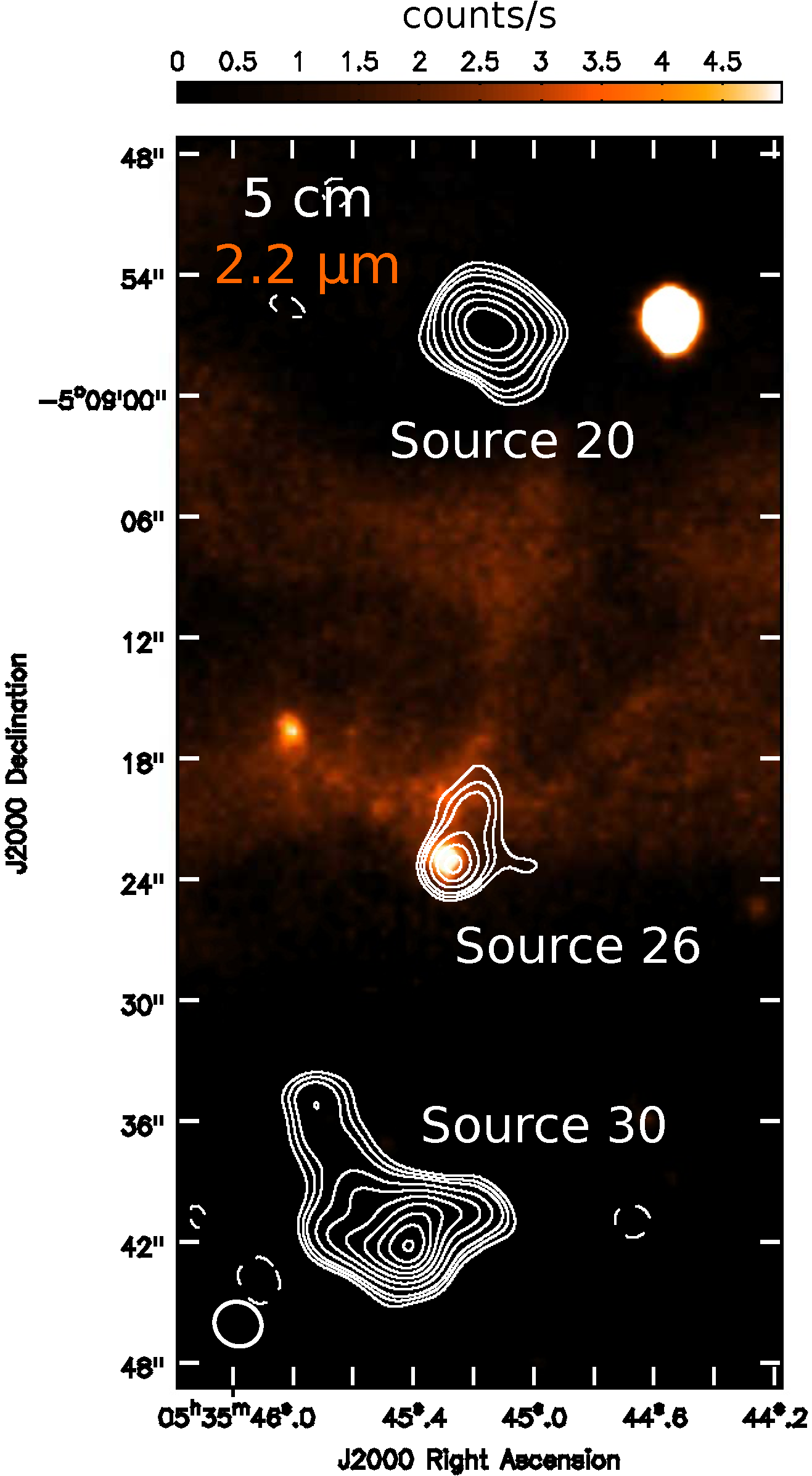}
	\includegraphics[width=0.81\hsize]{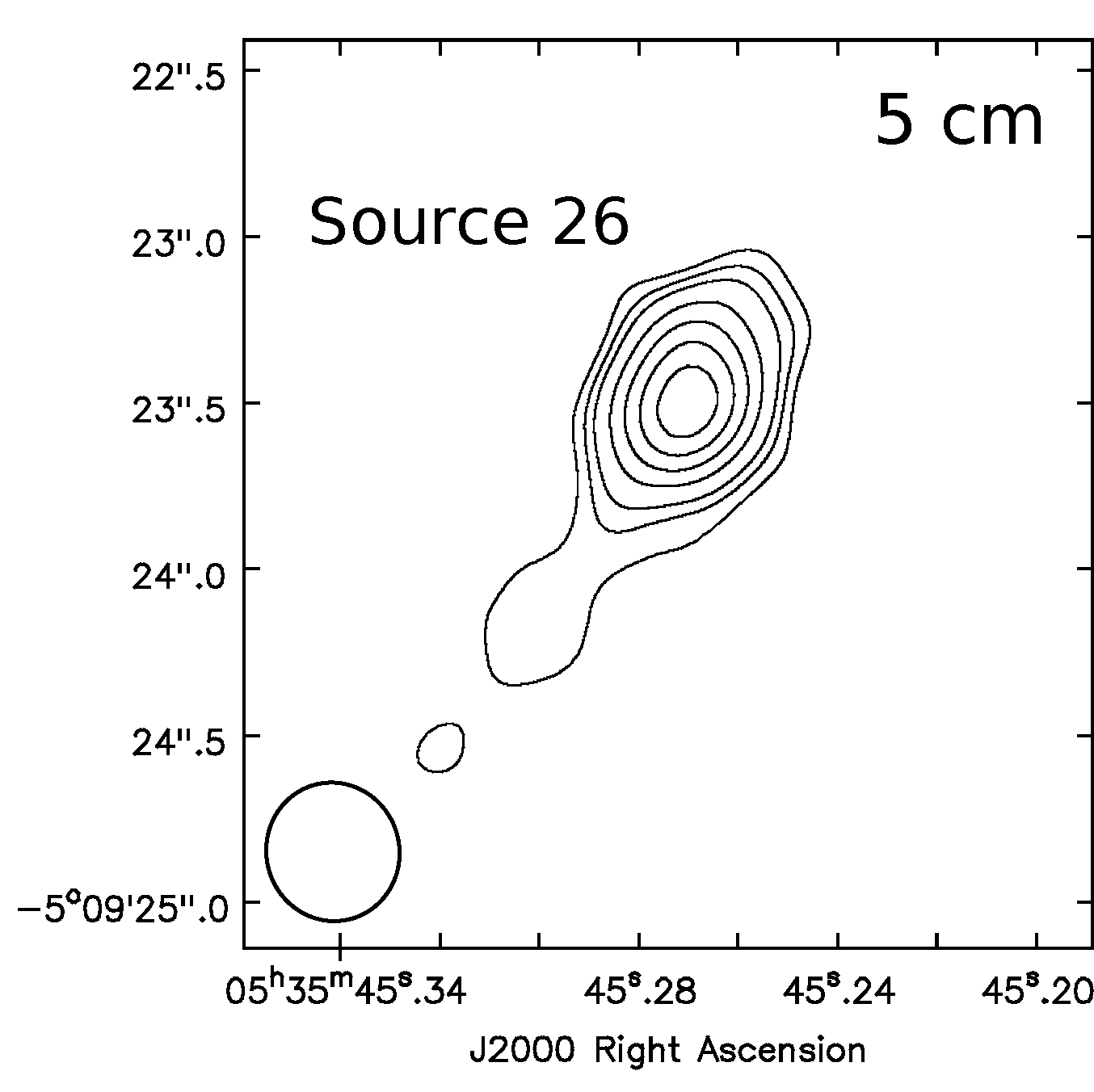}
	\caption{Sources 20, 26, and 30. {\it Top}: VLA 5 cm A-configuration map with natural weighting (contours) overlaid on the NEWFIRM 2.2 $\micron$ image (color scale, T. Megeath, private communication). To better detect the extended emission from sources 20 and 30, uvtaper = 50 k$\lambda$ was applied in the 5 cm data. {\it Bottom}:  VLA 5 cm A-configuration map with natural weighting of Source 26 with no uvtaper applied. The elongation of the source is clearly appreciated. In both panels, contour levels are $-$3, 3, 4, 5, 7, 9, 11, 13, 15, 17, 19, 21, and 23 times the rms of the maps, which are 30 $\mu$Jy beam$^{-1}$ (top) and 12 $\mu$Jy beam$^{-1}$ (bottom). The VLA synthesized beams are shown in the bottom left corner of the maps.
		\label{fig:20&26&30}}
\end{figure}

This is the first report at radio wavelengths of sources 20, 26, and 30. These three sources are located outside the ISF, $\sim4\farcm5$ from its ridgeline (see Fig. \ref{fig:filament}). They are roughly aligned in the NW-SE direction (Fig. \ref{fig:20&26&30}) and we propose that they are in fact a single object. This object would be a central, elongated jet (Source 26; see Fig.\ref{fig:20&26&30} bottom) that has two extended lobes (sources 20 and 30) with bow-shock morphology.

In our inspection of the 3 GHz VLASS map \citep{Lacy2020}, we find that the lobes are detected with a flux density of $2.3\pm0.4$ mJy (Source 20) and $3.4\pm0.7$ mJy (Source 30). These values, together with our 5 cm flux densities (Table \ref{tab:fluxes}), suggest a negative spectral index, indicative of non-thermal emission.

We favor that this is an extragalactic object since the central source (Source 26) has associated only weak IR emission shortward of 4.5 $\micron$ (Table \ref{tab:sources}), while its possible lobes have no counterparts at all (see Fig. \ref{fig:20&26&30}, top). We propose that we are detecting an extragalactic radio jet (Source 26) and terminal shocks (sources 20 and 30) where the jet impacts the environment. Additional data, including a reliable determination of the spectral indices and proper motion studies, should be obtained in order to confirm this prediction.

\subsubsection{Sources 24 and 25 (HOPS 66)}\label{24&25}

Sources 24 and 25, separated $2\farcs2$ ($849$ au) in projection, are located toward the densest part of the ISF and associated with the far-IR source HOPS 66 (\citealt{Adams2012,Furlan2016}; see Table \ref{tab:sources} for the counterparts).

\begin{figure}
	\includegraphics[width=0.93\hsize]{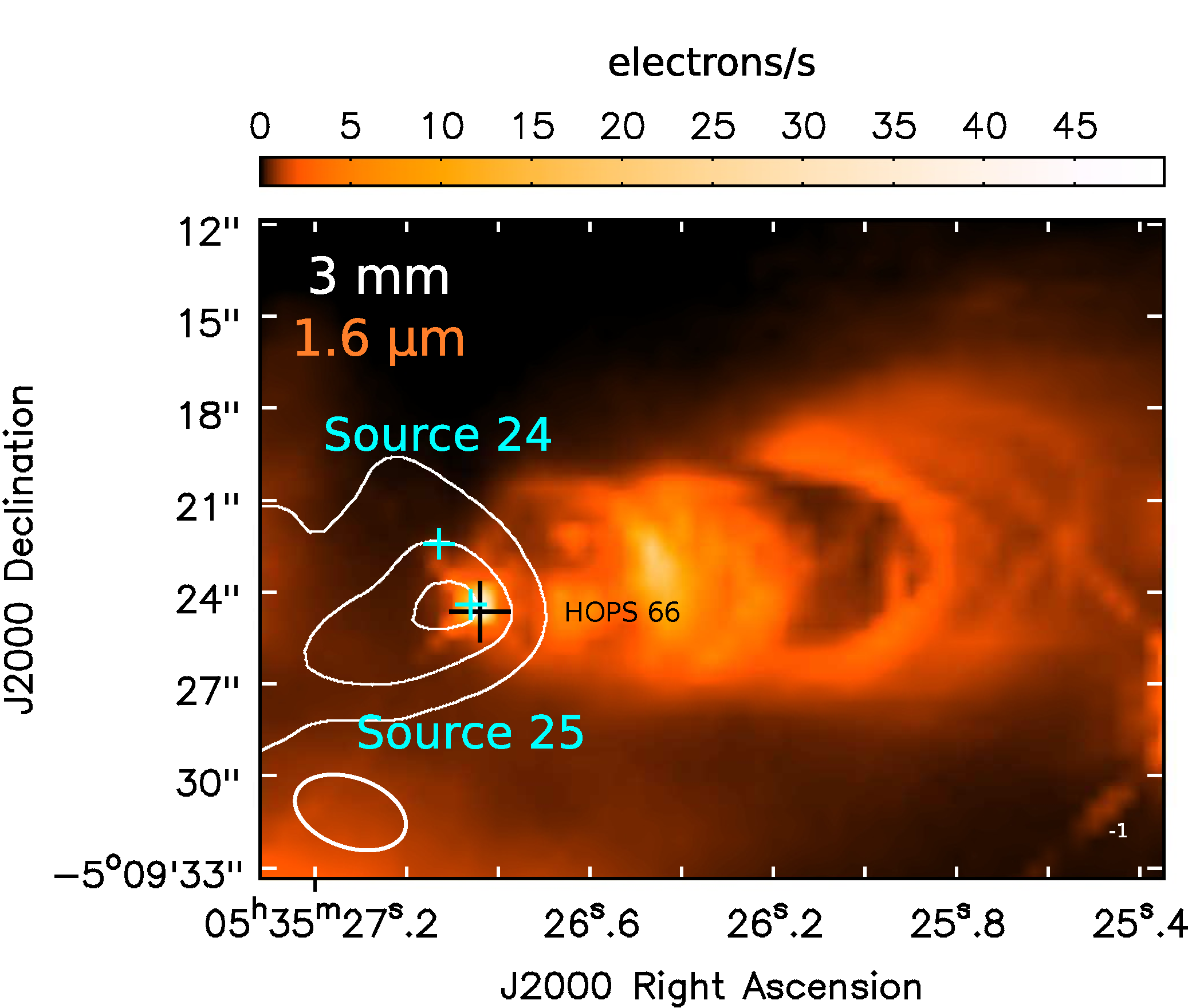}
	\includegraphics[width=0.92\hsize]{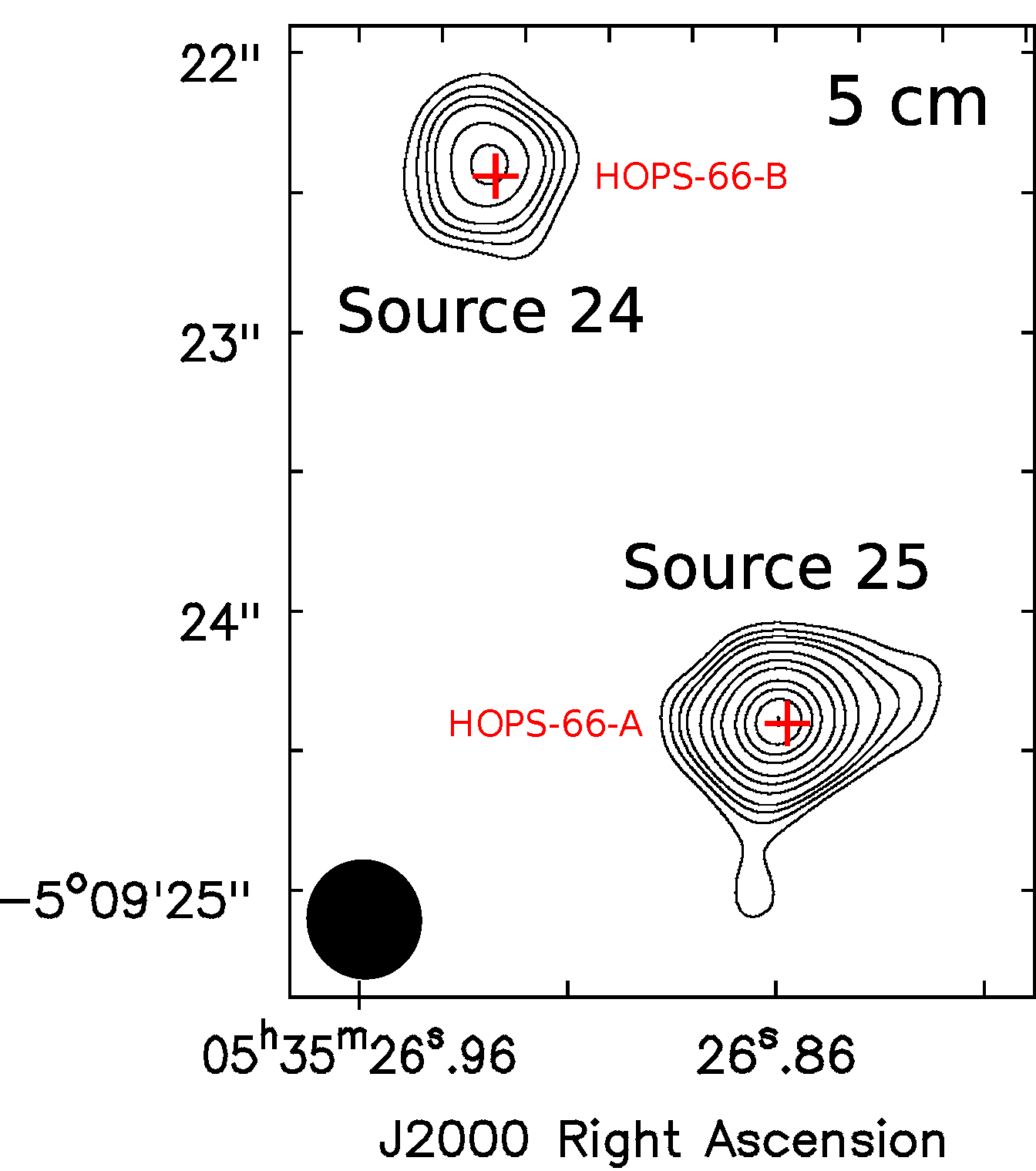}
	\caption{Sources 24 and 25. {\it Top}: ALMA 3 mm map (white contours, \citealt{Kainulainen2017}) overlaid on the HST 1.6 $\micron$ image (color scale, adapted from \citealt{Habel2021}). Plus signs mark the position of the radio sources (cyan, this work) and the HOPS source (black, \citealt{Furlan2016}, derived at 4.5 $\micron$) in the region. Contour levels are 0.3, 0.6, and 0.9 times 7.4 mJy beam$^{-1}$, the peak of the 3 mm core. {\it Bottom}: VLA 5 cm A-configuration map with natural weighting. Red plus signs mark the position of the 9 mm/870 $\micron$ sources \citep{Tobin2019}.  Contour levels are $-$3, 3, 4, 5, 6, 8, 10, 12, 15, 17, 19, 21, and 23 times 3.4 $\mu$Jy beam$^{-1}$, the rms of the map. In both panels, the synthesized beams (VLA: filled, ALMA: outlined) are shown in the bottom left corner. The E-W elongation of Source 25 is consistent with the direction of the monopolar cavity associated with it.
		\label{fig:24&25}}
\end{figure}

The 3 mm ALMA map \citep{Kainulainen2017} shows a compact dust core associated with both radio sources, suggesting they are embedded in the same parental material (see Fig. \ref{fig:24&25}, top). Source 25 is located $\sim1\arcsec$ from the peak of the core, and Source 24 is displaced about $\sim2\arcsec$ NE.
High angular resolution ALMA observations at 1.3 mm \citep {Sato2023} resolve the core into two peaks, each associated with one component of the binary system. We derive positive spectral indices for both sources consistent with thermal free-free emission, $\alpha=0.29\pm0.24$ for Source 24 and $\alpha=0.57\pm0.07$ for Source 25 (Table \ref{tab:fluxes} and Fig. \ref{fig:SEDs}). Both sources are also detected at 9 mm and 870 $\micron$ (see Fig. \ref{fig:sources}) with sizes $\sim$ 100 au \citep{Tobin2019}, indicating the presence of circumstellar discs.

In the HST map (\citealt{Habel2021}), a unipolar cavity opening to the west is clearly associated with Source 25 (Fig. \ref{fig:24&25}, top). Recently, a CO outflow centred on Source 25 with the blue lobe responsible for the IR cavity (\citealt{Tobin2020II}, \citealt{Sato2023}) has been imaged. In fact, the blue lobe exhibits a complex substructure that \citet{Sato2023} interpret as evidence of likely episodic ejection. In our 5 cm high-resolution map, the source is elongated at P.A. $= 106 \pm 2\degr$ (see Fig. \ref{fig:24&25}, bottom), consistent with the direction of the cavity and the outflow. Hence, we confirm this object as the radio jet driving the large scale outflow.

\citet{Furlan2016} classified HOPS 66 as a flat-spectrum luminous source with a high inclination angle. The latter would be consistent with our detection of Source 25 as an elongated radio jet. On the other hand, the far-IR emission peaks closer to Source 25 and, since both sources appear to be embedded in the same core, it is possible that the measured (broadband) SED has a contribution also from Source 24. In fact, the observed flattening of the SED (fig. 1, panel 3 in \citealt{Furlan2016}) could be due to flux coming from both sources, each one with the emission peak at a slightly different frequency.

\citet{Furlan2016} estimated for HOPS 66 a $L_{\rm bol} = 18~L_\odot$ (scaled to our adopted distance of 390 pc). Given the poor angular resolution of the far-IR data, this value probably includes the luminosity of both protostars, and the relative contribution of each one is unknown. However, given that the radio source 25 (likely a radio jet driving the outflow associated with the near-IR cavity) is the one with the higher 5 cm flux density, it seems plausible to attribute essentially all the bolometric luminosity to this radio source, with a much smaller contribution (of the order of a few tenths of $L_\odot$) to radio source 24. With this configuration, both sources would follow fairly well the Anglada correlation with $L_{\rm bol}$ (see discussion in Section \ref{subsec:nature}).

\subsubsection{Source 27}\label{27}

Source 27 is located toward the ISF, at $\sim24\arcsec$ from its ridgeline, in the region where the ISF and the E-W stripe apparently intersect (Fig. \ref{fig:filament}). The source is not associated with any 3 mm compact dust core \citep{Kainulainen2017}, and has been previously classified based on its near and mid-IR emission (see Table \ref{tab:sources} for its counterparts) as a  Class III YSO  \citep{Furesz2008,Getman2017} with M spectral type \citep{Hillenbrand1997}. According to \citet{Kounkel2018}, this source belongs to the ONC-16 cluster, which has a mean age of $0.4^{+1.2}_{-0.4}$ Myr and is located at a mean distance of $397\pm35$ pc. Recently, \citet{Gaia2023} reported another source at $0\farcs3$ of projected separation from our source 27 but at a distance of $909\pm246$ pc. It does not seem plausible that they belong to a binary system since they will be separated more than 300 pc.

Source 27 is reported at radio wavelengths for the first time here. We detect the source at the four bands (5, 3.6, 1.3, and 0.7 cm) in our 2014 C-configuration maps, but not in our more sensitive 2015 A-configuration 5 cm map, suggesting it is radio variable (see Fig. \ref{fig:SEDs}). Additionally, it has not been detected at 0.9 cm by \citep{Tobin2020I} in their 2016 A-configuration observations with a sensitivity of $\sim$ 8 $\mu$Jy beam$^{-1}$. We derive a positive spectral index $\alpha = 1.06\pm0.05$ (Table \ref{tab:fluxes} and Fig. \ref{fig:SEDs}), consistent with thermal free-free emission.

To explain the thermal free-free emission of this object, one possibility would be that we are detecting a radio jet, which is expected to be weak for an evolved YSO and more extended at longer wavelengths. Since the source is strong and resolved in one direction in our 1.3 cm and 0.7 cm maps (Tables \ref{tab:positions}, \ref{tab:fluxes} and Fig. \ref{fig:sources}), but remains unresolved at longer wavelengths, we can rule out this case. Another possibility would be that the positive spectral index is not real, but an artifact if we are measuring the combined flux of two components of an unresolved binary. In this case we would expect that, since the source is marginally resolved in one direction at 1.3 cm, the two components would be resolved in the 0.7 cm map with twice the higher resolution, and this does not happen. A viable option given its evolutionary stage would be that we are detecting free-free emission from disc photoevaporation. Further observations, such as the detection of optical and mid-IR forbidden line emission \citep{Font2004, Alexander2014}, should be carried out to better evaluate this possibility.

\subsubsection{Source 32 (HOPS 369)}\label{32}

\begin{figure}
	\includegraphics[width=0.99\hsize]{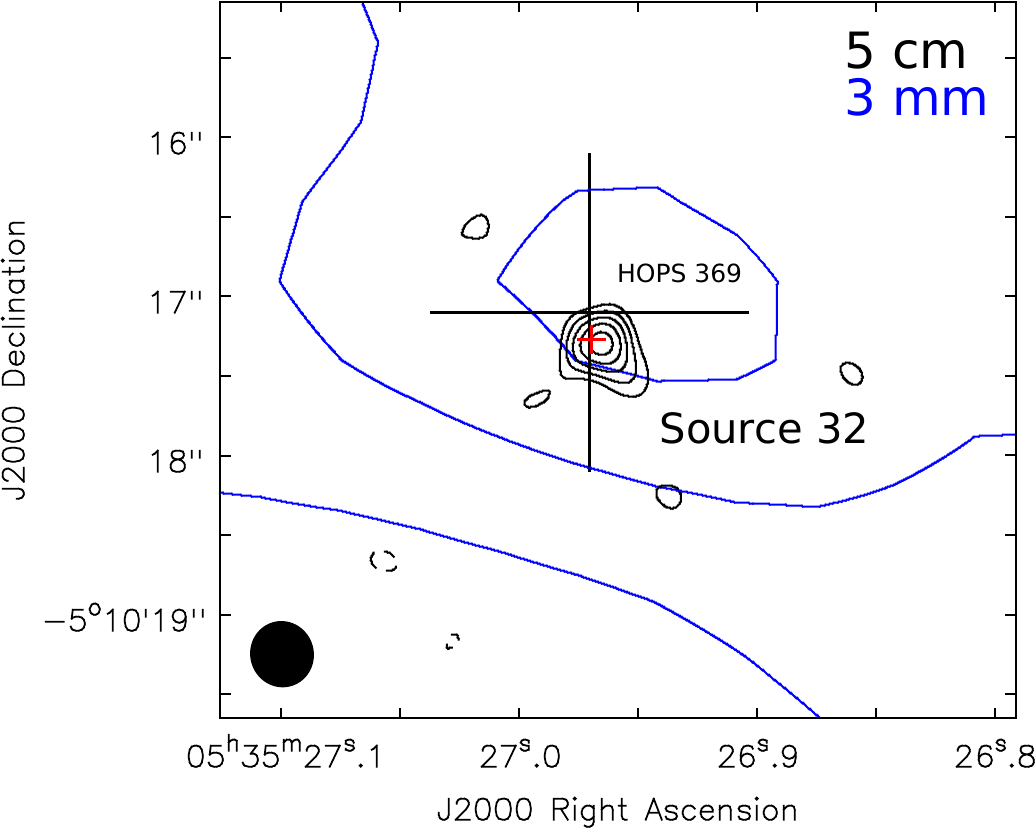}
	\caption{Source 32. VLA 5 cm A-configuration map with natural weighting (black contours) overlaid on the ALMA 3 mm map (blue contour, adapted from \citealt{Kainulainen2017}). Plus signs mark the position of the HOPS (black, \citealt{Furlan2016}, derived at 4.5 $\micron$) and the 9 mm/870 $\micron$ (red, \citealt{Tobin2020I}) sources in the region. Black contour levels are $-$3, 3, 4, 5, 6, and 7 times the rms of the 5 cm map, 3 $\mu$Jy beam$^{-1}$. Blue contour levels are 0.3, 0.6, and 0.9 times 1.8 mJy beam$^{-1}$, the peak of the 3 mm core. The VLA synthesized beam is shown in the bottom left corner of the image. The 3 mm ALMA synthesized beam (HPBW) is $3\farcs8\times2\farcs3$, P.A. = 71$\degr$.
		\label{fig:32}}
\end{figure}

This source is located toward the densest part of the ISF (Fig. \ref{fig:filament}), associated with the far-IR source HOPS 369 (see Table \ref{tab:sources} for the counterparts), and coincides with a faint 3 mm peak (see Fig. \ref{fig:32}). A compact bipolar CO(2-1) outflow toward this source has been recently reported (\citealt{Sato2023}; see their fig.13). The source has been previously classified using its broadband SED as a Class I/II YSO \citep{Furesz2008,Getman2017} and as a flat spectrum source with a high inclination angle (disc close to edge-on,  \citealt{Furlan2016}). On other hand, \citealt{Habel2021} report scattered light with an irregular morphology associated with this source.

We derived a positive spectral index, $\alpha=0.36\pm0.18$ (Table \ref{tab:fluxes} and Fig. \ref{fig:SEDs}) in the centimeter regime, consistent with thermal free-free emission. The radio source has been detected at 9 mm with a flux density of $0.051\pm0.007$ mJy \citep{Tobin2019}, in good agreement with our spectral index, suggesting that at that wavelength almost all the emission is due to free-free emission. It has also been detected at 870 $\micron$ by those authors (see Fig. \ref{fig:sources}) as a poorly resolved source that likely traces a disc of dust. Our 5 cm high resolution map shows a hint of an elongation in the SW direction (see  Figs. \ref{fig:sources} and \ref{fig:32}). Since the CO outflow shows a blueshifted lobe extending toward the NE \citep{Sato2023}, overall these results suggest the presence of a NE-SW outflow traced by radio continuum and the CO.
Higher angular resolution cm observations could be helpful to better resolve the radio jet.

\subsubsection{Sources 33 and 34 (VLA 13, HOPS 368)}\label{33&34}

\begin{figure*}
	\center\includegraphics[width=0.7\hsize]{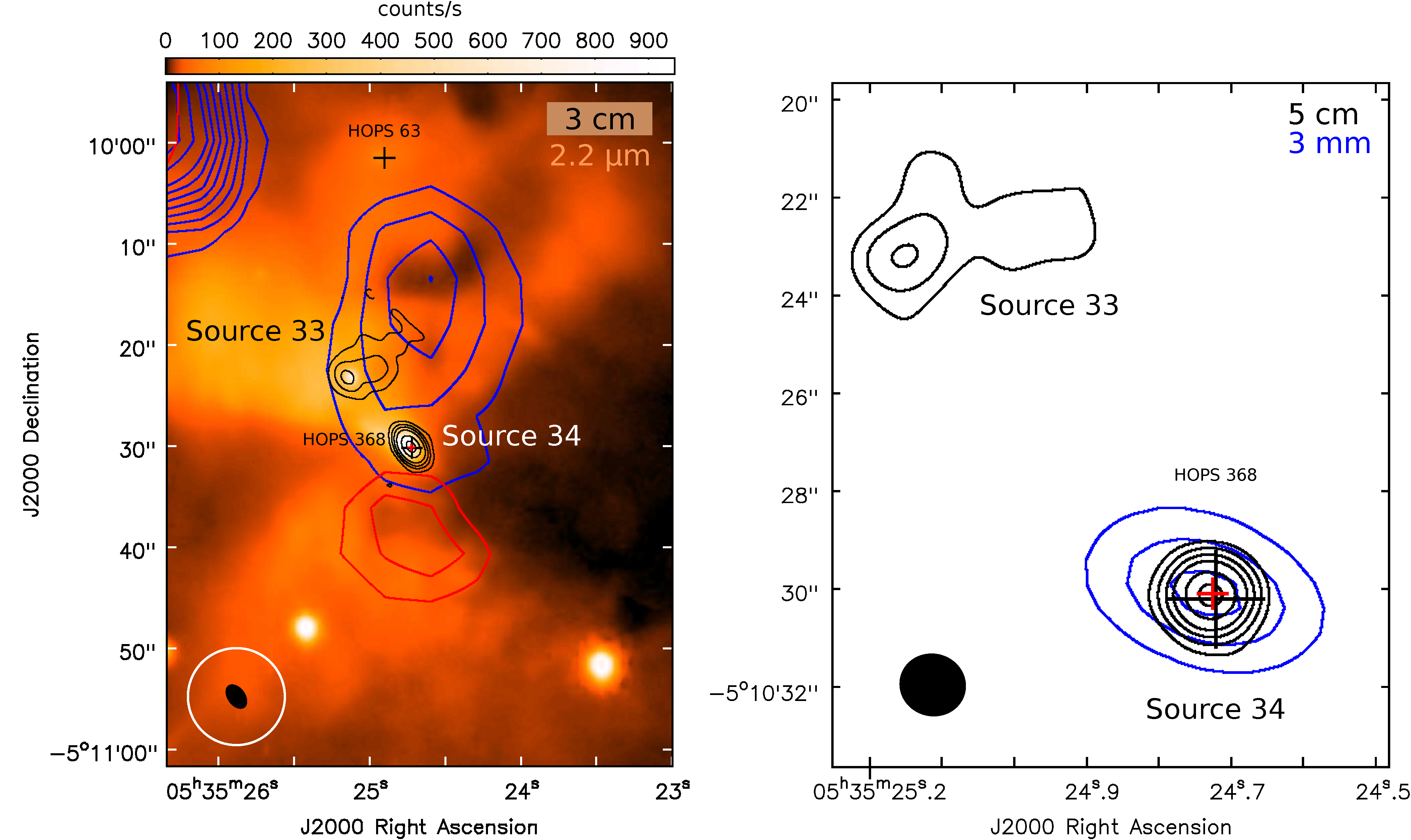}
	\caption{Sources 33 and 34. {\it Left}: APEX CO(6-5) blue- and red-shifted integrated intensity map (blue and red contours, respectively; B. Gonz\'alez-Garc\'\i a, private communication) and VLA 3 cm C-configuration map with robust 0 weighting (black contours) overlaid on the NEWFIRM 2.2 $\micron$ image (T. Megeath, private communication, see also \citealt{Habel2021}). Contour levels of the CO emission are 3, 4, 5, 6, 7, 8, 9, 10, 12, 14 and 16 times 8000 K, the rms of the map. Black contour levels are $-$3, 3, 7, 12, 20, 40, and 60 times 9.0 $\mu$Jy beam$^{-1}$, the rms of the 3 cm map.  {\it Right}: VLA 5 cm A-configuration map with natural weighting and uvtaper = 100 k$\lambda$ (black contours) overlaid on the ALMA 3 mm map (blue contours, adapted from \citealt{Kainulainen2017}). Contour levels are $-$3, 3, 6, 10, 15, 25, and 35 times 8.2 $\mu$Jy beam$^{-1}$, the rms of the 5 cm map. Blue contours are 0.3, 0.6, and 0.9 times 7.6 mJy beam$^{-1}$, the peak of the 3 mm core. In both maps, plus signs mark the position of the HOPS (black, \citealt{Furlan2016}, derived at 4.5 $\micron$) and the 9 mm/870 $\micron$ (red, \citealt{Tobin2020I}) sources in the region. The VLA (filled) and APEX (outlined) synthesized beams are shown in the bottom left corner of the maps. The 3 mm ALMA synthesized beam (HPBW) is $3\farcs8\times2\farcs3$, P.A. = 71$\degr$. The N-S outflow driven by Source 34 and the associated double cone-shaped nebula are clearly appreciated in the left panel.
		\label{fig:33&34}}
\end{figure*}

The radio source VLA 13, elongated at P.A. $\sim20\degr$, was reported by \citet{Reipurth1999} in their observations with angular resolution of $\sim8\arcsec$. Our higher angular resolution allows us to resolve that source into two, sources 33 and 34, separated $\sim9\arcsec$ ($\sim3500$ au) in projection (Fig. \ref{fig:33&34}). Source 34 was previously reported at 5 cm by \citet{Kounkel2014}, but this is the first report of Source 33 as an independent radio source. These sources are located toward the densest part of the ISF (Fig. \ref{fig:filament}), in a crowded region where different outflows overlap (see fig. 8 in \citealt{Takahashi2008}, fig. 7 in \citealt{Gonzalez2016}, and fig. 14 in \citealt{Sato2023}). Source 34 is detected toward the centre of a north-south bipolar CO outflow \citep{Takahashi2008, Gonzalez2016, Sato2023}, which is associated with a bright double cone-shaped K${_s}$ reflection nebula (e.g., \citealt{Takahashi2008}, \citealt{Habel2021}), likely tracing the outflow cavities (see Fig.~\ref{fig:33&34}). In the \citet{Sato2023} ALMA map the northern (blueshifted) outflow lobe shows a clumpy substructure with a distinct feature toward the position of source 33, suggestive of a shock with an obstacle (or maybe an independent outflow) associated with that source.

Source 34 is located toward the peak of a 3 mm compact dust core (\citealt{Kainulainen2017}, see Fig. \ref{fig:33&34}, right) and is associated with HOPS 368 (see Table \ref{tab:sources} for other counterparts), which has been modeled as a Class I YSO with a disc \citep{Adams2012,Furlan2016,Getman2017}. ALMA observations at 870 $\micron$ \citep{Tobin2019} reveal an elongated east-west structure, suggestive of an edge-on disc of dust perpendicular to the outflow axis (see Figs.~\ref{fig:sources} and \ref{fig:33&34}).
From our data (Table \ref{tab:fluxes}), we derive a positive spectral index $\alpha=0.73\pm0.14$ for this source, consistent with thermal free-free emission. The source follows the Anglada correlations with $L_{\rm bol}$ and \.{P} (see Section \ref{subsec:nature}) suggesting it is a radio jet. An elongation consistent with the direction of the outflow (and nearly perpendicular to the imaged circumstellar dust disc) is hinted at in our 5 cm high-resolution map (see Fig. \ref{fig:sources}). Higher angular resolution observations would be necessary in order to resolve the radio jet. We note that the flux density of Source 34 at 7 mm is lower than the extrapolated flux density from the fit with lower frequency data points (Fig. \ref{fig:SEDs}). This flattening of the spectrum could suggest that the radio emission from the jet becomes optically thin at high frequencies.

Source 33 is not associated with any 3 mm compact dust core \citep{Kainulainen2017} (see Fig. \ref{fig:33&34}, right). The radio source has an extended morphology and, in the 5 cm high-resolution map, it has two emission peaks, separated $\sim3\arcsec$ in projection (Fig. \ref{fig:33&34}, right). We derive a spectral index $\alpha=-0.05\pm0.01$ (Table \ref{tab:fluxes}), consistent with optically thin thermal free-free emission coming from an ionized region. The only star detected toward this radio source is V1343 Ori ($\sim1\arcsec$ to the NW of Source 33 eastern peak), at a distance of $396\pm30$ pc \citep{Gaia2023}, but its luminosity of only 0.1 $L_{\sun}$ \citep{DaRio2012} is not enough to yield the observed centimetre flux density. Since Source 33 appears located toward a clumpy feature in the blueshifted lobe of the CO outflow, as observed by \citet{Sato2023} with ALMA, we suggest that the ionization of the gas is occurring due to shocks of this outflow with the walls of the cavity.

In summary, we propose that Source 34 is the driving source of the outflow detected in its vicinity and that Source 33 is tracing ionized gas in the associated cavity.

\subsubsection{Source 38}\label{38}

Source 38 is reported at radio wavelengths for the first time here. It is located toward the densest part of the ISF but is not associated with any 3 mm compact dust core \citep{Kainulainen2017}, suggesting it is not a deeply embedded YSO.

\begin{figure}
	\center\includegraphics[height=0.89\textheight]{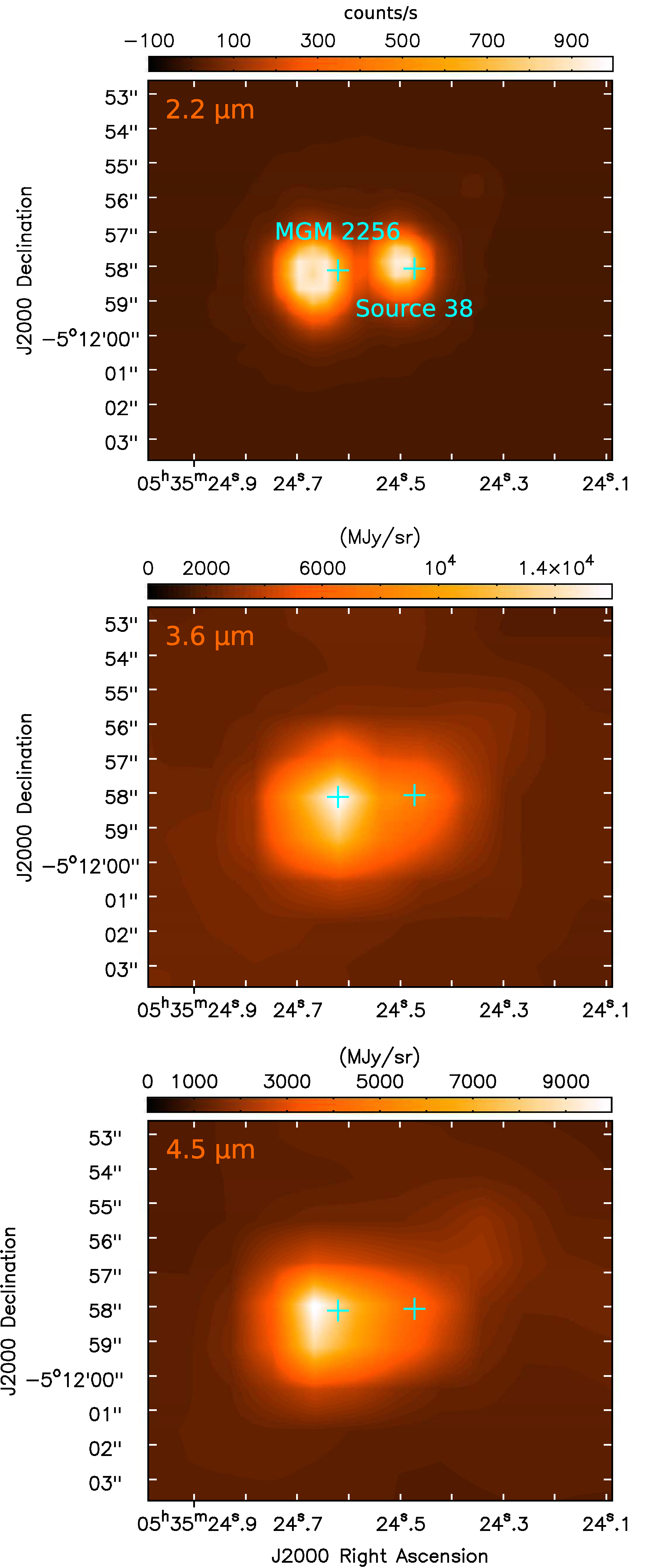}
	\caption{Source 38. Near-IR (NEWFIRM 2.2 $\micron$; T. Megeath, private communication) and mid-IR emission \citep[Spitzer 3.6 and 4.5 $\micron$]{Megeath2012}. Plus signs mark the position of the radio source (this work) and the mid-IR source MGM 2256 \citep[derived at 4.5 $\micron$]{Megeath2012}. Emission toward Source 38 is clearly observed in all panels. \label{fig:38}}
\end{figure}

The source has been detected in the optical and the near-IR (see Table \ref{tab:sources} for the counterparts) and classified with a spectral type M \citep{DaRio2009}. We report here emission associated with the radio source at 3.6 and 4.5 $\micron$ in the inspected {\it Spitzer} maps (Fig. \ref{fig:38}).

We determine a positive spectral index of $\alpha=1.3\pm0.8$ (Table \ref{tab:fluxes} and Fig. \ref{fig:SEDs}), consistent with thermal free-free emission. In fact, the source appears as a weak, marginally elongated source in the 5 cm (robust 0.5) map (see Fig. \ref{fig:sources}), which could indicate that we are detecting a weak radio jet from an evolved YSO (Class II/III).

Source 38 is located at $\sim2\farcs5$ ($\sim1000$ au) of projected separation from the mid-IR source MGM 2256 \citep{Megeath2012}, which we do not detect at radio wavelengths (Fig. \ref{fig:38}). \citet{Gaia2023} reported a distance of $755\pm90$ pc for MGM 2256 and $399\pm 11$ pc for Source 38, suggesting that they do not belong to a binary system since they are actually separated by more than 350 pc.

\subsubsection{Source 39 (HOPS 60)}\label{39}

\begin{figure}
	\includegraphics[width=\hsize]{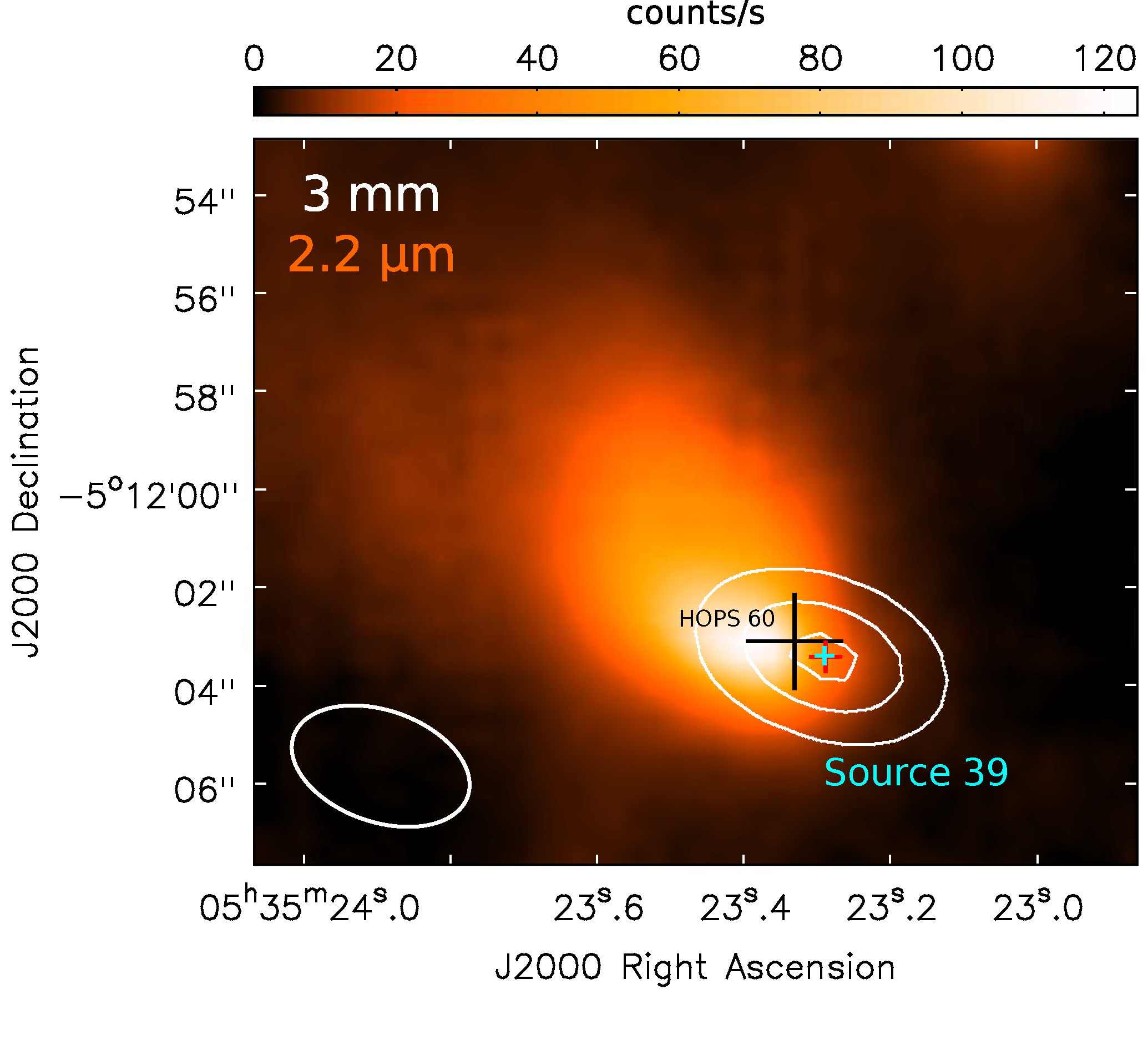}
	\caption{Source 39. ALMA 3 mm map (contours, \citealt{Kainulainen2017}) overlaid on the NEWFIRM 2.2 $\micron$ image (color scale, T. Megeath, private communication), in which the monopolar cavity associated with Source 39 is appreciated. Plus signs mark the position of the radio source (cyan, this work), the HOPS source (black, \citealt{Furlan2016}, derived at 4.5 $\micron$), and the 9 mm/870 $\micron$ source (red, \citealt{Tobin2020I}). Contour levels are 0.3, 0.6, and 0.9 times 18.9 mJy beam$^{-1}$, the peak of the 3 mm core. The ALMA synthesized beam is shown in the bottom left corner.
		\label{fig:39}}
\end{figure}

This source, located in the densest part of the ISF (Fig. \ref{fig:filament}), is associated with the far-IR source HOPS 60 \citep{Adams2012} and the 1.3 mm dust condensation named FIR 6b \citep{Chini1997}. The source is located toward the centre of a bipolar CO outflow at P.A. $= 65\degr$ \citep{Takahashi2008,Shimajiri2009, Gomez-Ruiz2019}, whose red lobe has opened a cavity to the NE seen in scattered light (\citealt{Habel2021}; Megeath, private communication, see Fig. \ref{fig:39}). \citet{Matsushita2021} reported evidence of rotation in the outflow. However, this result was later questioned by \citet{Soker2022}, who reanalyzing the data proposed that the molecular emission corresponds to a twin outflow. We detect only a single radio source; therefore, we cannot provide independent support to the binary hypothesis, unless the two radio sources are separated by less than the angular resolution of our observations.

The radio source is located near the peak of a compact dust core in the 3 mm ALMA image (\citealt{Kainulainen2017}, Fig. \ref{fig:39}), indicating it is deeply embedded. In our 5 cm high-resolution map, the source appears slightly elongated at P.A. $= 65\degr\pm17\degr$ (see Fig. \ref{fig:sources}), consistent with the orientation of  both the cavity and the outflow (P.A. $= 65\degr$) detected in its vicinity. Additionally, \citet{Tobin2020I} detected with ALMA at 870 $\micron$ a structure nearly perpendicular to the orientation of the outflow, which may correspond to a circumstellar dust disc (see Fig. \ref{fig:sources}). More recently,  (\citealt{Huang2024}) found  evidence of grain growth within this structure based on dust self-scattering measurements.

We derive a positive spectral index $\alpha=1.1\pm0.7$ (Table \ref{tab:fluxes} and Fig. \ref{fig:SEDs}), indicative of thermal free-free emission. Also, the source follows the correlations of the radio luminosity with $L_{\rm bol}$ and \.{P} (see Section \ref{subsec:nature}) suggesting it is a radio jet. Therefore, we propose that Source 39 is the driving source of the outflow seen at larger scales, but we cannot discriminate between the single and double outflow interpretation.

\subsubsection{Source 40}\label{40}

Source 40 is located toward the border of the ISF, $\sim38\arcsec$ from its ridgeline (Fig. \ref{fig:filament}), and is not associated with any 3 mm compact dust core \citep{Kainulainen2017}. The source was previously detected at 5 cm by \citet{Kounkel2014} and classified as a Class III YSO (\citealt{Tsujimoto2002,Getman2017}, see Table \ref{tab:sources} for the counterparts). It has a G spectral type according to \citet{Hillenbrand1997} and is located at a distance of $369\pm10$ pc \citep{Kounkel2017}.

The source is a close binary (separation $\sim$ 4 au) with both components detected at 6 cm in VLBA observations (milli-arcsec angular resolution, \citealt{Kounkel2017}), indicating that it has a very compact and therefore non-thermal core. We cannot constrain well the spectral index from our data (see Table \ref{tab:fluxes} and Fig. \ref{fig:SEDs}) nor taking into account 9 mm data ($0.28\pm0.02$ mJy, \citealt{Tobin2020I}).

In our 5 cm high-resolution maps, the source appears elongated along a P.A. $= 48\degr$, but is unresolved in the transversal direction (Table \ref{tab:positions}). This elongation is not related to the VLBA binary since we do not have enough angular resolution to resolve the binary. This elongation persists in maps made with higher resolution (from robust 0.5 to uniform weighting; see Fig. \ref{fig:sources}). It could be caused by extended emission that is resolved out with the VLBA or due to the presence of additional objects.

The source VLA 14, reported at 3.6 cm by \citet{Reipurth1999}, but undetected in our observations and those of \citet{Kounkel2014}, is located at a projected distance of $\sim5\farcs5$ ($\sim2100$ au) from Source 40. Since VLA 14 was detected with an angular resolution of $8\arcsec$, we cannot discard that VLA 14 and Source 40 are the same source.

\subsubsection{Sources 41 and 43 (HOPS 409)}\label{41&43}

\begin{figure}
	\includegraphics[width=0.99\hsize]{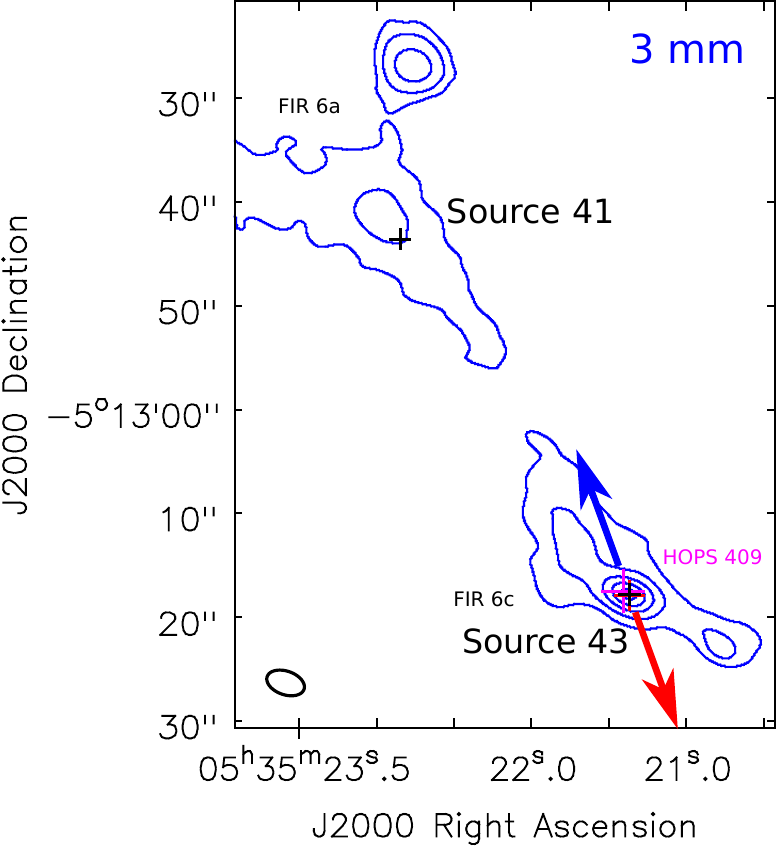}
	\caption{Sources 41 and 43. ALMA 3 mm map (contours, adapted from \citealt{Kainulainen2017}). Plus signs mark the radio sources (black, this work), the HOPS source (magenta, \citealt{Furlan2016}, derived at 4.5 $\micron$), and the 9 mm/870 $\micron$ source (red, \citealt{Tobin2020I}) in the region. FIR names are the millimeter peaks from \citet{Chini1997}. Contour levels are 0.1, 0.2, 0.4, 0.65, and 0.9 times 12.0 mJy beam$^{-1}$, the peak of the FIR 6c 3 mm core. The synthesized beam is shown in the bottom left corner of the image. Source 41 seems to fall in projection very close to the path of the blue-shifted lobe of Source 43 outflow, whose direction is indicated by arrows \citep{Shimajiri2008,Tobin2016outf}.
		\label{fig:41&43}}
\end{figure}

These sources are located toward the densest part of the ISF (Fig. \ref{fig:filament}). Source 43 is associated with the 1.3 mm dust condensation named FIR 6c \citep{Chini1997} and with the extremely young Class 0 object HOPS 409 (source PBR 135003 in \citealt{Tobin2015,Furlan2016}). The source is located toward the peak of a compact dust core at 3 mm (\citealt{Kainulainen2017}, see Fig. \ref{fig:41&43}), consistent with it being deeply embedded. \citet{Huang2024} studied the morphology and strength of the magnetic field of Source 43, among other Orion protostars, and reported a ``rotated hourglass'' morphology of the magnetic field in this source. The authors interpreted this morphology as evidence that the magnetic field strength in Source 43 is likely less dominant than gravity and angular momentum.
A large bipolar CO outflow, extending along a P.A. $\sim 20\degr$, is associated with this source \citep{Takahashi2008,Shimajiri2009,Tobin2016outf}. The outflow is also detected in SiO and complex organic molecules with ALMA (\citealt{Bouvier2025}). Additionally, \citet{Tobin2020I} detected with ALMA at 870 $\micron$ and 9 mm a structure nearly perpendicular (P.A.$\sim 100\degr$) to the direction of the outflow that could be a circumstellar dust disc (see Fig. \ref{fig:sources}). 

Although we do not resolve the radio source (Fig. \ref{fig:sources}), nor constrain its spectral index (Table \ref{tab:fluxes}), we propose that we are detecting the radio jet associated with the large scale outflow. We make this statement based on the facts that the outflow is centred on Source 43 and that this radio source follows the correlations with  $L_{\rm bol}$ and \.{P} (see Section \ref{subsec:nature}) suggesting it is a radio jet.
 
This is the first report at radio wavelengths of Source 41. It is located toward the border of a weak dust clump located $\sim40\arcsec$ NE of Source 43 (see Fig. \ref{fig:41&43}). It does not have known counterparts at other wavelengths and its spectral index is unconstrained (Table \ref{tab:fluxes}). We note that Source 41 seems to fall in projection close to the path of the blue-shifted lobe of the outflow driven by Source 43,  which extends toward a 1.3 mm dust condensation named FIR 6a \citep{Chini1997} (see Fig. \ref{fig:41&43}). As we have already mentioned, \citet{Shimajiri2009} and \citet{Bouvier2025} detected collimated SiO emission, a tracer of shocked gas, along this lobe (see their fig. 5). The former argued that interaction between the SiO jet and FIR 6a is probably taking place. The position of Source 41 in a region of strong shocks suggests its association with the outflow. It could be either a knot produced by the shocks or, if the source were indeed associated with the ISF, a possible case of induced star-formation. Further studies toward Source 41, such as proper motions or a reliable determination of its spectral index, should be carried out to address this question.

\subsubsection{Source 42 (HOPS 59)}\label{42}

\begin{figure}
	\includegraphics[width=0.99\hsize]{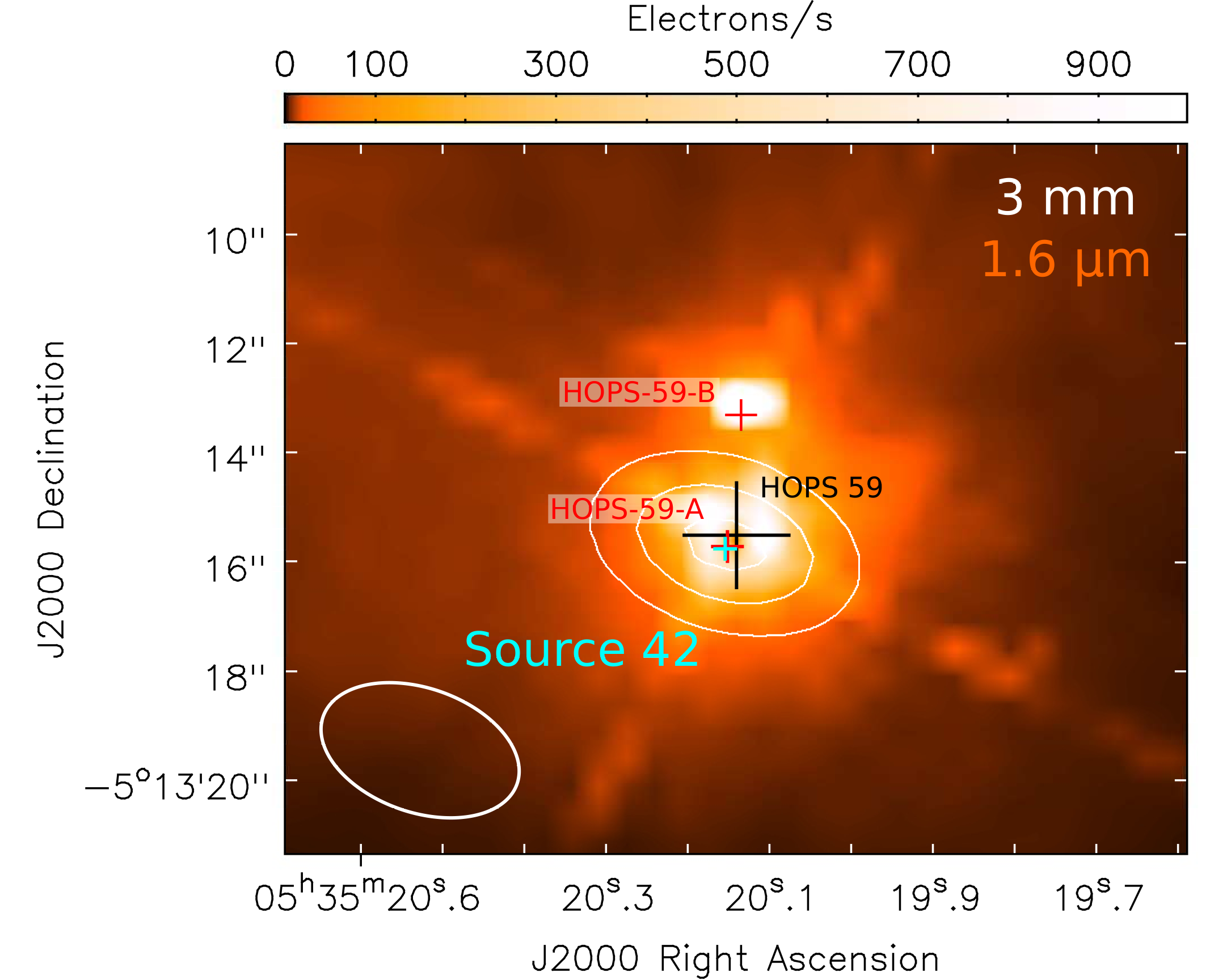}
	\includegraphics[width=0.97\hsize]{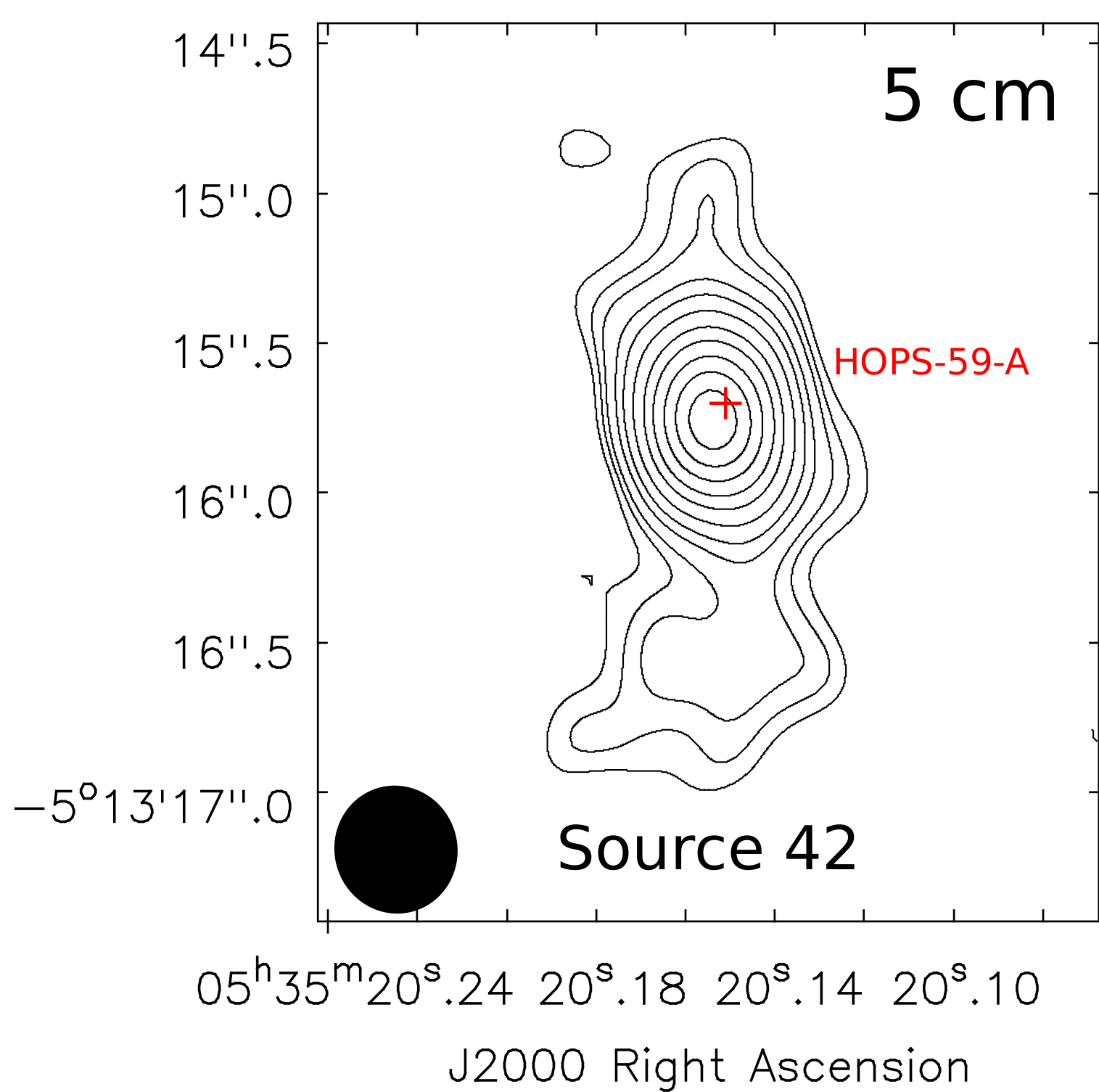}
	\caption{Source 42. {\it Top}: ALMA 3 mm map (contours, \citealt{Kainulainen2017}) overlaid on the HST 1.6 $\micron$ image (color scale, \citealt{Habel2021}). Plus signs mark the radio source (cyan, this work), the HOPS source (black, \citealt{Furlan2016}, derived at 4.5 $\micron$), and the 9 mm/870 $\micron$ sources (red, \citealt{Tobin2020I}) in the region. Contour levels are 0.3, 0.6, and 0.9 times 22.6 mJy beam$^{-1}$, the peak of the 3 mm map. {\it Bottom}: VLA 5 cm A-configuration map with natural weighting. Contour levels are $-$3, 3, 4, 5, 7, 9, 12, 16, 20, 25, 30, and 35 times 7.6 $\mu$Jy beam$^{-1}$, the rms of the map. In both panels, the synthesized beams (VLA: filled, ALMA: outlined) are shown in the bottom left corner. The elongation of Source 42 is consistent with the N-S outflow that has been associated with it \citep{Shimajiri2009}.
		\label{fig:42}}
\end{figure}

Source 42, located toward the densest part of the ISF (Fig. \ref{fig:filament}), is associated with the 1.3 mm dust condensation named FIR 6d \citep{Chini1997} and the far-IR source HOPS 59 \citep{Adams2012,Furlan2016}, to which a CO outflow in the N-S direction \citep{Shimajiri2009} and scattered light with an irregular morphology \citep{Habel2021} have been previously associated. The 3 mm ALMA map shows that Source 42 is located toward the peak of a compact dust core (\citealt{Kainulainen2017}, see white contours in Fig. \ref{fig:42}, top), indicating it is deeply embedded.

The source has a late K-M spectral type according to \citet{Hillenbrand1997}. Based on IR observations, Source 42 was classified as a YSO with a disc \citep{Tsujimoto2002,Getman2017}. This has been recently confirmed by \citet{Tobin2020I}, who imaged with ALMA at 870 $\micron$ a $\sim$100 au elongated structure suggestive of a dust disc (source HOPS-59-A in their nomenclature; see Fig. \ref{fig:sources}).

Our 5 cm high-resolution map shows a clearly elongated source at P.A. $= 11\degr\pm7\degr$ (see Fig. \ref{fig:42}, bottom), consistent with the direction of the outflow and nearly perpendicular to the proposed dust disc \citep{Tobin2020I}. We derive a spectral index $\alpha=-0.3\pm0.5$ (Table \ref{tab:fluxes}). Although with a large uncertainty, it is consistent with thermal free-free emission. Also, the source follows the Anglada correlation with $L_{\rm bol}$  (see Section \ref{subsec:nature}) suggesting that it is a radio jet. Therefore, we propose that Source 42 is the driving source of the outflow seen at larger scales.

The SED of the source has been modeled as a binary by \citet{Adams2012} and as a single object with a flat-spectrum by \citet{Furlan2016}. There is observational evidence supporting that Source 42 belongs to a binary system since it is located $\sim2\farcs5$ from the optical variable NSV 16490, which we do not detect at cm wavelengths (see Fig. \ref{fig:42}, top).  Furthermore, \citet{Gaia2023} reported consistent distances of $429\pm21$ pc for NSV 16490 and $384\pm10$ pc for Source 42. Nevertheless, we note that the star NSV 16490 has been detected from optical to mm wavelengths (\citealt{Gaia2023,Hillenbrand1997,Nielbock2003}; source HOPS-59-B in \citealt{Tobin2020I}), being always weaker than Source 42. Therefore, the observed SED \citep{Adams2012,Furlan2016}, probably with contributions from both components, is expected to be dominated by Source 42.

\subsubsection {Source 46 (HOPS 57)}\label{46}

\begin{figure}
	\includegraphics[width=0.99\hsize]{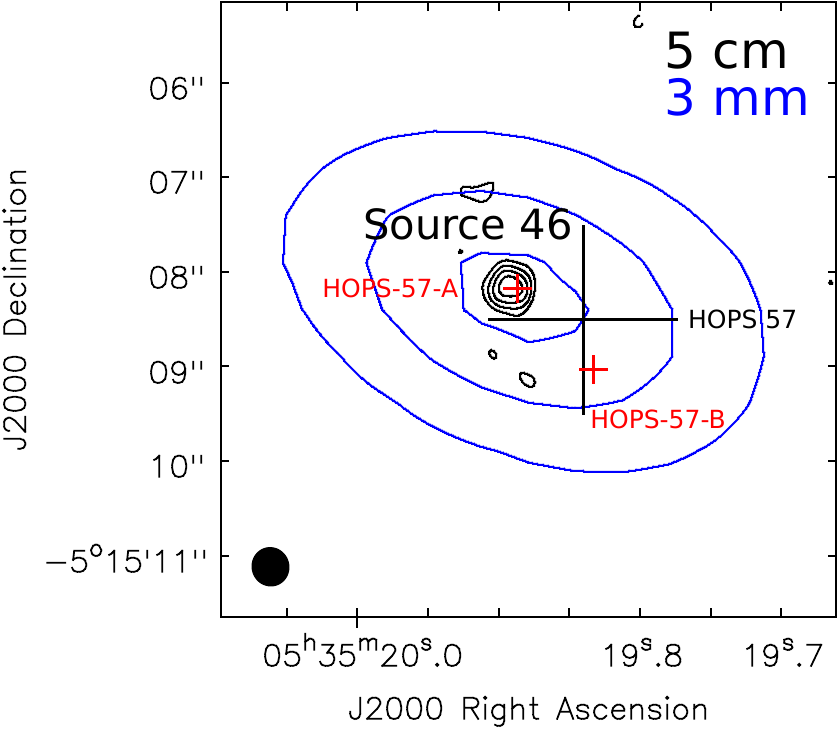}
	\caption{Source 46. VLA 5 cm A-configuration map with natural weighting (black contours) overlaid on the ALMA 3 mm map (blue contours, adapted from \citealt{Kainulainen2017}). Plus signs mark the HOPS source (black, \citealt{Furlan2016}, derived at 4.5 $\micron$) and the 9 mm/870 $\micron$ sources (red, \citealt{Tobin2020I}) in the region. Black contour levels are $-$3, 3, 4, 5, and 6 times 23 $\mu$Jy beam$^{-1}$, the rms of the 5 cm map. Blue contour levels are 0.3, 0.6, and 0.9 times 9.2 mJy beam$^{-1}$, the peak of the 3 mm core. The VLA synthesized beam is shown in the bottom left corner of the map. The 3 mm ALMA synthesized beam (HPBW) is $3\farcs8\times2\farcs3$, P.A. = 71$\degr$.
		\label{fig:46}}
\end{figure}

This source is located toward the densest part of the ISF (Fig. \ref{fig:filament}) and is associated with the far-IR and flat spectrum source HOPS 57 (\citealt{Furlan2016}; see Table \ref{tab:sources} for the counterparts) and with scattered light with an irregular morphology \citep{Habel2021}. Furthermore, the source coincides with the peak of a 3 mm compact dust core (\citealt{Kainulainen2017}, Fig. \ref{fig:46}), suggesting it is deeply embedded. Additionally, \citet{Tobin2020I} detected with ALMA at 870 $\micron$ with $\sim0\farcs11$ resolution, an elongated structure that appears to be a circumstellar dust disc (see Fig. \ref{fig:sources}). They also detected 9 mm emission with the VLA ($0.53\pm0.02$ mJy), which, together with our flux density at 5 cm (Table \ref{tab:fluxes}), allow us to derive a positive spectral index $\alpha=0.5\pm0.1$ for this source, consistent with thermal free-free emission. Source 46 could have a companion since it is separated $\sim1\arcsec$ ($\sim400$ au) in projection from the variable near-IR/optical star V2359 Ori \citep{Kounkel2016,Samus2017,Gaia2023,Tobin2020I}, reported a distance of $364\pm78$ pc \citep{Gaia2023}, which we do not detect in our observations.

\subsubsection{Source 47}\label{47}

Source 47, previously reported at 5 cm by \citet{Kounkel2014}, is located toward the border of the ISF, $\sim$$14''$ from its ridgeline (see Fig. \ref{fig:filament}), and is not associated with any 3 mm compact dust core \citep{Kainulainen2017}. Based on its near and mid-IR emission (see Table \ref{tab:sources} for the counterparts), the source has been classified as a Class III YSO \citep{Getman2017} with a spectral type K \citep{Hillenbrand1997}. \citet{Gaia2023} reported a distance of $394\pm4$ pc, confirming its association with the star-forming region.

\begin{figure}
	\includegraphics[width=0.99\hsize]{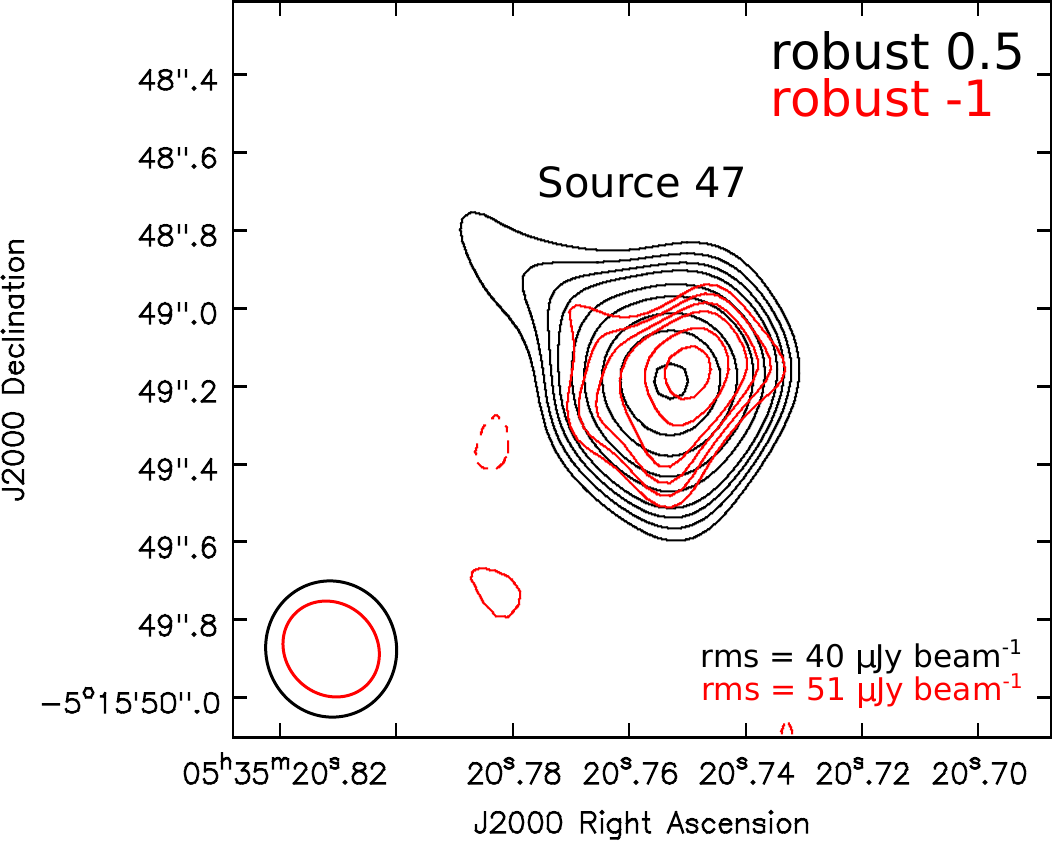}
	\caption{Source 47. VLA 5 cm A-configuration maps made with robust 0.5 (black contours) and robust -1 (red contours) weighting. Contour levels are $-$3, 3, 4, 5, 6, 8, 10, 13, 16 and 20 times the rms of each map, shown at the bottom right corner. The synthesized beams are shown in the bottom left corner of the maps.
		\label{fig:47}}
\end{figure}

We cannot constrain the spectral index from the 5 cm observations. However, the source was recently detected at 9 mm with a flux density of  $1.13\pm0.12$ mJy \citep{Tobin2020I}. Using this value, together with the flux densities shown in Fig. \ref{fig:SEDs}, we derive a spectral index $\alpha=-0.05\pm0.03$, which is consistent with optically thin thermal free-free emission. The source is marginally resolved in our high-resolution 5 cm natural weighting map, with an elongation in the NE-SW direction (Fig. \ref{fig:sources} and Table \ref{tab:positions}). The source is strong enough to be imaged at higher resolution (robust 0.5 and $-$1), but the main direction of elongation changes from NE-SW to NW-SE as the angular resolution increases (see Fig. \ref{fig:47}). This behavior suggests that a marginally resolved multiple system could be present. Higher angular resolution radio observations would be needed in order to resolve the possible components. Given that Source 47 is an evolved (Class III) YSO, free-free emission from a photoevaporating disc is a plausible alternative possibility to account for the relatively bright extended emission. Additional studies, including optical and mid-IR forbidden line emission observations \citep{Font2004, Alexander2014}, should be carried out to further investigate this possibility.

\section{Summary and conclusions}\label{sec:conclusions}

In this work we extended the study carried out by \citet{Osorio2017} to the whole field of view of those observations, a region of 12$\farcm$5 of diameter, comprising the southernmost 2$\farcm$5 of the OMC-3 region and the whole OMC-2 region.

We detect 47 radio sources, 23 of them for the first time, in addition to the 10 radio sources already reported by \citet{Osorio2017}. For these 47 radio sources, we report the positions, flux densities, and spectral indices, and discuss their nature by analyzing complementary information (i.e their radio spectral index, morphology, emission at other wavelength ranges, position relative to the ISF, association with dust clumps). Out of the 47 radio sources discussed in this work, 12 have previous high-resolution detection of thermal emission from dust attributed to the presence of a circumstellar disc \citep{Tobin2020I}. For these cases, our high-resolution detection of free-free emission allows us to establish a complete picture of these potential jet/disc systems.

Our main conclusions are:

\begin{enumerate}	
	\item Out of the 57 total radio sources in the field of view, 41 (72\%) are very likely associated with the star-forming region, and 6 are extragalactic background sources. We are unsure about the association with the star-forming region of the remaining 10 sources. Of the 41 sources associated with the star-forming region, 36 are tracing the position of YSOs, 3 are tracing non-thermal emission from jet knots, one is tracing the shock-ionized wall of an outflow cavity, and one could be either a YSO or a jet knot.
	
	\item  Most of the radio continuum sources are detected toward the Integral Shaped Filament (ISF), as expected if they were associated with YSOs. In general, the radio sources associated with YSOs previously classified as the youngest objects (Class 0/I) are located toward the densest part of the ISF, while the ones associated with more evolved YSOs (Class II/III) are located toward the border or outside the ISF.
	
	\item Interestingly, there are several radio sources distributed in a stripe nearly perpendicular to the ISF. Most of them have known counterparts at other wavelengths and approximately half of them have been previously classified as Class II/III objects. Since the ISF has been observed to be composed of velocity-coherent fibers, some of which deviate from the large-scale alignment, we suggest that the YSOs observed in this stripe have formed in perpendicular fibers whose material has already been depleted.
	
	\item Through their radio continuum emission, we have unambiguously identified the position of the driving source of all the previously reported molecular outflows in our field of view (namely sources 1, 9, 11, 25, 34, 39, 42, and 43, in addition to VLA 11, discussed by \citealt{Osorio2017}). We propose these sources as radio jets, and in the case of Source 1, Source 11, Source 25, and Source 42, we indeed see their well-collimated morphology.
	
	\item Using our radio data, complemented with other data from the literature, we verified that the observed radio emission of all but four of the 31 HOPS far-IR sources located within our field of view is consistent with the previously known empirical correlation between radio and bolometric luminosities. This correlation is usually interpreted as a consequence of the connection between accretion (traced by the bolometric luminosity) and outflow (traced by the radio continuum). For the four sources that are radio-underluminous with respect to this correlation, we suggest that either the ionization mechanism is less efficient or that they have important contributions to the bolometric luminosity other than accretion. Since three of these sources (HOPS 64, 108, and 369) are located in the proximity of the shocked region near the edge of the blue lobe of the HOPS 370 outflow, we suggest that external heating from this shocked region could have an important contribution to the bolometric luminosity of these objects.
	
	\item Only six of the HOPS sources within our field of view (HOPS 60, 71, 78, 368, 394, and 409) are associated with molecular outflows for which the momentum rate has been estimated. For all of these sources, the radio luminosity is roughly consistent with the established correlation between these two parameters.
	
	\item Although, in general, the nature of the radio emission from evolved YSOs (Class II/III) is non-thermal, we find two Class III YSOs (Source 27 and Source 47) where we favor a free-free nature for the observed radio emission. We suggest that this emission could result from disc photoevaporation.
	
	\item Using the VLA, with an angular resolution of a few tenths of an arcsecond, we have marginally resolved the radio emission from sources 3, 12, 14, 21, and 40, which were previously detected by the VLBA with milliarcsecond angular resolution. This suggests that these sources have a very compact—and therefore very bright—non-thermal core, as well as some extended emission. Source 3 appears to be an extragalactic object, while the remaining sources appear to be evolved YSOs, in which the compact non-thermal emission originates from a region within a few stellar radii, whereas the extended emission is due either to a surrounding envelope that is resolved by the VLBA or to the presence of additional objects.
	
	\item We find signs of radio variability for 10 radio sources, most of them evolved YSOs (Class II/III). Remarkably, Source 11 which is very young (Class I), shows a high variability in radio, which deserves further investigation. Interestingly, seven of these radio variables are also variables at shorter wavelengths (near-IR/optical/X-ray).
	
	\item We investigate the multiplicity of our sample in a range of projected distances between 154 (our angular resolution) and 3000 au. If we consider only our detected radio sources, we identify 7 possible binary systems and none higher-order system. Taking into account additional known objects in the field of view that are detected at other wavelengths, we detect radio components of 17 potential multiple systems: 11 binary systems, two triple systems, three quadruple systems, and one quintuple system.  Further analysis  (e.g., proper motion studies) will be necessary to determine if these are indeed gravitationally bound systems.
	
	\item The comparison of radio spectral-index distributions across seven star-forming regions suggests an increasing prevalence of negative spectral indices towards more evolved YSOs, particularly in Taurus and Ophiuchus, in agreement with previous suggestions \citep{Dzib2013, Dzib2015}. Although this trend is less evident in OMC-2/3, we find no statistically significant evidence that the evolutionary trends differ among regions. These results are consistent with evolutionary changes in the relative contributions of thermal and non-thermal radio emission mechanisms.
	
	\item We believe our study to be the most sensitive carried out until now in a significant fraction of Orion at wavelengths of a few centimeters, where free-free emission dominates. It shows the potential of this type of study at $\mu$Jy sensitivity, which will be routine for the next generation of radio interferometers, such as the ngVLA and SKA. In this sense, this work can be considered a pathfinder for the upcoming surveys, which will have even more sensitivity and angular resolution, further revealing the complexity of crowded star-forming regions.	
\end{enumerate}

\section*{Acknowledgements}

Most of the work presented in the paper was included in the PhD thesis of Ana K. Diaz-Rodriguez \citep{Diaz-Rodriguez2021}. A.K.D.R. acknowledges a Spanish MECD FPU fellowship FPU12/03181. G.A., A.K.D.-R., G.B.-C., A.F.P.-M., J.F.G., M.D.H.-E. and M.O. acknowledge support from grants PID2023-146295NB-I00 and CEX2021-001131-S, funded by MICIU/AEI /10.13039/501100011033 and by FEDER, EU. AS gratefully acknowledges support ANID BASAL project FB210003. The authors acknowledge V. Terr{\'o}n for his help with the code to find and classify counterparts of multiple SIMBAD sources. We would like to thank the editor and an anonymous referee for their thoughtful review of the manuscript and their valuable comments, which have helped to improve its clarity. The National Radio Astronomy Observatory is a facility of the National Science Foundation operated under cooperative agreement by Associated Universities, Inc.

\section*{Data availability}

The data underlying this article are available from the NRAO data archive (https://data.nrao.edu/portal/) under VLA Projects 14B-296 and 15A-369. Derived data will be shared on reasonable request to the corresponding author.




\bibliographystyle{mnras}
\bibliography{ms_final}

\clearpage
\appendix

\section{Observational properties of the detected radio sources}

\begin{table*}
\small
\setlength{\tabcolsep}{4.5pt}
\caption{Positions and sizes of the radio sources.}
\label{tab:positions}
\begin{tabular*}{\linewidth}{llccccccc}
\hline\hline
\multicolumn{9}{c}{}\\
{Source} & \multicolumn{2}{c}{Position$^a$} & {Deconvolved size$^a$} & P.A.$^b$ & {Column}  & {Previous}  \\
\cline{2-3}
{num.} & {R.A.(J2000)} & {Dec(J2000)} & {(\arcsec $\times$ \arcsec)} & (\degr) & {density$^c$} & {cm detections}  &Refs.  &  Notes$^d$\\
\hline
\multicolumn{9}{c}{OMC-3}\\
\cline{1-9}
1 & 05 35 25.9655$\pm$0.0011 &$-$05 05 43.369$\pm$0.016 & 0.50(3)$\times$0.23(7) & 99$\pm$5 & High & VLA 5$^e$, VLA 15$^f$, GBS$^g$ & 1, 2, 3  \\ 
2 & 05 35 25.775$\pm$0.003  &$-$05 05 43.84$\pm$0.05 & unresolved & ... & High & VLA 5$^e$ & 1 \\
3 & 05 35 29.2214$\pm$0.0007 &$-$05 05 44.150$\pm$0.011 & 0.23(3)$^h$ & 156$\pm$11 & Medium& VLA 6, VLA 17, GBS$^g$ & 1, 2, 3 \\ 
\cline{1-9}
\multicolumn{9}{c}{OMC-2}\\
\cline{1-9}
4 & 05 35 24.55$\pm$0.07 & $-$05 06 58.5$\pm$1.0 & 18(2)$\times$13(1) & ... & High  & VLA 7 & 1 & E\\
5 & 05 35 35.4843$\pm$0.0023 & $-$05 07 20.54$\pm$0.04 & unresolved & ... & Low\\ 
6 & 05 35 18.6135$\pm$0.0016 & $-$05 07 33.60$\pm$0.03 & 0.43(6)$\times$0.23(7) & 80$\pm$11 & Low \\ 
7  & 05 35 22.362$\pm$0.003 & $-$05 07 39.10$\pm$0.05 & unresolved & ... & High  \\
8 & 05 35 21.4827$\pm$0.0012 & $-$05 07 45.824$\pm$0.019 & 0.19(5)$\times$0.11(5) & 177$\pm$75 & Medium \\ 
9 & 05 35 24.0458$\pm$0.0009  & $-$05 07 52.074$\pm$0.014 & 0.35(6)$\times$0.20(6) & 16$\pm$17 & High  & VLA 8 & 1 & CE\\
10 & 05 35 25.897$\pm$0.003  & $-$05 07 56.4$\pm$0.4 & unresolved & ... & High  & VLA 9$^e$ & 1\\
11 & 05 35 25.5873$\pm$0.0007$^i$ & $-$05 07 57.596$\pm$0.011$^i$ & 0.45(4)$\times$0.18(5) & 24$\pm$5 & High & VLA 9$^e$ & 1\\ 
12 & 05 35 22.5628$\pm$0.0007 & $-$05 08 00.741$\pm$0.011 &  0.28(4)$\times$0.16(5) & 45$\pm$17 & High & GBS$^g$ & 3  \\
13 & 05 35 33.3778$\pm$0.0008 & $-$05 08 01.319$\pm$0.011 & 0.24(5)$\times$0.17(6) & 152$\pm$52 & Low& VLA 10 & 1 & CE \\ 
14 & 05 35 04.3023$\pm$0.0008 & $-$05 08 12.637$\pm$0.012 & 0.33(3)$\times$0.23(3) & 95$\pm$8 & Low& GBS$^g$ & 3 \\ 
15 & 05 35 13.812$\pm$0.003  & $-$05 08 30.16$\pm$0.04 & unresolved & ... & Low\\ 
16 & 05 35 32.881$\pm$0.003  & $-$05 08 31.35$\pm$0.04 & unresolved & ... & Low\\ 
17 & 05 35 26.2911$\pm$0.0016 & $-$05 08 40.037$\pm$0.024 & 0.42(7)$^h$ & 42$\pm$8 & High \\ 
18 & 05 35 33.041$\pm$0.003 & $-$05 08 50.91$\pm$0.04 & unresolved & ... & Low\\ 
19 & 05 35 39.079$\pm$0.003 & $-$05 08 56.39$\pm$0.04 & unresolved & ... & Low\\
20 & 05 35 45.130$\pm$0.016 & $-$05 08 56.88$\pm$0.24 & 4.8(3)$\times$4.1(3)$^j$& 65$\pm$23 & Low& & & BS\\ 
21 & 05 35 40.7757$\pm$0.0007 & $-$05 09 01.551$\pm$0.011 & 0.22(1)$\times$0.01(6) & 136$\pm$5 & Low& GBS$^g$ & 3 \\ 
22 & 05 35 43.2516$\pm$0.0011 & $-$05 09 16.943$\pm$0.016 & 0.27(3)$\times$0.19(3) & 19$\pm$20 & Low\\
23 & 05 35 19.620$\pm$0.006 & $-$05 09 20.36$\pm$0.09 &unresolved & ... & Medium \\
24 & 05 35 26.9293$\pm$0.0016 & $-$05 09 22.408$\pm$0.024 & 0.26(9)$\times$0.24(13) & 175$\pm$86 & High&HOPS 66-B&4\\
25 & 05 35 26.8588$\pm$0.0010 & $-$05 09 24.387$\pm$0.015 & 0.36(4)$\times$0.06(8) & 106$\pm$2 & High&HOPS 66-A&4  \\
26 & 05 35 45.2699$\pm$0.0015 & $-$05 09 23.485$\pm$0.023 & 0.56(6)$\times$0.26(5) & 157$\pm$6 & Low\\ 
27 & 05 35 25.2418$\pm$0.0010 & $-$05 09 27.393$\pm$0.014 & 0.38(15)$^h$ & 137$\pm$33 & Medium \\ 
28 & 05 35 31.9503$\pm$0.0017 & $-$05 09 27.79$\pm$0.03 & 0.31(9)$^h$ & 107$\pm$17 & Medium \\ 
29 & 05 35 31.971$\pm$0.003 &  $-$05 09 28.26$\pm$0.04 & unresolved & ... & Medium\\
30 & 05 35 45.417$\pm$0.019 & $-$05 09 41.4$\pm$0.3 & 6.7(7)$\times$4.2(4)$^j$& 90$\pm$8 & Low& & & BS \\ 
31 & 05 35 41.7768$\pm$0.0015 & $-$05 09 49.776$\pm$0.022 & 0.38(5)$\times$0.22(7) & 120$\pm$13 & Low\\
32 & 05 35 26.9643$\pm$0.0023 & $-$05 10 17.33$\pm$0.03 & unresolved & ... & High&HOPS 369&4 \\
33 & 05 35 25.148$\pm$0.006 & $-$05 10 23.02$\pm$0.09 & 2.1(3)$\times$0.5(5)$^k$& 161$\pm$5 & High & VLA 13$^e$ & 1 & CE\\
34 & 05 35 24.7274$\pm$0.0007 & $-$05 10 30.117$\pm$0.010 & 0.10(2)$\times$0.07(5) & 24$\pm$80 & High & VLA 13$^w$, GBS$^g$ & 1, 3\\
35 & 05 35 32.923$\pm$0.008 & $-$05 10 27.10$\pm$0.11 & unresolved$^k$& ... & Medium & & & E \\ 
36 & 05 35 47.7962$\pm$0.0022 & $-$05 10 30.83$\pm$0.03 & unresolved & ... & Low \\ 
37 & 05 35 26.866$\pm$0.003 & $-$05 11 07.53$\pm$0.04 & unresolved & ... & High  \\ 
38 & 05 35 24.472$\pm$0.004 & $-$05 11 58.06$\pm$0.06 & unresolved & ... & High  \\
39 & 05 35 23.2891$\pm$0.0011 & $-$05 12 03.398$\pm$0.017 & 0.24(4)$\times$0.09(6) & 65$\pm$17 & High & HOPS 60&4 \\
40 & 05 35 21.3186$\pm$0.0007 & $-$05 12 12.671$\pm$0.010 & 0.30(2)$^h$ & 48$\pm$3 & Medium & VLA 14?$^l$& 1\\ 
41& 05 35 22.8447$\pm$0.0015 & $-$05 12 43.554$\pm$0.023 & 0.19(9)$^h$ & 156$\pm$24 & High  \\ 
42 & 05 35 20.1534$\pm$0.0008 & $-$05 13 15.752$\pm$0.012 & 0.52(8)$\times$0.18(9) & 11$\pm$7 & High & HOPS 59-A &4 & CE\\
43 & 05 35 21.3660$\pm$0.0008 & $-$05 13 17.795$\pm$0.012 & unresolved & ... & High &HOPS 409&4 \\
44 & 05 35 15.7575$\pm$0.0011 & $-$05 13 32.305$\pm$0.017 &  0.27(3)$\times$0.18(3) & 175$\pm$17 & Medium \\ 
45 & 05 35 23.1236$\pm$0.0020 & $-$05 13 43.56$\pm$0.03 & 0.39(8)$\times$0.09(9) & 24$\pm$10 & Medium \\ 
46 & 05 35 19.893$\pm$0.003 & $-$05 15 08.16$\pm$0.04 & unresolved & ... & Medium&HOPS 57-A&4 \\
47 & 05 35 20.7530$\pm$0.0007 & $-$05 15 49.245$\pm$0.010 & 0.31$\times$0.16 & 21$\pm$5 & Medium& GBS$^g$ & 3  \\
\hline
\end{tabular*}
\begin{tablenotes}
\footnotesize
\item{$^a$Positions and sizes derived from elliptical Gaussian fits in the A-configuration C-band map, except for sources 4, 10, and 27, which are not detected in this map, and for whom the C-configuration C, X and K-band maps were used, respectively (see Table \ref{tab:maps}). Errors in deconvolved size are given in parenthesis. Position uncertainties correspond to those of absolute positions, and are calculated adding in quadrature a systematic error of 0\farcs01, accounting for uncertainties introduced by the phase calibration process \citep{Dzib2017}, to the formal error of the fit, $0.5\,\theta / {\rm SNR}$ \citep{Reid1988}, where $\theta$ is the source size and SNR is the signal-to-noise ratio.  $^b$Position angle of the major axis of the source measured from north to east. 
$^c$Derived from the column density map obtained by \citet{Stutz2018} from {\it Herschel} dust emission data (160 - 500 $\micron$), shown in Fig. \ref{fig:filament}. High: $>6 \times10^{22}$ cm$^{-2}$; Medium: $(2.3-6) \times10^{22}$ cm$^{-2}$; Low: $<2.3\times10^{22}$ cm$^{-2}$. 
$^d$E: Irregular extended source. CE: Source formed of a rather compact core and an extension elongated at a different P.A. The fit has been performed to the core. BS: Source with a bow-shock morphology. 
$^e$Split into two sources in our observations. 
$^f$From \citet{Reipurth2004}, different than source VLA 15 from \citet{Osorio2017}. 
$^g$Full name is GBS-VLA J$hhmmss.ss-ddmmss.s$, where J$hhmmss.ss-ddmmss.s$ are the J2000.0 coordinates of the source. 
$^h$Source resolved only in one direction. The major axis size is obtained using task IMFIT of AIPS. 
$^i$This position is displaced about $0\farcs1$ from the position of the protostar, see Section \ref{10and11}. 
$^j$From a map obtained with uvtaper = 50 k$\lambda$ (beam $\simeq 2\arcsec$) to better detect the extended emission. 
$^k$From a map obtained with uvtaper = 100 k$\lambda$ (beam $\simeq 1\arcsec$) to better detect the extended emission. 
$^l$Uncertain association. Position of source VLA 14, reported by \citet{Reipurth1999} from $8\arcsec$ angular resolution data, differs by $\sim$5$\farcs5$ from the position of our Source 40. }
{\bf References:} (1) \citealt{Reipurth1999}, (2) \citealt{Reipurth2004}, (3) \citealt{Kounkel2014}, (4) \citealt{Tobin2020I}.
\end{tablenotes}
\end{table*}

\clearpage

\begin{table*}
	\caption{Flux densities and spectral indices of the radio sources.}
	\label{tab:fluxes}
	\begin{tabular*}{\linewidth}{lcccc@{\extracolsep{1em}}ccc}
		\hline\hline
			\multicolumn{8}{c}{}\\
			&\multicolumn{4}{c}{ 2014 Flux Density$^a$}&{2015 Flux Density$^a$}&  &\\
			{Source} & \multicolumn{4}{c}{ (mJy)} &   {(mJy)} & {Spectral} &  {Notes$^c$}\\
			\cline{2-5}
			\cline{6-6}
			{Number} & {5 cm} & {3 cm} & {1.3 cm} & {0.7 cm}  & {5 cm} & {Index$^b$} \\
			\hline
			1 & $<$0.65$^d$ &... & ...& ...& 0.37$\pm$0.04 & $-$0.3$\pm$1.3$^e$ & \\
			2 & $<$0.65$^d$ & ... & ... & ...  & 0.062$\pm$0.012& ... &  \\
			3 & 0.67$\pm$0.08 & ... & ... & ... & 0.87$\pm$0.09 & $-$0.08$\pm$0.50$^e$ &  \\
			4 & 2.6$\pm$0.3 & 2.2$\pm$0.3$^f$ & ... & ... & ...$^g$ & $-$0.3$\pm$0.3 & E  \\
			5 & 0.082$\pm$0.018 & $<$0.13 & ... & ... & 0.057$\pm$0.012 & $\leq$2.0 &  \\
			6 & 0.094$\pm$0.018 & $$<$$0.12 & ... & ... & 0.104$\pm$0.016 & $\leq$0.6 &  \\
			7 & 0.055$\pm$0.015 & $<$0.07 & ... & ... & 0.034$\pm$0.009 & $\leq$1.8 &  \\
			8 & 0.057$\pm$0.014 & $<$0.06 & ... & ... & 0.066$\pm$0.008 & $\leq$0.2 &  \\
			9 & 0.27$\pm$0.03 & 0.23$\pm$0.03 & ... &  ... & 0.193$\pm$0.023 & 0.0$\pm$0.6 &  \\
			10 & 0.121$\pm$0.016$^h$ & 0.052$\pm$0.016 & ... & ... & $<$0.012 & $-$1.7$\pm$0.7$^i$ & V \\
			11 & 0.54$\pm$0.06$^h$ & 0.58$\pm$0.07 & ... & ... & 0.44$\pm$0.05 & 0.3$\pm$0.4& V \\
			12 & 0.063$\pm$0.013 & 0.066$\pm$0.016 & ... & ... & 0.26$\pm$0.03 & 0.1$\pm$0.6$^i$ & V \\
			13 & 0.36$\pm$0.04 & 0.32$\pm$0.05 & ... & ... & 0.32$\pm$0.04 & $-$0.13$\pm$0.20 &  \\
			14 & 0.81$\pm$0.16 & ... & ... & ... & 1.07$\pm$0.12 & $-$0.02$\pm$0.19$^e$ & V\\
			15 & $<$0.06 & $<$0.17 & ... & ... & 0.072$\pm$0.010 & ... &  \\
			16 & $<$0.04 & $<$0.05 & ... & ... & 0.026$\pm$0.005 & $\leq$2.3 &  \\
			17 & 0.049$\pm$0.011 & $<$0.03 & $<$0.08 & ... & 0.067$\pm$0.010 & $\leq$$-$0.6 &  \\
			18 & 0.037$\pm$0.012 & $<$0.04 & ... & ... & 0.036$\pm$0.008 & $\leq$0.24 &  \\
			19 & 1.36$\pm$0.14 & 2.16$\pm$0.23 & ... & ... & 0.039$\pm$0.012 & 0.9$\pm$0.3$^i$ & V\\
			20 & 1.93$\pm$0.21 & ... & ... & ... & 1.72$\pm$0.21$^j$ & ... & E \\
			21 & 1.04$\pm$0.11 & 0.76$\pm$0.13 & ... & ... & 0.39$\pm$0.04 & $-$0.6$\pm$0.4$^i$ & V\\
			22 & 0.26$\pm$0.05 & ... & ... & ... & 0.19$\pm$0.03 & ... &  \\
			23 & 0.065$\pm$0.020& $<$0.05& ... & ... & 0.075$\pm$0.016 & $\leq$$-$0.4 &  \\
			24 & $<$0.14$^d$ & $<$0.25$^d$ & 0.058$\pm$0.014 & $<$0.14 & 0.040$\pm$0.007 & 0.29$\pm$0.24 &  \\
			25 & $<$0.14$^d$ & $<$0.25$^d$ & 0.20$\pm$0.03 & 0.26$\pm$0.07 & 0.088$\pm$0.011 & 0.57$\pm$0.07 &  \\
			26 & 0.29$\pm$0.06 & ... & ... & ... & 0.26$\pm$0.04 & ... &  \\
			27 & 0.121$\pm$0.017 & 0.20$\pm$0.03 & 0.53$\pm$0.06 & 0.96$\pm$0.15 & $<$0.009 & 1.06$\pm$0.05 & V\\
			28 & $<$0.06$^d$ & $<$0.08$^d$ & $<$0.17$^d$ &  ... & 0.033$\pm$0.005 & 0.74$\pm$0.03$^d$&  \\
			29 & $<$0.06$^d$& $<$0.08$^d$ & $<$0.17$^d$ &  ... & 0.022$\pm$0.005 & 0.74$\pm$0.03$^d$ &  \\
			30 & 3.2$\pm$0.3 & ... & ... & ... & 3.2$\pm$0.4$^j$ & ... & E \\
			31 & 0.117$\pm$0.021 & $<$0.19 & ... & ... & 0.128$\pm$0.019 & $\leq$1.0 &  \\
			32 & 0.032$\pm$0.010 & $<$0.04 & 0.045$\pm$0.010 & $<$0.07 & 0.025$\pm$0.004 & 0.36$\pm$0.18 &  \\
			33 & 0.49$\pm$0.06 & 0.49$\pm$0.06 & 0.46$\pm$0.11$^k$& $<$2.5$^l$ & 0.49$\pm$0.07$^k$ & $-$0.05$\pm$0.01 & E \\
			34 & 0.44$\pm$0.05 & 0.58$\pm$0.06 & 0.99$\pm$0.10 & 0.69$\pm$0.14 & 0.34$\pm$0.04 & 0.73$\pm$0.14 &  \\
			35 & 0.19$\pm$0.03 & 0.077$\pm$0.016 & $<$0.15 & ... & 0.084$\pm$0.017$^k$& $-$1.7$\pm$0.5$^i$ & E, V \\
			36 & $<$0.12 & ... &  ... &  ... & 0.16$\pm$0.03 & 1.0$\pm$0.4$^e$& V\\
			37 & $<$0.04 & 0.033$\pm$0.009 & $<$0.07 & ... & 0.019$\pm$0.004 & 1.2$\pm$0.7 &  \\
			38 & $<$0.05 & 0.057$\pm$0.014 & ... & ... & 0.031$\pm$0.008 & 1.3$\pm$0.8 &  \\
			39 & 0.13$\pm$0.03 & 0.19 $\pm$0.03 & ... & ... & 0.089$\pm$0.012 & 1.1$\pm$0.7 &  \\
			40 & 0.56$\pm$0.06 & 0.53$\pm$0.07 & ... & ... & 0.84$\pm$0.08 & $-$0.1$\pm$0.3 & \\ 
			41 & $<$0.07 & $<$0.11 & ... & ... & 0.065$\pm$0.009 & $\leq$1.9 &  \\
			42 & 0.76$\pm$0.09 & 0.59$\pm$0.09 & ... & ... & 0.58$\pm$0.07 & $-$0.3$\pm$0.5 &  \\
			43 & 0.061$\pm$0.016 & $<$0.20 & ... & ... & 0.044$\pm$0.012 & $\leq$3.4 &  \\ 
			44 & 0.20$\pm$0.05 & ... & ... & ... & 0.23$\pm$0.03 & ... &  \\
			45 & $<$0.13 & ... & ... & ... & 0.073$\pm$0.016 & $-$0.06$\pm$0.69$^e$  & V \\
			46 & $<$0.20 & ... & ... & ... & 0.23$\pm$0.05 & ... &  \\ 
			47 & 1.17$\pm$0.11 & ... & ... & ... & 1.31$\pm$0.14 & $-$0.18$\pm$0.64$^e$ &   \\ 
			\hline
		\end{tabular*}
		\begin{tablenotes}
			\footnotesize
			\item{$^a$Flux densities at the representative wavelength of the observing bands, given in Table \ref{tab:maps}. Upper limits are 3$\sigma$, except when noted. No data (...) means the source is outside the imaged field of view, except when noted.
				$^b$The spectral index is obtained from a least-squares fit (in log-log scale) to the 2014 and 2015 detections, taking into account upper limits only when they are useful to constrain the fit (see Fig. \ref{fig:SEDs}).
				$^c$E: Source with extended emission. V: Suspected radio variable; the source presents changes by a factor greater than 1.5 (taking into account uncertainties, see Section \ref{subsec:variab}) in the flux density at a given frequency band (including data from the literature, see Fig. \ref{fig:SEDs}); in the case of source 34, the suspected variability is based on the minimum flux density from the observations of \citet{Kounkel2014}, derived indirectly from their table 2.
				$^d$The sources are blended. Upper limit corresponds to the total flux density of the two sources.
				$^e$Adopted from \citet{Kounkel2014}.
				$^f$Measured in a map with uvtaper = 20 k$\lambda$ in order to better detect the extended emission.
				$^g$Emission of this extended source is filtered out by the interferometer in the A configuration.
				$^h$The sources are blended. The flux density given for each source is obtained from a two-Gaussian fit.
				$^i$Derived using only 2014-epoch data since the 2015 flux density is significantly different.
				$^j$Measured in a map with uvtaper = 50 k$\lambda$ in order to better detect the extended emission.
				$^k$Measured in a map with uvtaper = 100 k$\lambda$ in order to better detect the extended emission.
				$^l$3$\sigma$ upper limit obtained according to the prescriptions given by \citet[Appendix A]{Beltran2001} and assuming the source is extended with a size $\sim 6\arcsec\times2\arcsec$.}
		\end{tablenotes}
	\end{table*}

\clearpage

\section{VLA maps at 5 cm of the detected radio sources}

	\begin{figure*} 
		{\includegraphics[height=0.18\textheight]{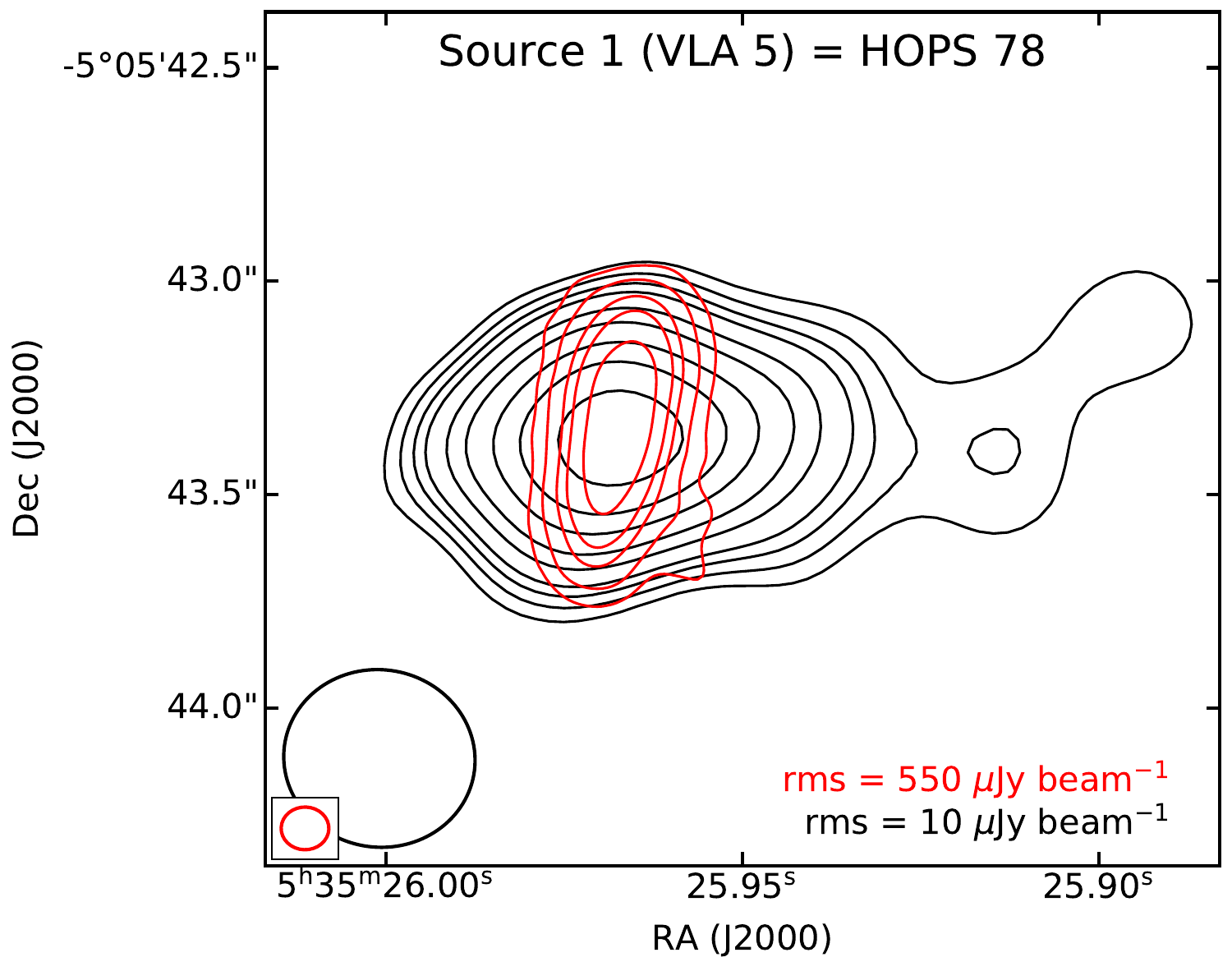}\hfill\includegraphics[height=0.18\textheight]{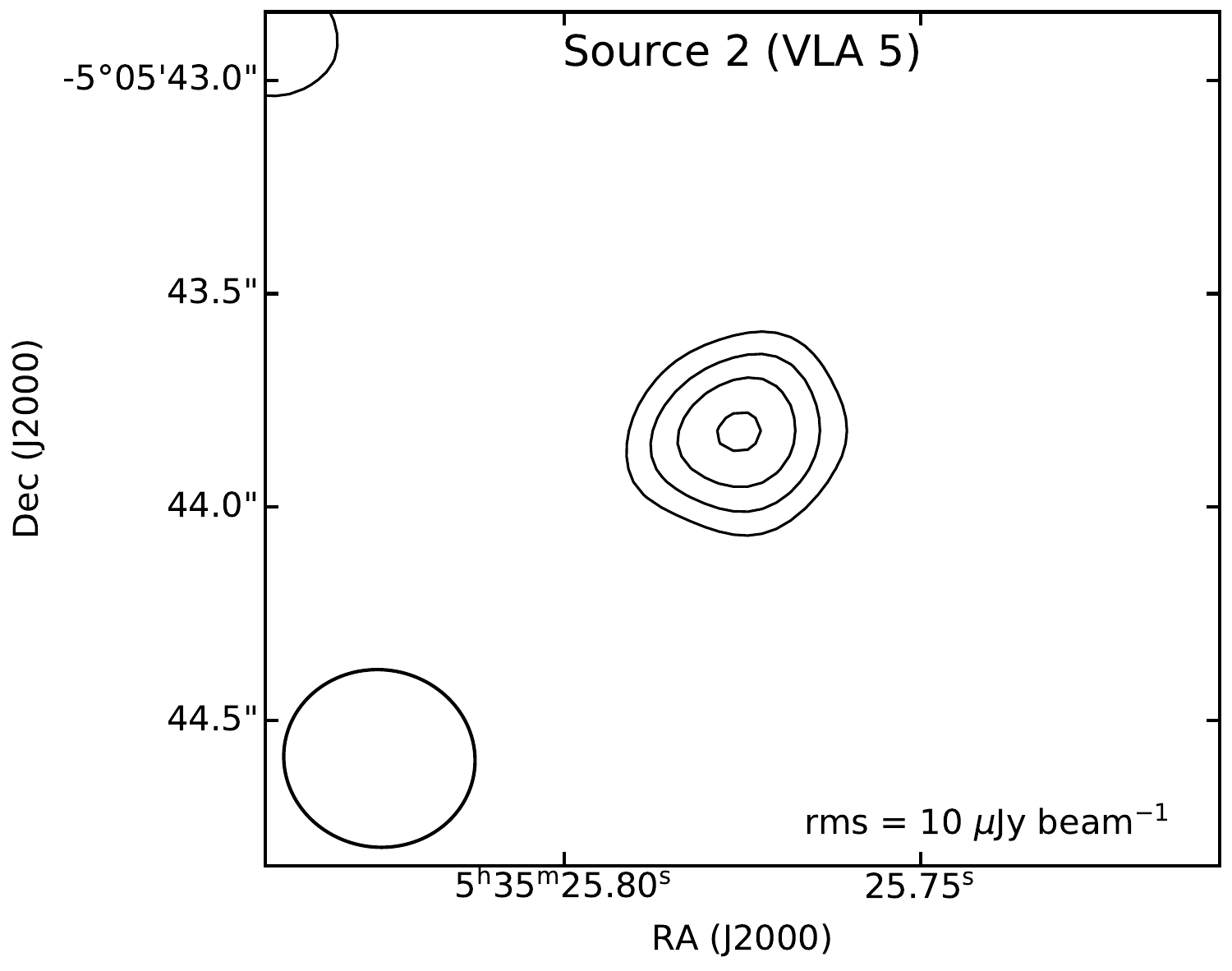}\hfill\includegraphics[height=0.18\textheight]{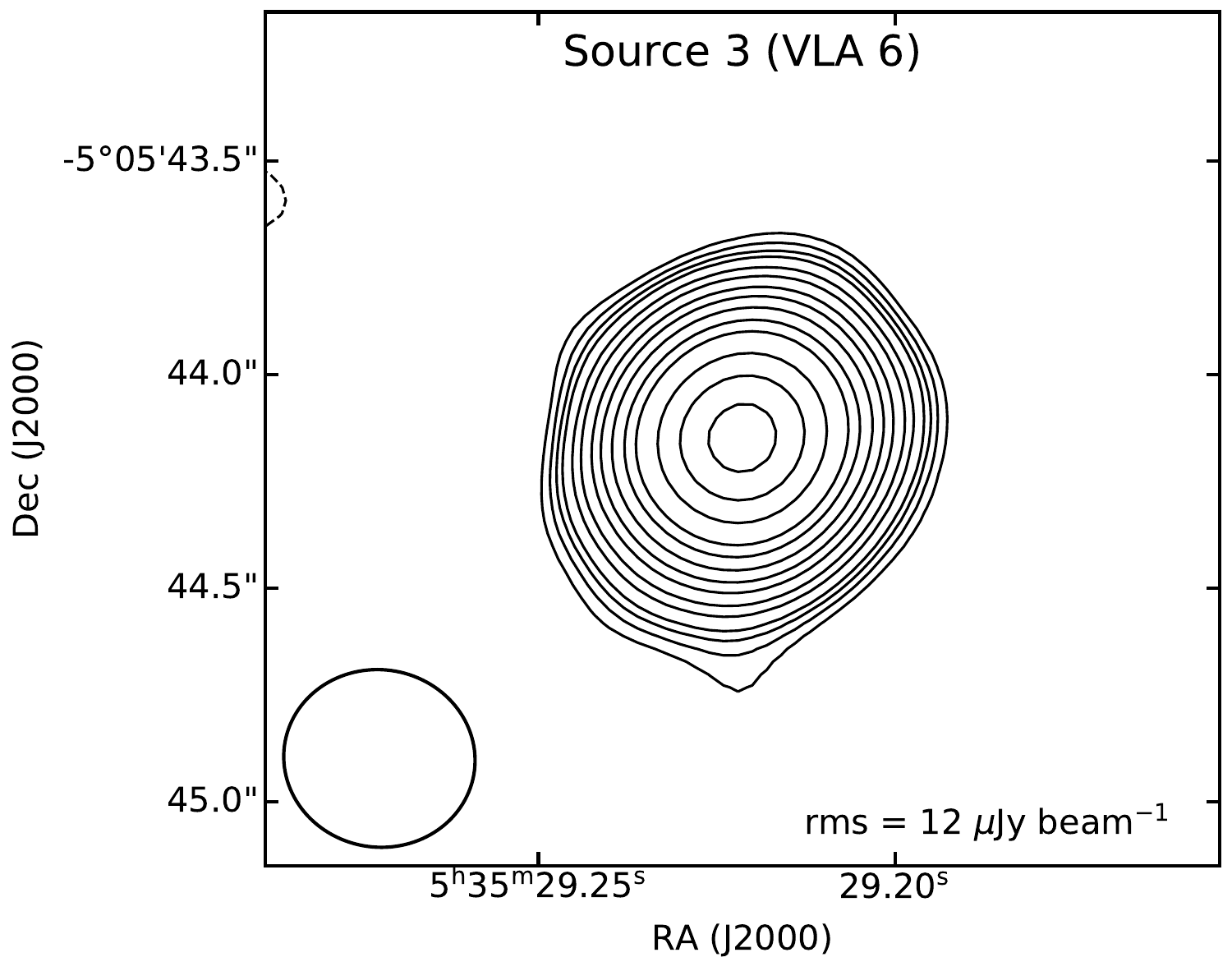}}\\
		{\includegraphics[height=0.18\textheight]{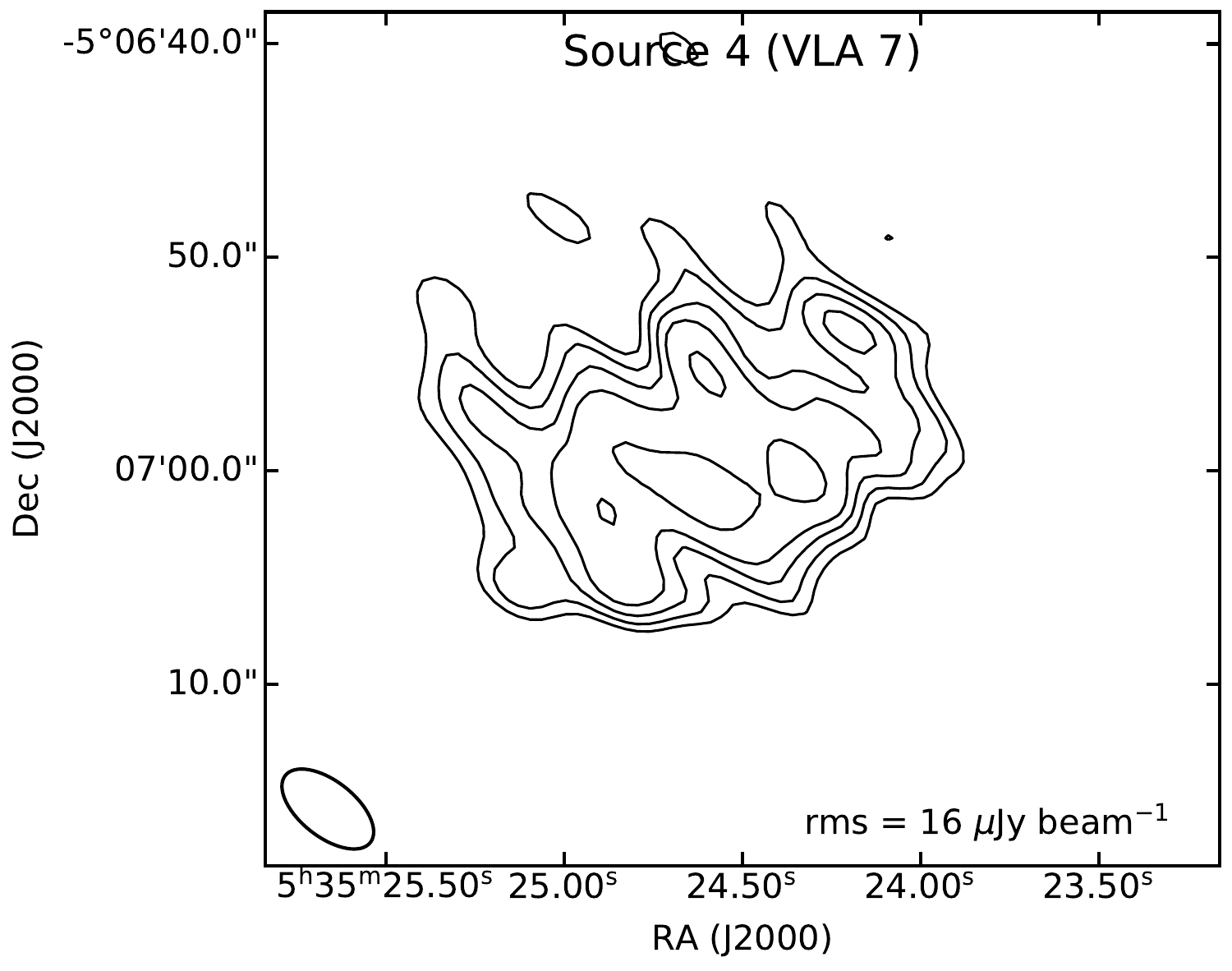}\hfill\includegraphics[height=0.18\textheight]{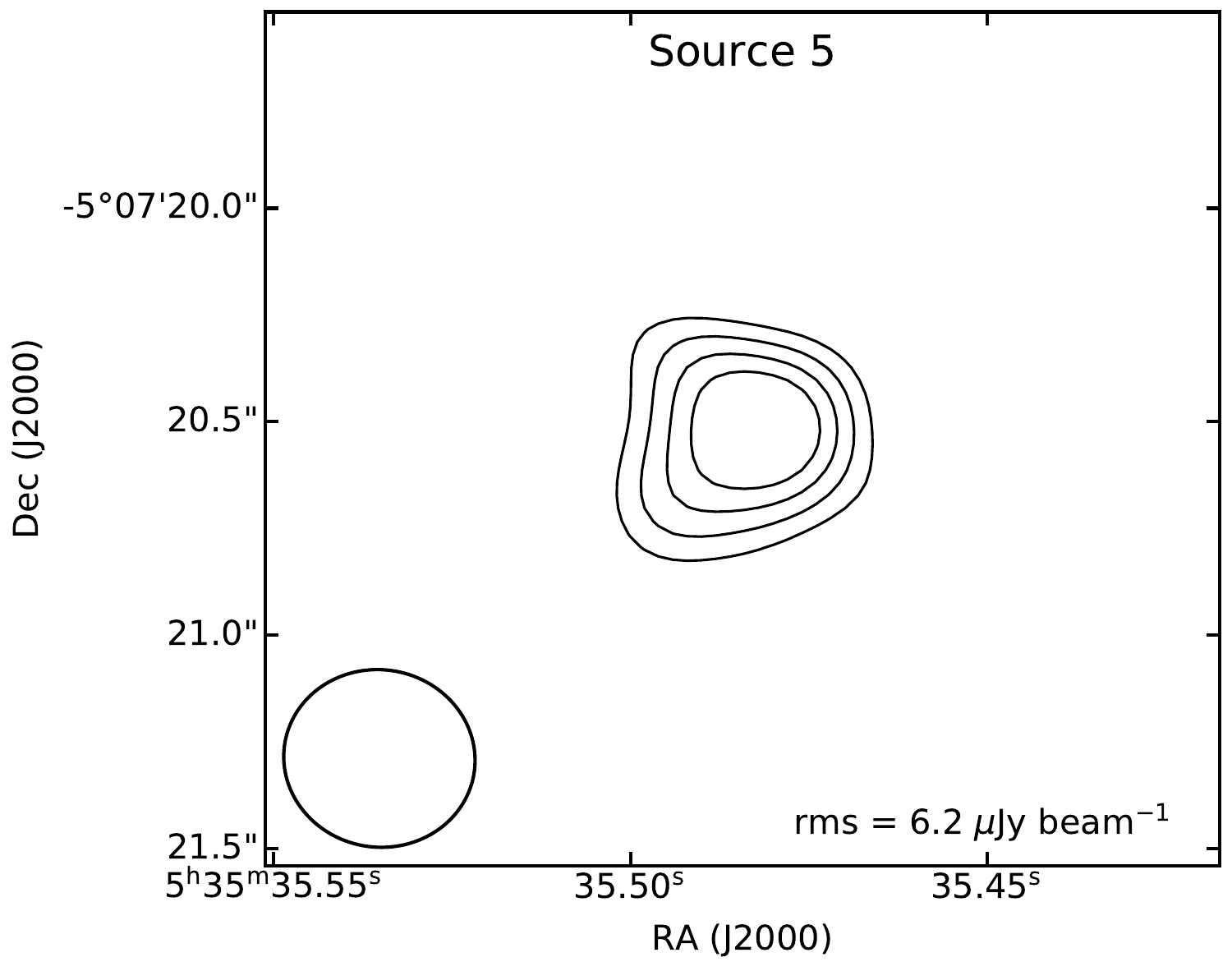}\hfill\includegraphics[height=0.18\textheight]{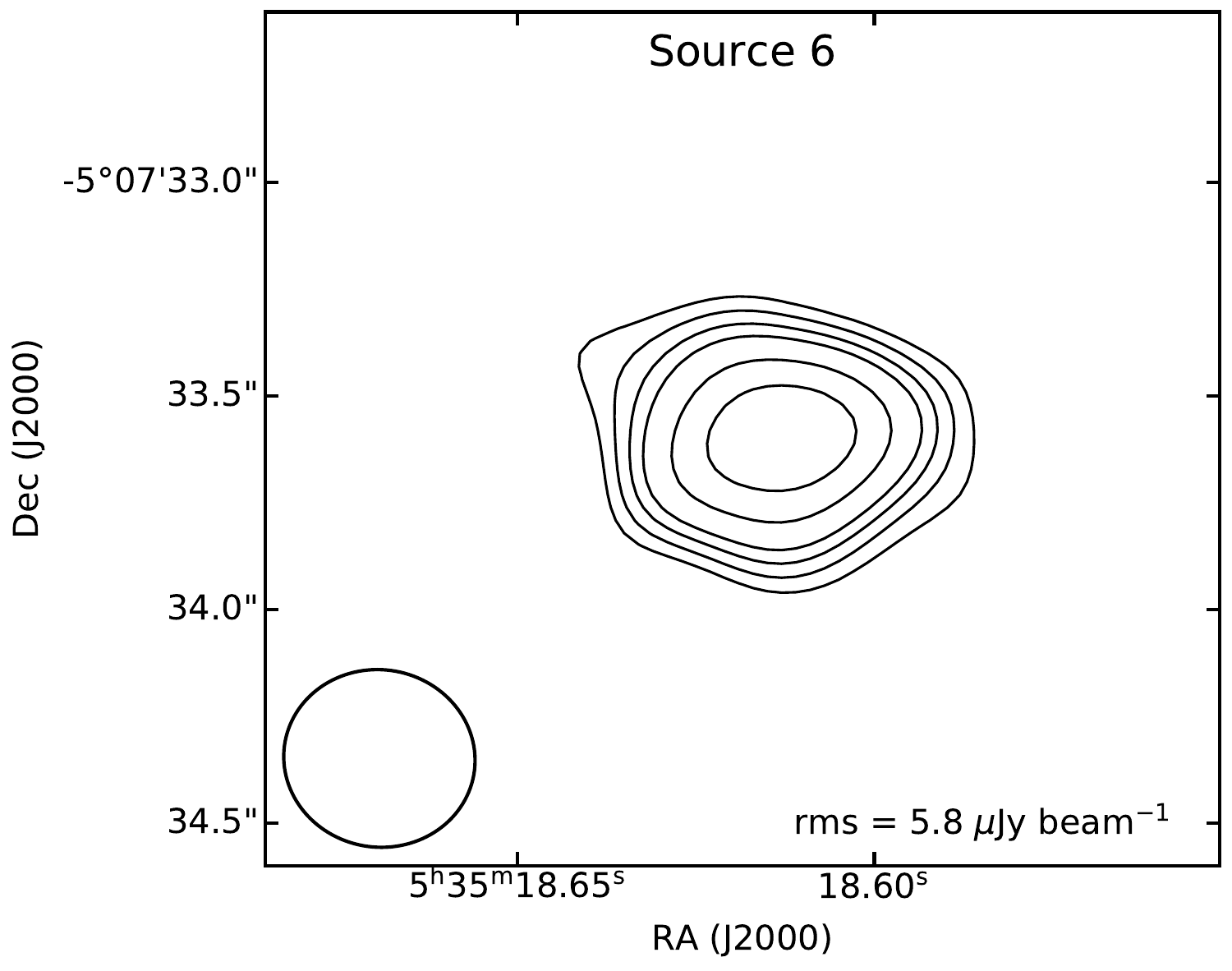}}\\
		{\includegraphics[height=0.18\textheight]{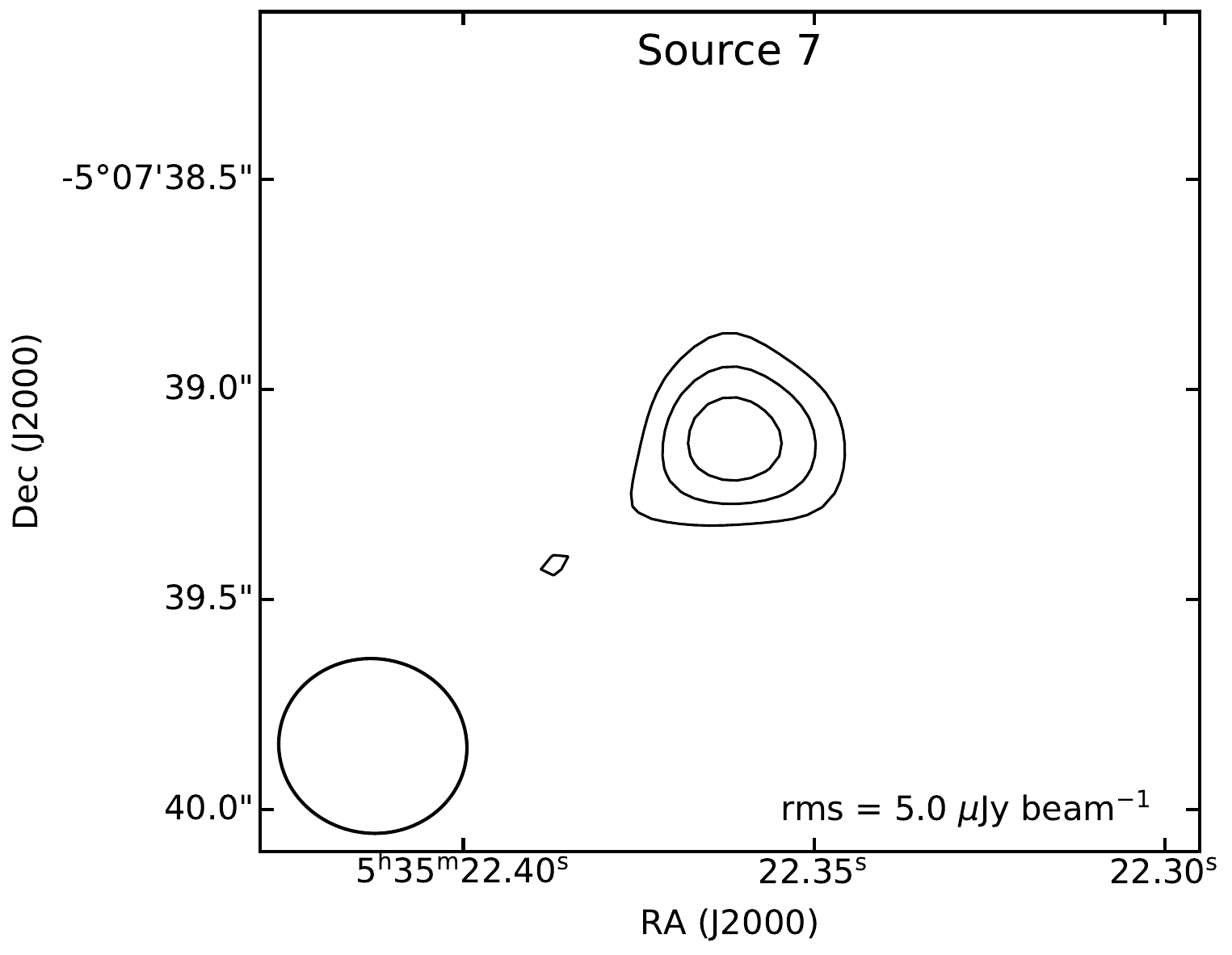}\hfill\includegraphics[height=0.18\textheight]{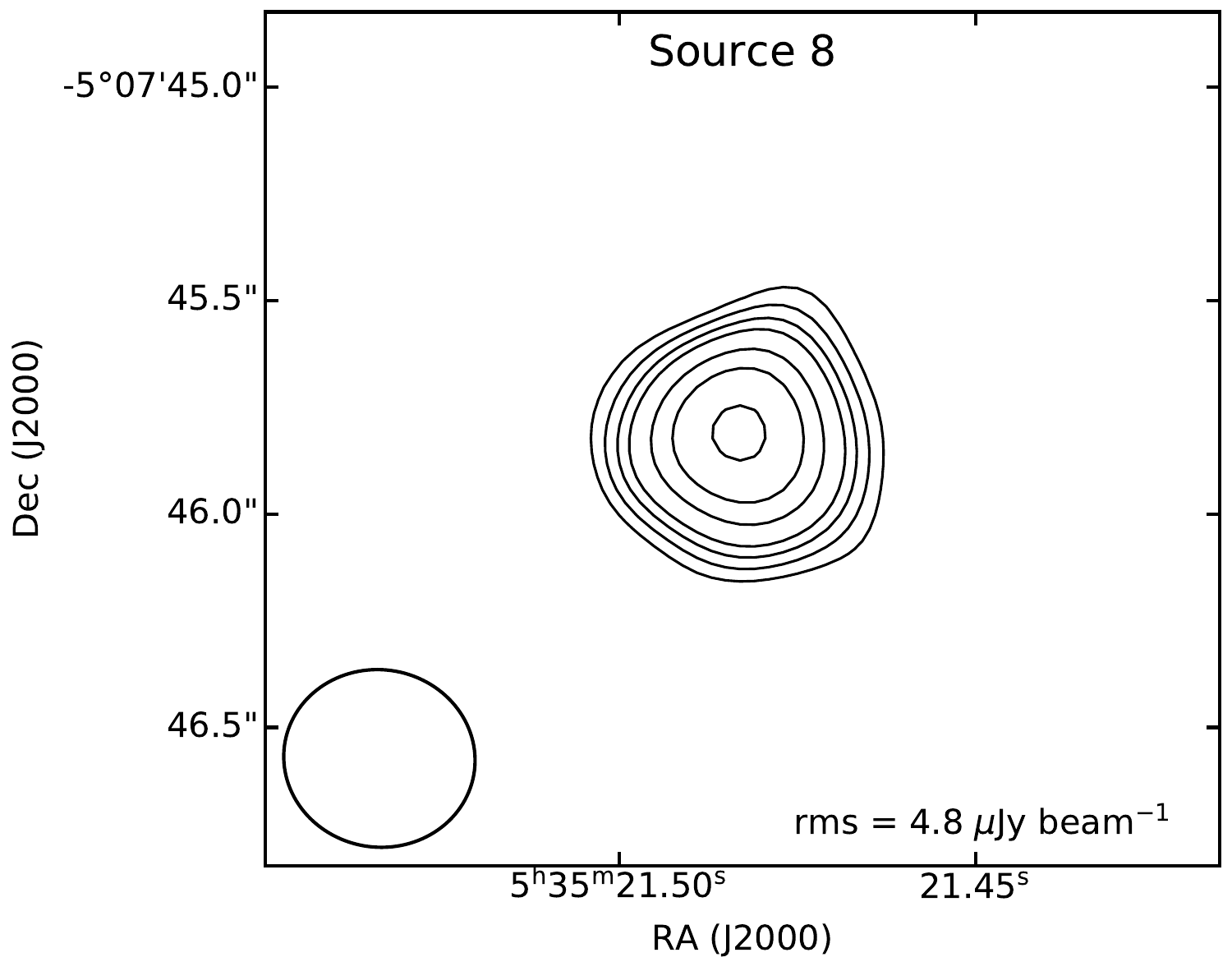}\hfill\includegraphics[height=0.18\textheight]{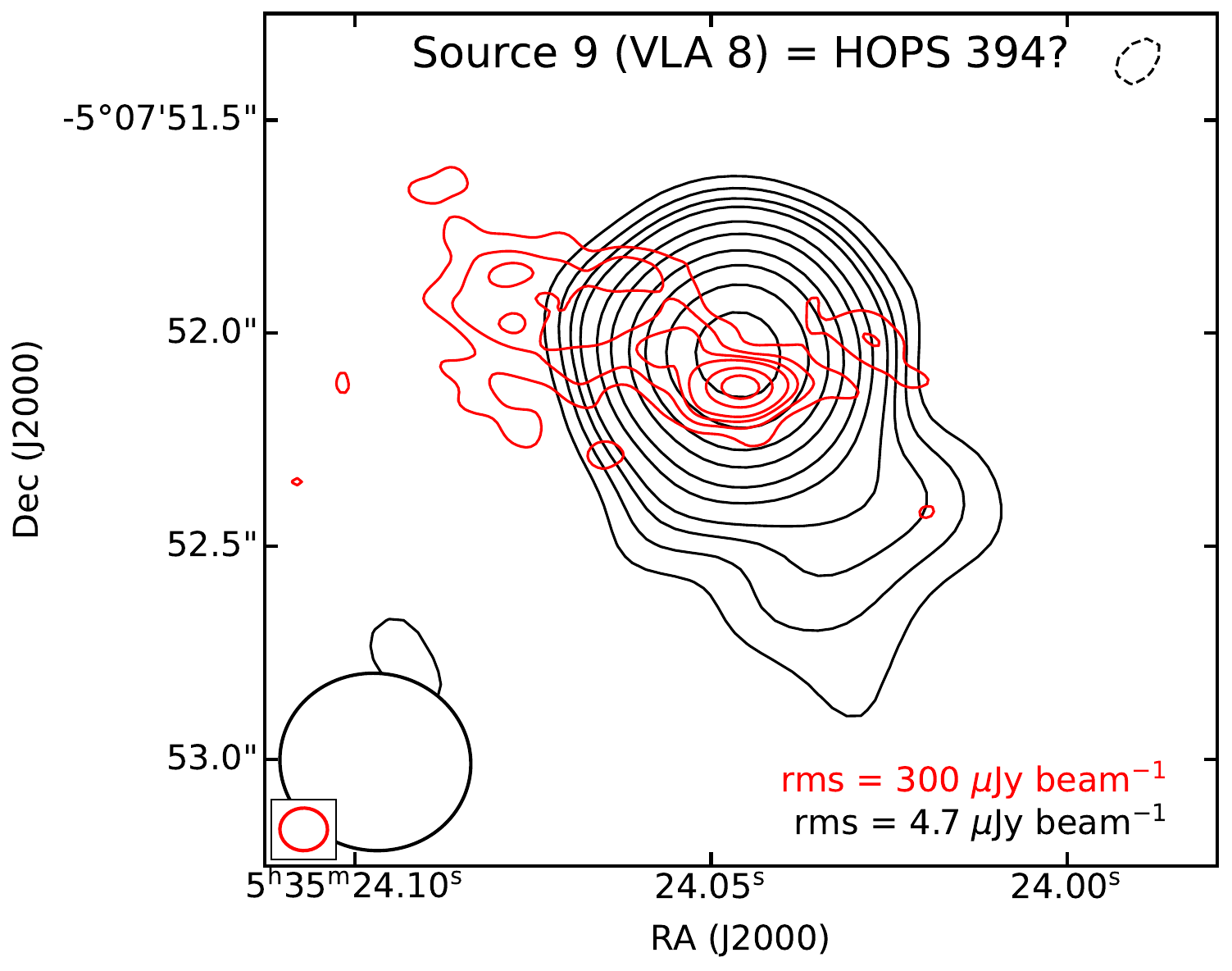}}\\
		{\includegraphics[height=0.18\textheight]{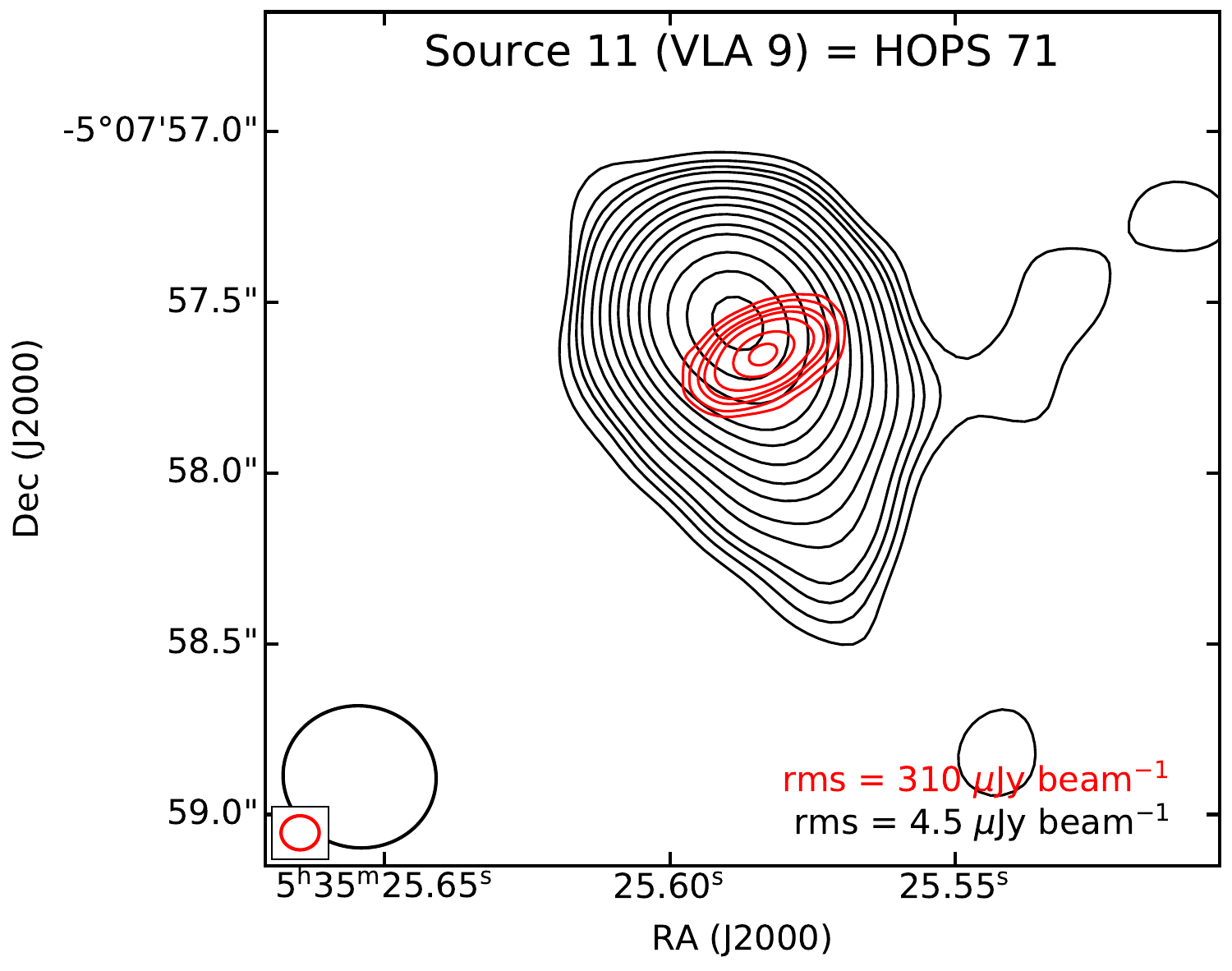}\hfill\includegraphics[height=0.18\textheight]{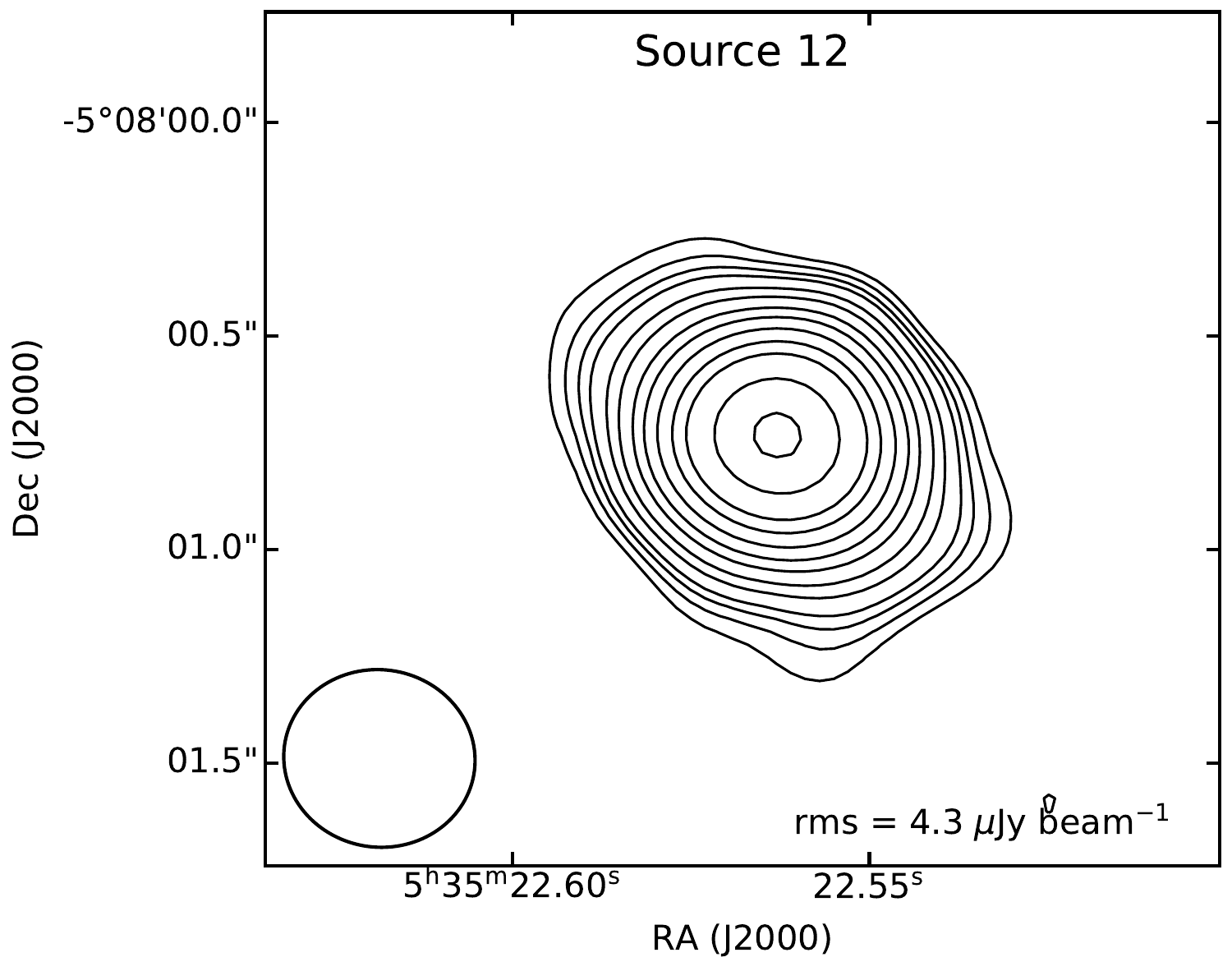}\hfill\includegraphics[height=0.18\textheight]{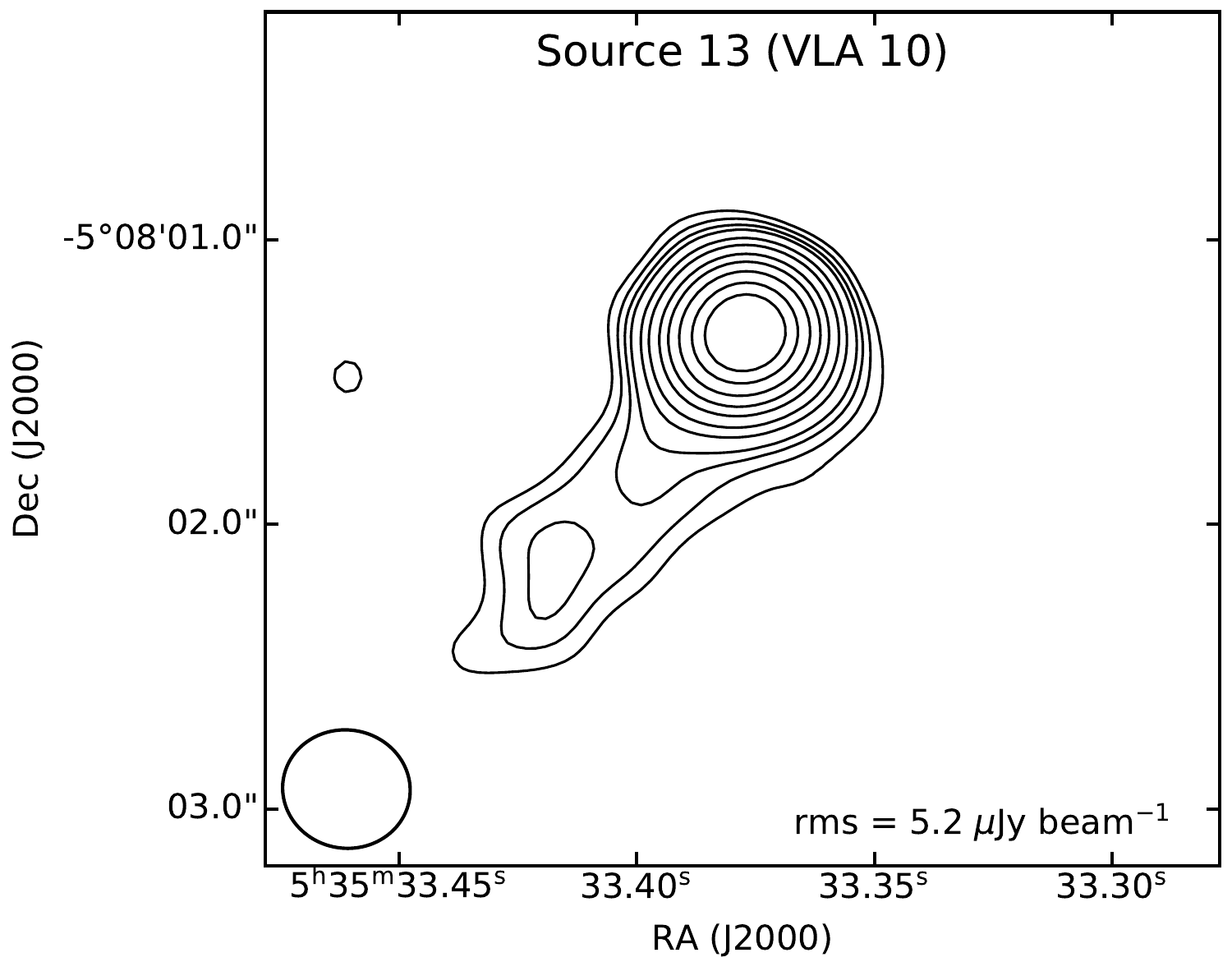}}\\
		{\includegraphics[height=0.18\textheight]{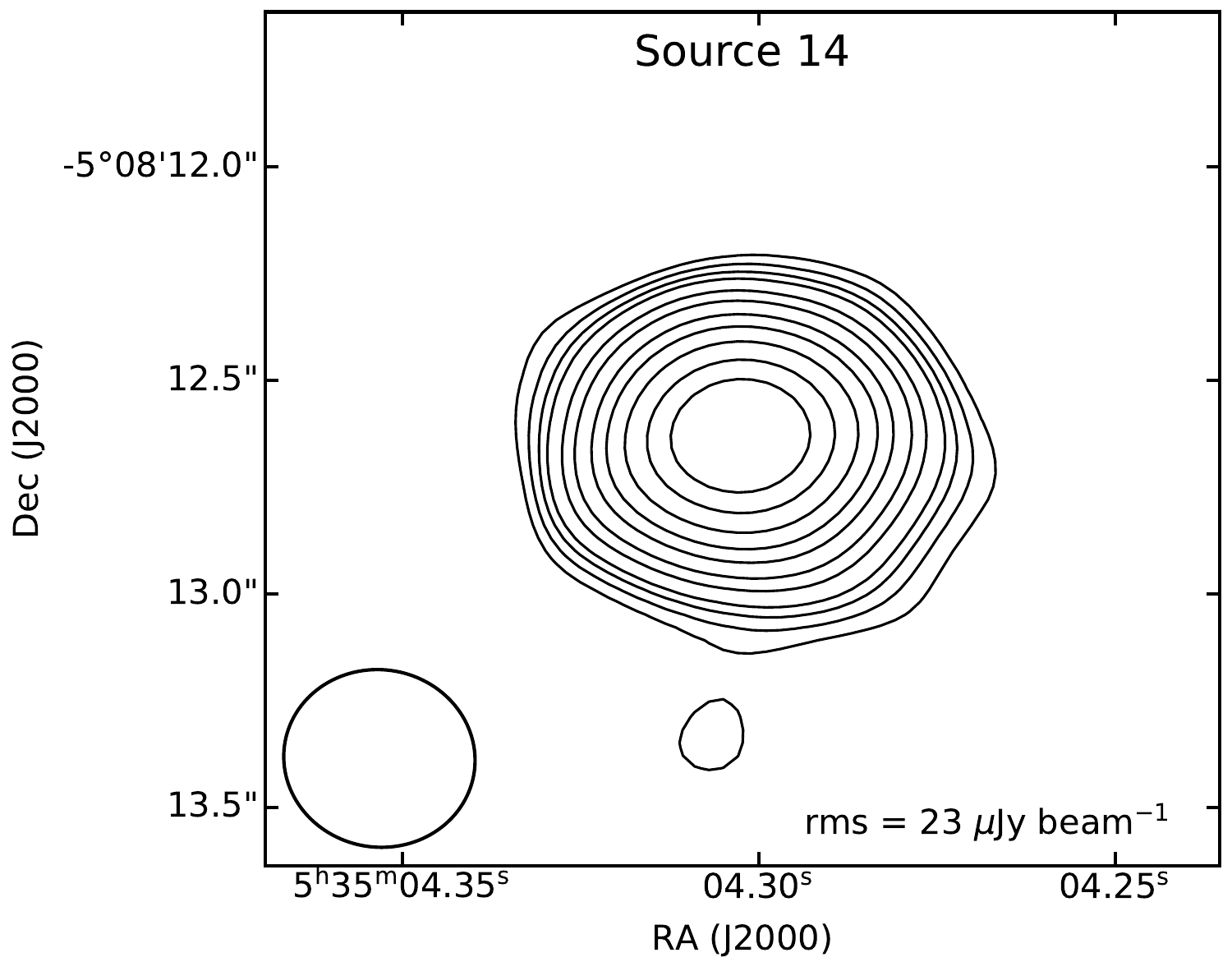}\hfill\includegraphics[height=0.18\textheight]{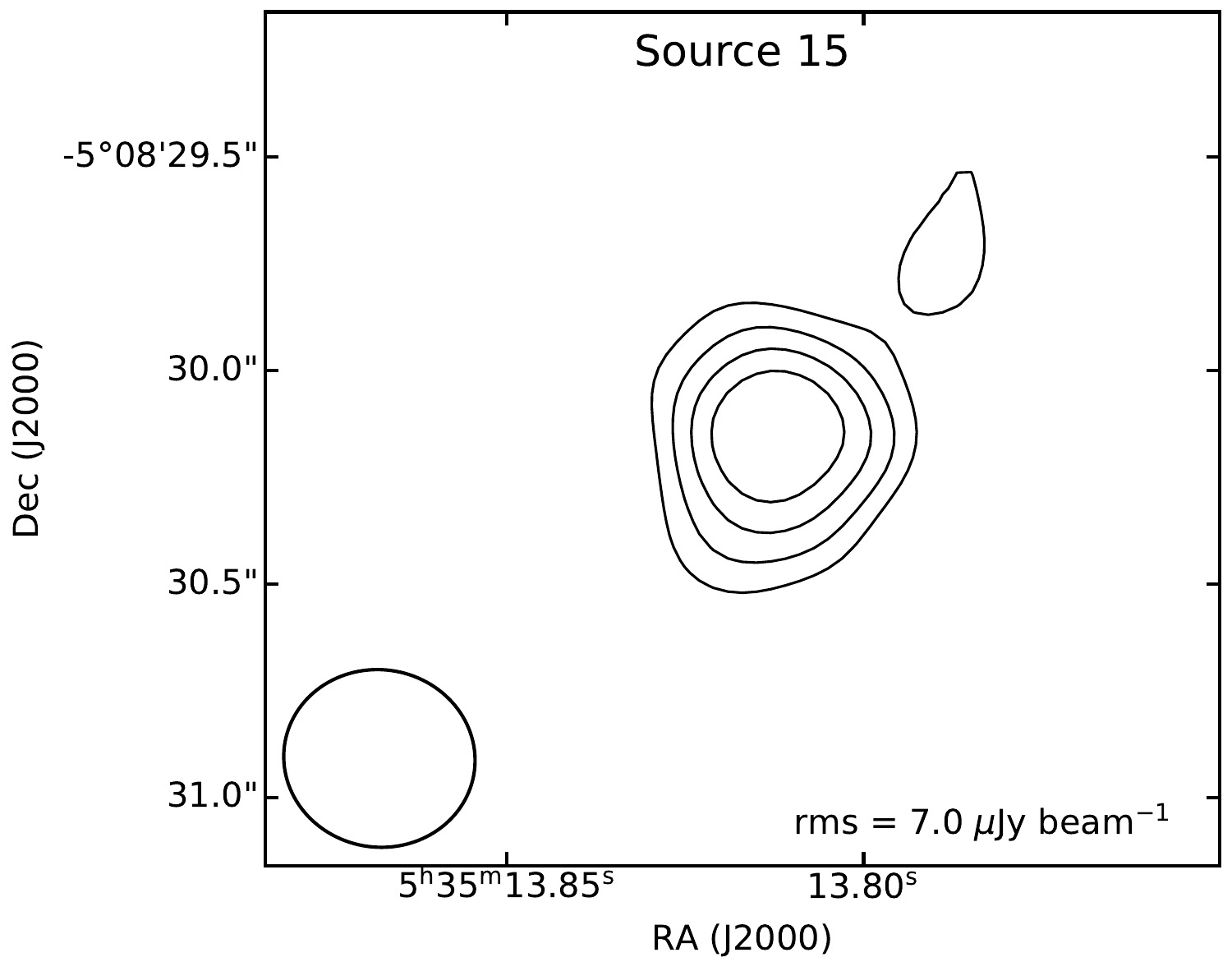}\hfill\includegraphics[height=0.18\textheight]{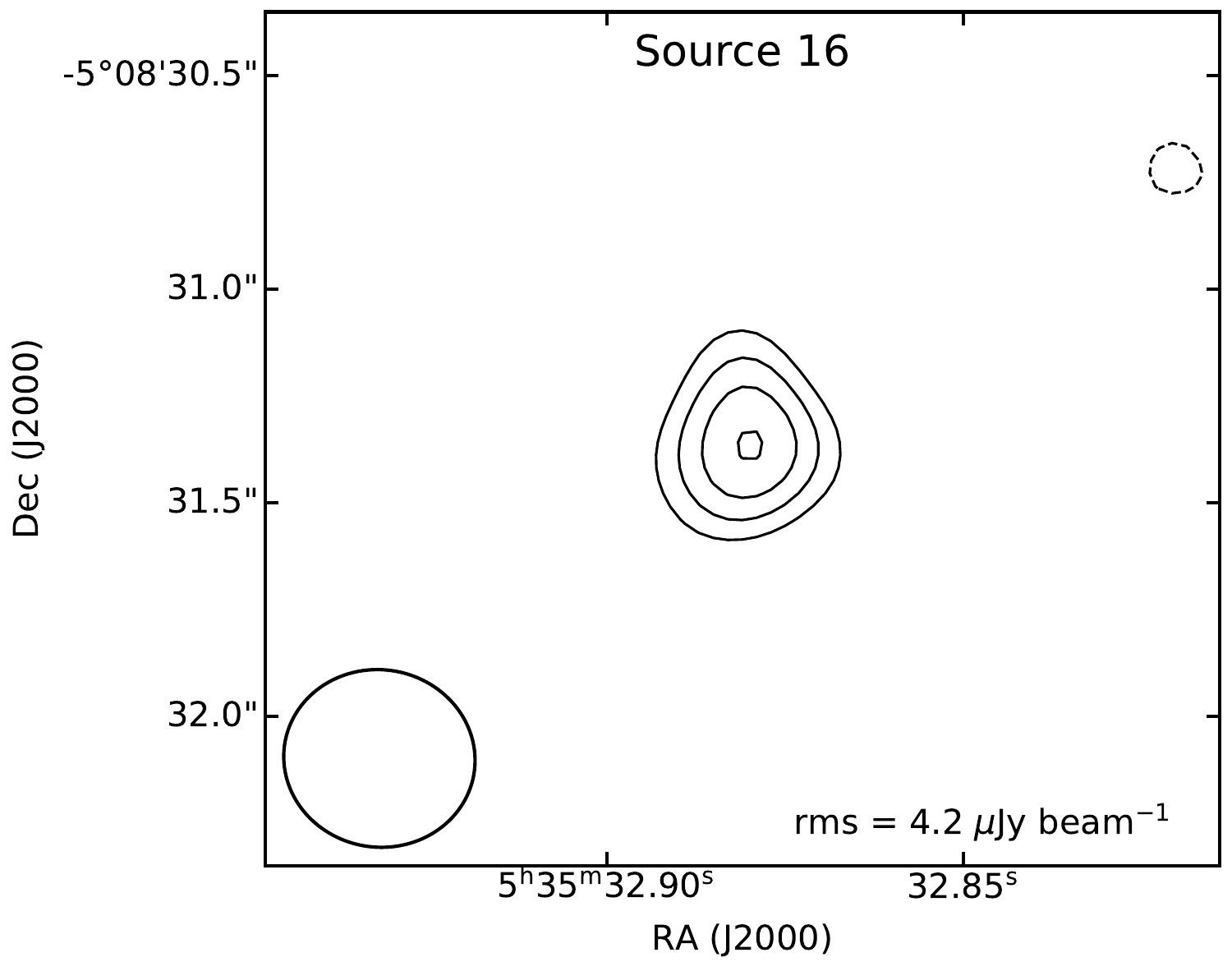}}
		\caption{VLA A-configuration maps of the detected radio sources obtained with natural weighting at 5 cm (black contours) with ALMA emission at 870 $\micron$ overlayed (red contours, adapted from \citealt{Tobin2020I}), except when noted. For sources 34, 38 and 40, A-configuration robust 0.75, 0.5, and $-$1 weighting maps, respectively, are presented; for sources 4 and 27, C-configuration robust 0 weighting maps at 5 cm and 1.3 cm, respectively, are presented; for Source 10, see Fig. \ref{fig:10&11}, top panel. To better detect the extended emission, maps with uvtaper 50 k$\lambda$ were used for sources 20 and 30, and with 100 k$\lambda$ for sources 33 and 35. VLA contours (black) are $-$3, 3, 4, 5, 6, 8, 10, 13, 16, 20, 25, 30, 40, 50, 60, 70, and 80 times the rms of the maps. ALMA contours (red) are 5, 10, 20, 30, 50, 100, and 150 times the rms of the maps, except for sources 9 and 34, where contours are 4, 6, 8, 10, 15, 20, 30, 50 and 70 times the rms of the map. The beam and the rms are shown in the bottom left and right corners, respectively, of each map.}
		\label{fig:sources}
	\end{figure*}
	
	\begin{figure*} 
		{\includegraphics[height=0.18\textheight]{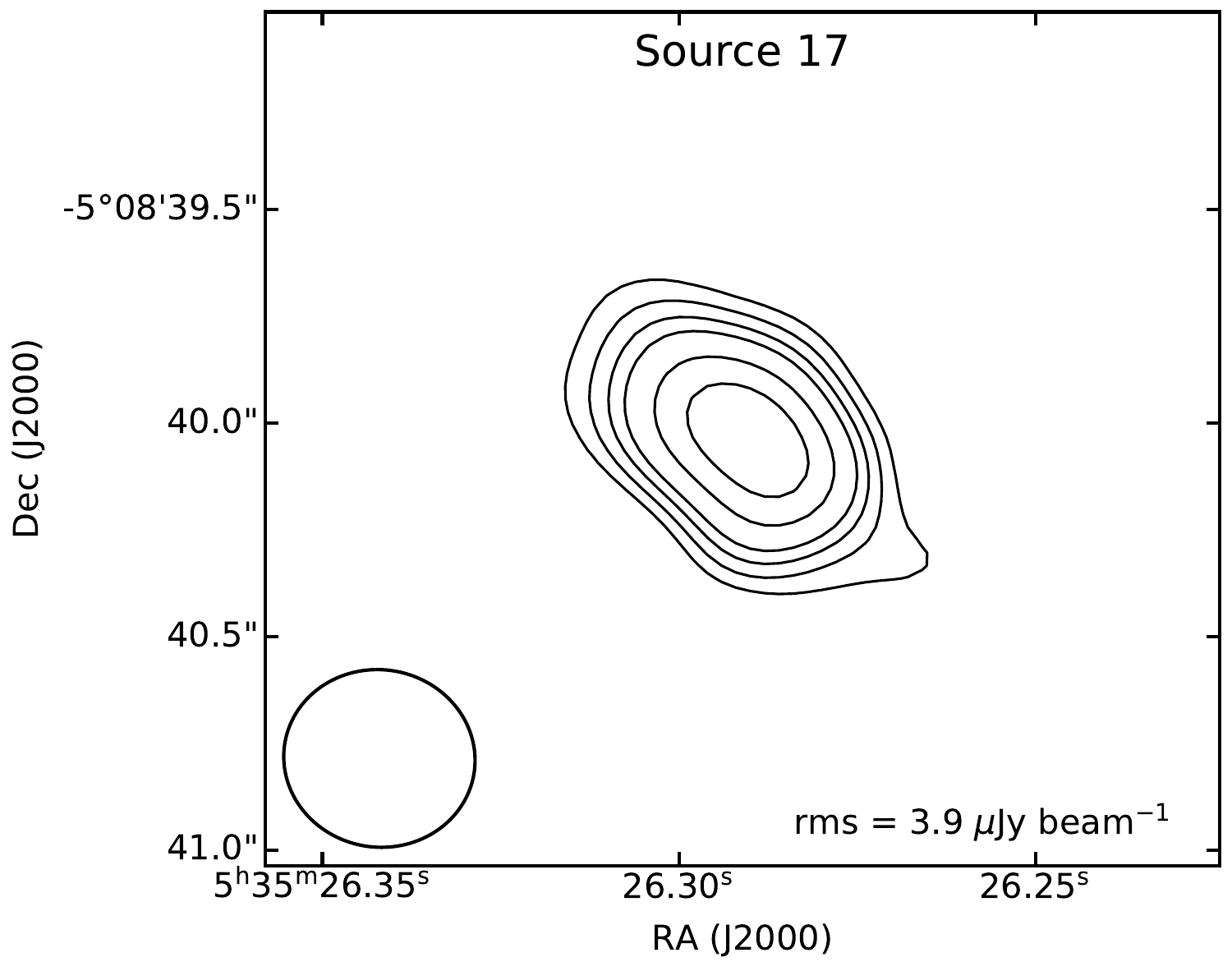}\hfill\includegraphics[height=0.18\textheight]{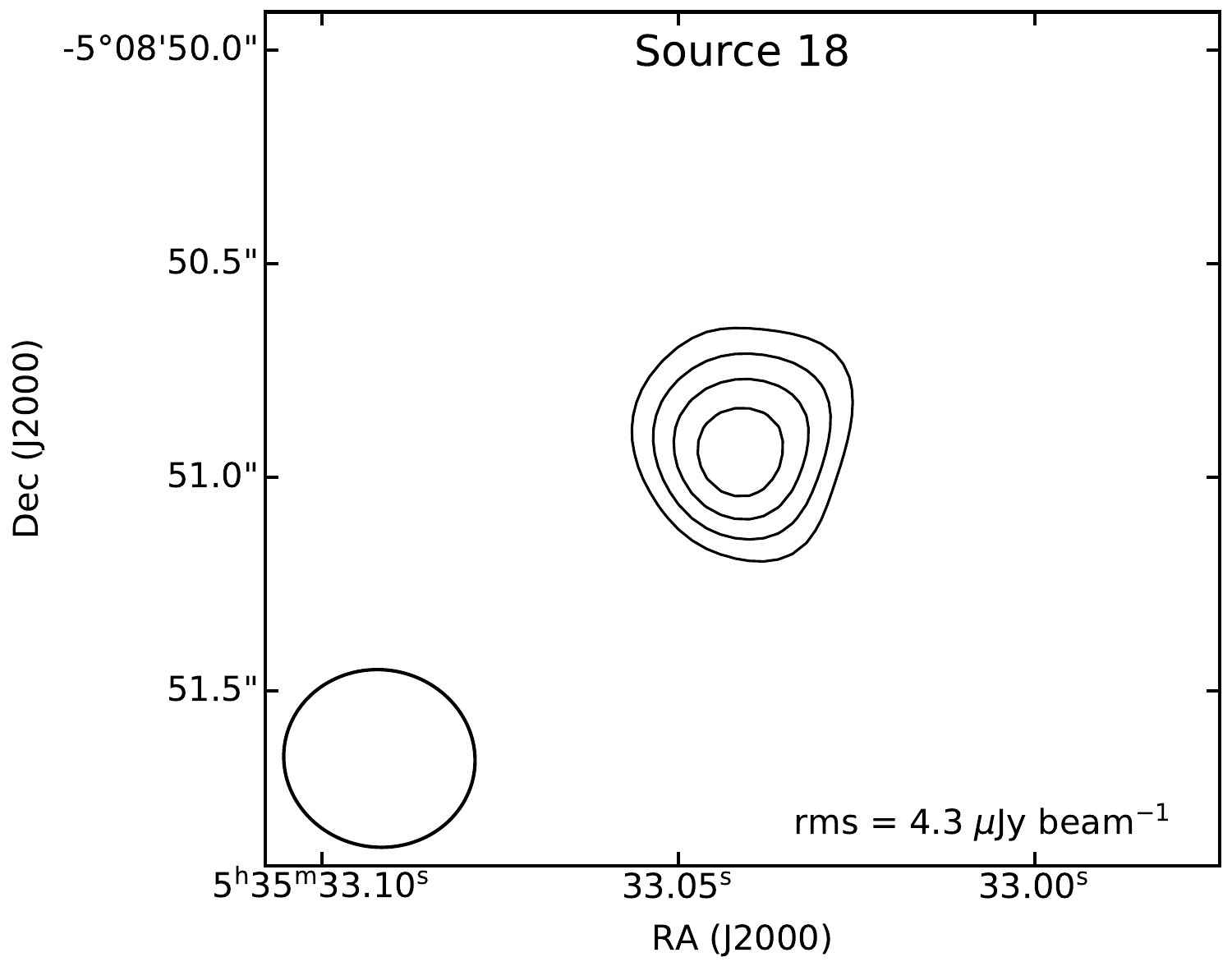}\hfill\includegraphics[height=0.18\textheight]{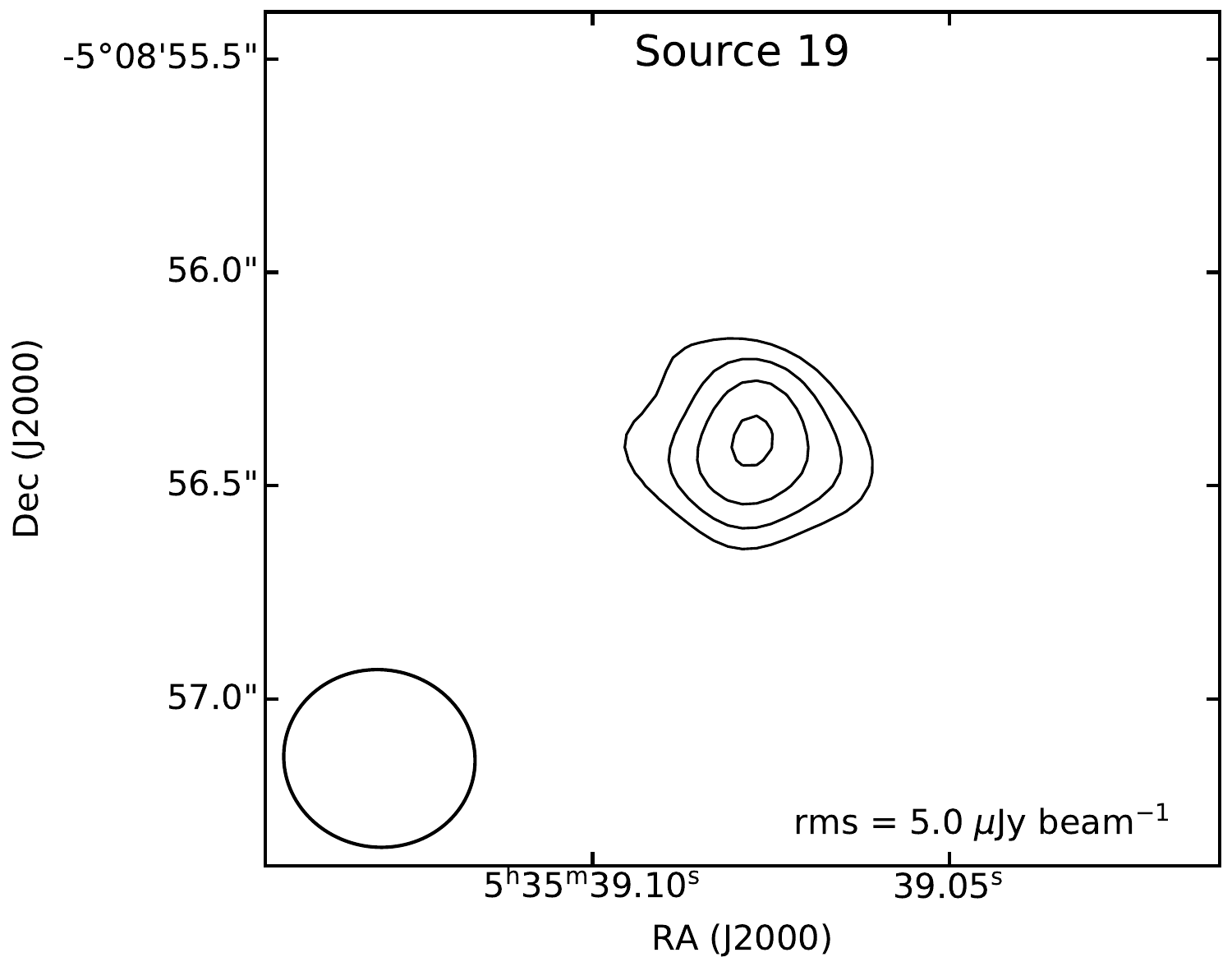}}\\
		{\includegraphics[height=0.18\textheight]{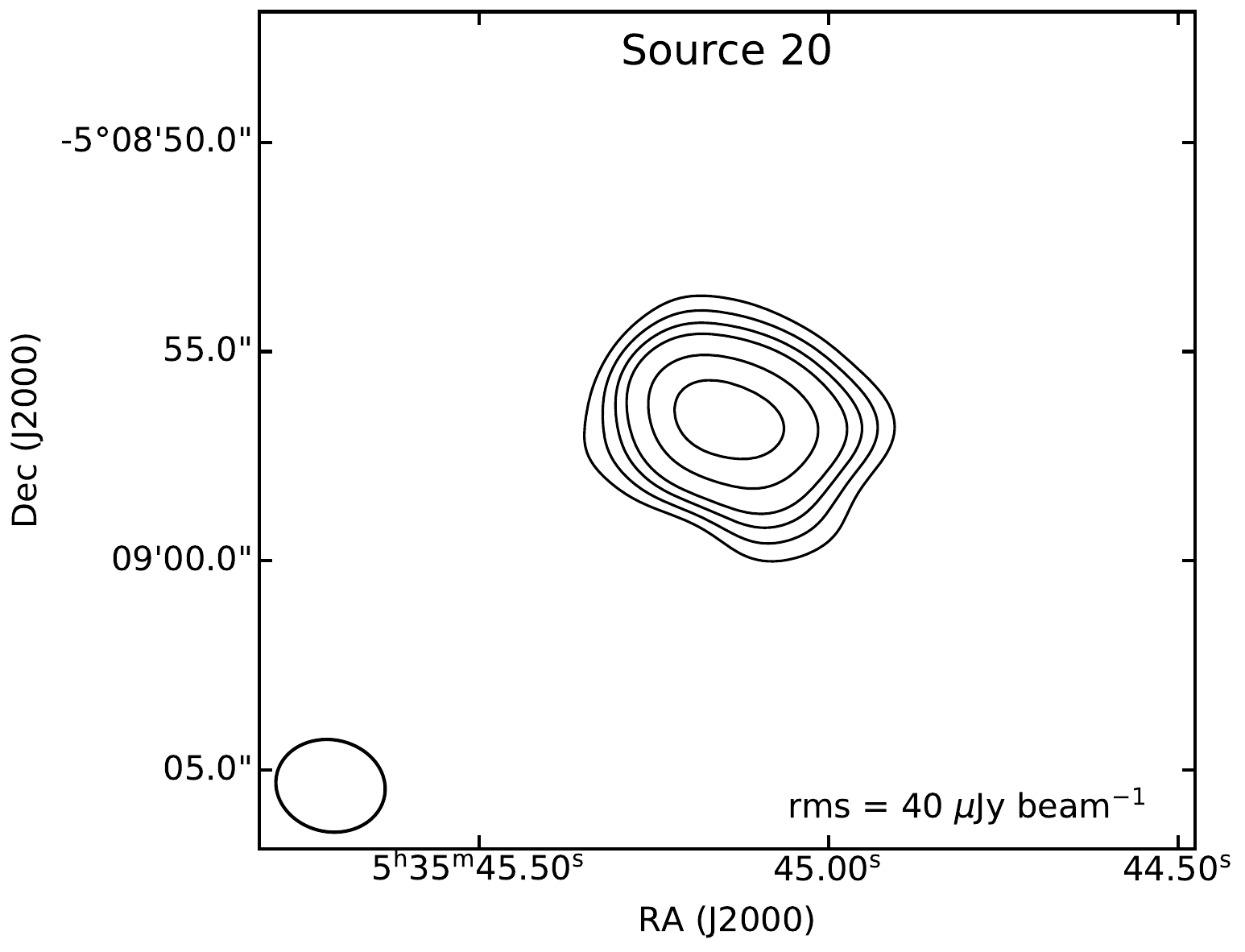}\hfill\includegraphics[height=0.18\textheight]{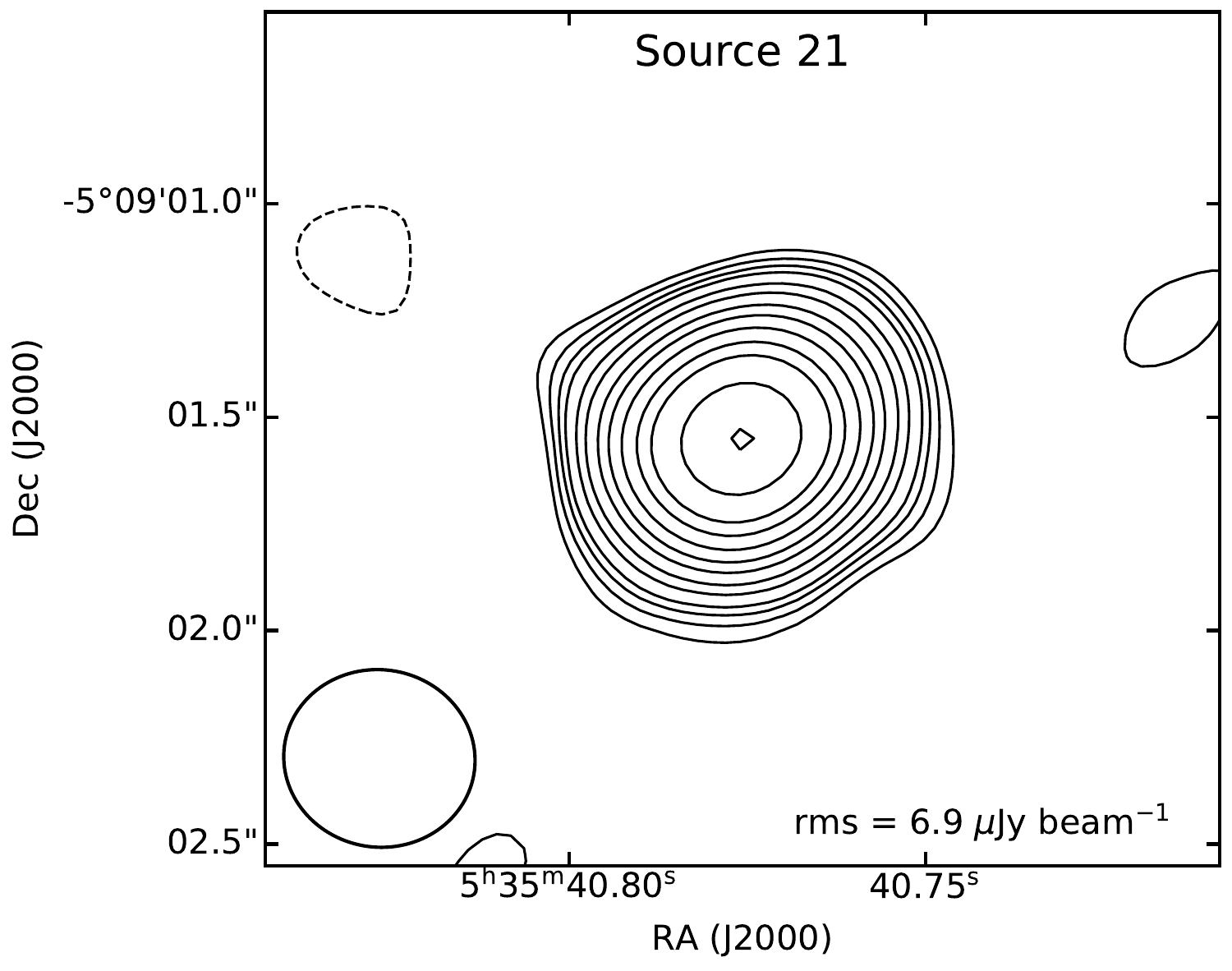}\hfill\includegraphics[height=0.18\textheight]{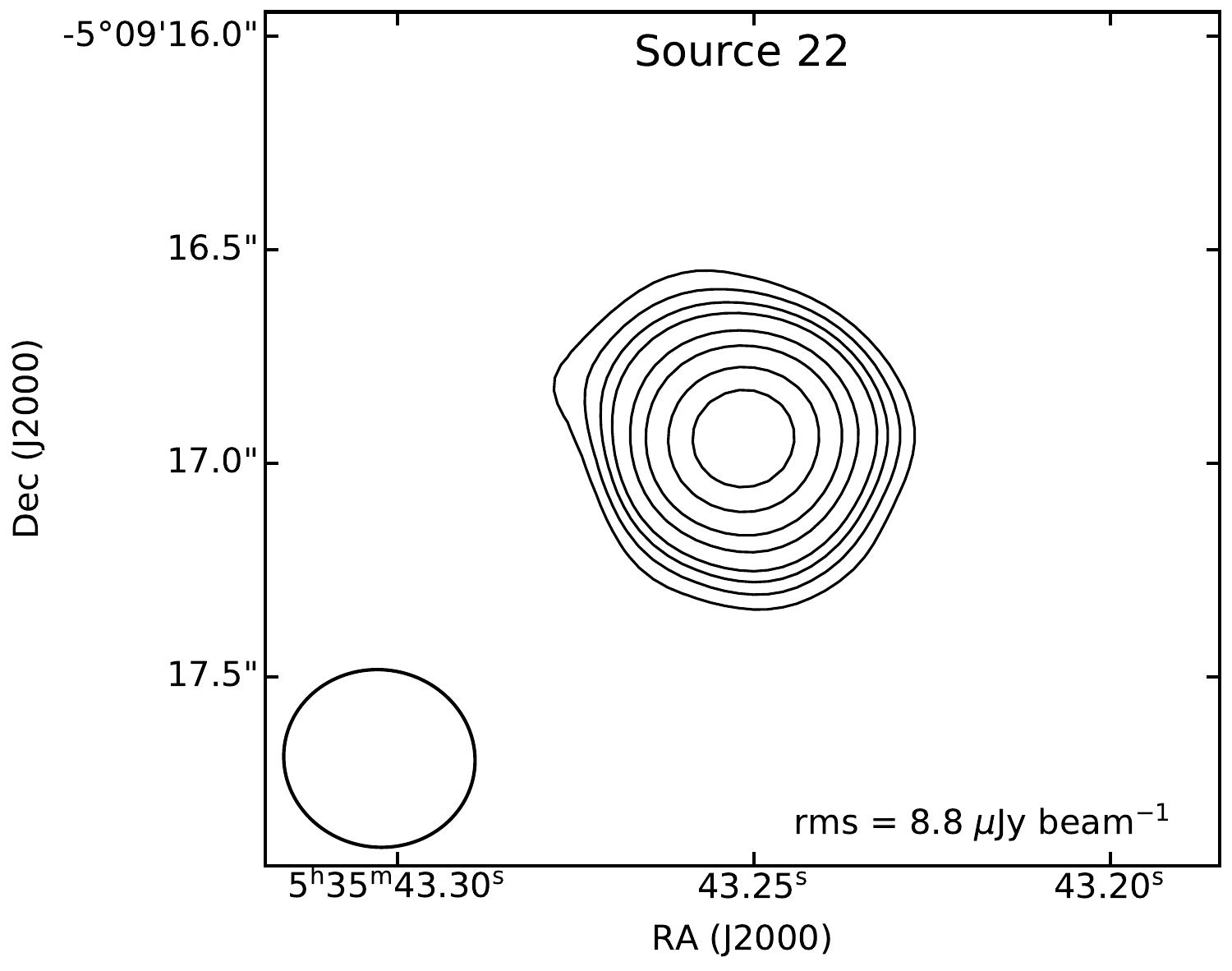}}\\
		{\includegraphics[height=0.18\textheight]{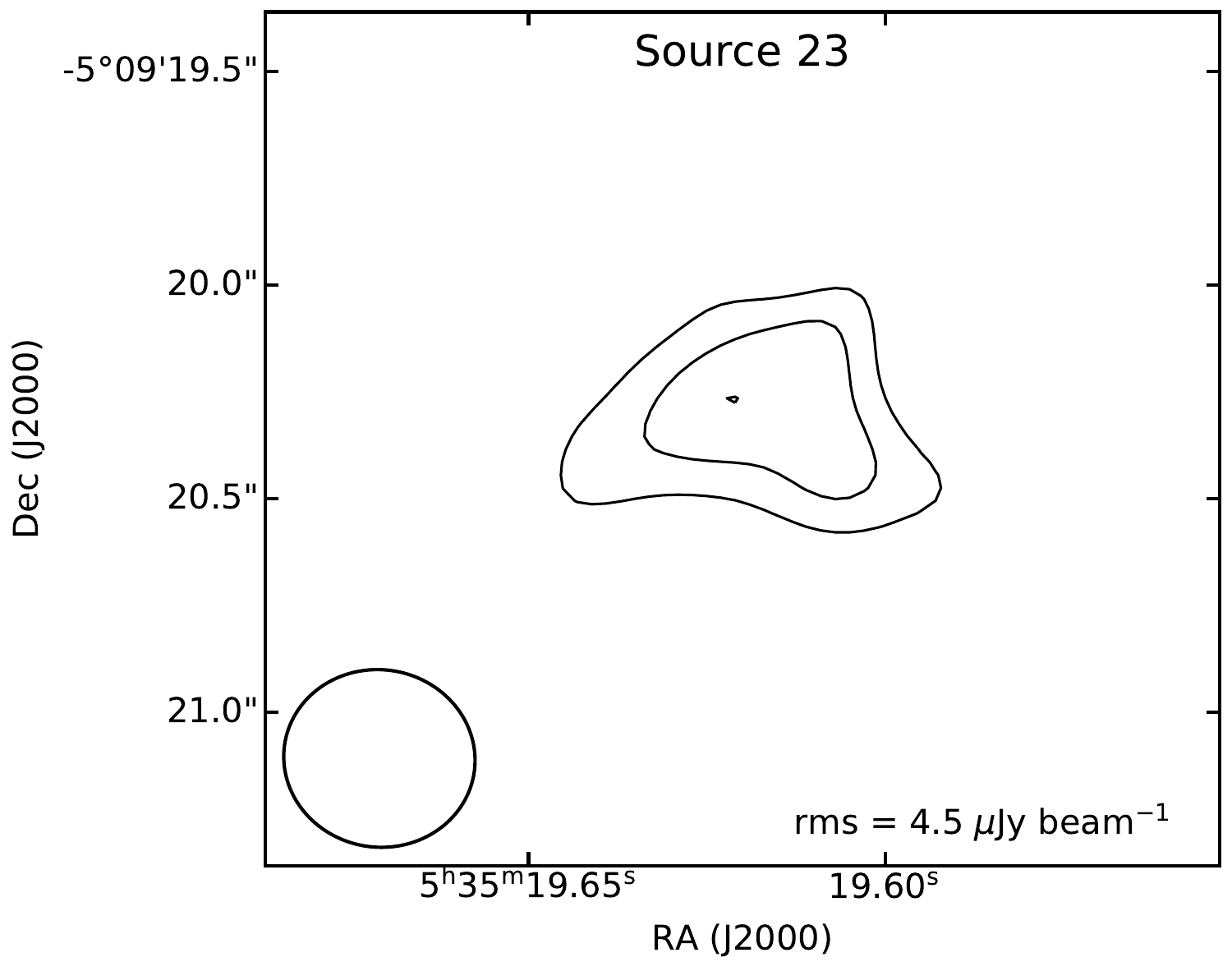}\hfill\includegraphics[height=0.18\textheight]{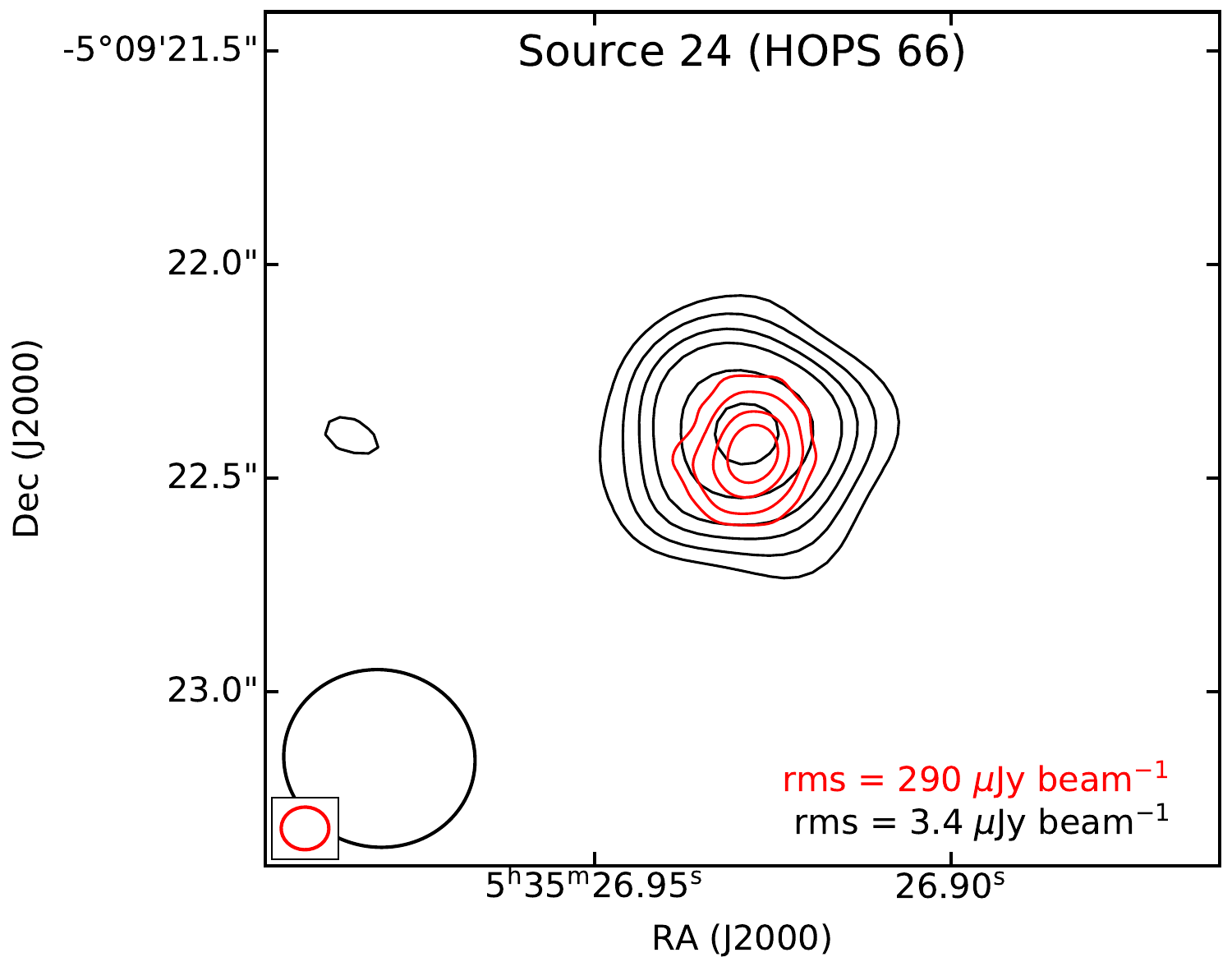}\hfill\includegraphics[height=0.18\textheight]{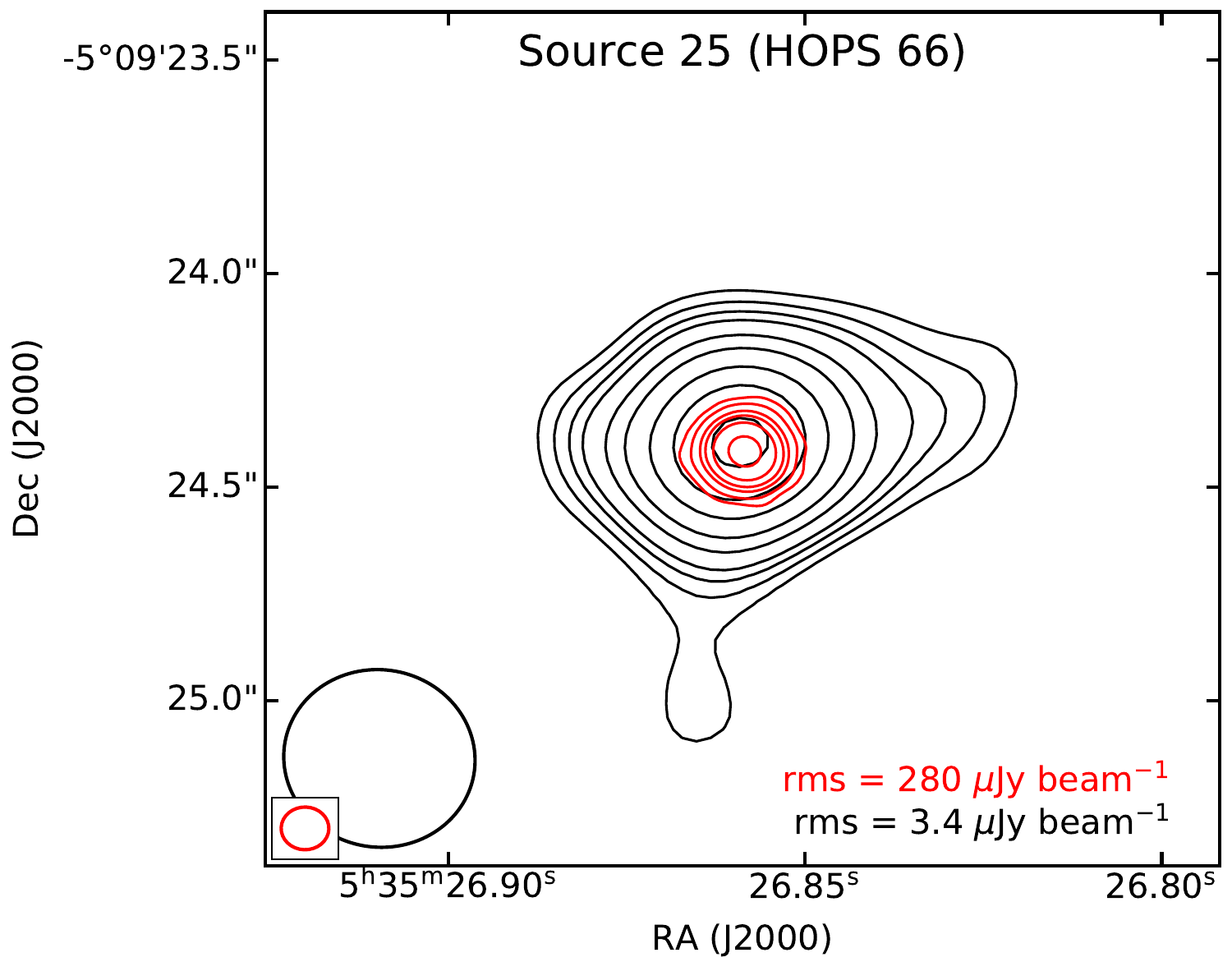}}\\
		{\includegraphics[height=0.18\textheight]{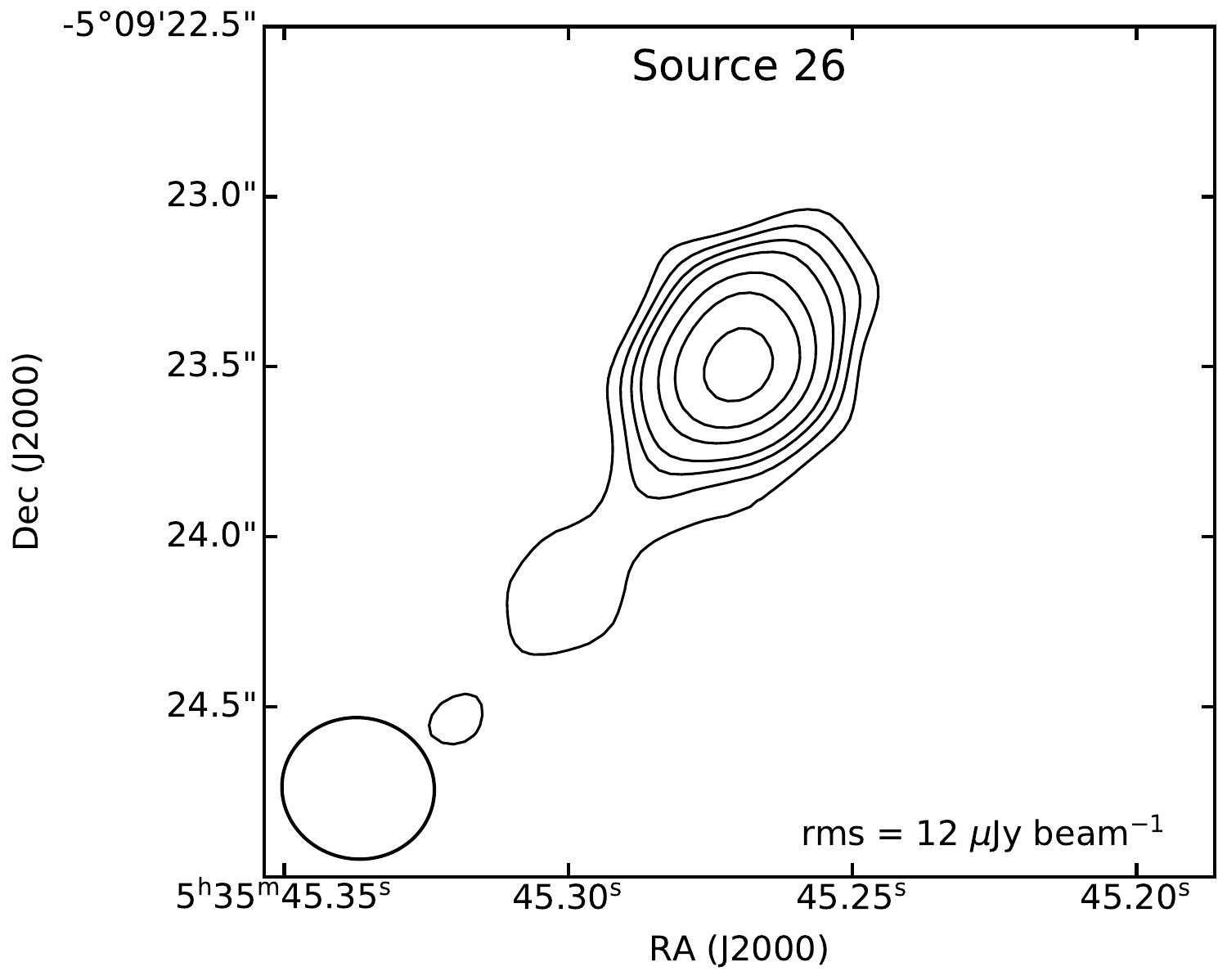}\hfill\includegraphics[height=0.18\textheight]{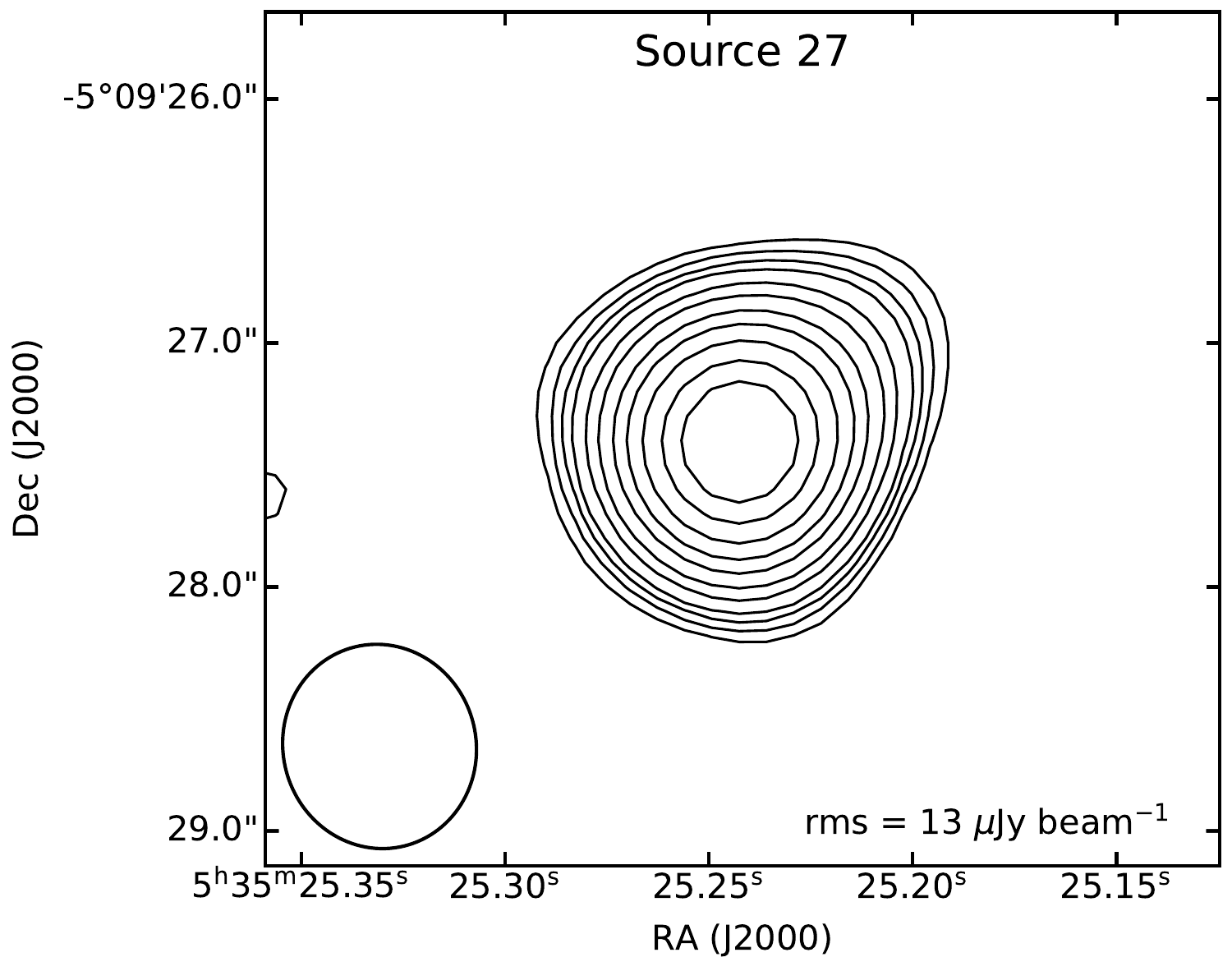}\hfill\includegraphics[height=0.18\textheight]{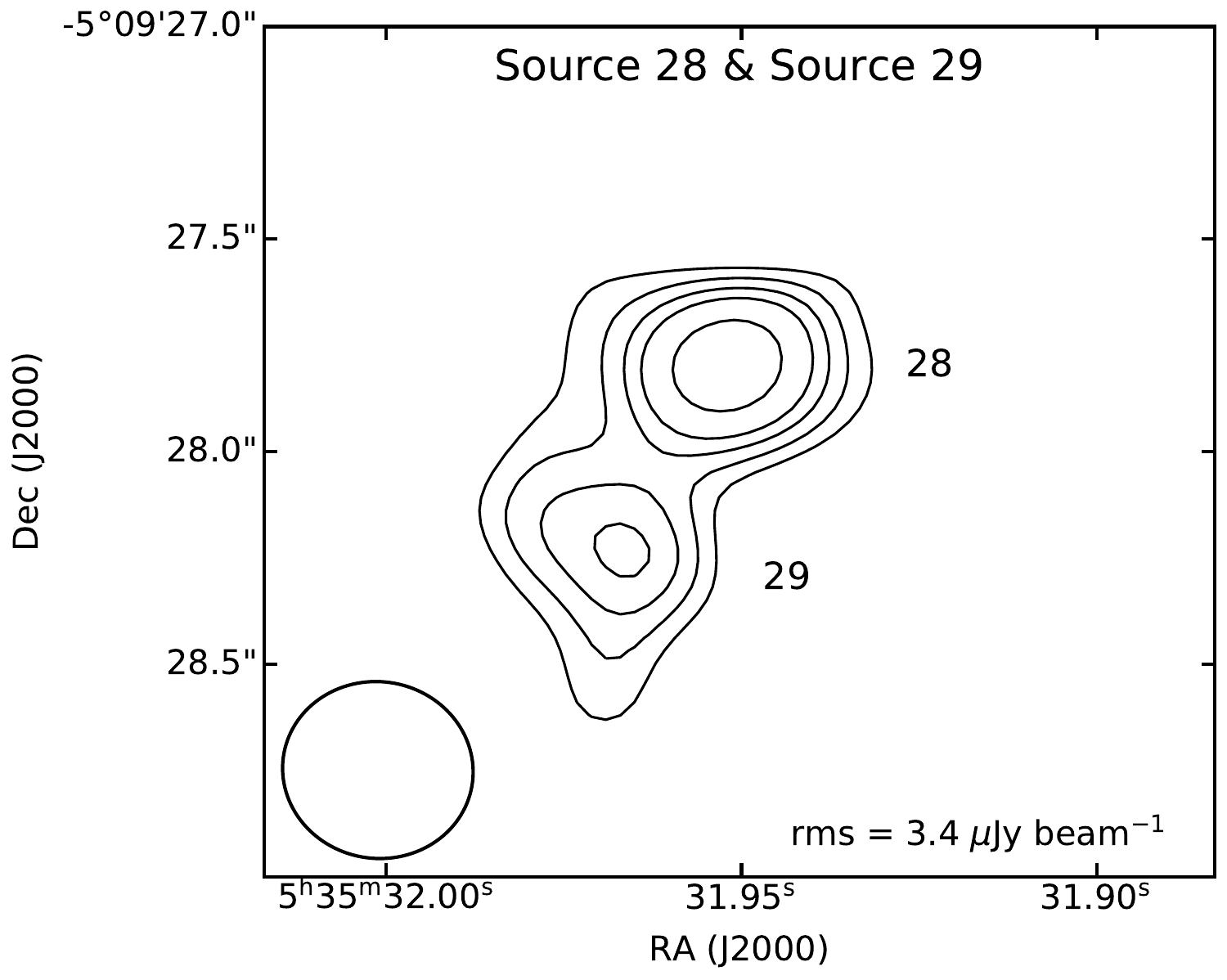}}\\
		{\includegraphics[height=0.18\textheight]{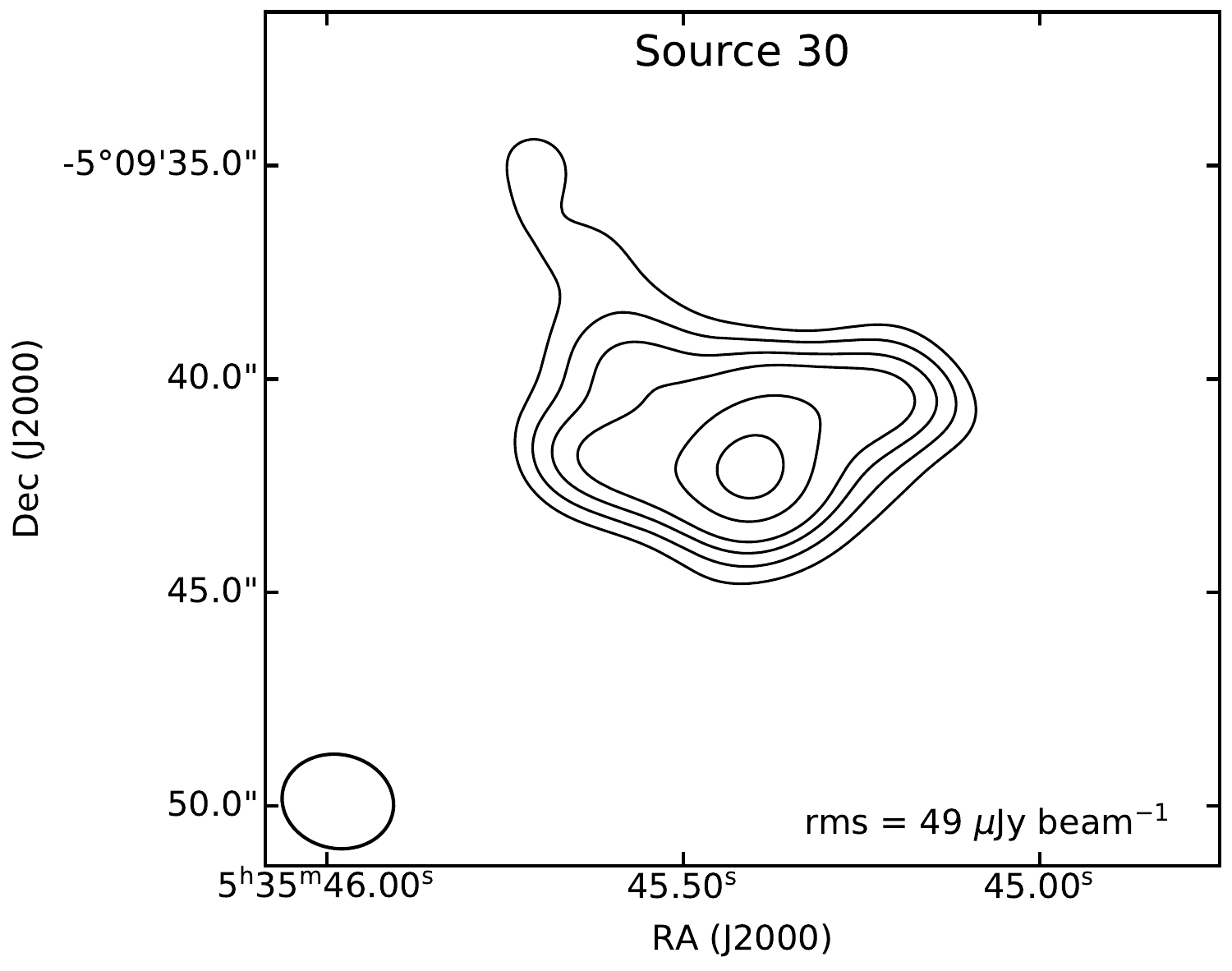}\hfill\includegraphics[height=0.18\textheight]{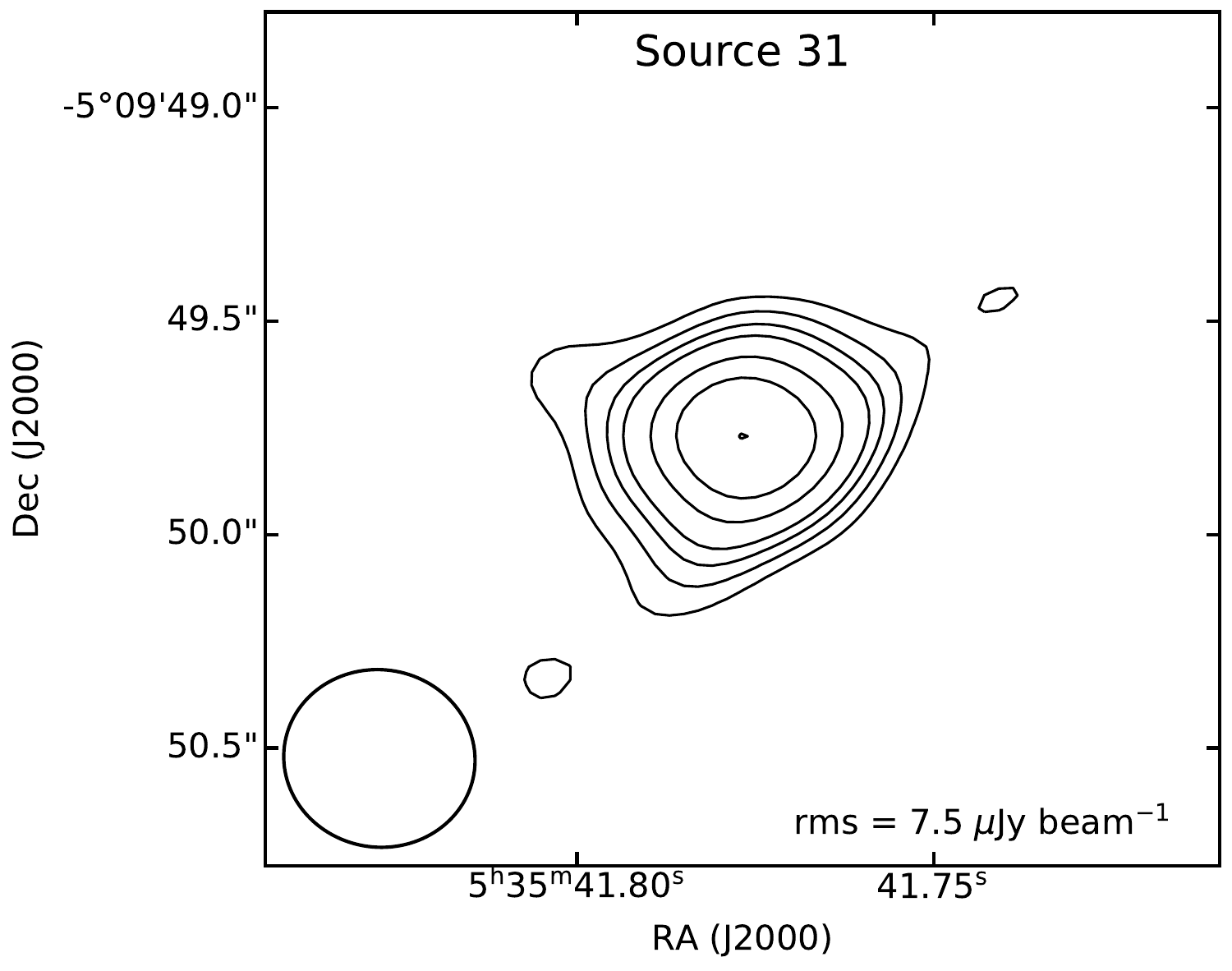}\hfill\includegraphics[height=0.18\textheight]{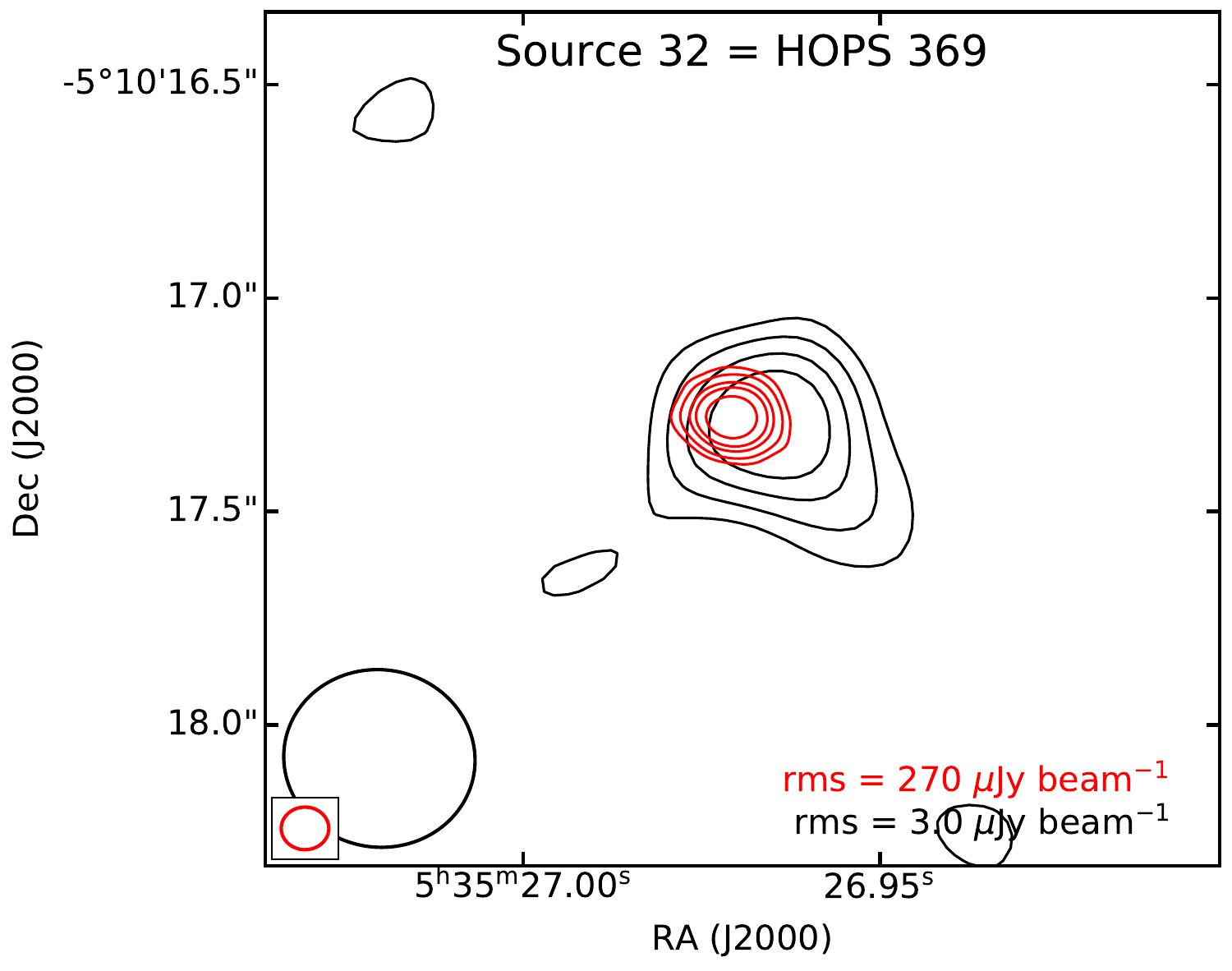}}
		\contcaption{}
		\label{fig:sources2}
	\end{figure*}	
	
	\begin{figure*} 
		{\includegraphics[height=0.18\textheight]{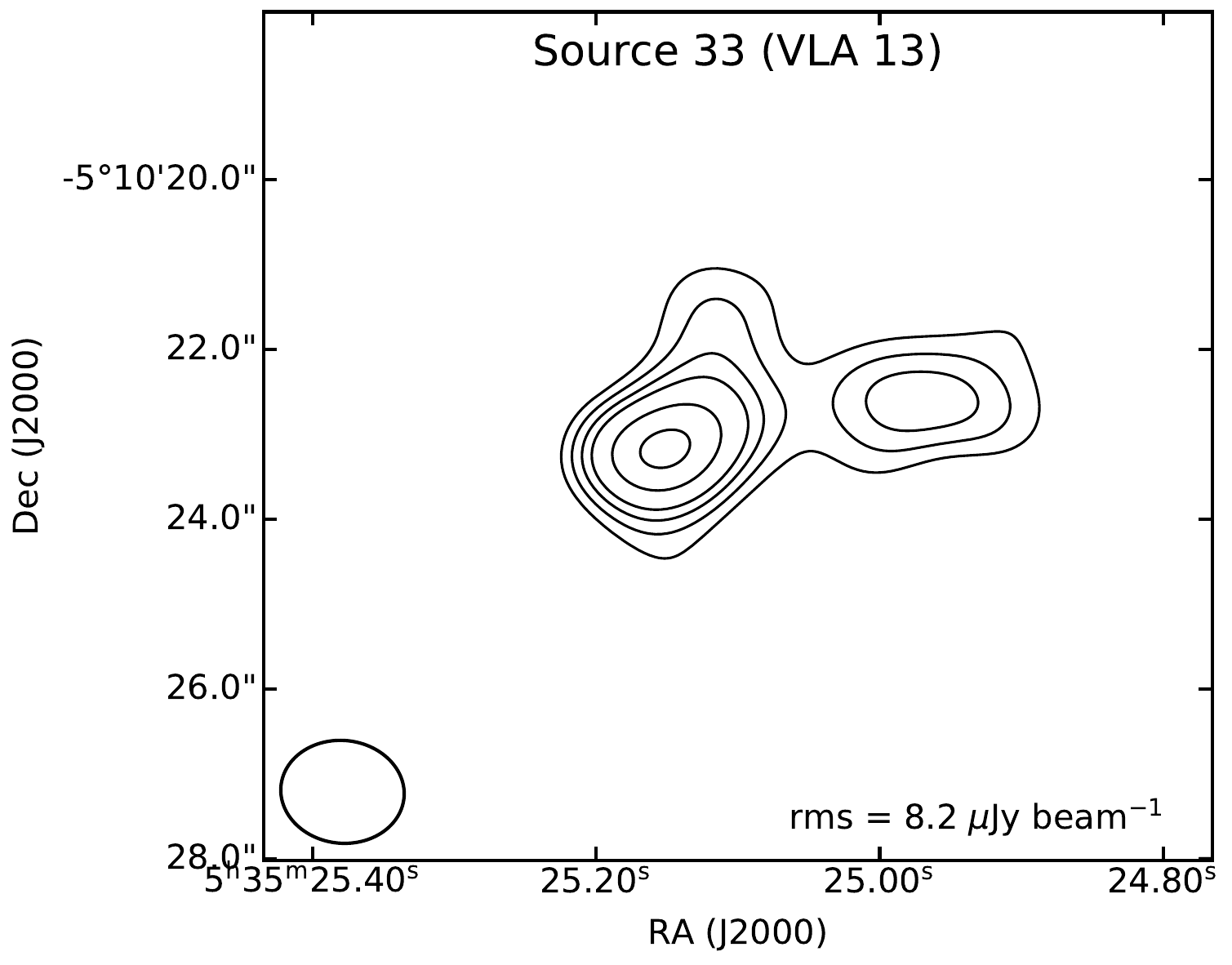}\hfill\includegraphics[height=0.18\textheight]{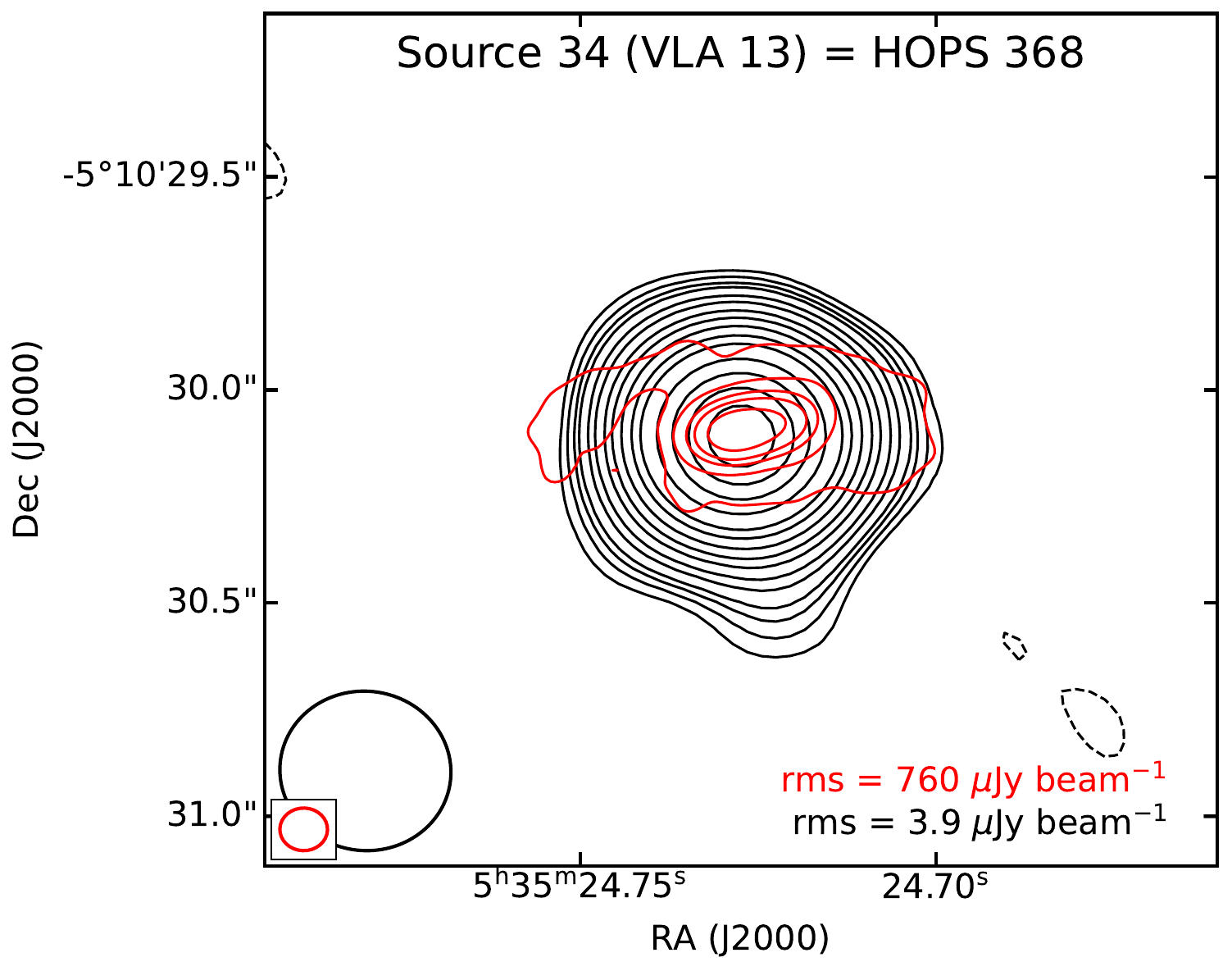}\hfill\includegraphics[height=0.18\textheight]{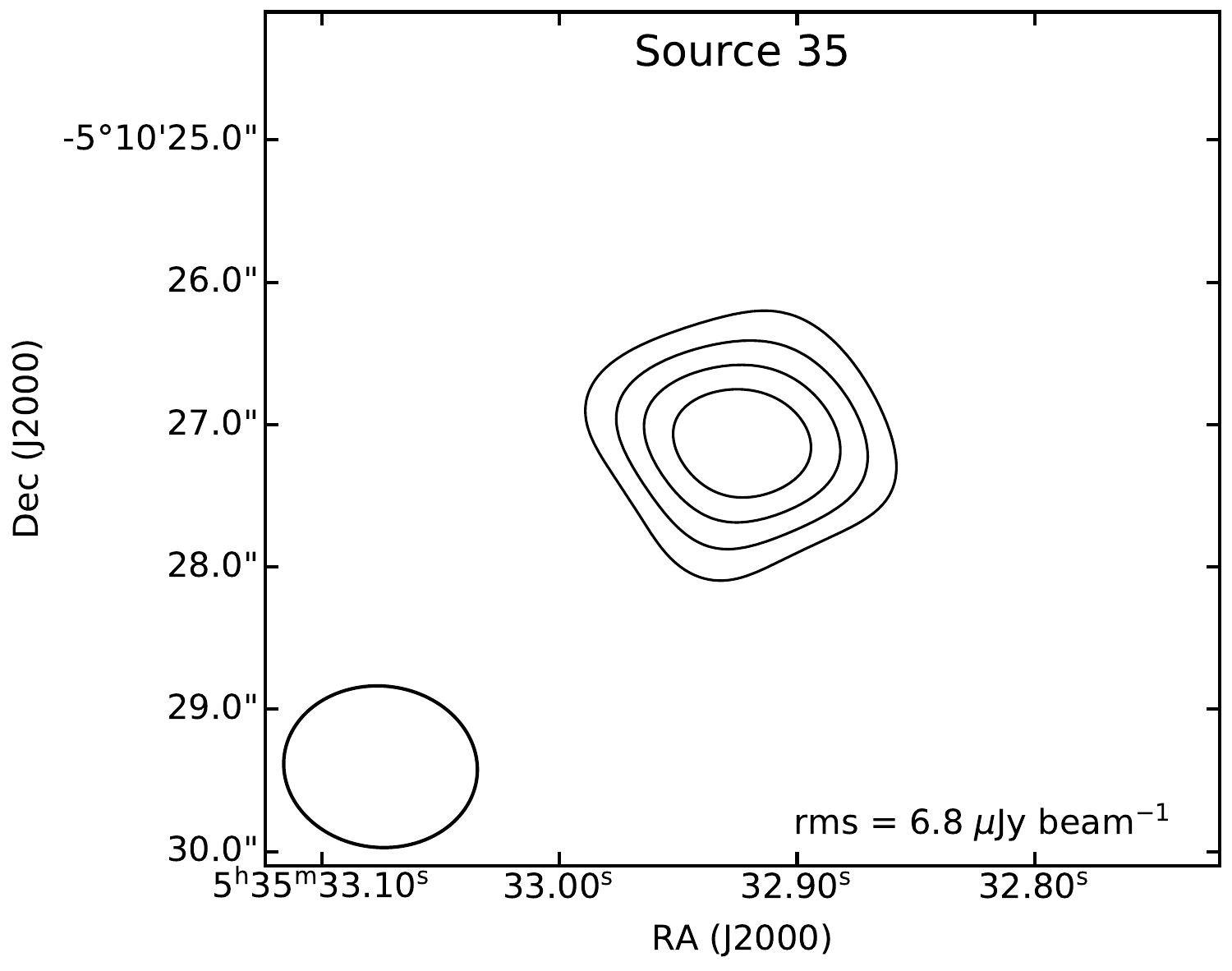}}\\
		{\includegraphics[height=0.18\textheight]{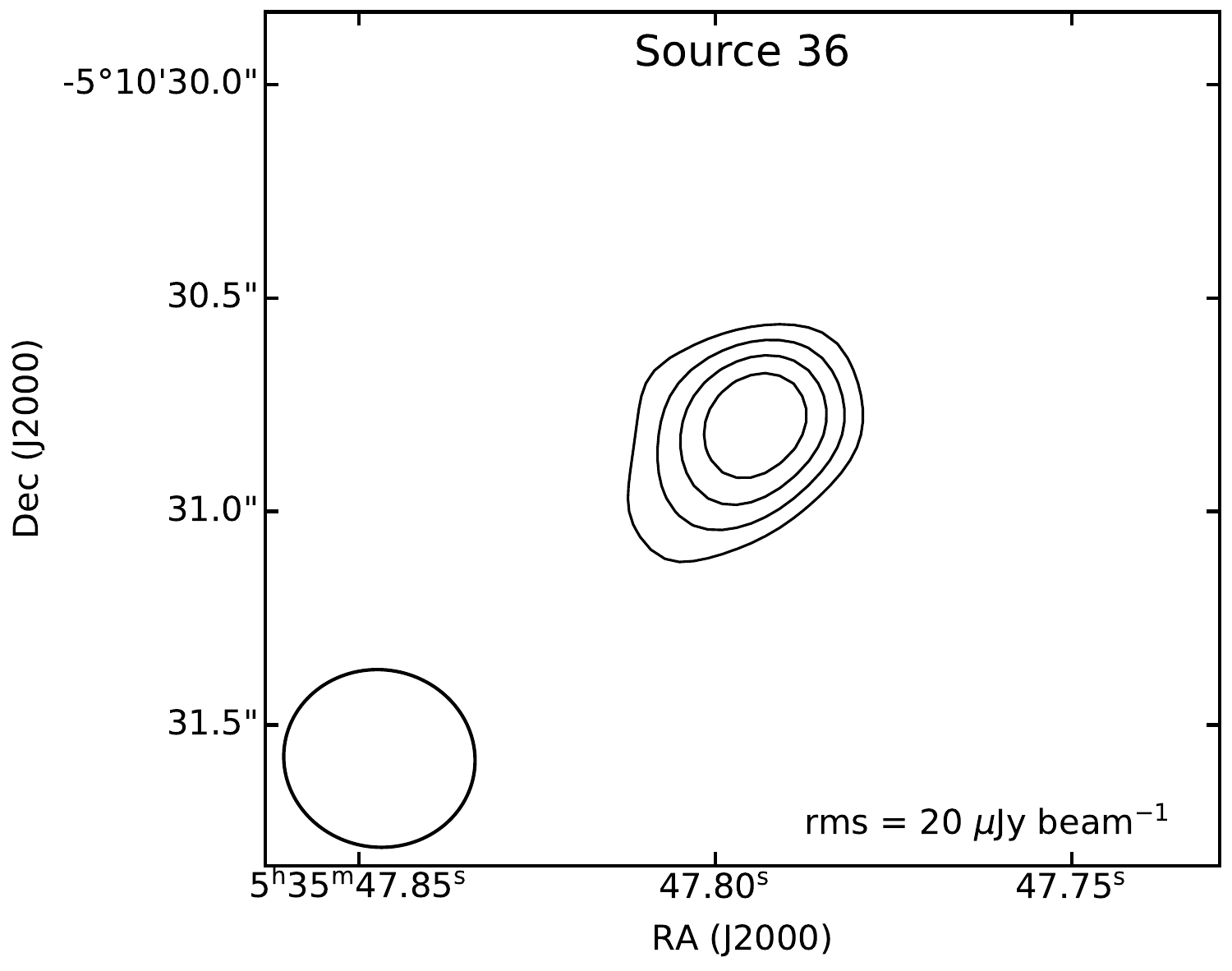}\hfill\includegraphics[height=0.18\textheight]{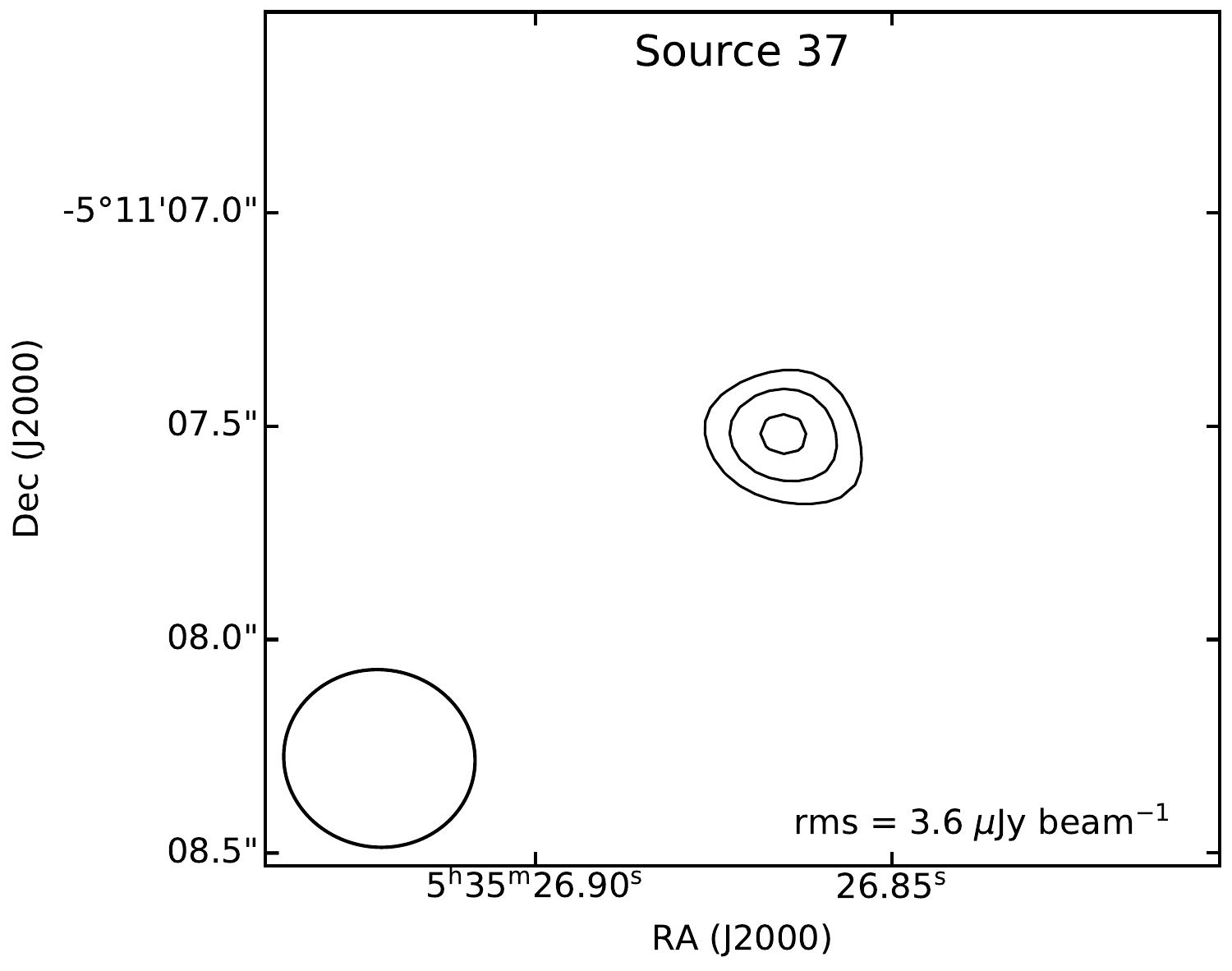}\hfill\includegraphics[height=0.18\textheight]{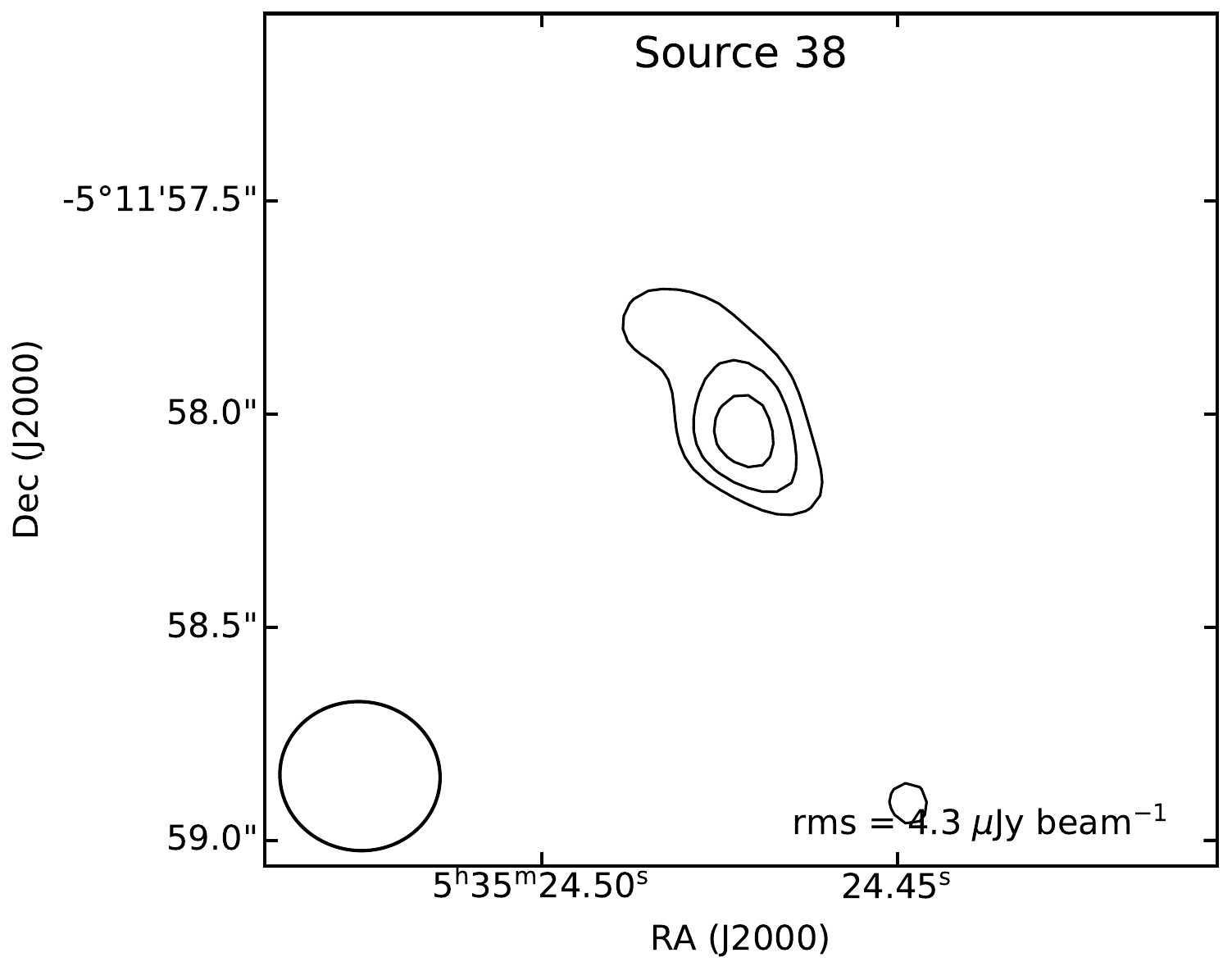}}\\{\includegraphics[height=0.18\textheight]{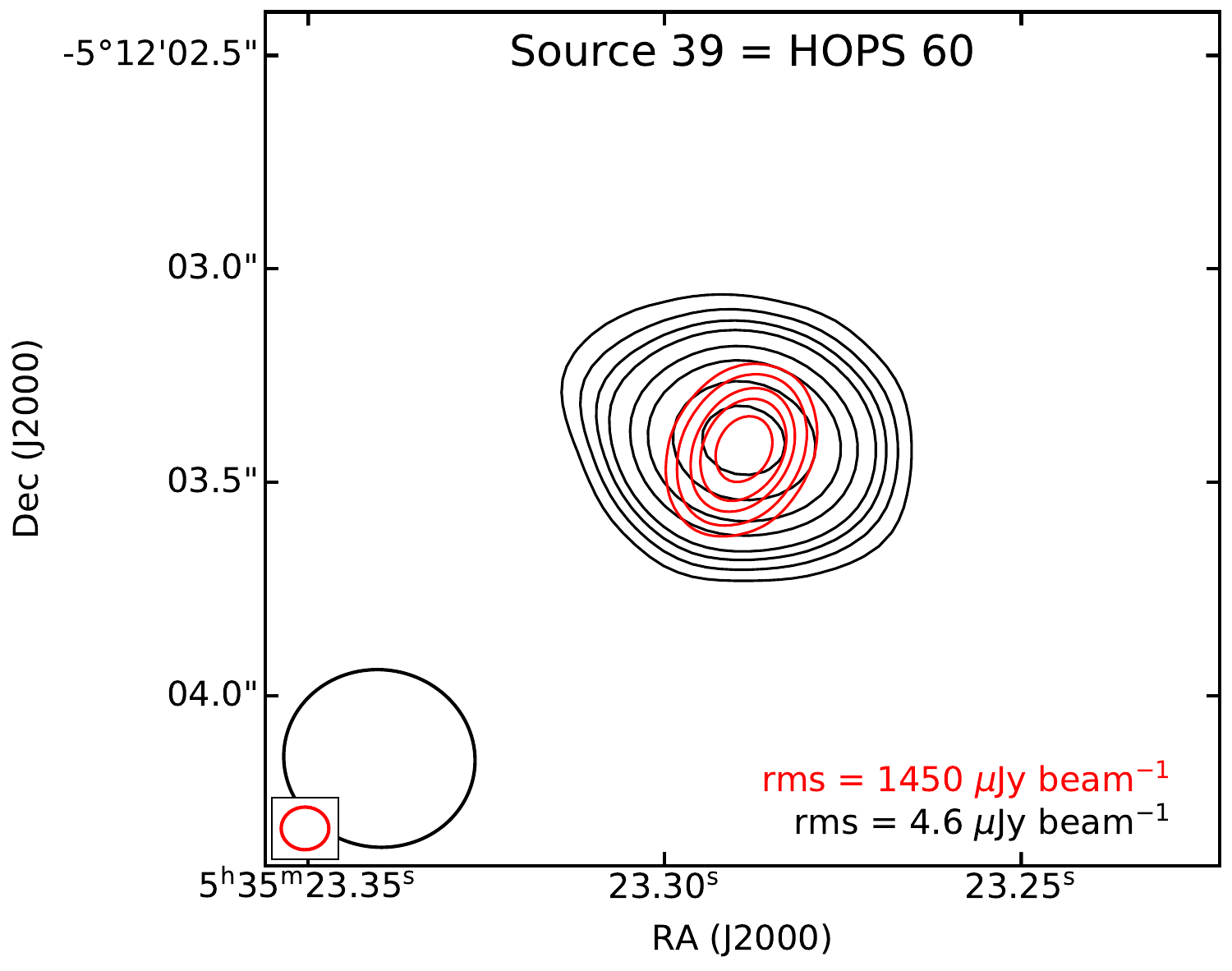}\hfill\includegraphics[height=0.18\textheight]{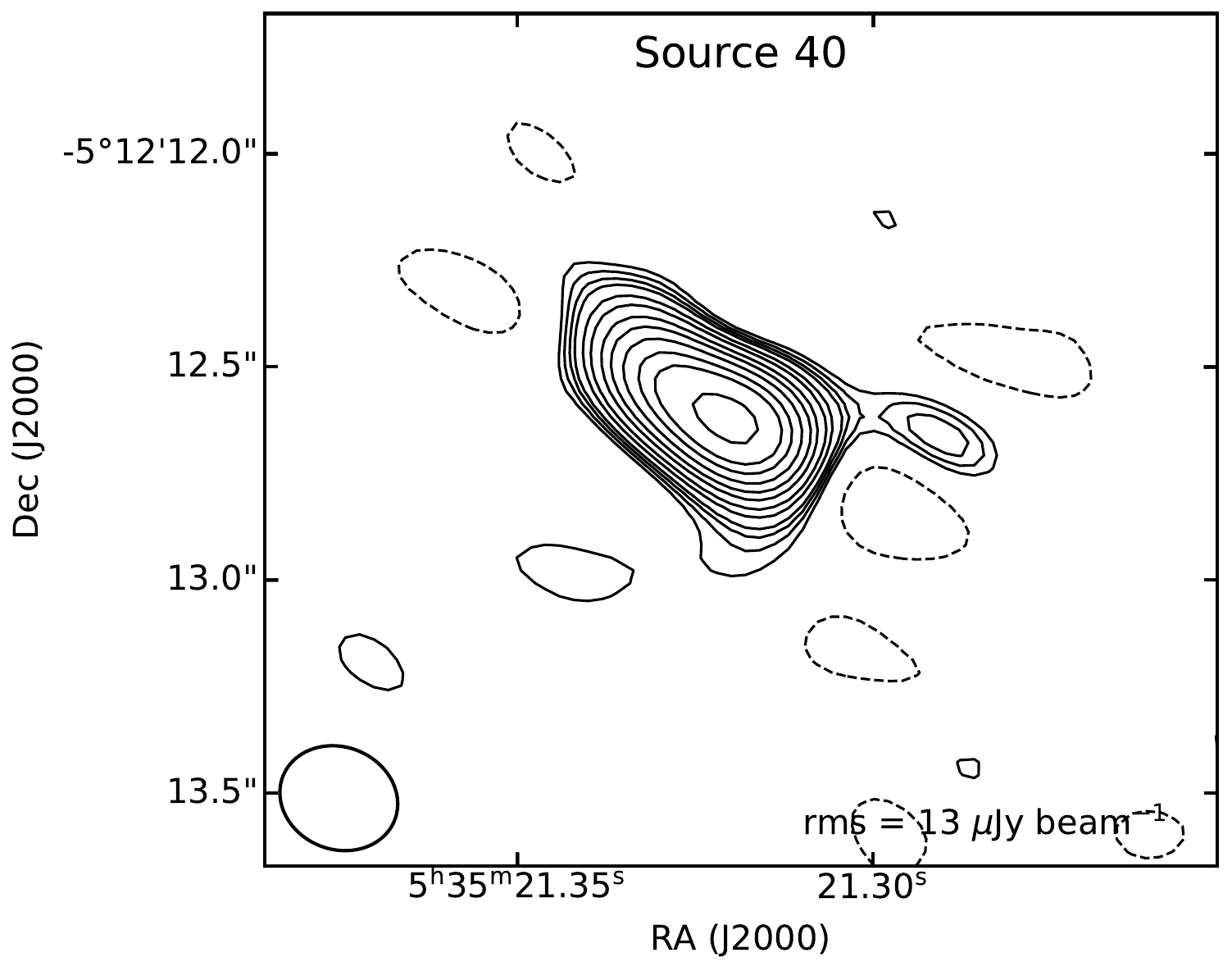}\hfill\includegraphics[height=0.18\textheight]{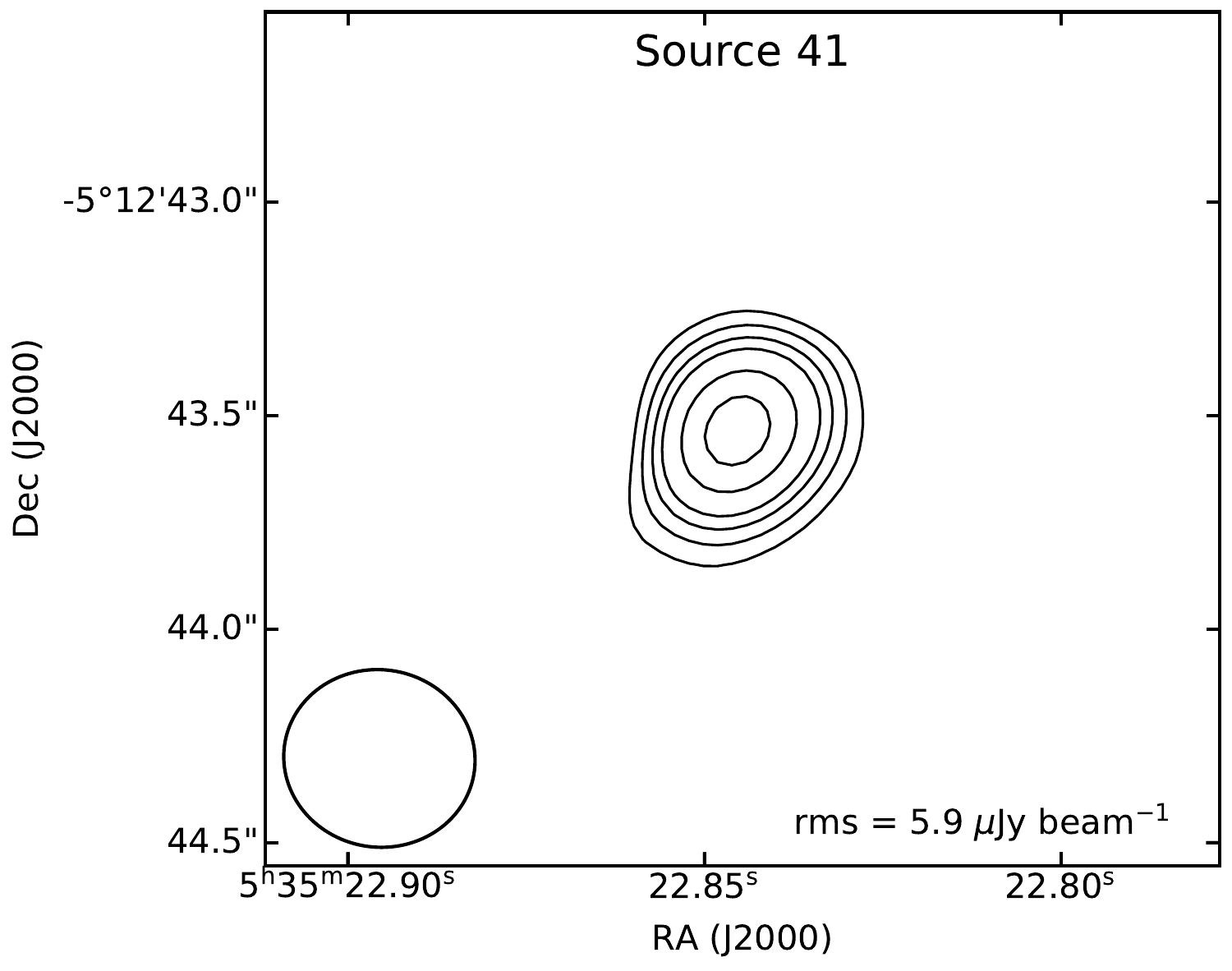}}\\
		{\includegraphics[height=0.18\textheight]{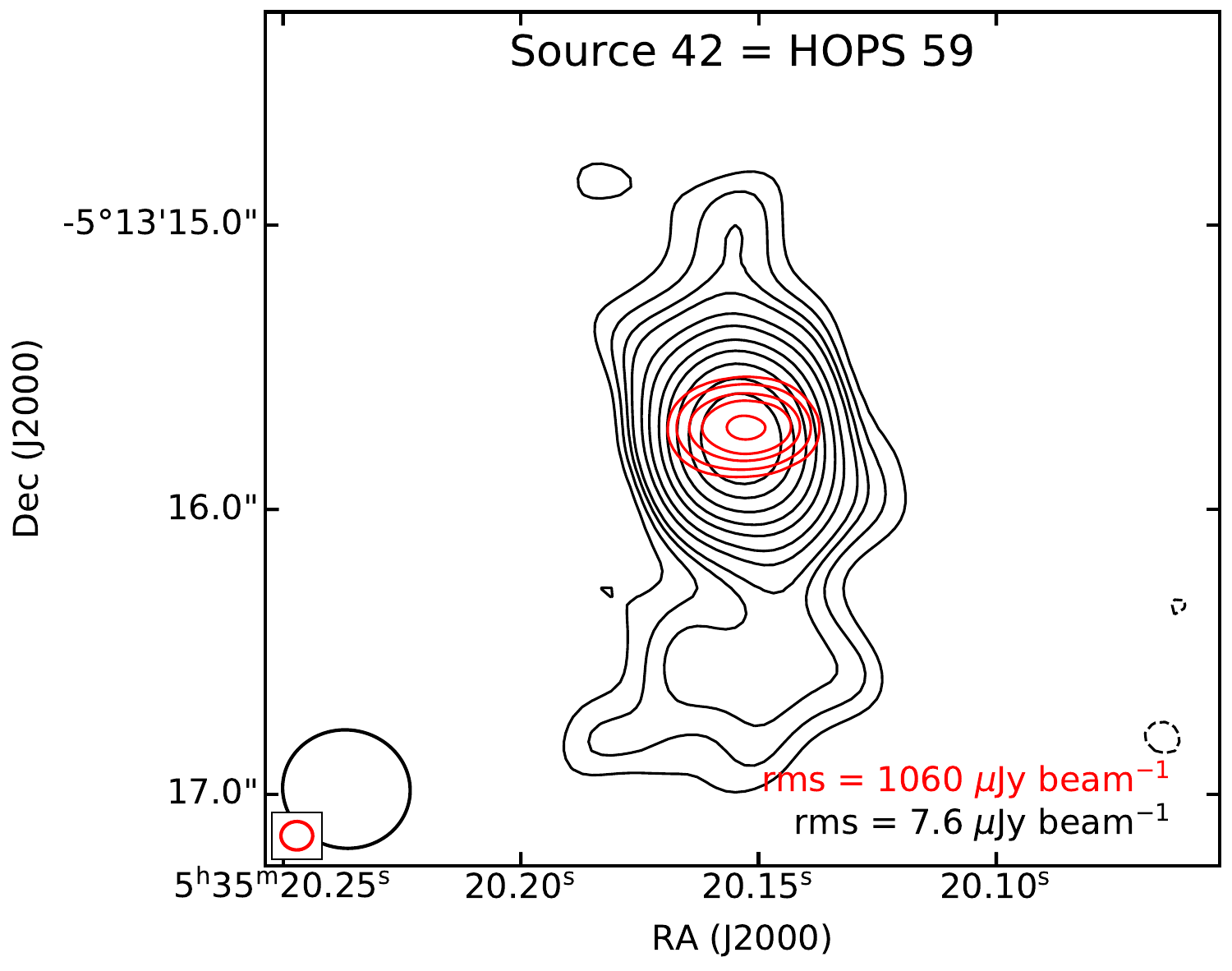}\hfill\includegraphics[height=0.18\textheight]{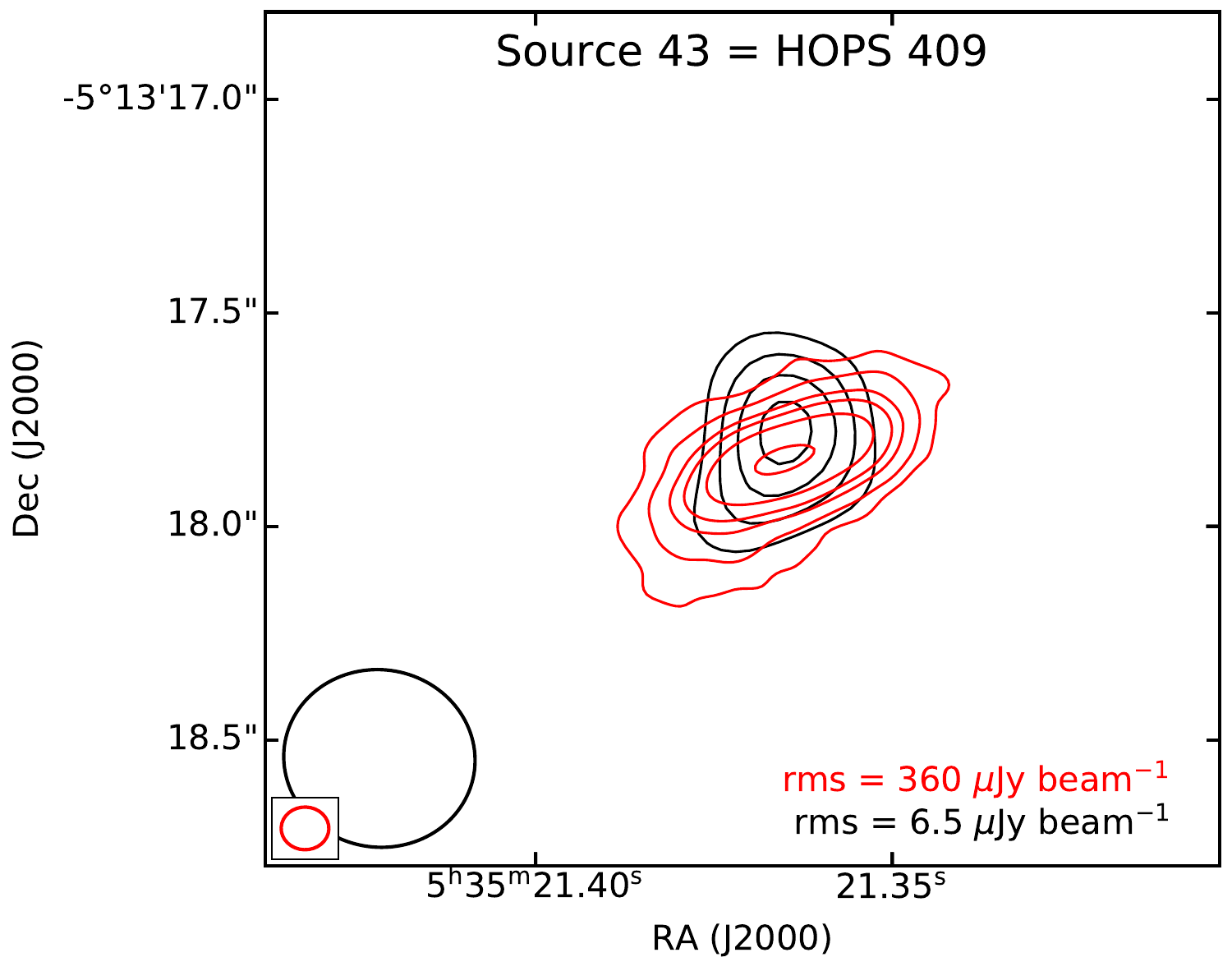}\hfill\includegraphics[height=0.18\textheight]{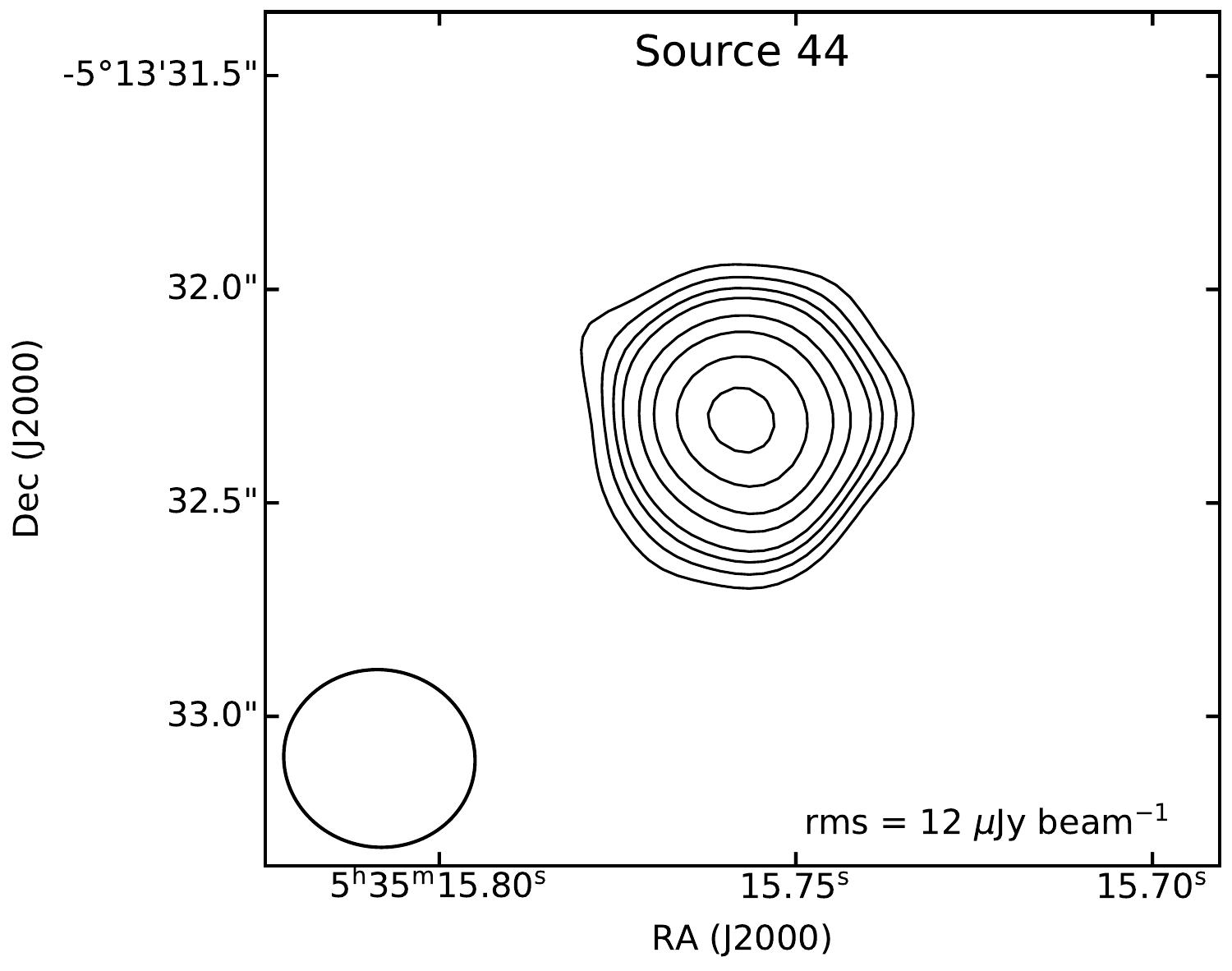}}\\
		{\includegraphics[height=0.18\textheight]{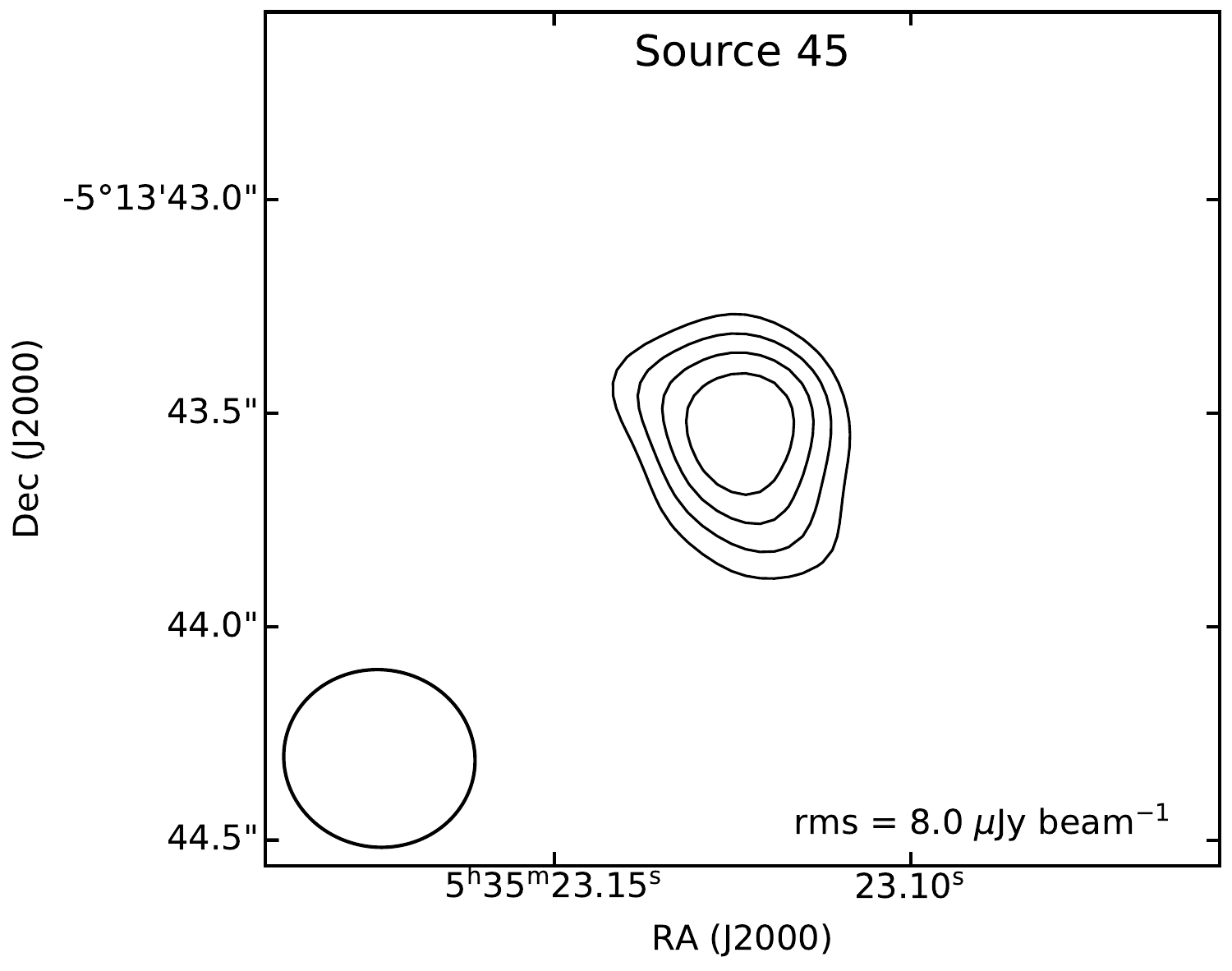}\hfill\includegraphics[height=0.18\textheight]{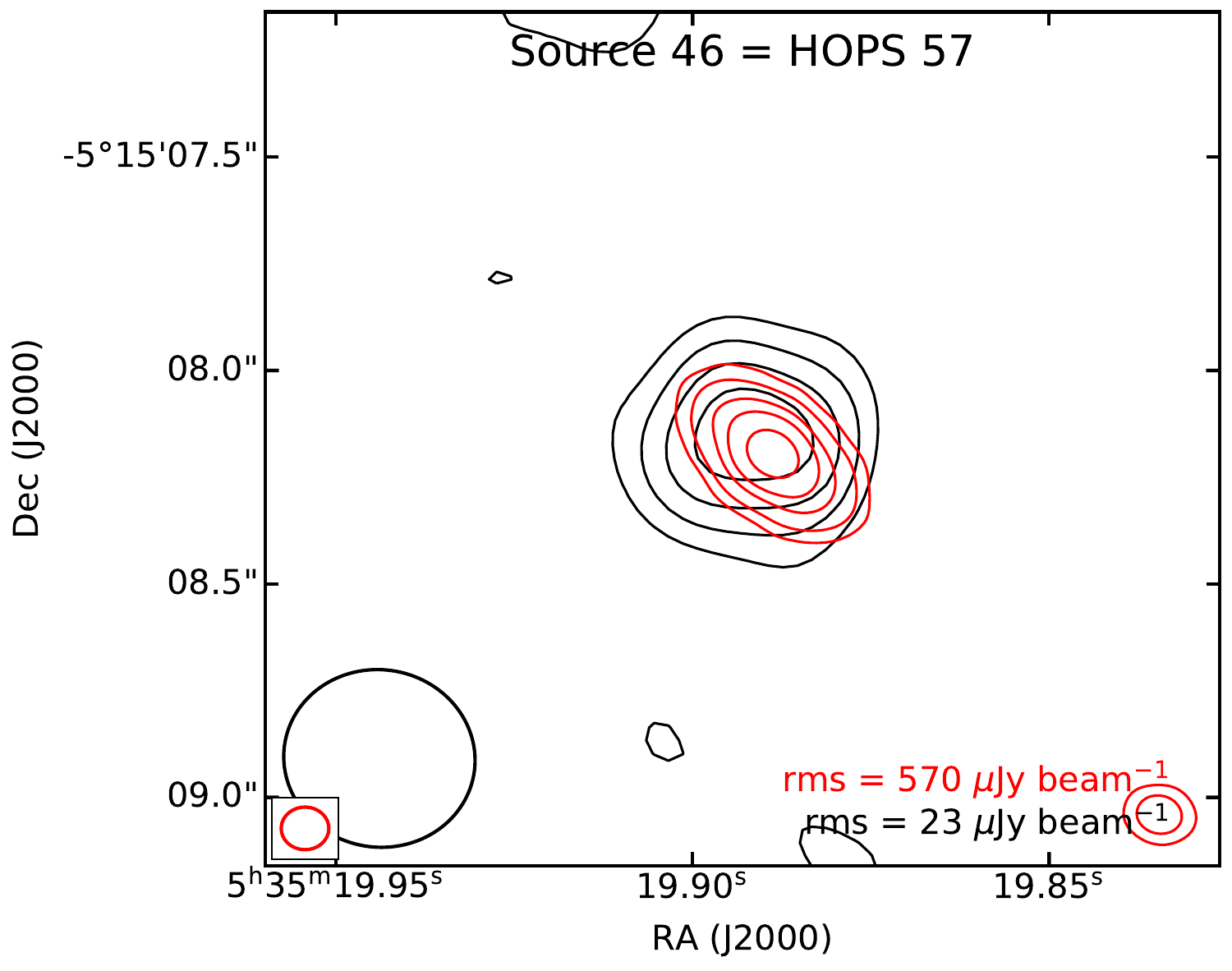}\hfill\includegraphics[height=0.18\textheight]{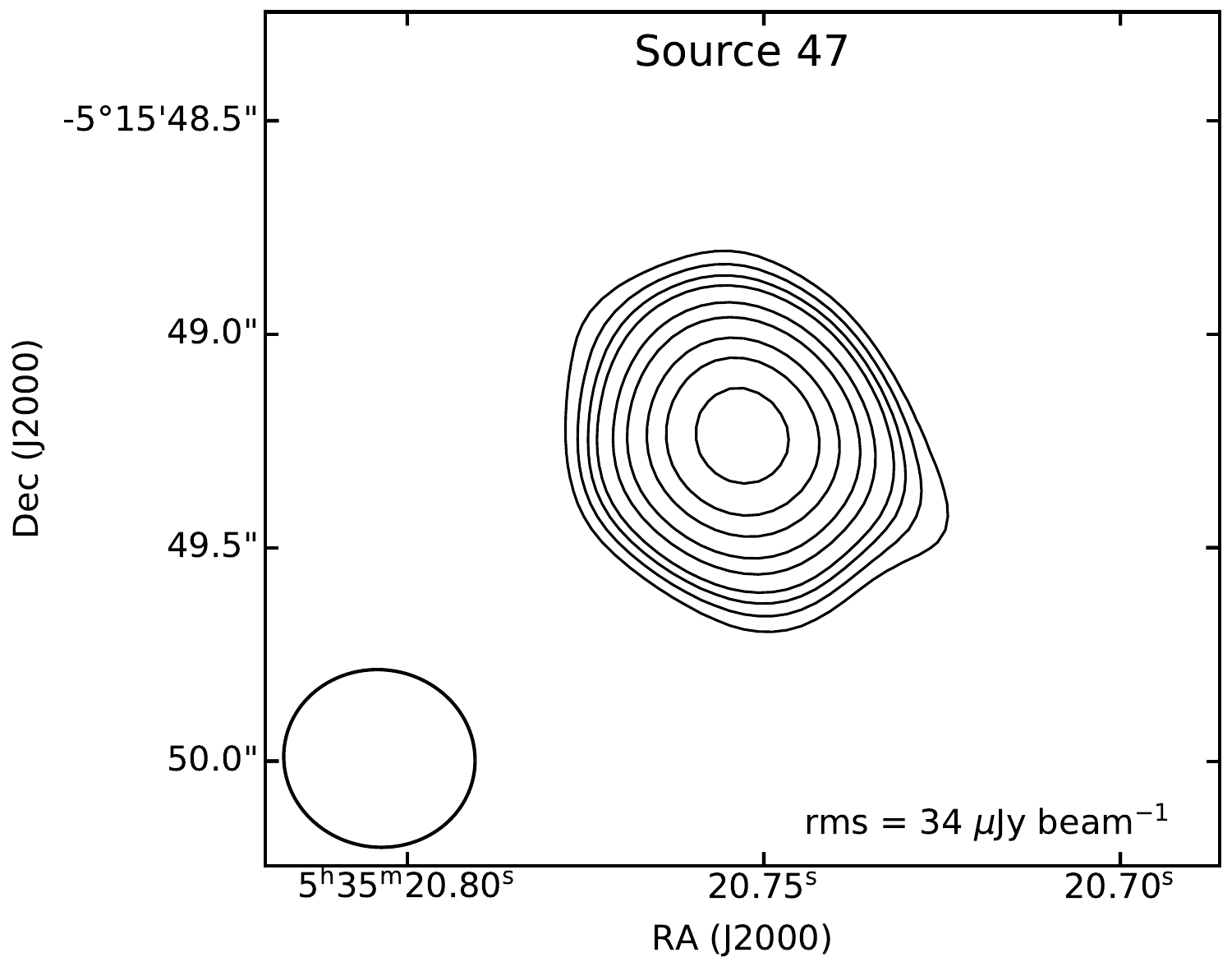}}
		\contcaption{}
		\label{fig:sources3}	
	\end{figure*}

\clearpage

\section{Spectra of the detected radio sources}

	\begin{figure*}
		{\includegraphics[width=0.33\textwidth]{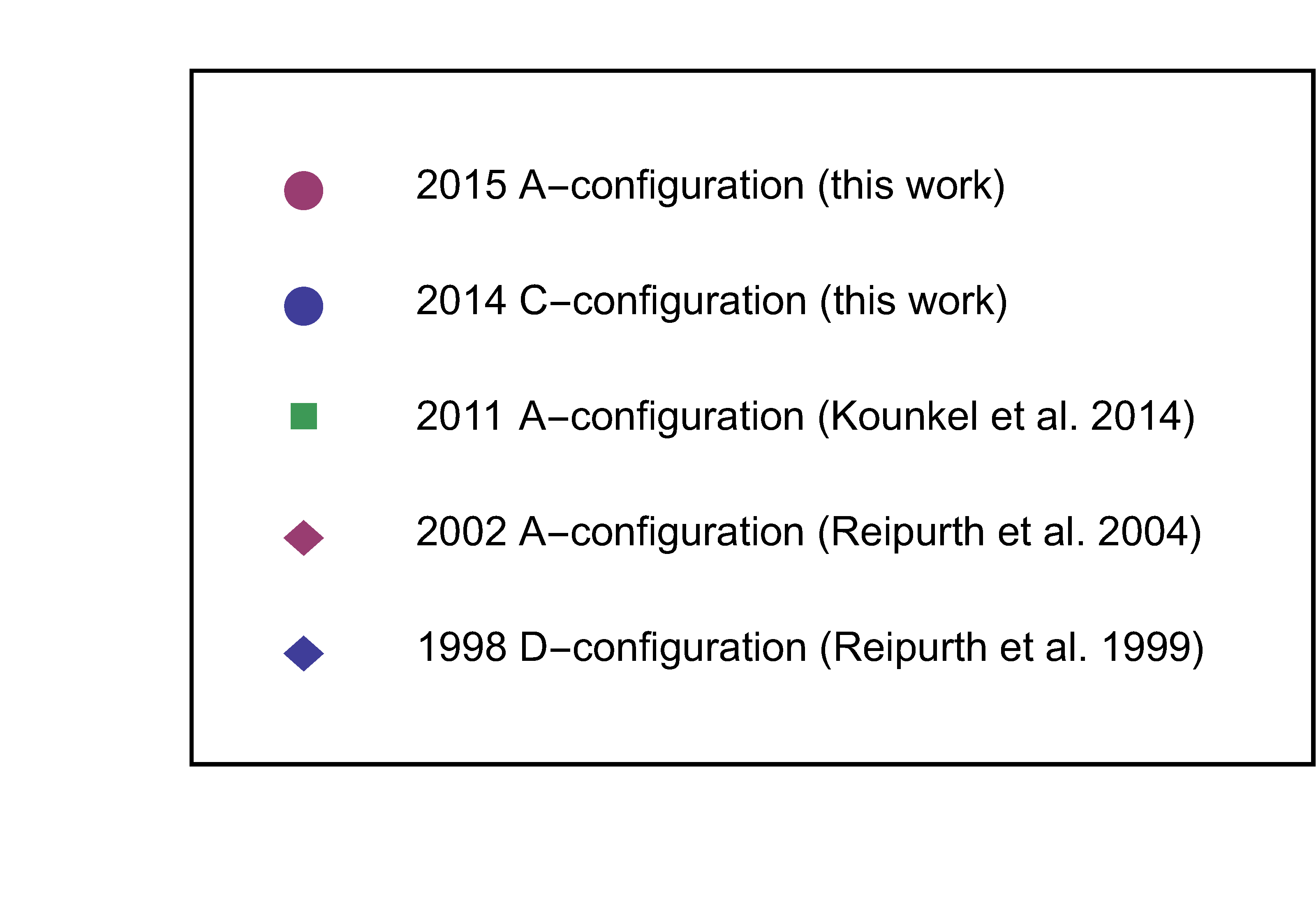}\includegraphics[width=0.33\textwidth]{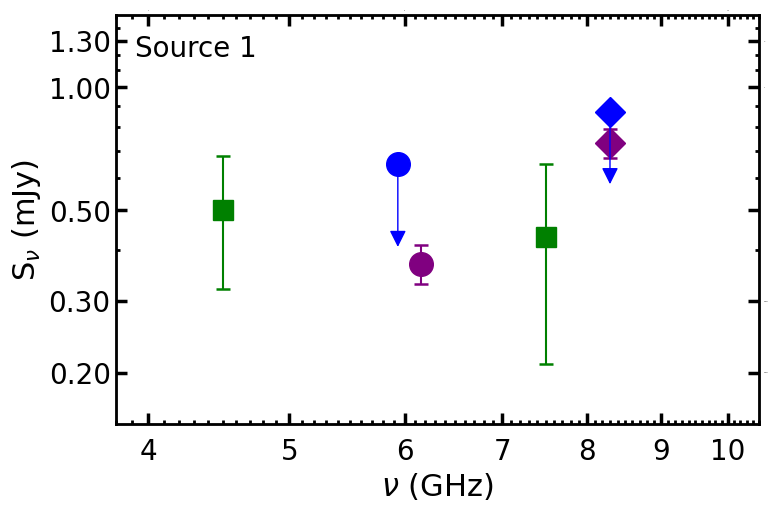}\hfill\includegraphics[width=0.33\textwidth]{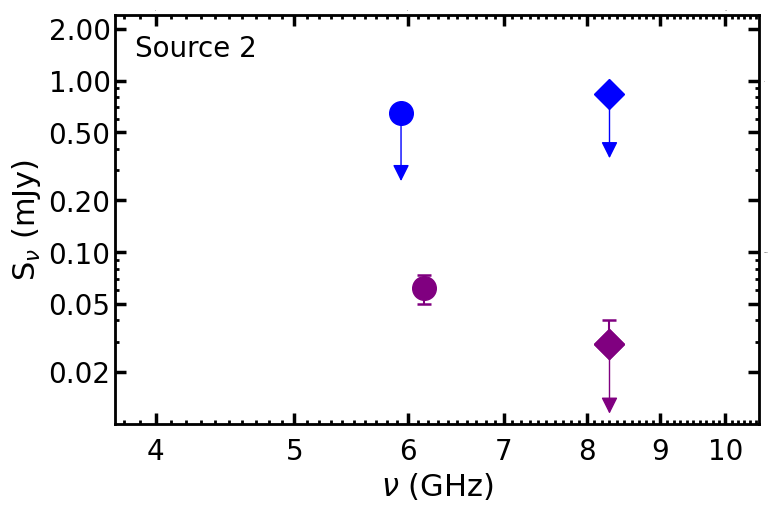}}
		{\includegraphics[width=0.33\textwidth]{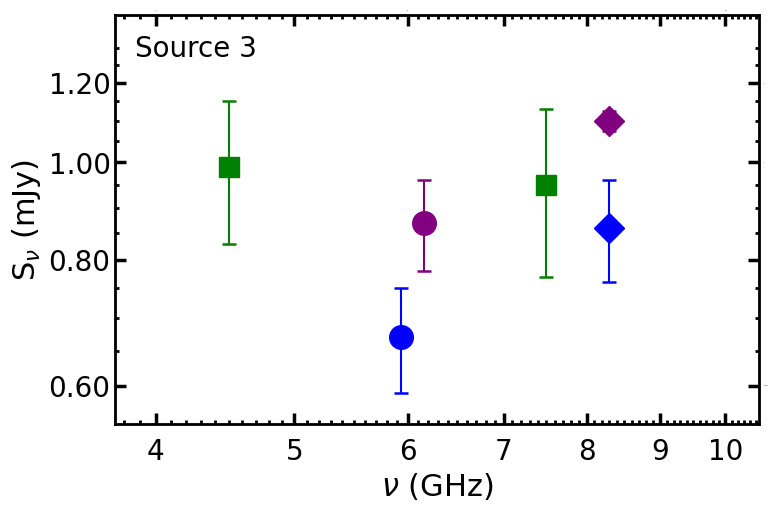}\hfill\includegraphics[width=0.33\textwidth]{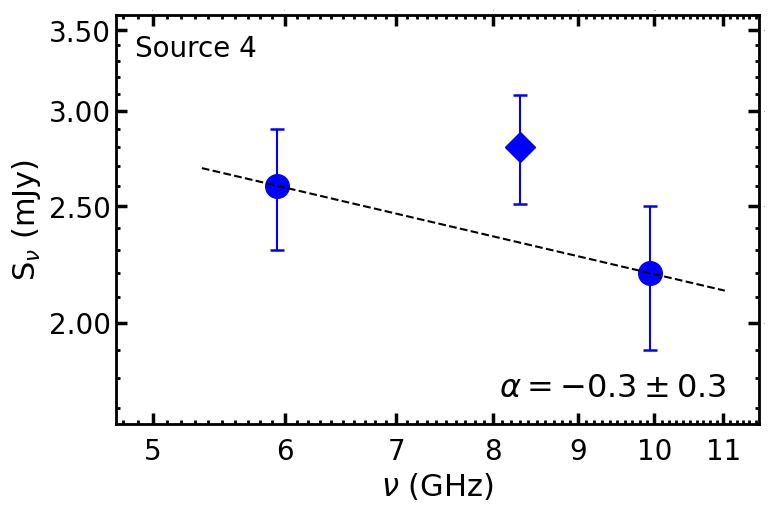}\hfill\includegraphics[width=0.33\textwidth]{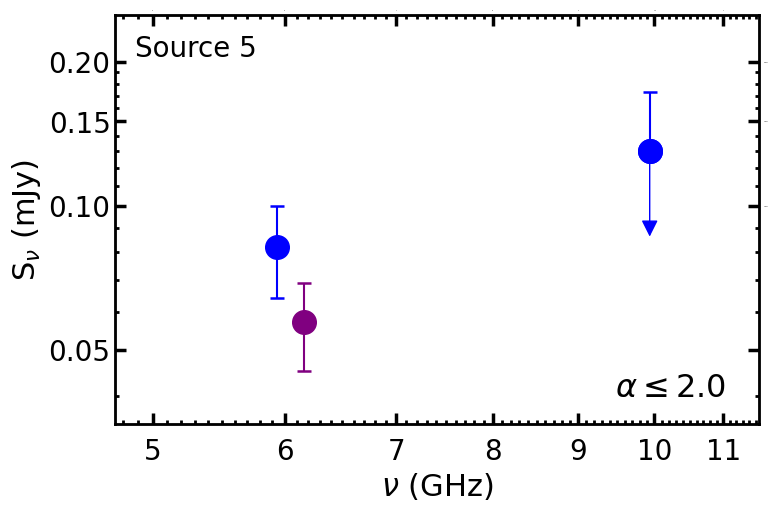}}\\
		{\includegraphics[width=0.33\textwidth]{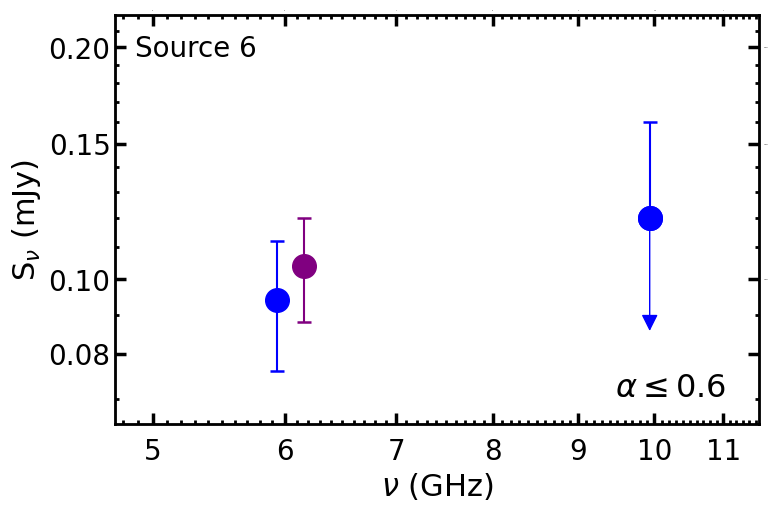}\hfill\includegraphics[width=0.33\textwidth]{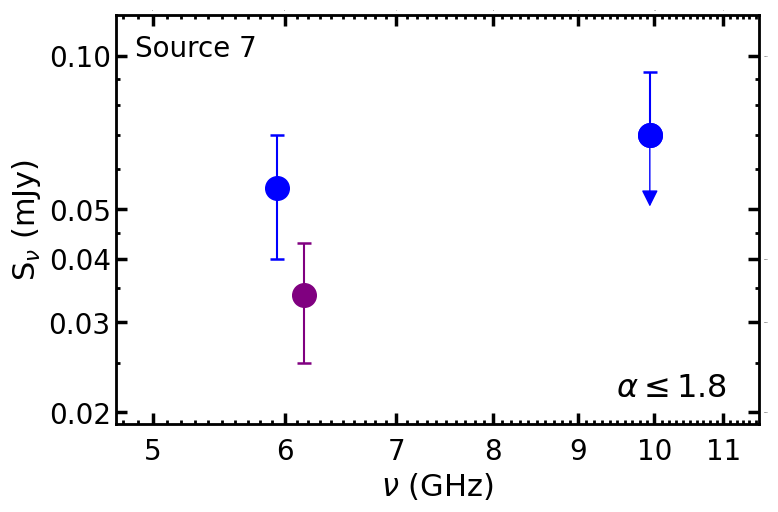}\hfill\includegraphics[width=0.33\textwidth]{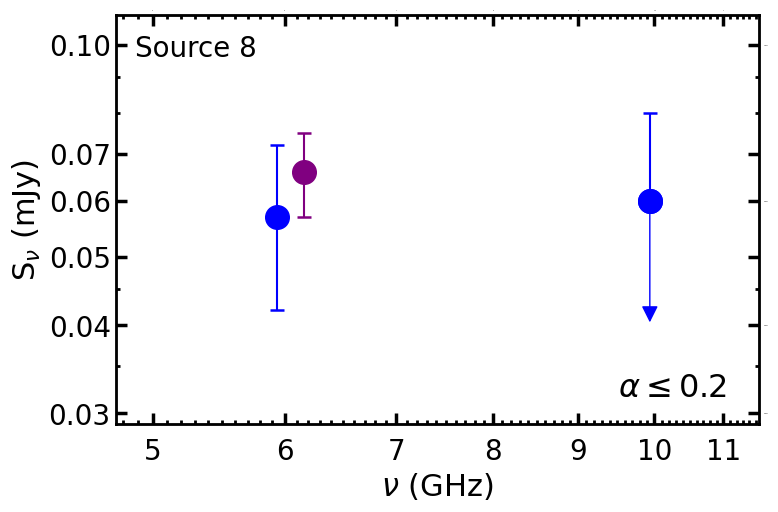}}\\
	{\includegraphics[width=0.33\textwidth]{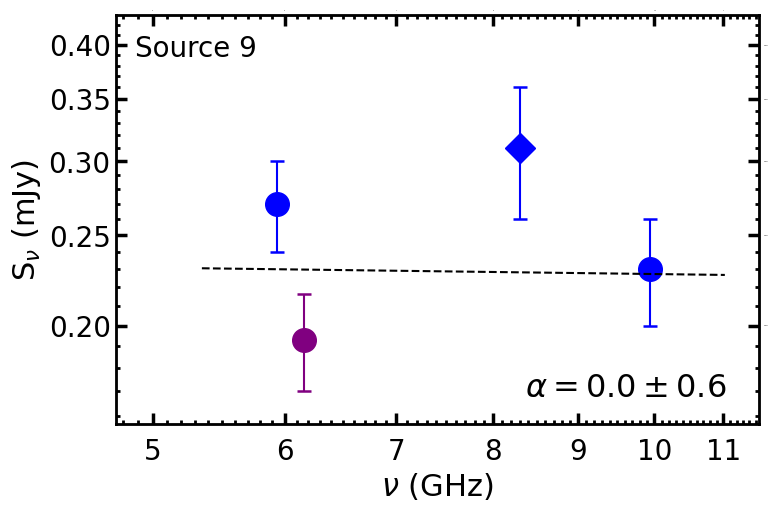}\hfill\includegraphics[width=0.33\textwidth]{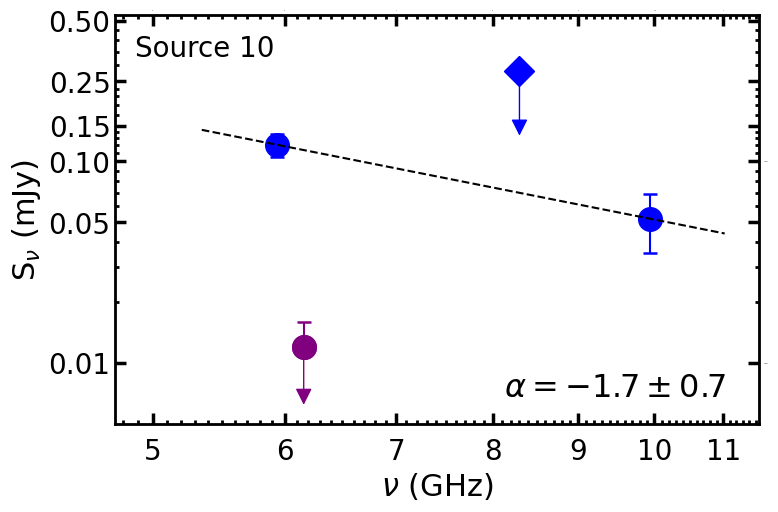}\includegraphics[width=0.33\textwidth]{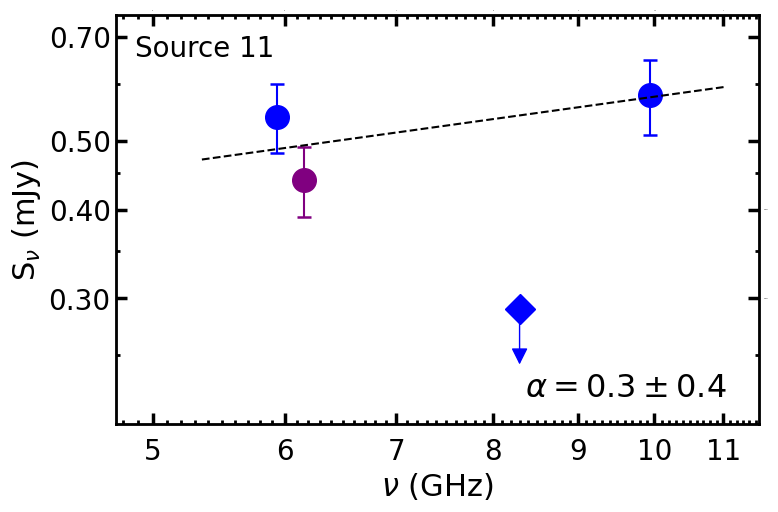}}\\
	{\includegraphics[width=0.33\textwidth]{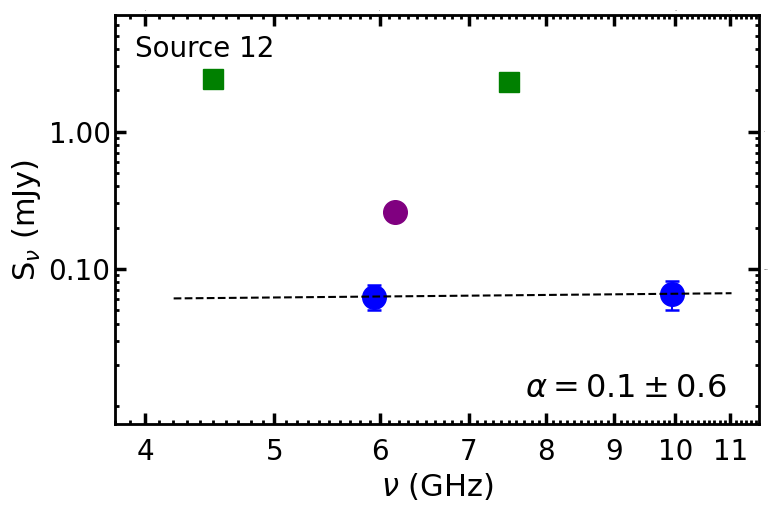}\hfill\includegraphics[width=0.33\textwidth]{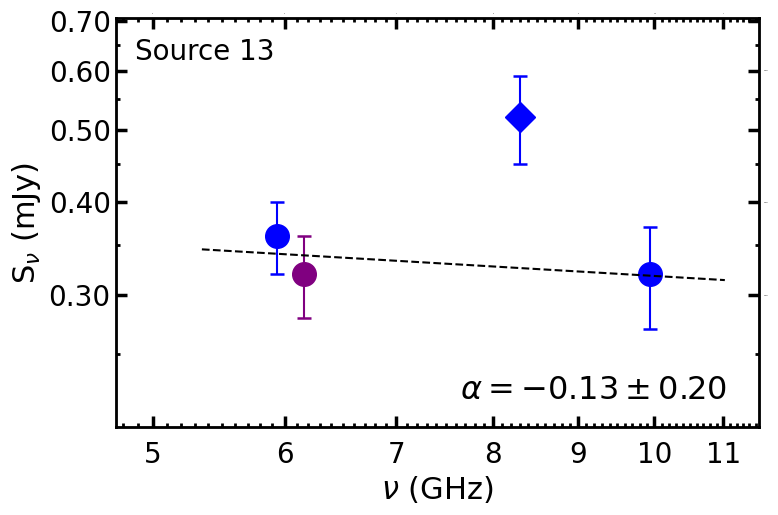}\hfill\includegraphics[width=0.33\textwidth]{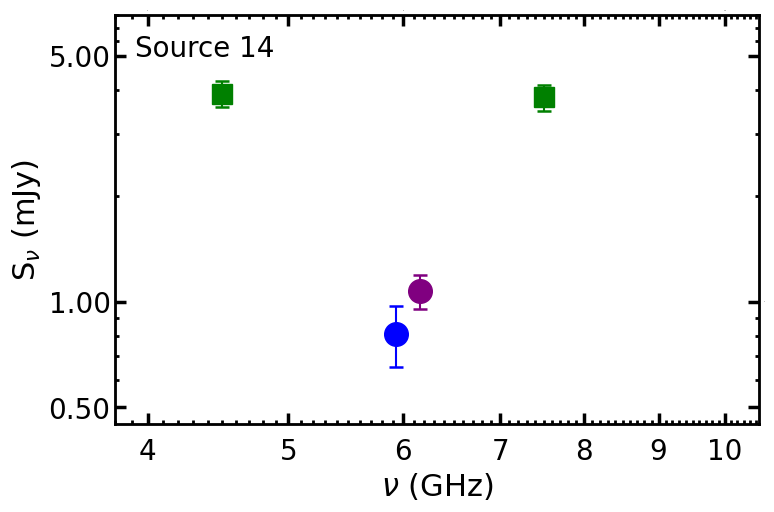}}\\
		\caption{Spectra of the radio sources in the centimeter range. Error bars are $\pm1\sigma$. In general, upper limits are represented by arrows where the central symbols are at 3$\sigma$. For sources blended at low-resolution observations (sources 1 and 2, 10 and 11, 24 and 25, 33 and 34) the total flux density measured for the two sources is given as an upper limit for each one. Least-squared fits (dashed lines) were made using only data from our work (2014 and 2015 epochs), taking into account upper limits when they constrain the fit. In the case of sources 10, 12, 19, 21, and 35, only 2014 data are used since their 2015 flux density is significantly different. The data represented (see first panel) are: red circles, 2015 A-configuration (red circles; this work); blue circles, 2014 C-configuration (this work); green squares, 2011 A-configuration \citep{Kounkel2014}; red diamonds, 2002 A-configuration \citep{Reipurth2004}, recalibrated; blue diamonds, 1998 D-configuration \citep{Reipurth1999}.}
	\label{fig:SEDs}
	\end{figure*}
	
	\begin{figure*} 
	{\includegraphics[width=0.33\textwidth]{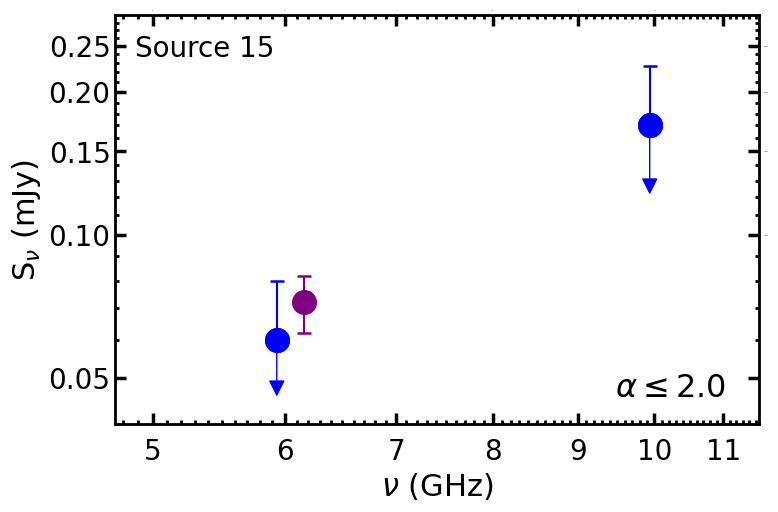}\hfill\includegraphics[width=0.33\textwidth]{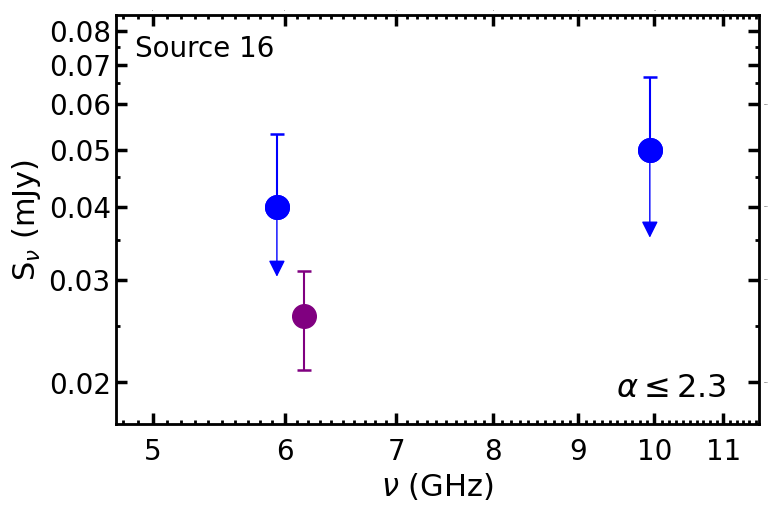}\hfill\includegraphics[width=0.33\textwidth]{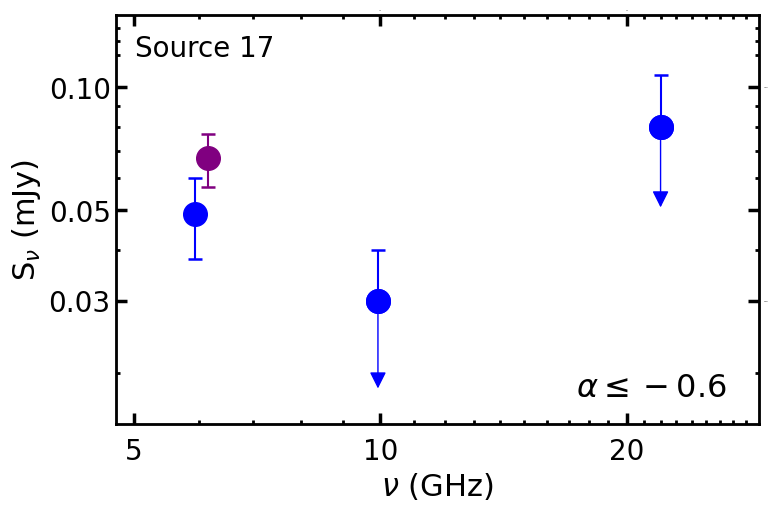}}\\
	{\includegraphics[width=0.33\textwidth]{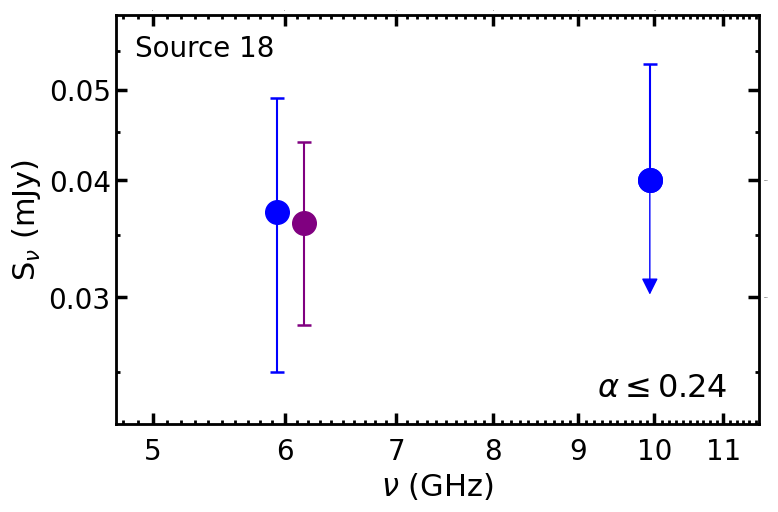}\hfill\includegraphics[width=0.33\textwidth]{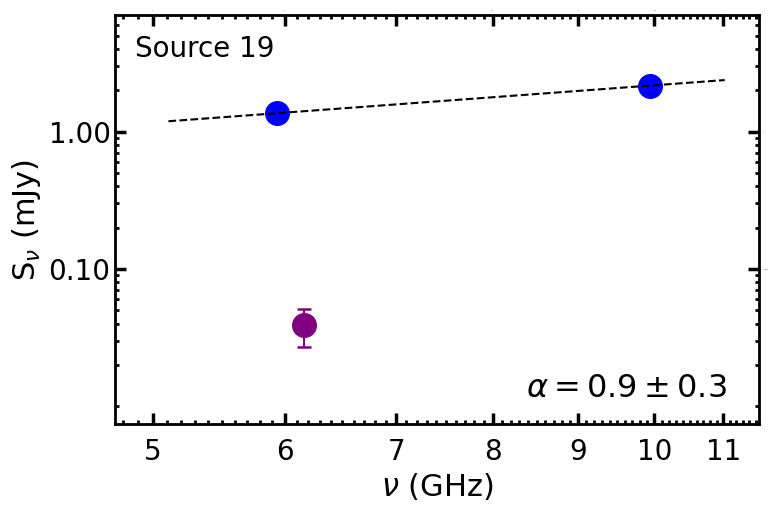}\hfill\includegraphics[width=0.33\textwidth]{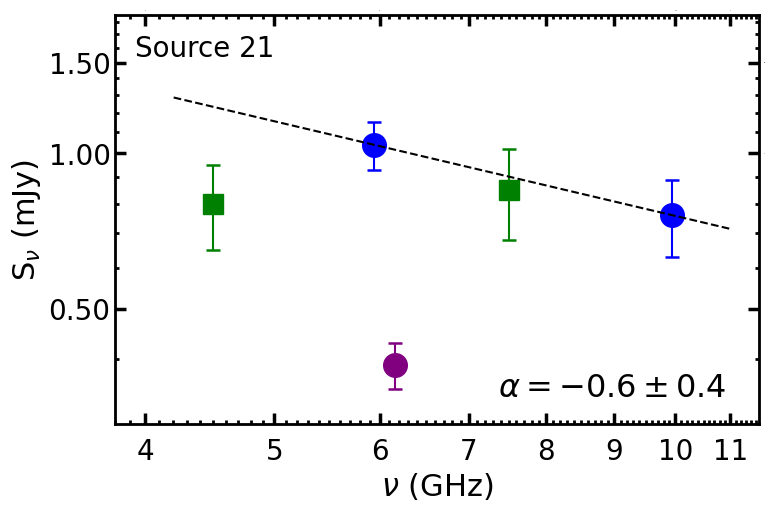}}\\
	{\includegraphics[width=0.33\textwidth]{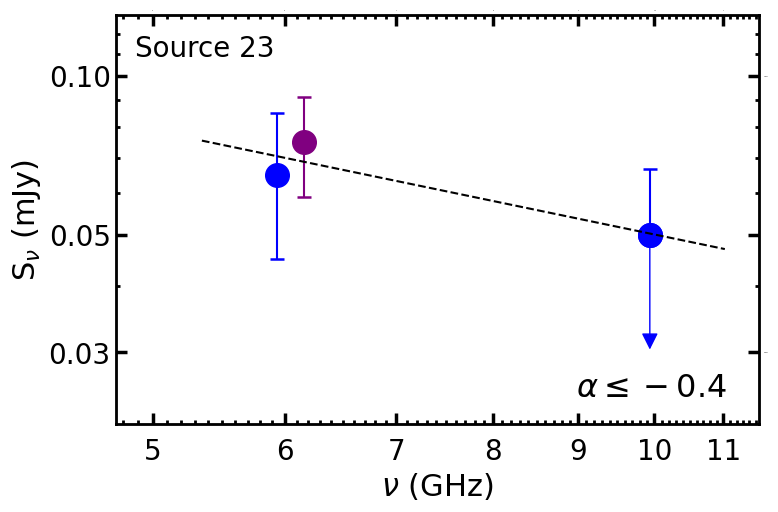}\hfill\includegraphics[width=0.33\textwidth]{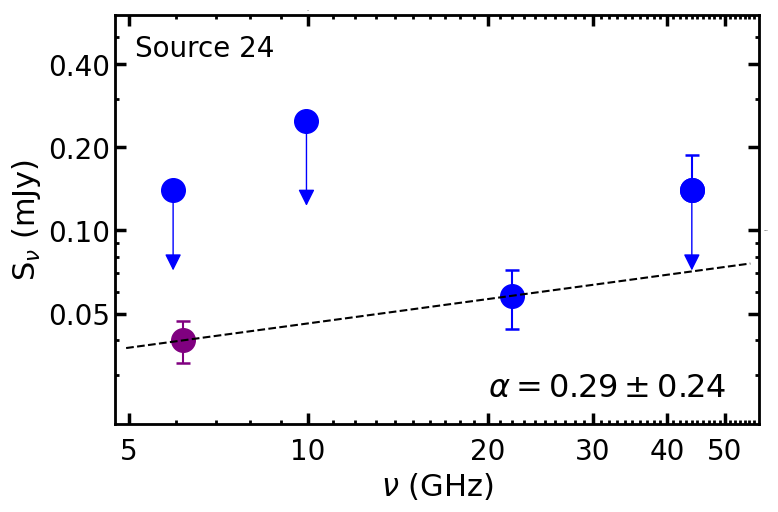}\hfill\includegraphics[width=0.33\textwidth]{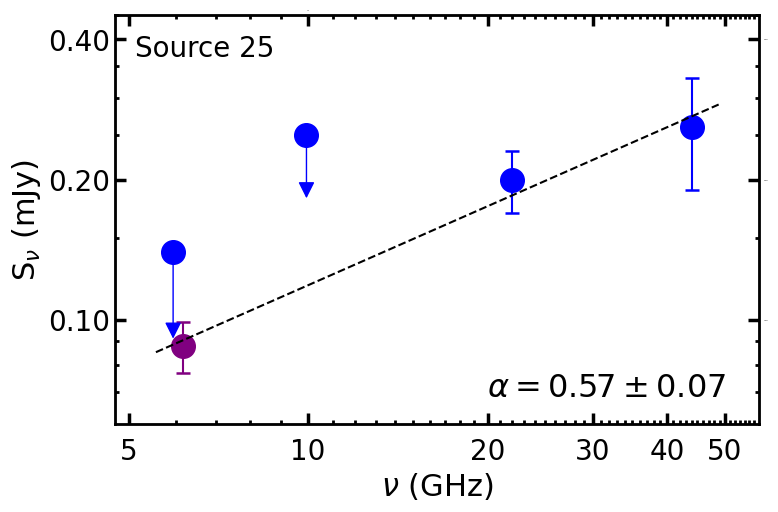}}\\
	{\includegraphics[width=0.33\textwidth]{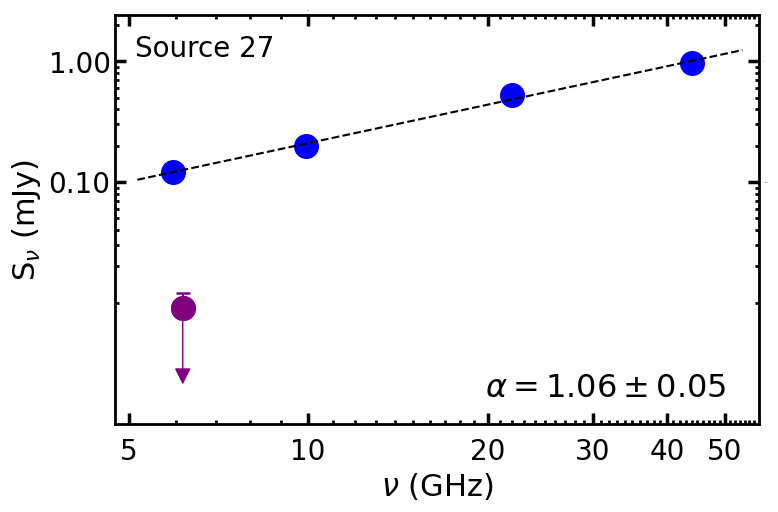}\hfill\includegraphics[width=0.33\textwidth]{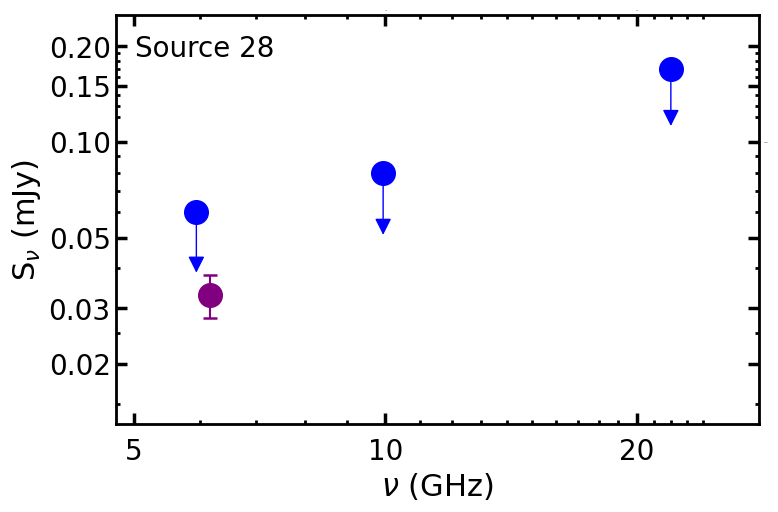}\hfill\includegraphics[width=0.33\textwidth]{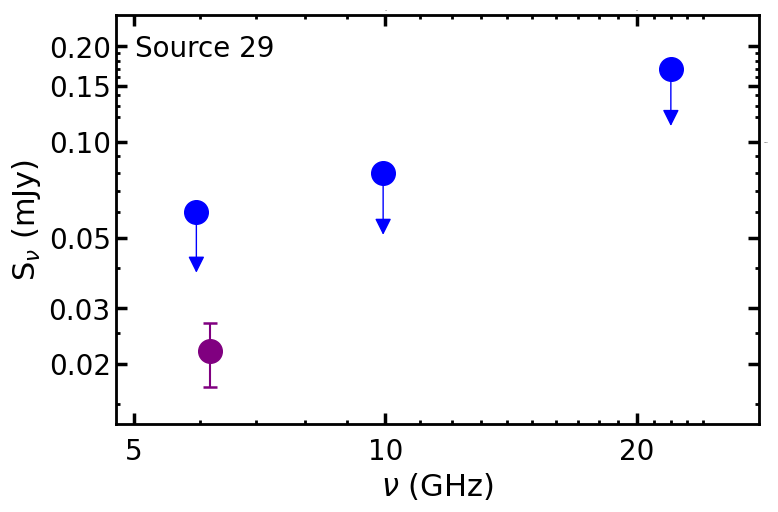}}\\
	{\includegraphics[width=0.33\textwidth]{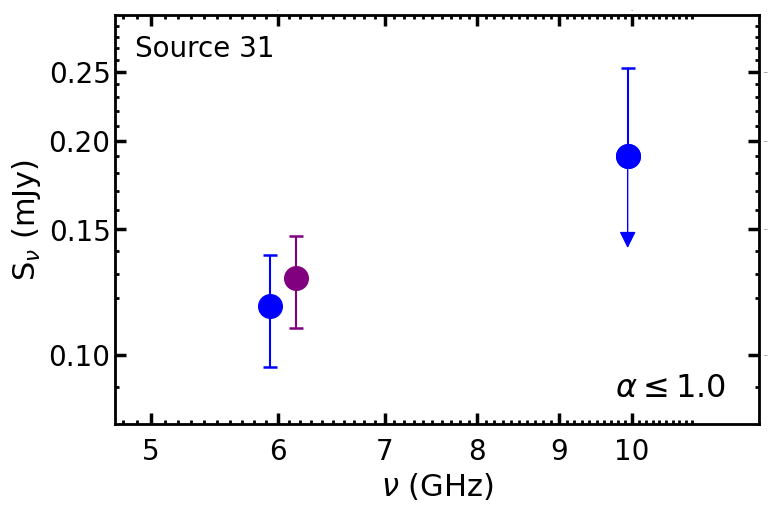}\hfill\includegraphics[width=0.33\textwidth]{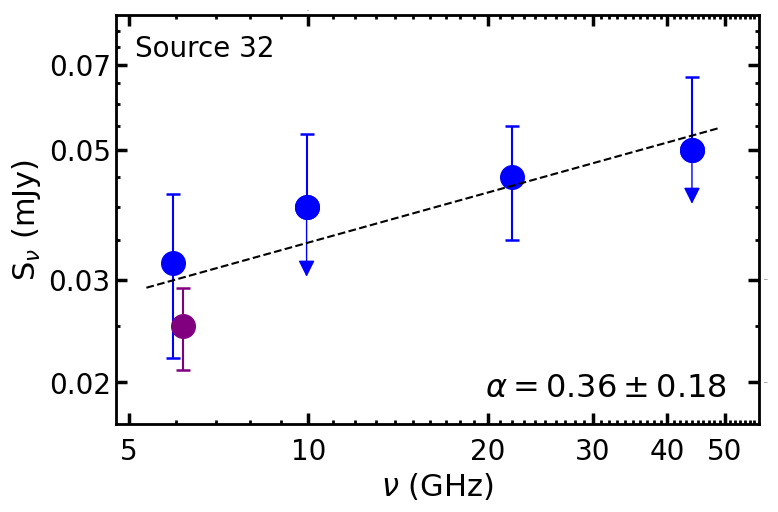}\hfill\includegraphics[width=0.33\textwidth]{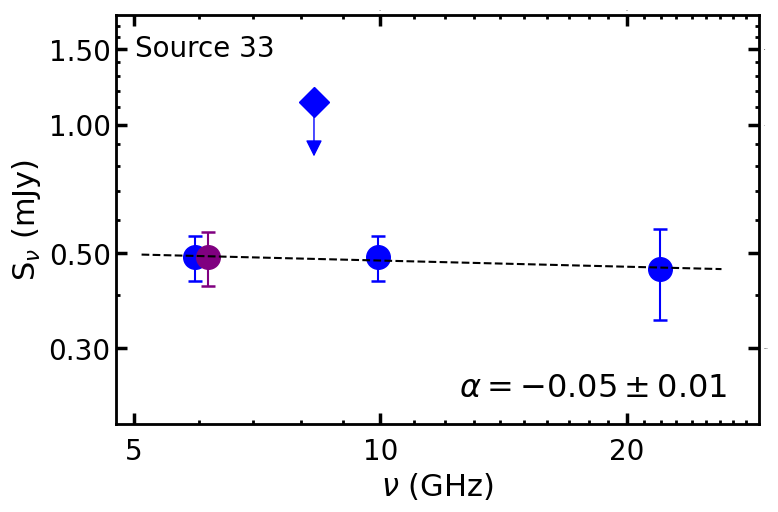}}\\
	{\includegraphics[width=0.33\textwidth]{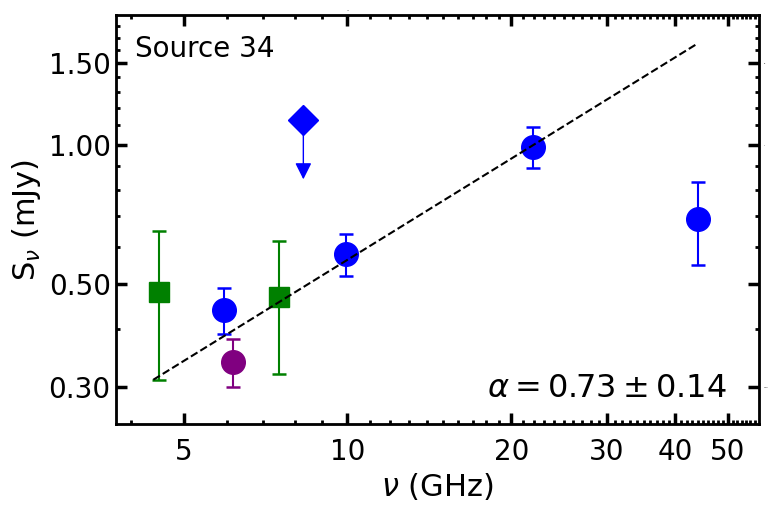}\hfill\includegraphics[width=0.33\textwidth]{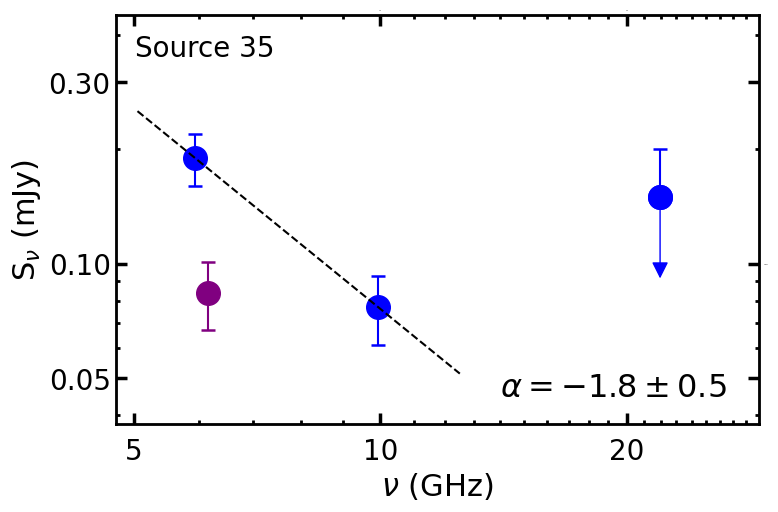}\hfill\includegraphics[width=0.33\textwidth]{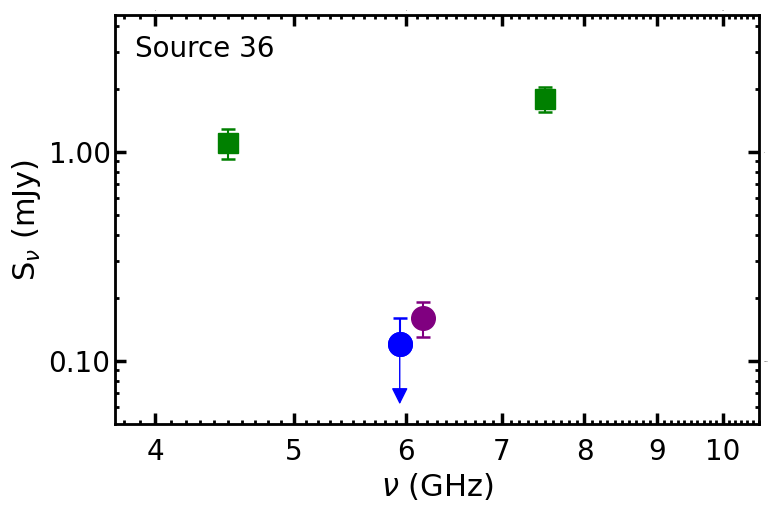}}
	\contcaption{}
	\label{fig:SEDs2}
	\end{figure*}
	
	\begin{figure*} 
	{\includegraphics[width=0.33\textwidth]{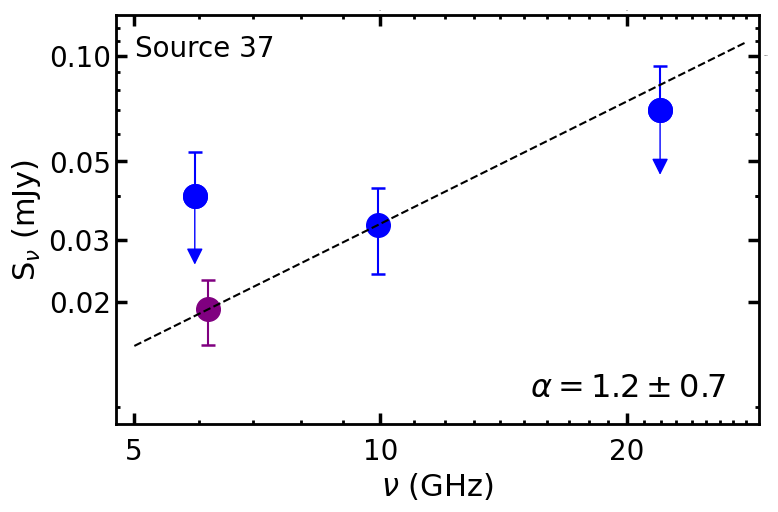}\hfill\includegraphics[width=0.33\textwidth]{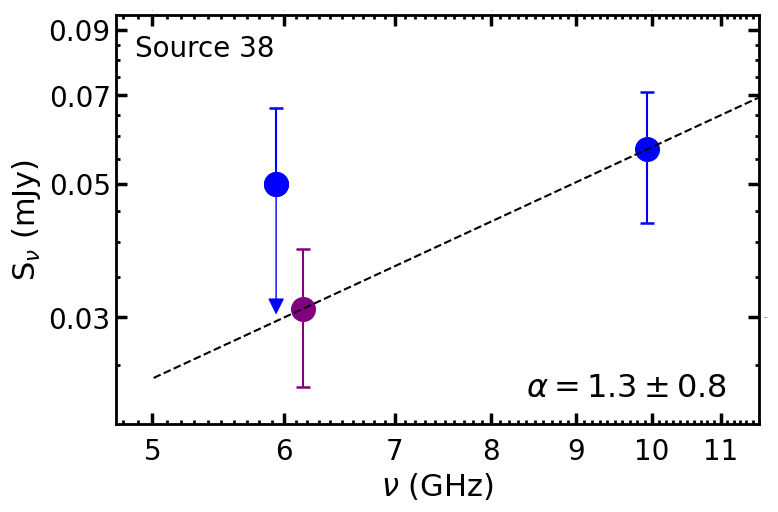}\hfill\includegraphics[width=0.33\textwidth]{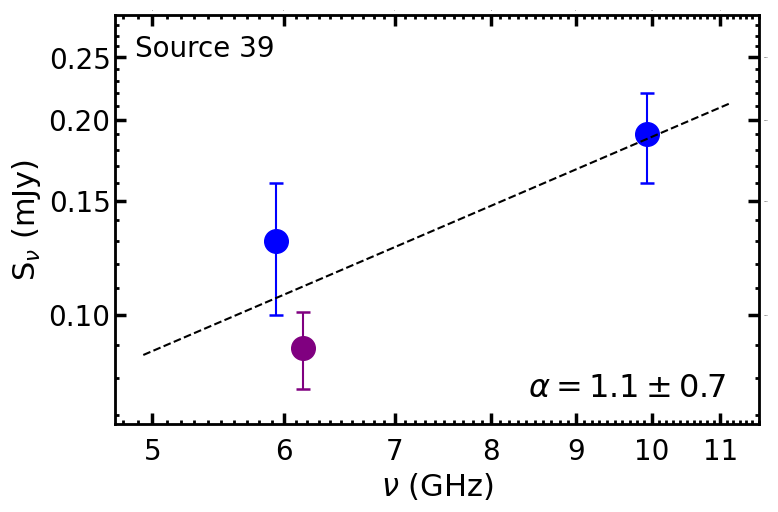}}\\
	{\includegraphics[width=0.33\textwidth]{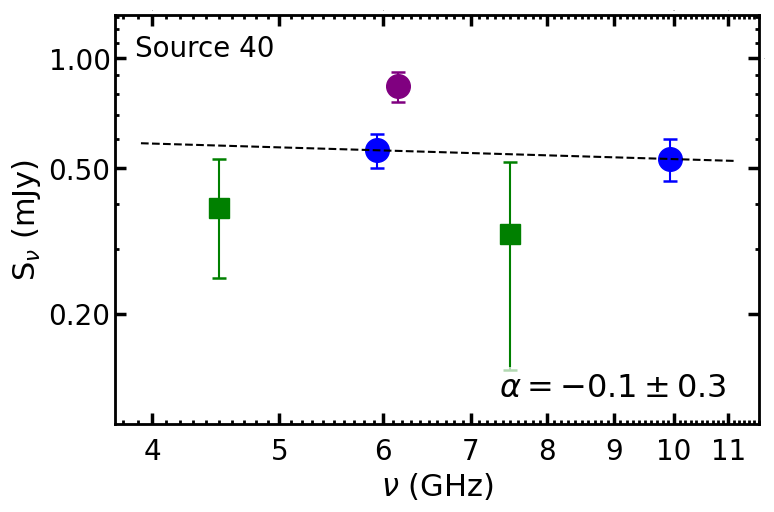}\hfill\includegraphics[width=0.33\textwidth]{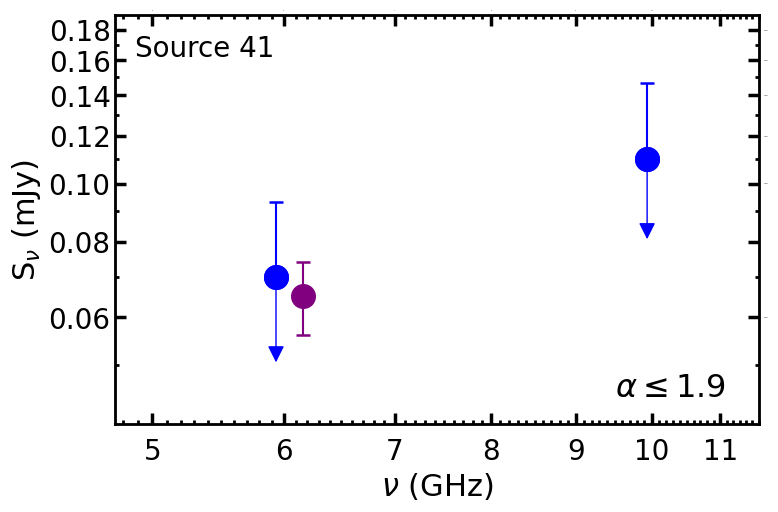}\hfill\includegraphics[width=0.33\textwidth]{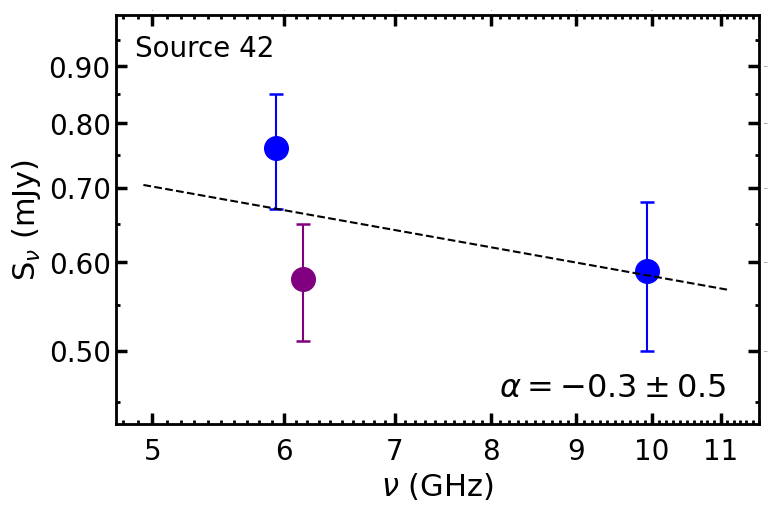}}\\
	{\includegraphics[width=0.33\textwidth]{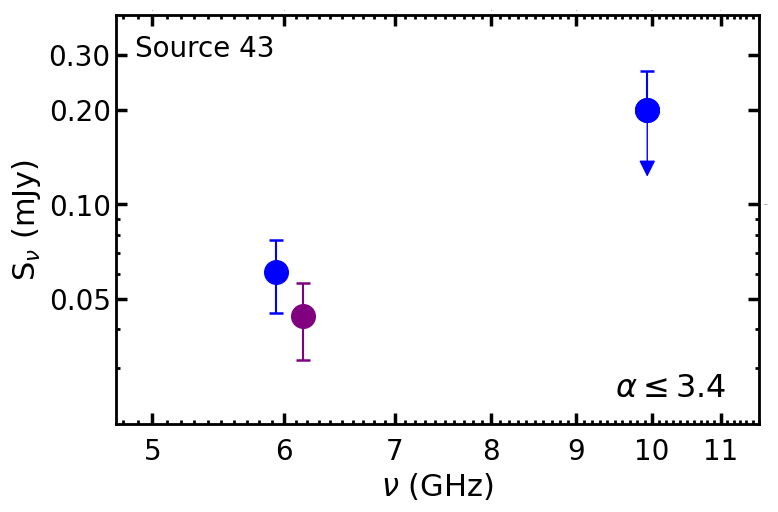}\hfill\includegraphics[width=0.33\textwidth]{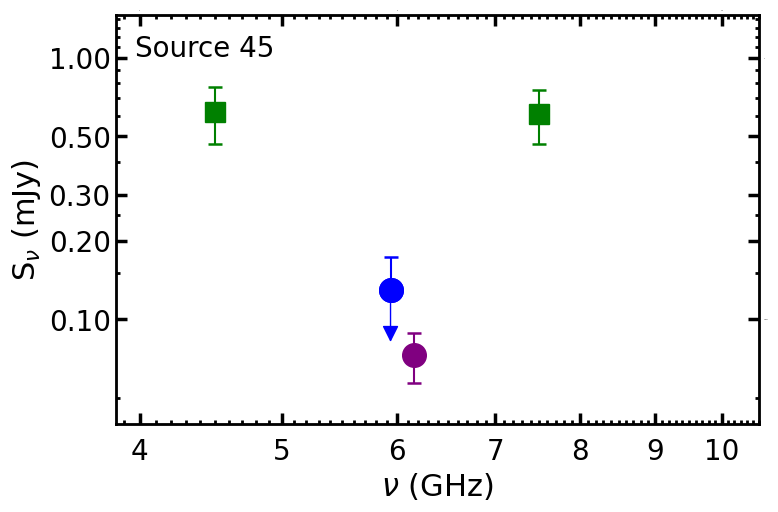}\hfill\includegraphics[width=0.33\textwidth]{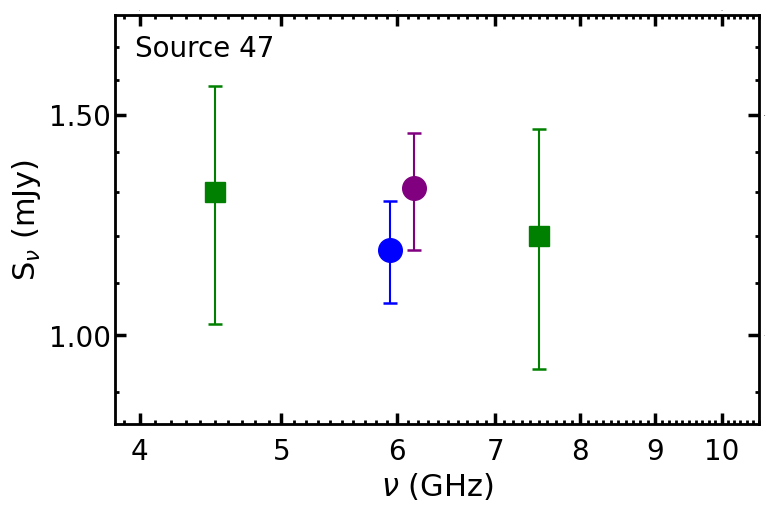}}
	\contcaption{}
	\label{fig:SEDs3}
	\end{figure*}

\clearpage

\section{Statistical treatment of spectral indices}
\label{app:spectral_statistics}

To account for measurement uncertainties and censored data, we performed 10,000 Monte Carlo realizations of the radio spectral indices, restricting their values to the physically motivated interval $-2.5\leq\alpha\leq2.5$. For detections, spectral indices were drawn from truncated Gaussian distributions centred on the measured values, with standard deviations given by their uncertainties. For upper and lower limits, values were drawn uniformly from the allowed intervals defined by the corresponding limits and the adopted boundaries. These realizations were generated using a custom Python function, \texttt{sample\_matrix}, based on the random-number routines of \texttt{NumPy} and the Gaussian cumulative distribution and inverse cumulative distribution functions \texttt{scipy.special.ndtr} and \texttt{scipy.special.ndtri}.

We used two approaches depending on the statistical quantity being evaluated. In the first approach, the source catalogue was kept fixed, and only the spectral indices were varied between realizations. This procedure was used to construct the cumulative distribution functions (CDFs), their Monte Carlo percentile bands, and the Kolmogorov–Smirnov (KS) comparisons between samples. The CDFs were calculated using \texttt{numpy.mean}, while the KS statistics were computed using a custom Python function, \texttt{ksdist}, based on the maximum difference between the empirical cumulative distributions. In the second approach, Monte Carlo sampling was combined with bootstrap resampling, in which sources were randomly selected with replacement using \texttt{numpy.random.Generator.integers}. This procedure was used to estimate the median spectral indices and their 95 per cent percentile intervals, as well as the intervals for the fractions of sources with $\alpha<-0.1$ and their evolutionary contrasts. The central estimates of the fractions were instead calculated as the mean fractions over the fixed-catalogue Monte Carlo realizations.

The reported intervals correspond to the 2.5th and 97.5th percentiles of the relevant simulated distributions, calculated using \texttt{numpy.percentile}. For the median spectral indices, the central values were taken as the medians of the Monte Carlo–bootstrap distributions, calculated using \texttt{numpy.median}. The percentile bands of the CDFs and the ranges of the KS statistics reflect uncertainties associated with the spectral-index measurements and the treatment of censored data, whereas the Monte Carlo–bootstrap intervals additionally account for sampling variability. All results involving censored sources depend on the assumed uniform distributions within their allowed intervals.

\begin{table}
\centering
\caption{Kolmogorov--Smirnov distances between OMC-2/3 and other star-forming regions for each evolutionary class.}
\label{tab:ks_omc_regions}
\begin{tabular}{llrrc}
\toprule
Comparison region & Class & $N_{\rm OMC}$ & $N_{\rm region}$ & $D_{\rm KS}$ \\
\midrule
Perseus & 0/I & 12 & 10 & $0.333^{+0.217}_{-0.133}$ \\
Perseus & II & 6 & 6 & $0.333^{+0.167}_{-0.167}$ \\
Perseus & III & 8 & 12 & $0.375^{+0.250}_{-0.125}$ \\
\addlinespace
Serpens & 0/I & 12 & 8 & $0.542^{+0.208}_{-0.208}$ \\
Serpens & II & 6 & 8 & $0.375^{+0.167}_{-0.167}$ \\
Serpens & III & 8 & 12 & $0.375^{+0.167}_{-0.167}$ \\
\addlinespace
Ophiuchus & 0/I & 12 & 14 & $0.321^{+0.226}_{-0.155}$ \\
Ophiuchus & II & 6 & 12 & $0.333^{+0.167}_{-0.167}$ \\
Ophiuchus & III & 8 & 27 & $0.463^{+0.213}_{-0.162}$ \\
\addlinespace
Taurus & 0/I & 12 & 10 & $0.267^{+0.217}_{-0.100}$ \\
Taurus & II & 6 & 18 & $0.389^{+0.167}_{-0.167}$ \\
Taurus & III & 8 & 22 & $0.375^{+0.205}_{-0.136}$ \\
\addlinespace
G14.2-N & 0/I & 12 & 7 & $0.750^{+0.167}_{-0.250}$ \\
G14.2-N & II & 6 & 3 & $0.667^{+0.333}_{-0.333}$ \\
G14.2-N & III & 8 & 6 & $0.708^{+0.292}_{-0.292}$ \\
\addlinespace
G14.2-S & 0/I & 12 & 5 & $0.400^{+0.200}_{-0.167}$ \\
G14.2-S & II & 6 & 5 & $0.333^{+0.167}_{-0.100}$ \\
G14.2-S & III & 8 & 6 & $0.375^{+0.292}_{-0.167}$ \\
\bottomrule
\end{tabular}
\par\smallskip
\begin{minipage}{0.48\textwidth}\footnotesize
\textit{Notes.} Values of $D_{\rm KS}$ are medians with asymmetric offsets to the 2.5th and 97.5th percentiles from 10\,000 fixed-catalogue Monte Carlo realisations. $N_{\rm OMC}$ and $N_{\rm region}$ are the number of sources involved in the comparison.
\end{minipage}
\end{table}

\begin{table}
\centering
\caption{Pairwise within-region Kolmogorov--Smirnov distances between evolutionary classes.}
\label{tab:ks_within_regions}
\begin{tabular}{llrrc}
\toprule
Region & Classes & $N_1$ & $N_2$ & $D_{\rm KS}$ \\
\midrule
OMC-2/3 & 0/I--II & 12 & 6 & $0.417^{+0.250}_{-0.167}$ \\
OMC-2/3 & 0/I--III & 12 & 8 & $0.333^{+0.208}_{-0.167}$ \\
OMC-2/3 & II--III & 6 & 8 & $0.375^{+0.167}_{-0.167}$ \\
\addlinespace
Perseus & 0/I--II & 10 & 6 & $0.300^{+0.167}_{-0.100}$ \\
Perseus & 0/I--III & 10 & 12 & $0.367^{+0.183}_{-0.150}$ \\
Perseus & II--III & 6 & 12 & $0.333^{+0.250}_{-0.083}$ \\
\addlinespace
Serpens & 0/I--II & 8 & 8 & $0.375^{+0.250}_{-0.125}$ \\
Serpens & 0/I--III & 8 & 12 & $0.292^{+0.125}_{-0.125}$ \\
Serpens & II--III & 8 & 12 & $0.292^{+0.167}_{-0.125}$ \\
\addlinespace
Ophiuchus & 0/I--II & 14 & 12 & $0.309^{+0.155}_{-0.131}$ \\
Ophiuchus & 0/I--III & 14 & 27 & $0.460^{+0.177}_{-0.161}$ \\
Ophiuchus & II--III & 12 & 27 & $0.315^{+0.157}_{-0.120}$ \\
\addlinespace
Taurus & 0/I--II & 10 & 18 & $0.289^{+0.156}_{-0.111}$ \\
Taurus & 0/I--III & 10 & 22 & $0.518^{+0.200}_{-0.209}$ \\
Taurus & II--III & 18 & 22 & $0.465^{+0.141}_{-0.131}$ \\
\addlinespace
G14.2-N & 0/I--II & 7 & 3 & $0.429^{+0.428}_{-0.143}$ \\
G14.2-N & 0/I--III & 7 & 6 & $0.429^{+0.286}_{-0.214}$ \\
G14.2-N & II--III & 3 & 6 & $0.500^{+0.333}_{-0.167}$ \\
\addlinespace
G14.2-S & 0/I--II & 5 & 5 & $0.400^{+0.200}_{-0.200}$ \\
G14.2-S & 0/I--III & 5 & 6 & $0.333^{+0.267}_{-0.100}$ \\
G14.2-S & II--III & 5 & 6 & $0.333^{+0.300}_{-0.100}$ \\
\bottomrule
\end{tabular}
\par\smallskip
\begin{minipage}{0.48\textwidth}\footnotesize
\textit{Notes.} Values of $D_{\rm KS}$ are medians with asymmetric offsets to the 2.5th and 97.5th percentiles from 10\,000 fixed-catalogue Monte Carlo realisations. $N_1$ and $N_2$ are the number of sources involved in the comparison.
\end{minipage}
\end{table}

\begin{figure*}
\centering
\includegraphics[width=0.46\textwidth] {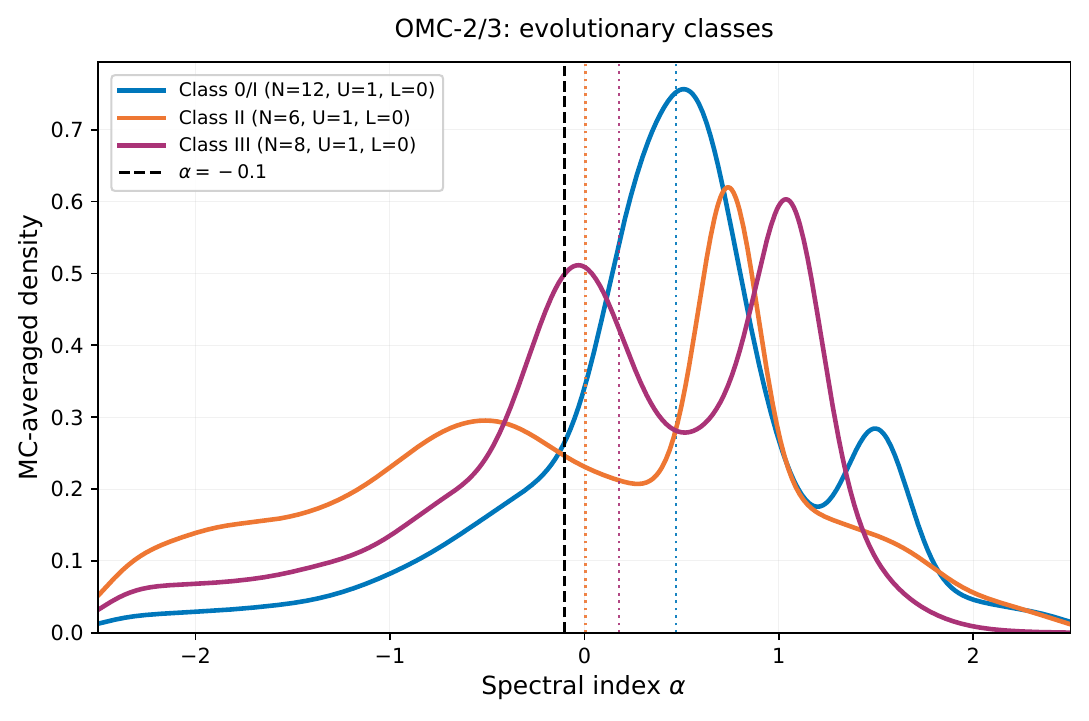}\hfill
\includegraphics[width=0.46\textwidth] {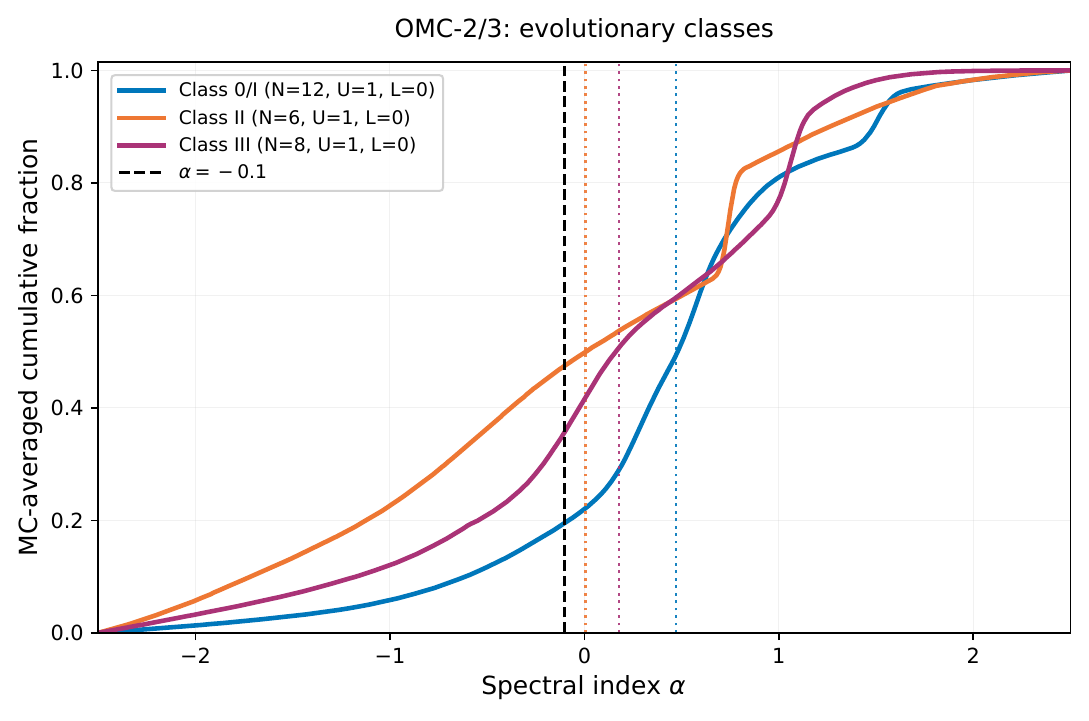} \\
\includegraphics[width=0.46\textwidth] {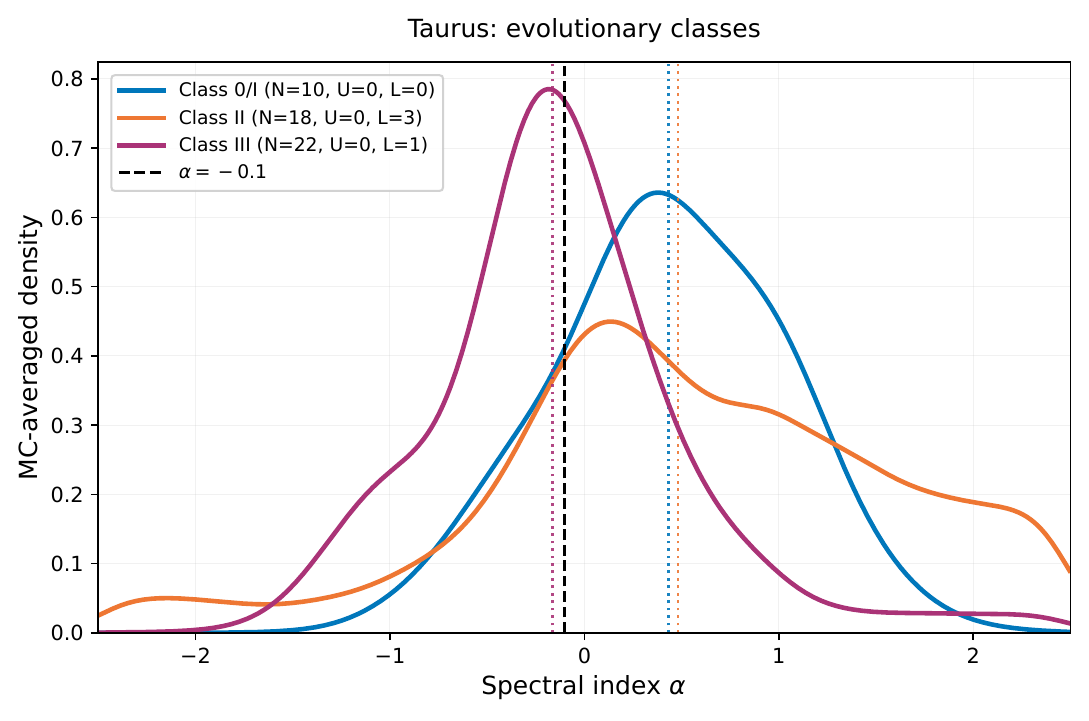}\hfill
\includegraphics[width=0.46\textwidth] {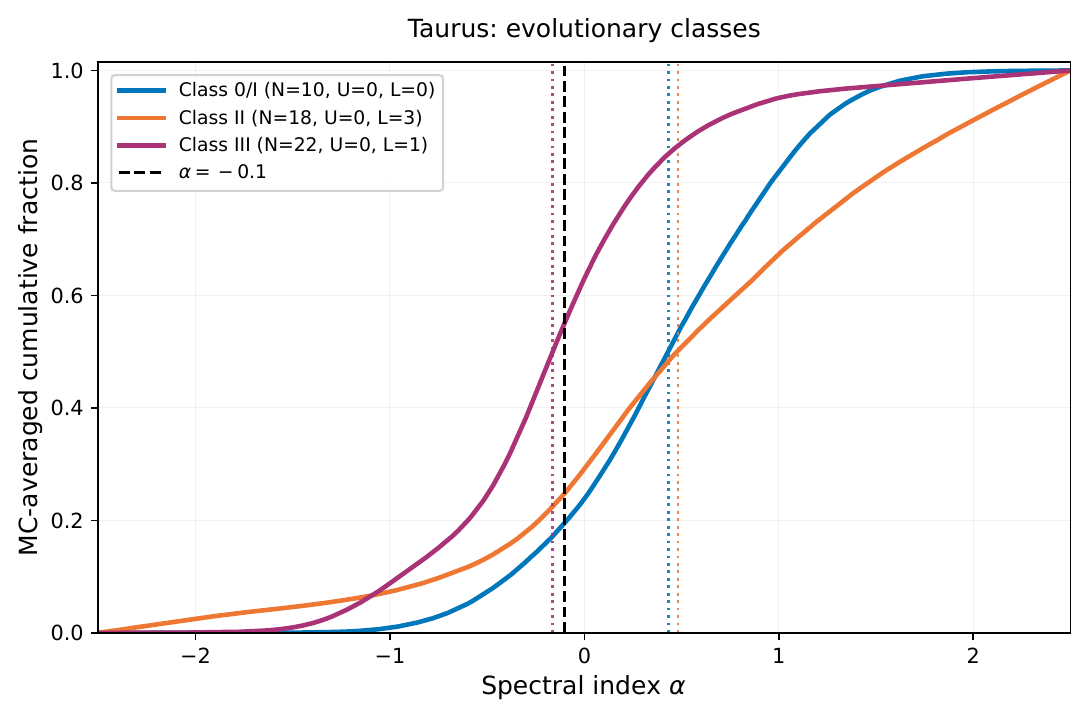} \\
\includegraphics[width=0.46\textwidth] {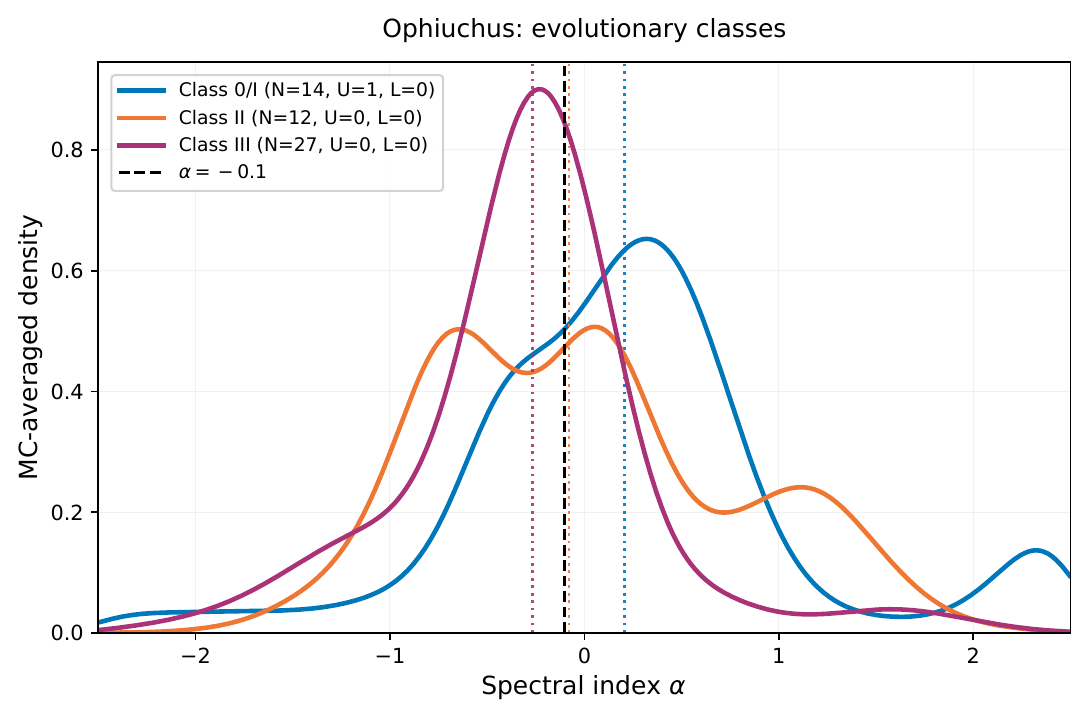}\hfill
\includegraphics[width=0.46\textwidth] {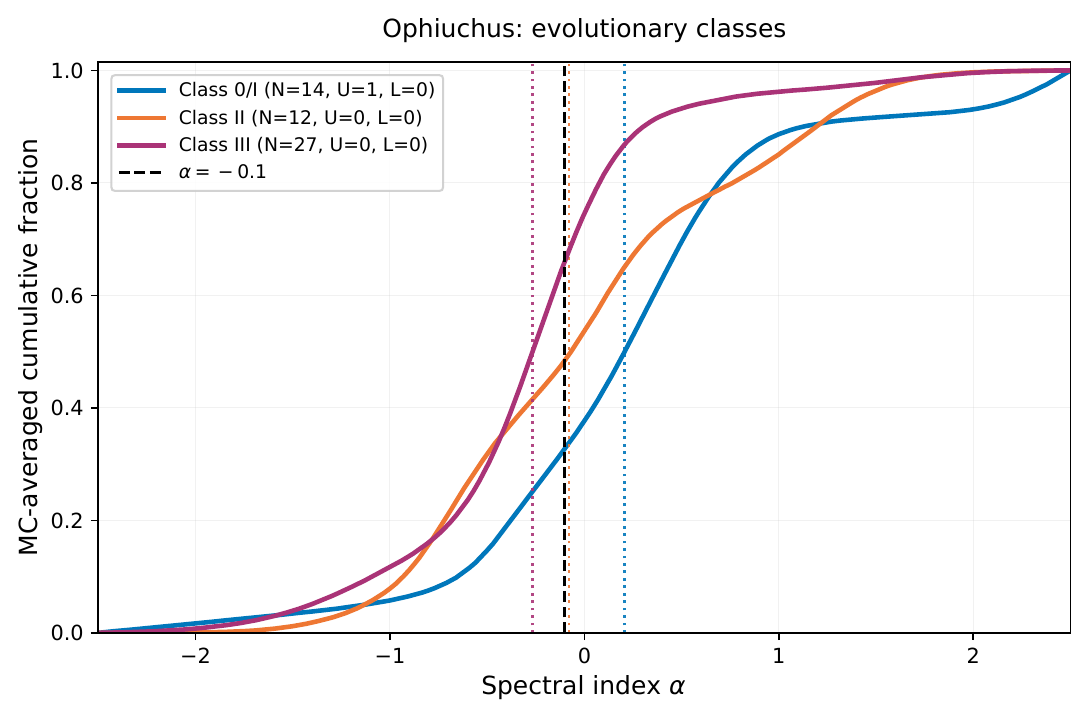} 
\caption{Monte Carlo-averaged kernel density estimates (KDEs; left) and cumulative distribution functions (CDFs; right) of the radio spectral indices for Class 0/I, II, and III sources in OMC-2/3 (top), Taurus (middle), and Ophiuchus (bottom). The distributions were obtained from 10,000 Monte Carlo realizations, accounting for measurement uncertainties and censored data. The density estimates were constructed using the  \texttt{numpy.histogram} function in Python and smoothed with a Gaussian kernel of bandwidth 0.15 using \texttt{scipy.ndimage.gaussian\_filter1d}. The coloured dotted lines indicate the median spectral indices of each evolutionary class, while the black dashed line marks $\alpha=-0.1$, below which spectral indices are consistent with non-thermal emission. }
\label{fig:regions_kde_cdf}
\end{figure*}

\subsection{Evolutionary trends in the fraction of sources with negative radio spectral indices}\label{app:negative_fractions}

To investigate possible evolutionary trends within individual star-forming regions, we calculated the change in the fraction of sources with $\alpha<-0.1$ between Class 0/I and Class III, defined as $\Delta f_{\rm Region}=f_{\mathrm{III}}-f_{\mathrm{0/I}}$. We obtained $\Delta f_{\rm Region}=0.167^{+0.375}_{-0.417}$ for OMC-2/3, $0.354^{+0.318}_{-0.345}$ for Taurus, and $0.344^{+0.291}_{-0.323}$ for Ophiuchus (Table~\ref{tab:class_fraction_differences}). The Monte Carlo–bootstrap 95\% percentile intervals marginally exclude zero for Taurus and Ophiuchus, providing tentative evidence for an increase in the fraction of sources with negative radio spectral indices towards more evolved YSOs in these regions. In contrast, the corresponding interval for OMC-2/3 includes zero, indicating that the observed increase is not statistically significant.

\begin{table}
\centering
\caption{Difference in the fractions of Class III and Class 0/I radio sources with $\alpha$$<$$-0.1$, for three regions. }
\label{tab:class_fraction_differences}
\begin{tabular}{lrrc}
\toprule
Region & $N_{\rm 0/I}$ & $N_{\rm III}$ & $\Delta f_{\rm Region}$ \\
\midrule
OMC-2/3 & 12 & 8 & $0.167^{+0.375}_{-0.417}$ \\
Taurus & 10 & 22 & $0.354^{+0.318}_{-0.345}$ \\
Ophiuchus & 14 & 27 & $0.344^{+0.291}_{-0.323}$ \\
\bottomrule
\end{tabular}
\par\smallskip
\begin{minipage}{\linewidth}\footnotesize
\textit{Notes.} $\Delta f_{\rm Region}$ = $f_{\rm III}(\alpha$$<$$-0.1)$ $-$ $f_{\rm 0/I}(\alpha$$<$$-0.1)$, where $ f_{\rm III}(\alpha$$<$$-0.1)$ and $f_{\rm 0/I}(\alpha$$<$$-0.1)$ are, respectively, the fractions of Class III and Class 0/I sources with $\alpha$$<$$-0.1$ in the region, obtained from Table~\ref{tab:spectral_medians_fractions_boot}. Values are medians with 95\% percentile offsets from 10\,000 Monte Carlo-bootstrap draws. The 95\% Monte Carlo-bootstrap percentile intervals account for measurement uncertainties and sampling variability, but depend on the assumptions adopted for censored data. $N_{\rm 0/I}$ and $N_{\rm III}$ are the number of sources involved in the comparison.
\end{minipage}
\end{table}

To assess whether these evolutionary trends differ among star-forming regions, we also compared the values of $\Delta f_{\rm Region}$ between regions. Although the estimated increase is larger in Taurus and Ophiuchus than in OMC-2/3, the Monte Carlo–bootstrap percentile intervals for the differences between regions include zero in all cases (Table~\ref{tab:between_region_fraction_contrasts}. Therefore, we find no statistically significant evidence that the evolutionary trends differ among regions.

These results should be interpreted with caution because of the limited sample sizes and the model dependence introduced by the treatment of censored spectral indices.

\begin{table}
\centering
\caption{Between-region contrasts of the class-dependent fraction of sources with $\alpha$$<$$-0.1$.}
\label{tab:between_region_fraction_contrasts}
\begin{tabular}{llc}
\toprule
Region 1 & Region 2 & $\Delta f_{\rm Region\, 1}-\Delta f_{\rm Region\, 2}$ \\
\midrule
Taurus & OMC-2/3 & $0.195^{+0.497}_{-0.519}$ \\
Ophiuchus & OMC-2/3 & $0.175^{+0.489}_{-0.509}$ \\
Taurus & Ophiuchus & $0.018^{+0.443}_{-0.442}$ \\
\bottomrule
\end{tabular}
\par\smallskip
\begin{minipage}{\linewidth}\footnotesize
\textit{Notes.} $\Delta f_{\rm Region\, 1}$ and $\Delta f_{\rm Region\, 2}$ are the differences in the fractions of Class III and Class 0/I radio sources with $\alpha$$<$$-0.1$ in each region, given in Table~\ref{tab:class_fraction_differences}. 
Values of $\Delta f_{\rm Region\, 1}-\Delta f_{\rm Region\, 2}$ are medians with 95\% percentile offsets from 10\,000 Monte Carlo-bootstrap draws. 
\end{minipage}
\end{table}

\subsection{Robustness and observational limitations}
\label{app:validation}

We assessed the sensitivity of selected KS comparisons to the treatment of censored measurements by considering alternative imputations that preferentially place the unknown spectral indices closer to, or farther from, their reported limits. The median KS statistics were relatively stable under these assumptions, remaining
at $D_{\rm KS}=0.267$ for OMC-2/3 versus Taurus (Class 0/I), and varying between $D_{\rm KS}=0.454$ and $0.463$ for Ophiuchus (Class III), and between $D_{\rm KS}=0.500$ and $0.542$ for Serpens (Class 0/I).

We also performed 10\,000 bootstrap resamplings, propagating measurement and imputation uncertainties.
Under the uniform-imputation model, the median KS statistics and their 95\% percentile intervals
were $D_{\rm KS}=0.333^{+0.267}_{-0.166}$ for Taurus (Class 0/I), $D_{\rm KS}=0.477^{+0.324}_{-0.250}$ for Ophiuchus (Class III), and $D_{\rm KS}=0.542^{+0.333}_{-0.250}$ for Serpens (Class 0/I). These broad intervals indicate that finite sample sizes contribute appreciably to the uncertainty in the inferred distributional differences.

The comparisons are also affected by differences in observational sensitivity, frequency coverage,
source-selection criteria, and censoring mechanisms among the surveys. Since the samples comprise
radio-detected sources with available evolutionary classifications rather than complete YSO populations,
observational selection effects may influence the spectral-index distributions. Furthermore, standard
KS $p$-values obtained from individual imputed data sets cannot be interpreted as calibrated
significance levels for the original censored samples. We therefore use the KS statistic as a descriptive
measure and do not claim statistically significant differences between the underlying distributions.

\section{Notes on the nature of other individual sources}\label{subsec:individual2}

\subsubsection{Source 3 (VLA 6)}\label{3}
	
	This source is located toward a tenuous part of the ISF, at $\sim40\arcsec$ from its ridgeline (see Fig. \ref{fig:filament}) and is not associated with any 3 mm compact dust core \citep{Kainulainen2017}. The source, also known as VLA 6, was previously detected with the VLA at 3.6 cm by \citet{Reipurth1999,Reipurth2004} and at 5 cm by \citet{Kounkel2014}, and does not have known counterparts at shorter wavelengths.
	
	The source has also been detected at 6 cm with the VLBA (milli-arcsec angular resolution) with a maximum flux of $1.7\pm0.5$ mJy \citep{Kounkel2017}, indicating that it is very compact and therefore non-thermal. These authors proposed the source as extragalactic, as previously suggested by \citet{Reipurth1999}, since it does not have proper motions like the YSOs of the region.
	
	However, in our inspection of the 3 GHz VLA Sky Survey (VLASS, \citealt{Lacy2020}) map, we find that this source is detected with a flux density of $0.85\pm0.16$ mJy, which, together with the flux densities shown in Fig. \ref{fig:SEDs}, results in a flat spectral index $\alpha = 0.08\pm0.18$, consistent with thermal, optically thin free-free emission. In our 5 cm high-resolution map the source is marginally resolved in the N-S direction (Fig. \ref{fig:sources}). This suggests that the source could be composed of a non-thermal compact core (detected by the VLBA) surrounded by a more extended thermal envelope (detected by us with the VLA). Follow up observations at shorter radio wavelengths (e.g., 0.7 cm) could better constrain the spectral index and shed light on the nature of this object.

\subsubsection{Sources 5, 16, 18, 23, 31, 35}\label{weakKs}

\begin{figure*}
	\center\includegraphics[width=1.01\hsize]{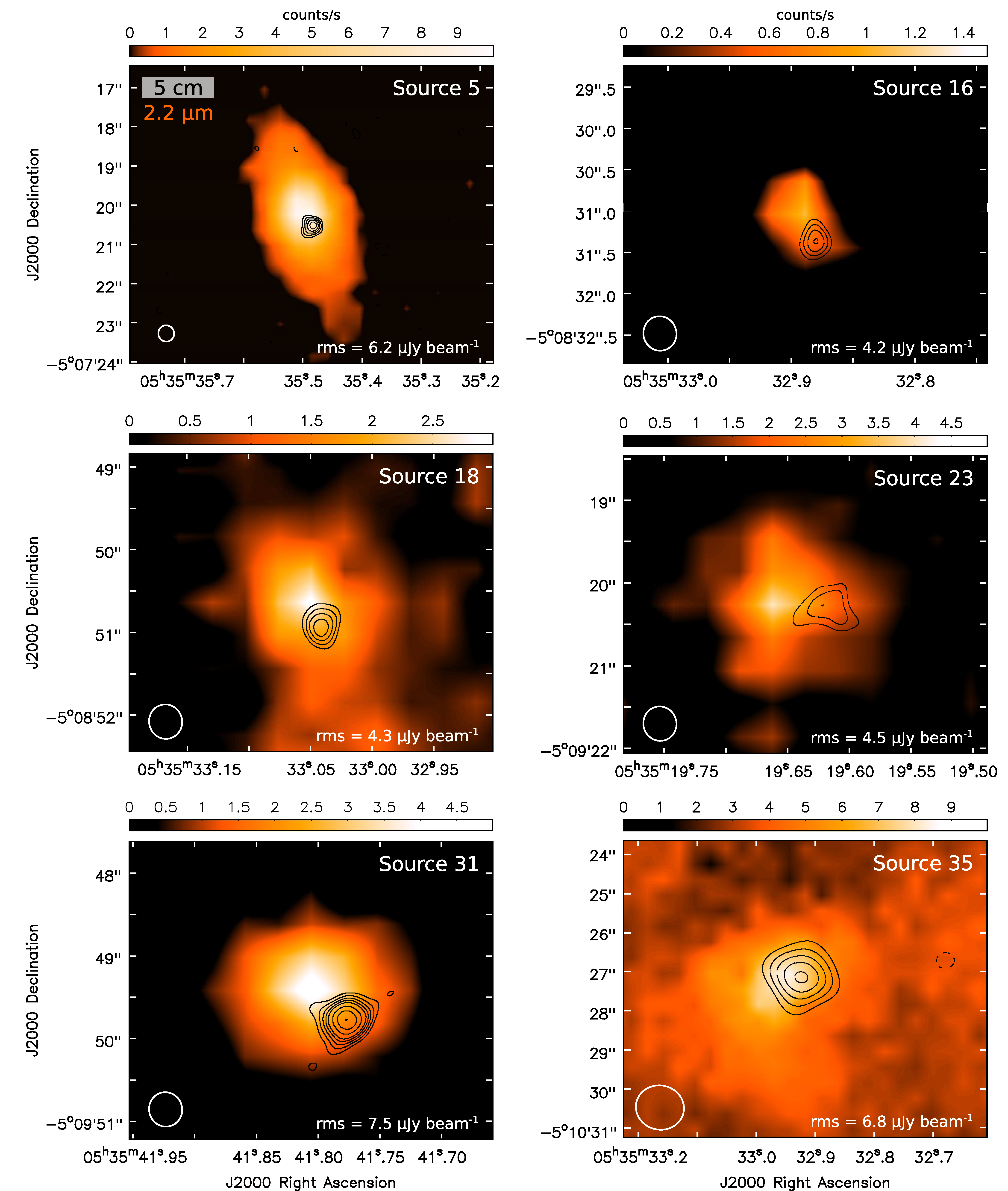}
	\caption{Sources 5, 16, 18, 23, 31 and 35. VLA 5 cm A-configuration map with natural weighting (black contours) overlaid on the NEWFIRM 2.2 $\micron$ image (color scale, T. Megeath, private communication). In the case of Source 35, uvtaper = 100 k$\lambda$ was applied to better detect the extended emission. Contour levels are $-$3, 3, 4, 5, 6, 7, 9, 11, and 13 times the rms of the 5 cm maps, shown at the bottom right corner. The VLA synthesized beam is shown in the bottom left corner of the maps.
		\label{fig:Ks}}
\end{figure*}

These sources are reported here for the first time at radio wavelengths. They are detected toward the border or outside the ISF, more than $1\farcm4$ from its ridgeline (see Fig. \ref{fig:filament}), and are not associated with any 3 mm compact dust core \citep{Kainulainen2017}. We find that all of them are associated with weak ($\le$ 10 counts/s) emission in the inspected NEWFIRM K${_s}$ image (see Fig. \ref{fig:Ks}).

In the case of Source 5, its near-IR emission is extended, with morphology and angular size consistent with that of a disc galaxy. The source also has an X-ray counterpart with a hardness ratio suggesting an extragalactic nature according to \citet{Tsujimoto2002} (see Table \ref{tab:sources} for its counterparts), although its radio spectral index is unconstrained.

We determine a negative spectral index for sources 23 and 35 (see Table \ref{tab:fluxes} and Fig. \ref{fig:SEDs}), suggesting non-thermal emission. Therefore they could be either extragalactic sources or evolved YSOs. On the other hand, the spectral indices of sources 5, 16, 18, and 31 are unconstrained, and the nature of their radio emission is uncertain.

\subsubsection{Sources 6, 8, 15, and 44}\label{?}

These sources are reported here for the first time at radio wavelengths and do not have any known counterpart at other wavelengths. They are detected toward the border or outside the ISF (see Fig. \ref{fig:filament}), with distances to its ridgeline ranging between 37$\arcsec$ (Source 8) and $2\farcm7$ (Source 15). Their radio spectral indices are unconstrained (see Table \ref{tab:fluxes} and Fig. \ref{fig:SEDs}) and, therefore, their nature is uncertain.

\subsubsection{Source 7}\label{7}

This source is reported here for the first time at radio wavelengths. It is located toward the densest part of the ISF, $\sim23\arcsec$ from its ridgeline (see Fig. \ref{fig:filament}), although it is not associated with any 3 mm compact dust core \citep{Kainulainen2017}. Based on IR observations (see Table \ref{tab:sources} for the counterparts), the source has been previously classified as a YSO with a disc \citep{Megeath2012,Getman2017}. Interestingly, \citet{Kounkel2016} reported a near-IR counterpart (MGM 2341) of Source 7 and a second near-IR source (MGM 2341-b), undetected at radio wavelengths, at a projected separation of $0\farcs5$ ($\sim$ 190 au), than could be its binary companion.

\subsubsection{Source 13 (VLA 10). }\label{13}

This source, previously detected at 3.6 cm and named VLA 10 \citep{Reipurth1999}, is located toward the border of the ISF, $\sim2\arcmin$ from its ridgeline (see Fig. \ref{fig:filament}). The source has an X-ray counterpart (Table \ref{tab:sources}) whose hardness ratio, along with the lack of near-IR counterpart, suggests an extragalactic nature according to \citet{Tsujimoto2002}. \citet{Reipurth1999} also suggested it is probably a background radio galaxy based on the lack of a clear association with dust.

We cannot constrain with our observations the spectral index of Source 13 (Table \ref{tab:fluxes} and Fig. \ref{fig:SEDs}). However, in our inspection of the 2-4 GHz VLASS map \citep{Lacy2020} we find that the source is marginally detected at $\sim$ 2$\sigma$ with a flux density of $\sim$ 0.4 mJy. This value (although with a large uncertainty) together with our flux densities, suggests a flat spectral index. 

In our 5 cm high-resolution map, the source is resolved as a compact core with an elongation to the SE (see Fig. \ref{fig:sources}). This morphology suggests that our observations could be resolving the jet of a radio galaxy, which would be in agreement with previous proposals for its nature. Proper motion studies and a reliable determination of the spectral index should be done to clarify the nature of this source.

\subsubsection{Source 37}\label{37}

\begin{figure}
	\includegraphics[width=0.99\hsize]{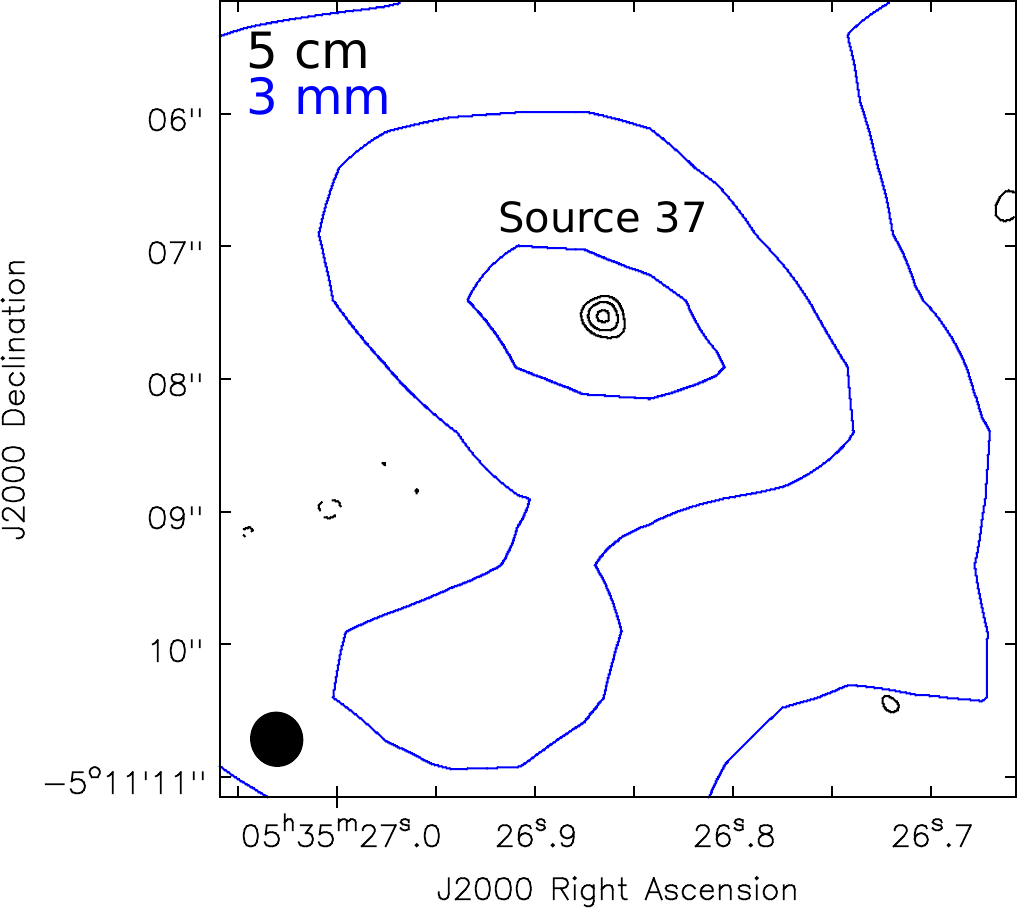}
	\caption{Source 37. VLA 5 cm A-configuration map with natural weighting (black contours) overlaid on the ALMA 3 mm map (blue contours, adapted from \citealt{Kainulainen2017}). Black contour levels are $-$3, 3, 4, and 5 times 3.6 $\mu$Jy beam$^{-1}$, the rms of the 5 cm map.  Blue contour levels are 0.3, 0.6, and 0.9 times 1.2 mJy beam$^{-1}$, the peak of the 3 mm core. The VLA synthesized beam is shown in the bottom left corner of the map. The 3 mm ALMA synthesized beam (HPBW) is $3\farcs8\times2\farcs3$, P.A. = 71$\degr$.
		\label{fig:37}}
\end{figure}

Source 37 is located toward the densest part of the ISF (see Fig. \ref{fig:filament}), coincident with a faint 3 mm peak (see Fig. \ref{fig:37}). It has mid-IR, near-IR, X-ray, and variable optical counterparts (Table \ref{tab:sources}) and was classified as a YSO with disc based on IR emission \citep{Megeath2012,Getman2017}. Its spectral type is G according to \citet{Penston1975} and is a spectroscopic binary according to \citet{Tobin2009}.

This is the first report at radio wavelengths of Source 37, for which we determine a positive spectral index $\alpha=1.2\pm0.7$ (Table \ref{tab:fluxes} and Fig. \ref{fig:SEDs}), consistent with thermal free-free emission. \citet{Gaia2023} reported a distance of $378\pm3$ pc, confirming its membership to the star-forming region.

\subsubsection{Source 45}\label{45}

Source 45, previously reported at 5 cm by \citet{Kounkel2014}, is located toward the border of the ISF, $\sim21\arcsec$ from its ridgeline (see Fig. \ref{fig:filament}), and is not associated with any 3 mm compact dust core \citep{Kainulainen2017}.  We detect this source at 5 cm with a slight NE-SW elongation and, comparing our flux density with the one reported by \citet{Kounkel2014} from A-Configuration VLA  observations, we see signs of radio variability (see Table \ref{tab:fluxes} and Fig. \ref{fig:SEDs}).

Based on its near and mid-IR emission (see Table \ref{tab:sources} for the counterparts), \citet{Tsujimoto2002} and \citet{Getman2017} classified this source as a Class II YSO. The source has an M spectral type according to \citet{Hillenbrand2013} and could be a member of a multiple system since the brown dwarf TKK 738 and the IR variable V2401 Ori (none of them detected in our observations) lie within $5\arcsec$ ($1900$ au) in projection from the radio source. \citet{Gaia2023} reported a distance of $365\pm30$ pc for this source. On the other hand,  \citet{Kounkel2018} classified the source as a field star that does not belong to any of the clusters identified by them in Orion A. Therefore, there is a possibility that this source could be a foreground star.


\bsp	
\label{lastpage}
\end{document}